%% file: mlam_thesis.tex
\documentclass[a4paper]{mlam_thesis}
\usepackage{indentfirst}
\usepackage{tikz}
\usepackage{tablefootnote}
\usepackage{caption}
\usepackage{natbib}
\usepackage{tocloft}
\usepackage{longtable}

\newcommand*\circled[1]{\tikz[baseline=(char.base)]{
            \node[shape=circle,draw,inner sep=1.5pt] (char) {#1};}}
\newcommand*\rectangled[1]{\tikz[baseline=(R.base)]
            \node[draw,rectangle,inner sep=1.5pt](R) {#1};\!}
\newcommand*\msolar{\mathop{}\!\,\mathcal{M}_{\odot}}
\newcommand*\mwd{\mathop{}\!\mathcal{M}_{\mathrm{WD}}}
\newcommand*\kmps{\mathop{}\!\,\mathrm{km\,s}^{-1}}
\newcommand*\gyr{\mathop{}\!\,\mathrm{Gyr}}
\newcommand*\FeH{\mathop{}\!\mathrm{[Fe/H]}}

\newcommand*\magnitude{\mathop{}\!\,\mathrm{mag}}
\newcommand*\pc{\mathop{}\!\,\mathrm{pc}}
\newcommand*\upd{\mathop{}\!\mathrm{d}}
\newcommand{\angstrom}{\,\textup{\AA}}

\begin{document}
\pagenumbering{roman}
\include{0_cover}
\cleardoublepage
\pagenumbering{arabic}
\setcounter{page}{1}
\include{1_introduction}
\cleardoublepage
\include{2_ps_3pi}
\cleardoublepage
\include{3_theoretical_wdlf}
\cleardoublepage
\include{4_vmax_pm_limited}

\cleardoublepage
\include{5_wdlf_galaxy}

\cleardoublepage
\include{6_conclusion}

\cleardoublepage
\appendix
\include{appendix_A}
\cleardoublepage
\include{appendix_B}

\cleardoublepage
\include{appendix_C}

\cleardoublepage
\include{appendix_D}
\bibliography{mlam_thesis}

\end{document}

%% file: 0_cover.tex
\begin{titlepage}
\begin{center}
\begin{minipage}{0.75\linewidth}
    \centering
    {\uppercase{\Large White Dwarf Luminosity Functions from the Pan--STARRS1 3$\pi$ Survey\par}}
    \hrulefill\\
    \vspace{1cm}
    {\Large Marco Cheuk-Yin Lam\par}
    \vspace{2cm}
    {\Large Institute for Astronomy}\\
    {\Large School of Physics and Astronomy\par}
    \vspace{1cm}
    \includegraphics[width=0.5\linewidth]{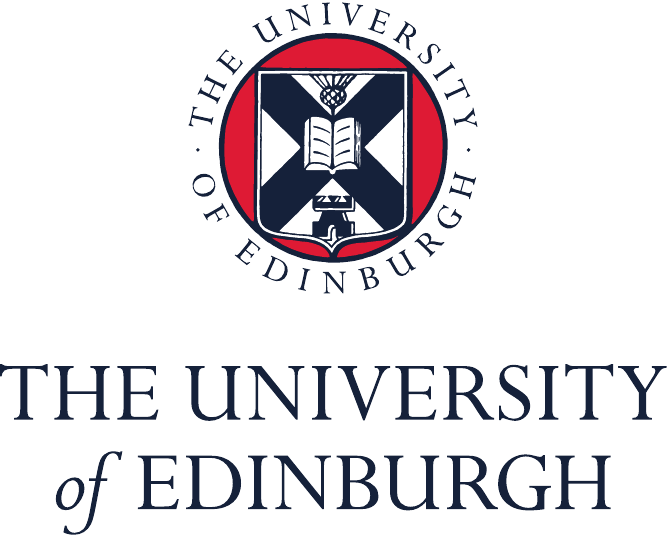}\\
    \vspace{1cm}
    {\Large Doctor of Philosophy\par}
    \vspace{3cm}
    {\Large 4th May 2016 \par}
\end{minipage}
\end{center}
\end{titlepage}

\cleardoublepage
\begin{center}
Lay Summary\vspace*{0.1in}\\
\end{center}

When a star less massive than 8 solar masses run out of hydrogen in the core for nuclear fusion, it expands to a red giant during which it fuses helium to carbon and oxygen in the core. The core temperature can never reach that required for carbon fusion so an inert mass of carbon and oxygen builds up at the center. At the end stage of the stellar evolution, a star sheds the outer layers and forms a planetary nebula, leaving behind the core as white dwarf. Therefore, white dwarfs are mostly composed of carbon and oxygen. A white dwarf is very dense: it has a mass comparable to that of the Sun, while it is of the size of the Earth.

White dwarfs are among the most common objects in the oldest visible part of the Milky Way -- the stellar halo; however, due to their low luminosity and low number density compared to the stars in the younger structures, the discs, of the Galaxy, they are scarce in the solar neighbourhood which is in the middle of the discs. Hence, they are still poorly understood one hundred years after their discovery as relatively few have been observed. They are crucial to the understanding of the geometry, kinematics and star formation history of the Galaxy, as well as to the study of the end-stage of stellar evolution for low- and intermediate-mass stars.

White dwarfs were traditionally identified by their brightness in the ultraviolet; however, if they have cooled for a long time, they become faint at those wavelengths that they cannot be seen by the most sensitive modern detectors. Because of the small radii, white dwarfs are much fainter than stars that are at the same temperatures. When the intrinsic brightness and the temperatures of objects are compared, white dwarfs stand out from other stars. However, distance to an object is extremely difficult to measure, hence it is usually not possible to measure the intrinsic brightness directly. Instead, proper motion, which is the measure of the observed changes in apparent positions of stars in the sky as seen from the center of mass of the Solar System, is much easier to measure. Since we can see the motion of an object from a close distance much easier than an object that is far away, it is possible to estimate the distance, hence the intrinsic brightness. Thus, proper motion was used as a means to identify white dwarf candidates. The use of proper motion as a selection criterion has proven effective and has yielded large samples of candidates with previous work, for example the SuperCOSMOS Sky Survey and Sloan Digital Sky Survey.

This thesis presents the sample of white dwarfs identified by their large proper motion as measured with the Panchromatic Synoptic Telescope And Rapid Response System 1~(Pan--STARRS1). Their number densities at different luminosities are measured to study the possible star formation scenario.

\cleardoublepage
\begin{abstract}

White dwarfs are among the most common objects in the stellar halo; however, due to their low luminosity and low number density compared to the stars in the discs of the Milky Way, they are scarce in the observable volume. Hence, they are still poorly understood one hundred years after their discovery as relatively few have been observed. They are crucial to the understanding of several fundamental properties of the Galaxy -- the geometry, kinematics and star formation history, as well as to the study of the end-stage of stellar evolution for low- and intermediate-mass stars.

White dwarfs were traditionally identified by their ultraviolet~(UV) excess, however, if they have cooled for a long time, they become so faint in that part of the spectrum that they cannot be seen by the most sensitive modern detectors. Proper motion was then used as a means to identify white dwarf candidates, due to their relatively large space motions compared to other objects with the same colour. The use of proper motion as a selection criterion has proven effective and has yielded large samples of candidates with the SuperCOSMOS Sky Survey and Sloan Digital Sky Survey. In this work I will further increase the sample size with the Panchromatic Synoptic Telescope And Rapid Response System 1~(Pan--STARRS1).

To construct luminosity functions for the study of the local white dwarfs, I require a density estimator that is generalised for a proper motion-limited sample. My simulations show that past works have underestimated the density when the tangential velocity was assumed to be a constant intrinsic parameter of an object. The intrinsically faint objects which are close to the upper proper motion limits of the surveys are most severely affected because of the poor approximation of a fixed tangential velocity. The survey volume is maximised by considering the small/intermediate scale variations in the observation properties at different epochs. This type of volume maximisation has not been conducted before because previous surveys did not have multi-epoch data over a footprint area of this size. The tessellation of the 3$\pi$ Steradian Survey footprint is so complex that the variations are strong functions of position. I continue to demonstrate how a combination of a galactic model and the photometric limits as a function of position can give a good estimate of the completeness limits at different colour and different line-of-sight directions. Finally, I compare the derived white dwarf luminosity function with previous observational and theoretical work. The effect of interstellar reddening on the luminosity functions is also investigated.
\end{abstract}

\cleardoublepage
\noindent {\normalfont \Huge \bfseries \textcolor{darkmidnightblue}{Declaration}}\vspace*{0.1in}\\

This thesis describes work carried out at the Institute for Astronomy, University of Edinburgh between September 2012 and March 2016. No part of this thesis has been submitted for any other degree or professional qualification, and all the work contained within is my own, unless specifically stated.\\

Chapter 4 was published in the Monthly Notices of the Royal Astronomical Society, written with co-authors N. Rowell and N. C. Hambly. Figure 4.4 and the subsection ``\textit{Survey volume generalised for kinematic selection}'' were prepared by N. Rowell.\\

\vspace{3cm}
\begin{flushright}
Marco Cheuk-Yin Lam\\
4th May 2016
\end{flushright}

\cleardoublepage
\noindent {\color{darkmidnightblue} \normalfont \Huge \bfseries Acknowledgement}\vspace*{0.1in}\\

I would not have been able to survive the past three and a half year without the help of many people who deserve special mention.

First of all, I would like to thank my parents, who have provided limitless support to my choice of such an unconventional career path for people from a financial city. Not to mention their initial decision in providing for my sixth-form and tertiary eduction in England. Mr. Michael Mak and Miss Josephine Ho definitely deserve my gratitude for their guidance to both me and my parents in making that decision.

My supervisor Dr. Nigel Hambly has provided an enormous amount of help and guidance, he definitely has the best patience in the World. I would also like to thank Prof. Annette Ferguson, Prof. Ross McLure, Prof. Philip Best, Dr. Edouard Bernard, Dr. Nicholas Rowell, Dr. Jorge Pe\~{n}arrubia for their scientific/mathematical input at various stages of my PhD life; Dr. Peter Jones, Dr. Dave Green, Prof. Richard de Grijs, Dr. Christopher Tout, Dr. Simon Hodgkin and Dr. Sergey Koposov who have influenced me one way or another during my sixth-form and undergraduate years. Special mention Ms. Paula Wilkie -- the most efficient secretary; Dr. Horst Meyerdierks -- who saved 31TB of my data from a RAID\,6 with three concurrent disc failures and; Dr. Eugene Magnier and Dr. Bertrand Goldman who have been helpful in the KP3 monthly telecon.

Finally, many thanks to my friends who have been beside me through my highs and lows: Gerard and Josh for the amazing holidays; Jacky, Stephen, Kenneth, Perry, Leo, Chris, CY and Chong-Jin for their long-distance emotional support; my lovely flatmates Sam T., Sam P. and Mike who didn't once set the flat on fire; Mike and David for the bike trips despite my tortoise speed; Derek, Jo\~{a}o, Rapha\"{e}l and Richard for the nights in and the nights out; Emma and Duncan for the ROE Coders outreach team; Lee and Graham for ultimate frisbee and board games; various people from the PIPC for organising the annual Firbush trip; and the OPTICON for the NEON Observing School in Asiago.\\

\begin{flushright}
Marco\\
4th May 2016
\end{flushright}

\cleardoublepage
\makeatletter
\gdef\ttl@savemark{\chaptermark{Contents}}
\makeatother
\renewcommand{\contentsname}{\textcolor{darkmidnightblue}{\Huge Contents}\vspace*{-0.25in}}
\tableofcontents

\cleardoublepage
\makeatletter
\gdef\ttl@savemark{\chaptermark{List of Figures}}
\makeatother
\renewcommand{\listfigurename}{\textcolor{darkmidnightblue}{\Huge List of Figures}\vspace*{-0.25in}}
\listoffigures

\cleardoublepage
\makeatletter
\gdef\ttl@savemark{\chaptermark{List of Tables}}
\makeatother
\renewcommand{\listtablename}{\textcolor{darkmidnightblue}{\Huge List of Tables}\vspace*{-0.25in}}
\listoftables

%% file: 1_introduction.tex
\chapter{Introduction}
White dwarfs are the most common remnants of stars, and they are among the oldest objects in the Milky Way. Although these ancient relics are far from rare, their intrinsic faintness has made them extremely difficult to detect. They were first used as cosmic clocks, or cosmochronometers, 50 years ago, but it is only in this high speed automated digital era that white dwarfs are discovered in bulk and have become useful in high resolution cosmochronology. In this chapter, I present a brief picture of the Milky Way, and review some of the basic mathematical constructions and recent developments in white dwarf science. I will finish with a brief discussion on the use of the white dwarf luminosity function in the context of Galactic archaeology.

\section{The Milky Way Galaxy}
Milky Way, the Galaxy, was first recorded in Meteorologica by Aristotle~(384--322\,BC), whom believed the Galaxy was ``the ignition of the fiery exhalation of some stars which were large, numerous and close together''. Fast forward two thousand years, in 1610, Galileo Galilei used a telescope to study the Milky Way and discovered that it is composed of a large number of faint stars.


\subsection{Galaxy Formation}
The formation and evolution of galaxies is one of the greatest outstanding problems in Astrophysics. The earliest attempt to explain the formation scenario of the Galaxy was that of \citeauthor{1962ApJ...136..748E}~(\citeyear{1962ApJ...136..748E}, hereafter ELS). Through a comprehensive study of 221 dwarf stars, they found that (1) stars with the largest ultraviolet~(UV) excess~(i.e.\ lowest metallicity) were moving in highly eccentric orbits while stars with little or no UV excess were moving in nearly circular orbits; (2) UV excess was correlated with the velocity perpendicular to the plane; (3) stars with large UV excess have small angular momenta. Because the orbital eccentricity and angular momentum are slowly changing ``adiabatic invariants'', they deduced that the proto-galactic cloud had to contract rapidly in a monolithic collapse within a Galactic rotation period of $\sim 2 \times 10^{8}$ years. As the materials fell inwards, condensations formed and seeded the nowadays globular clusters. The angular momentum of the proto-galaxy is conserved, so the angular velocity increased as the collapse continued while massive stars from the very early epoch enriched the gas cloud through supernovae. The collapse in the radial direction ceased when it became rotationally supported, but it continued in the direction perpendicular to the plane to form the disc. However, when a larger dataset became available, the collapse timescale was found to be $2-3\gyr$~(\citealp{1974A&A....36..333I}; \citealp{1979PASP...91..553S}) rather than $\sim0.1\gyr$, contradicting the ELS picture.

In order to explain the existence of both spherical and disc-like components in the Galaxy, which the ELS model has difficulty to do, \citet{1976MNRAS.176...31L} proposed that the formation of a spiral galaxy had to proceed in two stages with very different star formation rates: the first episode of infall was that of the ELS model forming a spheroidal component, followed by a much slower star formation process that allowed the remaining gas to settle to a disc. Simulations from \citet{1986ApJ...309..472Q} showed that the heating of the disc can be caused by the accretion of satellites, which suggests that there was a primordial thin disc which was dynamically heated by the second infall forming the thick disc, embedding a new thin disc formed within.

Independently, \citet{1977egsp.conf..219S} brought up a few observations that could not be explained by the ELS model: (1) there was no detectable spread in metal-abundance among the stars within a globular cluster; and (2) there was no statistically significant difference in the distribution of metal abundance for clusters located at different Galactocentric distance, within the range covered by the survey. Although these were consistent with the results of ELS, they were unexpected from the model of slow collapse. He reinforced his argument with the work by \citet{1968ApJ...154..891P} who proposed that globular clusters might have originated in gravitationally bound gas clouds before galaxies formed. However, this is only possible if the protoglobular gas clouds were massive enough to permit several generations of cluster formation and the associated chemical enrichment. These led the author to consider a new model that the halo was formed from the infall of a number of more or less isolated fragments. A year later, a detailed description of the model was published regarding how the stellar halo was built up from the debris of satellites~(\citealp{1978ApJ...225..357S}; hereafter SZ). This picture is close to what is predicted by the hierarchical galaxy formation scenario where galaxies are formed by accretion of smaller building blocks~\citep{1978MNRAS.183..341W}. 

There are, however, always some observables that cannot be explained by any single model. For example, the ELS picture cannot explain the existence of young halo field horizontal branch~(HB) stars and RR Lyrae stars compared to their inner halo counterparts~\citep{1991ApJ...375..121P,1999AJ....118.1373L} or the large age spread in the outer halo field stars~\citep{1996AJ....112..668C,1999AJ....118.2306R}. A number of works suggested the $\FeH$-eccentricity correlation in ELS was due to proper-motion bias~(e.g. \citealp{1991ApJ...380..403N} and reference therein; \citealp{1995ApJS...96..175B,1998AJ....115..168C,2000AJ....119.2843C}). On the other hand, the SZ model cannot explain the origin of the thick disc~\citep{1992ApJS...78...87M} and the outer halo clusters were found to be as old as those in the inner regions and there was no net age gradient in the halo clusters. Metal poor clusters were formed throughout the entire halo at approximately the same epoch and on a shorter timescale predicted by SZ~\citep{1996ApJ...463..602R}. The most obvious candidates for the fragments are either the dwarf spheroidal or dwarf irregular galaxies. However, their chemical composition abundance patterns are not compatible with those in the Milky Way~\citep{2007PASP..119..939G}. 

Currently, the accepted picture of galaxy formation is the hierarchical structure formation scenario in which galaxies are built from accretion of smaller components. The formation process relies on a hierarchical process driven by the gravitational forces of the large-scale distribution of cold dark matter~($\Lambda$CDM). This large-scale structure is the remnant of the quantum fluctuations during the epoch of inflation. These fluctuations were initially inflated to super-horizon scales by the exponential expansion, but the re-entering to the horizon after the inflation had frozen these inhomogeneities and thus seeded the large-scale structure of the Universe. At the epoch of recombination~($z\sim1100$), baryons decoupled from the thermal bath of particles and slowly flow to these local density enhancements due to gravity. By $z\sim10$, sufficient dark matter and baryons were able to fall out of the Hubble flow to form modern day galaxies. According to \citet{2005ARA&A..43..247R}, at a look back time, $\tau$, of $\sim12\gyr$, the Galactic halo was formed in a dissipational collapse of a protogalactic cloud with a dynamical time scale of a few hundred million years, carrying a weak prograde net rotation. The collapse was then followed by minor mergers of satellites which continued into the present day~(e.g. Sagittarius, \citealp{2002ApJ...569..245N}), but the rate was highest at early times. At $\tau\sim10-11\gyr$, gas clouds, which were slightly metal enriched by supernova ejecta from the halo, collided dissipationally with angular momentum conserved, forming a rotating disc which could be compared to a metal poor version of the current day thin disc. At $\tau\sim9-10\gyr$, a major merger event occurred and heated the then-thin disc to a few times thicker than it was, forming the current impression of the thick disc. The merger induced a brief period of increased star formation in the disc. After $\tau\sim8\gyr$, the disc settled to a thin rotating structure, the thin disc, and maintained a roughly constant star formation rate.

The properties of the dark matter halos are well understood within the $\Lambda$CDM paradigm. However, simulations of how baryons produce the observable galaxies in this framework are far from realistic. Bridging the gap between simulations and observations is the next step to construct a unified model of galaxy formation. Many of the observables in the Galaxy are related to events that occurred when the Galaxy was still in its infancy, which would, by assuming the Galaxy is not peculiar, provide a link to the distant Universe where we can observe a large number of high-redshift objects.

\subsection{The Galactic Populations of White Dwarfs}
The Galaxy is a superposition of multiple populations. With regard to white dwarfs, only the populations within the few hundred parsecs of the Sun are of relevance\footnote{With the exception of a few pencil beam surveys with open clusters, globular clusters and the bulge.}, they are the thin disc, thick disc and stellar halo~(hereafter, halo). Because of the large surface gravity of white dwarfs, metals settle beneath the photosphere rapidly in $10^{5}-10^{6}$\,years, depending on the mass and temperature~\citep{2009A&A...498..517K}. Any observed metal features would be attributed from very recent accretion events rather than the signature of their progenitor metallicities. Large distances from the Galactic plane can tell us they are likely halo members, but for objects that are close to the plane it is impossible to assign the population. On the other hand, the Galactic relaxation time is of the order of magnitude of the Hubble time so the information is preserved in the velocity phase space, thus population assignment has to be done solely with kinematics.

\subsubsection*{Local Number Density and Distribution}
The number density of white dwarfs in the solar neighbourhood is dominated by the thin disc population. Its density at the Galactic plane is roughly one and two orders of magnitude higher than that of the thick disc and the halo. The scale-height of the thin disc is, however, only a fraction of the older populations; as the distance from the plane increases, the number of thin disc objects drops rapidly. In real terms, the number count from SuperCOSMOS Sky Survey~(SSS, \citealp{2001MNRAS.326.1279H,2001MNRAS.326.1295H,2001MNRAS.326.1315H})found that thin disc white dwarfs outnumber the thick disc and halo counterparts by roughly 5 and 20 times respectively~(\citealp{2011MNRAS.417...93R}, hereafter RH11). See Table~\ref{table:number_density} for the complete listing of white dwarf number density found in refereed publications from the last 20 years. The reported values are still gradually increasing as fainter objects are discovered and the improvement in completeness with digital imaging as compared to photographic plates. For the thin disc, recent studies report roughly constant density, meaning we are approaching a complete sample of the thin disc white dwarfs in the solar neighbourhood. However, for the halo, the first ultracool high velocity white dwarf, WD~0346+286, was only found two decades ago~\citep{1997ApJ...489L.157H}, and a dozen more were found since then. From the shape of the luminosity function~(see Chapter \ref{ch:wdlf}), it is expected that more of these ultracool objects are yet to be found in the solar neighbourhood. Gaia satellite will be the discovery machine of these extreme objects~\citep{2014A&A...565A..11C}.

The scale-height for the white dwarf population in the solar neighbourhood was determined by the best fit observed number of white dwarfs as a function of distance from the Galactic plane to models of exponential discs with different scale-height. Due to a small available sample, it was traditionally fitted with a single model for all white dwarfs, for example \citet{1980ApJ...238..685G} finds a scaleheight of $245\pm25\pc$, \citet{1982PASJ...34..381I} -- $300\pc$, \citet{1986ApJS...61..569D} -- $270\pc$, \citet{1989MNRAS.240..533B} -- $275\pm50\pc$, and \citet{2002MNRAS.335..673V} -- $260\pm40\pc$. However, older objects experienced more kinematic heating so the cooler white dwarf population should have larger scale-height. In view of this, \citeauthor{2006AJ....131..571H}~(\citeyear{2006AJ....131..571H}, hereafter H06) studied the change in the scale-height in the range $9.0\magnitude<\mathrm{M}_{\mathrm{bol}}<13.0\magnitude$ in bins of $0.5\magnitude$. However, they were not able to decouple the different Galactic components. At brighter magnitudes, the sample was dominated by thin disc objects; their large luminosity meant that they could be seen at large distances where the thick disc and halo contributions became important. In contrast, at the faint end, the population was dominated by thick disc or halo objects. It is enlightening to see the trend of increasing scale-height as a function of magnitude, from $\sim300\pc$ at $9.0\magnitude$ to $\sim900\pc$ at $12.0\magnitude$, but the result is not of practical use just yet and the authors themselves adopted the conventional scale-height of $250\pc$ for their analysis of the solar neighbourhood. The lack of metallicity indicator in white dwarf atmospheres has made it almost impossible to distinguish thick disc white dwarf from the thin disc counterpart; even in the velocity space, the two populations share similar kinematics, and in most cases where 3D velocities are not available, the velocity distribution in the projected plane has made the task even more difficult. Hence, no work has been conducted on the white dwarf thick disc scale-height. For the halo, their extreme rareness has made this task impossible. The extended structure of the halo means when the distance limit of the surveys is only a few hundred parsecs, it is safe to approximate an infinite scale-height for the halo. This argument also applies to the scale-lengths of all components where studies of the main sequence stars found scale-lengths in the order of thousands of parsecs.

\subsubsection*{Velocity Distribution}
Before the era of the Hipparcos astrometric catalogue~\citep{1997ESASP1200.....E}, where milliarcsecond astrometry was available for $10^{5}$ stars brighter than $\sim12\magnitude$\footnote{The survey was complete to V$\sim7-9\magnitude$ depending on spectral type.}, representative samples to study the velocity distribution of the Galaxy were limited to nearby stars from proper-motion surveys where they were likely to be biased towards high velocity stars~(e.g.~\citealp{1993IAUS..156..107J}). However, the astrometric machine lacked accurate photometric and spectroscopic measurements. One way to overcome this problem is to use the statistical analysis of the projected velocities to derive the local stellar kinematics~\citep{1998MNRAS.298..387D}. A more thorough way is to combine with follow-up spectra to obtain radial velocities to study the sample in the six-dimensional phase space~(e.g. \citealp{2004A&A...418..989N}). While in the Hipparcos catalogue there were only 20 white dwarfs, it was impossible to study the velocity properties with only white dwarfs. However, field white dwarfs are subjected to the same gravitational perturbations as other stars, the kinematic structure should be similar to that exhibited by the low mass main sequence stars which were formed at similar time. For example, the white dwarfs from \citet{2001Sci...292..698O} and the F and G dwarfs from \citet{2004A&A...418..989N} show similar kinematic structure~\citep{2005ARA&A..43..247R}.

The two discs have similar kinematics, as they share similar formation scenarios. However, the older thick disc has experienced more kinematic heating, most significantly from the major merger event that formed the thin disc. The two components were then subjected to the heatings through scattering by molecular clouds, gravitational perturbation of the spiral arms and interactions with infalling satellites to give the current profile. The mean velocity in the direction of the Galactic centre, $<\mathbf{\mathcal{U}}>$, and the North Galactic Pole, $<\mathbf{\mathcal{W}}>$, are roughly the same for the two components: $\left\langle\mathbf{\mathcal{V}}\right\rangle\sim\left\langle\mathbf{\mathcal{W}}\right\rangle\sim-10\kmps$. In the direction of Galactic rotation, $\mathbf{\mathcal{V}}$, the thick disc is lagging the thin disc by $\sim20\kmps$ at $<\mathbf{\mathcal{V}}>\sim-40\kmps$. The velocity dispersion of the thick disc is roughly twice that of the thin disc, which is expected from the formation scenario. In comparison, the halo is a pressure-supported system. The velocity properties can be studied with a number of standard tracers, for example metal poor subdwarfs, HB stars, RR Lyraes stars and globular clusters~\citep{1993ARA&A..31..575M} and is always found to carry small $<\mathbf{\mathcal{U}}>\sim-20\kmps$ and large $<\mathbf{\mathcal{V}}>\sim-200\kmps$. This set of velocities indicates little rotation of the halo, and the stellar orbits are typically much more eccentric than those in the discs. When projecting this velocity ellipsoid onto the plane of observation, halo objects always carry large tangential velocities. This property is very important in the kinematic selection of white dwarfs and will be discussed in Chapter~\ref{sec:vtan_distribution}.

\subsection{White Dwarfs in the context of Galactic Archaeology}
White dwarfs are good chronometers because their cooling rates are well understood in most temperature ranges. The total cooling time can be well approximated with as few as two parameters at high temperature: mass and luminosity. At lower temperatures, the atmospheric hydrogen/helium ratio is also important. The use of the white dwarf luminosity function as cosmochronometer was first introduced by \citet{1959ApJ...129..243S}. Given a finite age of the Galaxy, there is a minimum temperature below which no white dwarfs can reach in a limited cooling time. This limit translates to an abrupt downturn in the white dwarf luminosity function. Evidence of such behaviour was observed by \citet{1979ApJ...233..226L}, however, it was not clear at the time whether it was due to incompleteness in the observations or to some defect in the theory~(eg. \citealp{1984ApJ...282..615I}). A decade later, \citet{1987ApJ...315L..77W} gathered concrete evidence for the downturn and estimated the age of the disc to be $9.3 \pm 2.0\gyr$~(see also \citealp{1988ApJ...332..891L}). In order to obtain accurate ages for individual objects, it is desirable to have good quality trigonometric parallaxes and low resolution spectra for spectral line fitting to derive the luminosity, surface gravity and atmospheric composition. However, it is also possible to do it statistically with a larger sample of WDs that only have broadband photometry to achieve practical accuracy. The total progenitor stellar lifetime is found from a stellar evolution model as a function of zero age main sequence~(ZAMS) mass and metallicity, where the mass can be found by applying the initial--final mass relation of main sequence stellar mass and white dwarf mass, and the metallicity is tested with different values since no observable of isolated white dwarfs can reveal the progenitor metallicity. The total time can then be found by adding the cooling time and the stellar life time. 

\section{White Dwarfs}
\begin{quote}
\textit{We learn about the stars by receiving and interpreting the messages which their light brings to us. The message of the Companion of Sirius when it was decoded ran: ``I am composed of material 3,000 times denser than anything you have ever come across; a ton of my\ material would be a little nugget that you could put in a matchbox.'' What reply can one make to such a message? The reply which most of us made in 1914 was -- ``Shut up. Don't talk nonsense.''} -- Stars and Atoms, Sir Arthur Eddington 1927
\end{quote}

The pair of 40 Eridani B/C was first resolved by William Herschel in 1783, and then later re-observed by various astronomers. However, the strange nature of its small size and being hot but faint were not realised until 1910 by H. N. Russell, E. C. Pickering and W. Fleming. W. Adams obtained the spectrum of Sirius B in 1914 and used the term ``white dwarf'' for the first time. Together with its luminosity, the first estimate of its radius was made. The mass of this star was known from its orbit to be approximately $1\msolar$, which gives a mean density of $5\times10^{4}$\,g\,cm$^{-3}$. Sir Arthur Eddington pointed out in 1924 that objects with such densities would cause gravitational redshift of its own radiation, which was confirmed by Adams in 1925. This also provided independent evidence of its high density. One should note that the current effective temperature measurement would give a mean density about two orders of magnitude higher than the first result.

The rise of quantum mechanics in the 1920s provided a solution to the then ``impossible'' density. In 1926, it was shown that electrons obey Fermi-Dirac statistics for fermions, as a consequence of Pauli's exclusion principle. R. H. Fowler realised immediately that this allows completely degenerate electrons to be the source of pressure to support the interior of white dwarf against gravity. This pressure remains even at zero temperature, so it was established that white dwarfs are very stable final configurations and opened up the field of the study of zero-temperature configurations. Subsequent improvements on white dwarf structural models were made in the following decade. \citet{1929ZPhy...56..851A} and \citet{1930LEDPM...9..944S} corrected Fowler's equation of state (EoS) for relativistic effects due to the extreme densities. In 1931, by coupling relativity and quantum mechanics, S.~Chandrasekhar found that there exists an upper mass limit at which a white dwarf can remain stable. This is now known as the Chandrasekhar limiting mass. L. D. Landau arrived at the same conclusion independently a year later. This provided the first clue to a fundamental difference in the final stage of stellar evolution of medium and low mass stars. In the next eight years, Chandrasekhar formulated the complete set of EoSs at all densities, including the effects of finite temperatures and special relativity, and the resulting structure of zero-temperature models and their mass-radius relations.

\subsection{Stellar Evolution of the Progenitors of White Dwarfs}
\label{sec:wd_progenitor}
Main sequence~(MS) stars with initial mass less than $\sim8\msolar$ end up as white dwarfs~(WDs) at the end of their lives. Since this mass range encompasses the vast majority of stars in the Galaxy, these degenerate remnants are the most common final product of stellar evolution. There is little nuclear burning and gravitational contraction provides negligible amounts of energy, so WDs cannot replenish the energy they radiate away. As a consequence, the luminosity and temperature decrease monotonically with time. The electron degenerate nature means that a WD with a typical mass of $\sim0.6\msolar$ has a similar size to the Earth which gives rise to high density, large surface gravity and low luminosity. Their intrinsic faintness has made them one of the least detected objects.

\begin{figure}
\includegraphics[width=\textwidth]{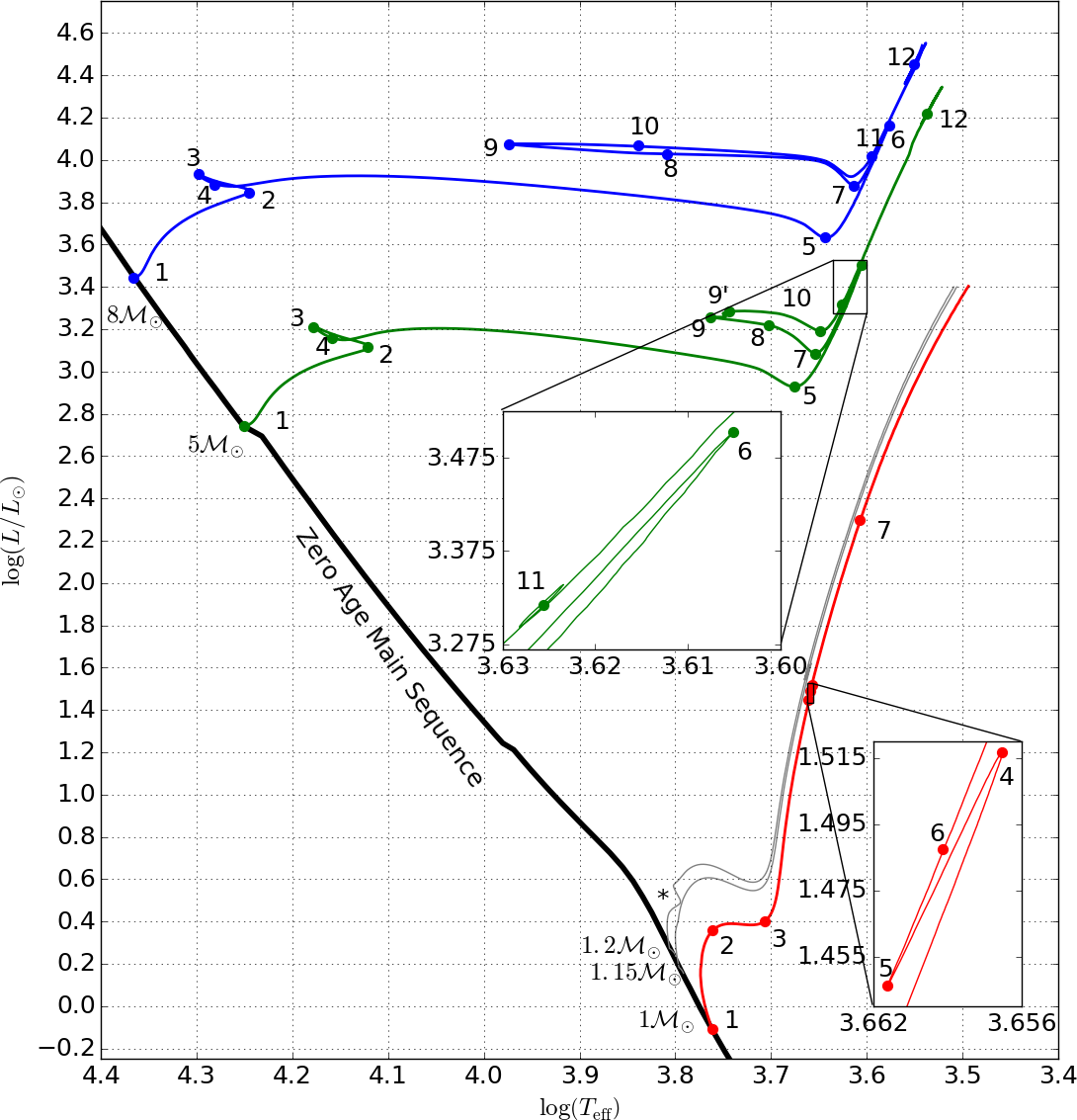}
\caption[HR diagram of the evolutionary tracks of $1$, $5$ and $8\msolar$ MS stars with solar metallicity]{An HR diagram showing the evolutionary tracks of $1$, $5$ and $8\msolar$ MS stars with solar metallicity marked in red, green and blue respectively. The numbered dots along the path denote some of the major changes during the evolution~(see Chapter \ref{sec:wd_progenitor} for the explanation). The $1.15$ and $1.2\msolar$ are to demonstrate the effect of the onset of a convective core as it is evolving to a subgiant on a HR diagram at $\log(L/L_{\odot}) \sim 0.5$}
\label{fig:ms_evolution}
\end{figure}

Even though all low mass stars will end up as WDs, the evolution process and the composition of WDs are different depending on their initial states, most heavily on mass. Fig.~\ref{fig:ms_evolution} shows the evolutionary tracks of $1$, $5$ and $8\msolar$ MS stars with solar metallicity. The numbered dots along the path denote some of the major changes during the evolution. The following is simplified and rearranged from the text of \citet{2012sse..book.....K}.

\subsubsection*{$\mathbf{1\mathcal{M}_{\odot}}$ \textbf{Evolution}}
\begin{enumerate}
\item[\circled{1}] Hydrogen burning predominantly via proton-proton chain reaction.
\item[\circled{2}] Reached the Sch\"{o}nberg--Chandrasekhar limit, which is the maximum mass of a non-fusing isothermal core that can support the overlying envelope, the core becomes degenerate such that it can support the outer layer with degenerate pressure.
\item[\circled{3}] Core grows as the convective envelope expands, the opacity of the photosphere decreases with temperature~(reached the Hayashi limit). 
\item[\circled{4}] Core helium ignites degenerately, known as the helium flash, $T_{\mathrm{c}}$ rises and runaway occurs, reducing degeneracy without thermostatic control. $\mathcal{M}_{\mathrm{c}}=0.476\msolar$ and $T_{\mathrm{c}}=7.9\times10^{7}$\,K, where the subscript c denotes the core.
\item[\circled{5}] Star settles to the HB with core helium burning.
\item[\circled{6}$\rightarrow$\circled{7}] Convective envelope overlay on the hydrogen burning shell. Helium core gradually catches up with the hydrogen burning shell.
\item[\circled{7}] Core helium depleted as the helium burning core catches up the hydrogen burning shell, star ascends onto the asymptotic giant branch~(AGB), the early AGB~(E-AGB) phase.
\item[\circled{8}] (Not shown in diagram)~Thermal pulses occur at the thermally pulsing AGB~(TP-AGB) phase which lead to extensive mass loss.
\end{enumerate}
Final product: \textbf{Carbon--Oxygen WD}

\subsubsection*{$\mathbf{1.15-1.20\mathcal{M}_{\odot}}$ \textbf{Evolution}}
\begin{itemize}
\item[\circled{*}] This feature is due to the transition of a radiative core to a convective core at this mass range.
\end{itemize}

\subsubsection*{$\mathbf{>2.2\mathcal{M}_{\odot}}$ \textbf{Evolution}}
\begin{itemize}
\item[\circled{6}] Core helium ignites gently when the core temperature reaches $1.2\times10^{8}$\,K.
\end{itemize}

\subsubsection*{$\mathbf{5\mathcal{M}_{\odot}}$ \textbf{Evolution}}
\begin{enumerate}
\item[\circled{1}] Hydrogen burning is via the Carbon--Nitrogen--Oxygen~(CNO) cycle which began before reaching the ZAMS when the star was still forming. Because of the strong dependence of temperature in the energy generation rate, $\epsilon_{\mathrm{CNO}} \propto \rho \, T^{16}$, the luminosity rises steeply at the centre of core, so it is convective~($\mathcal{M}_{\mathrm{conv, c}} \approx 1.2\msolar$). Remainder of the star is radiative. At the core, the chemical composition increases with time. Temperature is thermostatically fixed, so density must increase to maintain pressure. Hence, luminosity and radius slowly increase and convective core shrinks to $0.5\msolar$.
\item[\circled{2}] Release of gravitational energy at the core becomes important, $L_{\mathrm{grav}} \approx 0.1 L_{*}$, star contracts and core temperature increases.
\item[\circled{3}] Core burning ceases and becomes isothermal as shell hydrogen burning begins.
\item[\circled{4}] Helium core mass reaches $0.6\msolar$, the Sch\"{o}enbog--Chandraskhar mass limit of a $5\msolar$ star, the maximum mass of an isothermal gas-pressure supported core. The core collapses in a thermal timescale.
\item[\circled{4}$\rightarrow$\circled{5}] Crossing the Hertzsprung gap in the Hertzsprung--Russell~(HR) diagram. The envelope expands.
\item[\circled{5}] Hydrogen ionisation at the surface~(Hayashi limit), deep convective envelope develops due to increased opacity. Core contracts and heats up in order to maintain thermal equilibrium, as a consequence the luminosity increases.
\item[\circled{6}] Helium ignites smoothly.
\item[\circled{7}$\rightarrow$\circled{9}] Helium burns in a growing convective core via triple-alpha process to form carbon, which also burns with helium to produce oxygen.
\item[\circled{8}] The star crosses the Cepheid instability strip when overstable pulsations are driven by helium ionisation, in the following A--B--C--A cycle:
\begin{enumerate}
\item[\rectangled{A}] Energy is absorbed by He$^{+}$ to form He$^{++}$ and e$^{-}$.
\item[\rectangled{B}] Star contracts and heats up as the opacity rises.
\item[\rectangled{C}] All helium is fully ionised, so the helium shell cannot store any more energy through ionisation, star relaxes and deionises.
\end{enumerate}
\item[\circled{9}] Helium is exhausted in the core so shell helium is ignited. The star ascends onto the AGB.
\item[\circled{10}] The star crosses Cepheid instability strip from the opposite direction, following the A--B--C--A cycle again.
\item[\circled{11}] Second dredge up occurs when the carbon--oxygen core becomes degenerate. The size of the helium burning shell increases, star expands and cools. The hydrogen shell is extinguished which causes the convective envelope to deepen. More CNO products are dredged up to the surface so the hydrogen burning shell reignites. At this point, there is a temperature inversion in the core as the bottom of the helium shell has the highest temperature.
\item[\circled{12}] Thermal pulses happen at the envelope. This process has no thermostatic control and the reactions can run away. At this point, small amount of neon is produced in the core, which comes from alpha processes during core helium burning.
\item[\circled{13}] (Not shown in diagram)~Carbon burning begins degenerately at $5\times10^{8}$\,K. Since degenerate pressure does not depend on temperature, it becomes a runaway burning. The star experience massive mass loss and the core grows rapidly and eventually ceases and becomes a WD.
\end{enumerate}
Final product: \textbf{Carbon--Oxygen WD}

\subsubsection*{\textbf{$8\mathcal{M}_{\odot}$ Evolution}}
The evolution of a $8\msolar$ star is similar to that of a $5\msolar$ until \circled{11}.
\begin{enumerate}
\item[\circled{11}] After core He-burning, $\mathcal{M}_{\mathrm{He}}=1.9\msolar$ and $\mathcal{M}_{\mathrm{C/O}}=1.2\msolar$. Second dredge up occurs and carbon ignites degenerately in a shell in carbon--oxygen core. The burning shell propagates inwards, raising the degeneracy of the inner layers. Once central burning complete, minor flashes occur to convert carbon--oxygen core to oxygen--neon--magnesium core.
\end{enumerate}
Final Product: \textbf{Oxygen--Neon WD or Supernova}

\subsection{Structure of White Dwarfs}
A simple structural model developed by \citet{1939isss.book.....C}, and a cooling model developed by \citet{1952MNRAS.112..583M} capture the most essential physics of WDs, giving a good approximate description to the most common types of WDs. They were far from ideal but they set very good frameworks for further development. Most of the following derivations and discussions in this section are adapted from \citet{1990RPPh...53..837K, 2012sse..book.....K} and \citeauthor{2001PASP..113..409F}~(\citeyear{2001PASP..113..409F}, hereafter F01).

\subsubsection{Equation of State for Electron Gas}
\label{sec:eos}
\begin{figure}[hbt]
\begin{center}
\includegraphics[width=6cm]{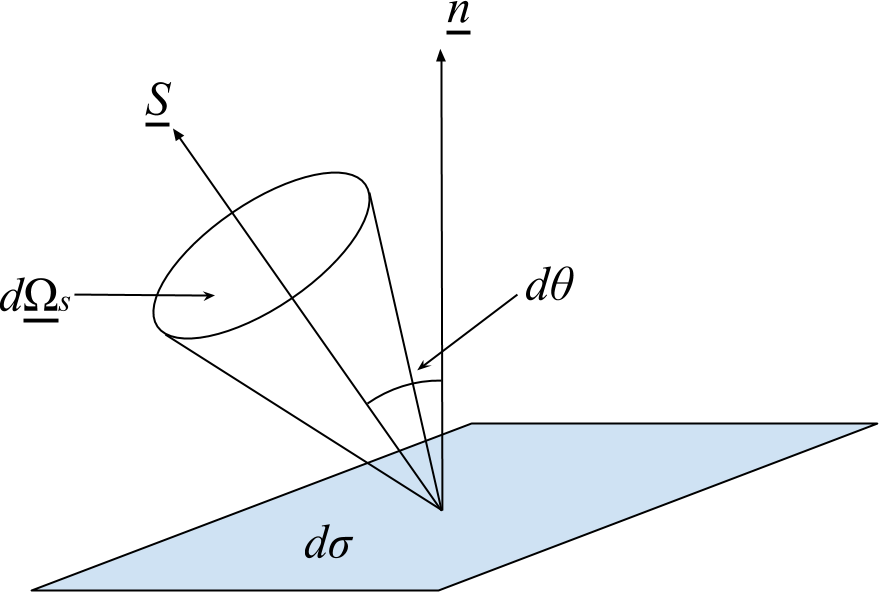}
\end{center}
\caption[Momentum flux through an area element]{Illustration for the transport of momentum in the direction \underline{\textbf{\textit{S}}} across an area element $\upd $\textbf{\textit{$\sigma$}} in the direction \underline{\textbf{\textit{n}}}}
\label{fig:momentumflux}
\end{figure}

The EoS is the relation between the state variables. The simplest model can be described by three variables: pressure\,($P$), number density\,($n$) and energy density\,($u$). Derivation of the EoS for ideal, non-interacting degenerate electron gas starts from Pauli's exclusion principle. An electron is a fermion, so each quantum cell of the six-dimensional phase space cannot hold more than two electrons~(2 spin states). The volume of such a quantum cell is $\upd x^{3}\upd p^{3} = h^{3} $, where $x$ is position, $p$ is momentum and $h$ is Planck's constant. Therefore in the shell $ [\,p,\,p+\upd p\,] $ of the momentum space, there are $ 4 \pi p^{2}\upd p / h^{3} $ quantum cells, which in total cannot contain more than $ 8 \pi p^{2}\upd p / h^{3} $ electrons. Hence, quantum mechanics requires the density of states, $g(p)$, to follow
\begin{equation}
g(p)\upd p\upd V = \frac{8 \pi p^{2}}{h^{3}} \upd p\upd x^{3}
\end{equation}
The occupation of states follows Fermi-Dirac statistics which, as a function of energy $\epsilon$, is given by
\begin{equation}
f(\epsilon) = \left[1+\exp\left(\frac{\epsilon-\mu}{kT}\right)\right]^{-1}
\end{equation}
where $\mu$ is the chemical potential, $k$ is the Boltzmann constant and $T$ is the temperature. By definition, pressure is the momentum flux through a unit area. Consider a surface element $\upd \sigma$ with a normal vector $\underline{\textit{\textbf{n}}}$, the pressure can be found by determining the number of electrons going through $\upd \sigma$ into a solid angle element $\upd \underline{\Omega_{S}}$ in the direction $\underline{\textit{\textbf{S}}}$ per second. Each electron carries a momentum $p\cos(\theta)$ in the direction $\underline{\textit{\textbf{n}}}$, and crosses the surface element $\upd \sigma$ with a velocity $v(p) \cos(\theta)$. Therefore, the number of electrons going through the surface element $\upd \sigma$ into the solid-angle element $\upd \underline{\Omega_{S}}$ per second is $g(p)\upd p\upd \underline{\Omega_{S}} v(p) \cos(\theta) \upd \sigma /(4 \pi)$~(Fig.~\ref{fig:momentumflux}). The total momentum flux in direction $\textit{\textbf{n}}$ can hence be obtained by integrating over all directions $\underline{\textit{\textbf{S}}}$ of a hemisphere and over all absolute values $p$, under the assumption that the distribution function is isotropic in all directions for a population of non-relativistic electrons. The pressure $P_{e}$ of the electrons is therefore given by
\begin{equation}
P_{e} = \int_{\Omega_{s}=0}^{2\pi} \int_{p=0}^\infty g(p)f(\epsilon)p\cos^{2}(\theta)v(p)\upd p\,\frac{\upd \Omega_{s}}{4\pi}
\end{equation}
This integral is the most general form for this simple treatment of WDs. It has no analytic solution when the degeneracy is not complete. The solution can be found in chapter 15 of~\citep{2012sse..book.....K} where a table of numerical results is provided. I am only showing the fully degenerate case where this integral is analytical.
In the case of completely degenerate electron gas, all phase cells up to the Fermi momentum, $p_{F}$, are occupied by two electrons, while all cells above $p_{F}$ are empty. The corresponding energy is the Fermi energy, $\epsilon_{F}$. This simplifies the distribution function to
\[
 f(\epsilon) =
  \begin{cases}
 1 & \text{if } \epsilon \leq \epsilon_{F} \text{ ,}\\
 0 & \text{if } \epsilon > \epsilon_{F} \text{ .}
  \end{cases}
\]
Hence, the electron pressure integral becomes
\begin{equation}
P_{e} = \frac{4\pi}{3 h^{3}} \int_{0}^{p_{F}} p^{3}v(p)\upd p
\end{equation}
From special relativity, the mass $m$, the velocity $v$ and the momentum $p$ are related by $ p = \gamma(v)mv $ where $\gamma(v) = [1-(v/c)^{2}]^{-\frac{1}{2}} $ and $c$ is the speed of light in vacuum. This can be rearranged to express $v$ as a function of $p$,
\begin{equation}
v(p) = \frac{\frac{p}{m_{e}}}{\left[(\frac{p}{m_{e}c})^{2}+1\right]^{\frac{1}{2}}} \text{ .}
\end{equation}
Using a substitution of $ \xi = p/m_{e}c $ and $ x = p_{F}/m_{e}c $, the integral becomes
\begin{align} \label{eq:pressure}
P_{e} &= \frac{8\pi m_{e}^{4} c^{5}}{3 h^{3}} \int_{0}^{x} \frac{\xi^{4}}{(1+\xi^{2})^{\frac{1}{2}}}\upd \xi \\
  &= \frac{\pi m_{e}^{4} c^{5}}{3 h^{3}} f(x) ,
\end{align}
with $ f(x) = x(2x^{2}-3)(x^{2}+1)^{\frac{1}{2}} + 3 \sinh^{-1}(x) $. The number density of electrons can be found by integrating the density of states in phase space over the range of momentum from 0 to $p_{F}$. This gives
\begin{align}
n_{e} &= \frac{8\pi }{3 h^{3}} \int_{0}^{p_{F}} p^{2}\upd p\\
  &= \frac{8\pi m_{e}^{3} c^{3}}{3 h^{3}} x^{3}\\
  &= \frac{\rho}{\mu_{e}m_{u}} , \label{eq:numberdensity}
\end{align}
where $\rho$ is the matter density, $\mu_{e}$ the ionisation fraction and $m_{u}$ the atomic mass.
The relativistic energy of an electron is given by $ E = m_{e}c^{2}(\sqrt{1+p^{2}/m_{e}^{2}c^{2}}-1) $, so the energy density of an electron gas can be found by integrating this energy over the range of momentum from 0 to $p_{F}$
\begin{align}
u_{e} &= \frac{8\pi }{3h^{3}} \int_{0}^{p_{F}} E(p)p^{2}\upd p\\
  &= \frac{\pi m_{e}^{4} c^{5}}{3 h^{3}} \left[8x^{3}[(x^{2}+1)^{\frac{1}{2}}-1]-f(x)\right]
\end{align}
Once again, the general form does not allow the formulation of an analytical EoS, I can only consider the limiting cases in the non-relativistic and ultra-relativistic regimes such that equation (\ref{eq:pressure}) can be simplified to
\[
 P_{e} \rightarrow
  \begin{cases}
 \frac{8 \pi m_{e}^{4} c^{5}}{15 h^{3}}x^{5} & \text{if } x \ll 1 \text{ ,}\\
 \frac{2 \pi m_{e}^{4} c^{5}}{3 h^{3}}x^{4} & \text{if } x \gg 1 \text{ .}
  \end{cases}
\]
Now, it is possible to relate between pressure, number density and energy density,
\begin{alignat}{2}
P_{e, non-rel} & =\, \frac{1}{20} \left(\frac{3}{\pi}\right)^{\frac{2}{3}}\frac{h^{2}}{m_{e}}n_{e}^{\frac{5}{3}}&\, = \, \frac{2}{3} \, u_{e} \label{eq:pnonrel} \text{ ,}\\
P_{e, rel} &=\, \frac{1}{8} \left(\frac{3}{\pi}\right)^{\frac{1}{3}}hc\,n_{e}^{\frac{4}{3}} &\, = \, \frac{1}{3} \, u_{e} \text{ .}\label{eq:prel}
\end{alignat}
These expressions give approximations to the EoSs for completely degenerate stellar configurations in the limits of non-relativistic and ultra-relativistic degenerate cases. They are essentially the upper and lower boundaries for the Chandrasekhar WDs.

\subsubsection{Mass-Radius Relation and Chandrasekhar Mass Limit}
\label{sec:chandrasekharmass}
A spherically symmetric stellar structure can be determined completely by four basic equations: the conservation of energy, mass and momentum, and the energy transport. In the case of a zero-temperature Chandrasekhar WD, only the equations of mass conservation
\begin{equation}
\frac{\upd \mathcal{M}}{\upd r} = 4 \pi r^{2} \rho
\end{equation}
and hydrostatic equilibrium~(ie. the conservation of momentum),
\begin{equation}
\frac{\upd P}{\upd r} = - \frac{G\mathcal{M}\rho}{r^{2}}
\end{equation}
are relevant. To understand the basic features of the mass-radius relation, it is possible to do so with a simple dimensional analysis, rather than solving the above equations. The pair above can be re-written as
\begin{align}
P_{c}    &\propto \frac{G\mathcal{M}^{2}}{R^{4}} \textrm{ and}\\
\rho_{c} &\propto \frac{\mathcal{M}}{R^{3}} \textrm{.}
\end{align}
Combining the density relation with equation (\ref{eq:numberdensity}), (\ref{eq:pnonrel}) and (\ref{eq:prel}) leads to
\begin{align}
P_{e, non-rel} &\propto n_{e}^{\frac{5}{3}} \propto \rho^{\frac{5}{3}} \propto \frac{\mathcal{M}^{\frac{5}{3}}}{R^{5}} \textrm{ and} \label{eq:degeneracy_pressure_nonrel}
\\
P_{e, rel}     &\propto n_{e}^{\frac{4}{3}} \propto \rho^{\frac{4}{3}} \propto \frac{\mathcal{M}^{\frac{4}{3}}}{R^{4}}
\end{align}
in an equilibrium. They must balance with the gravity so they are proportional to $ \mathcal{M}^{2}/R^{4} $ in both cases, giving
\begin{align}
\mathcal{M}_{non-rel} &\propto R^{-3} \textrm{and}\\
\mathcal{M}_{rel}     &\propto \text{constant.}
\end{align}
These two relations show that in the limit of zero mass non-relativistic WDs, the radius decreases with an increasing mass. However, as the mass increases, the EoS moves into the relativistic regime and the mass of a WD approaches a constant value, which means there is a maximum mass at which a WD can exist. This limiting mass is now known as the Chandrasekhar mass. One simple estimation of this mass is to combine the polytropic model\footnote{A solution of Lane-Emden equation in which the pressure depends only on density in the form $P = K \rho^{\frac{n+1}{n}}$, where $n$ is the polytropic index and $K$ is an appropriate constant.} with a polytropic index $n=3$ and equation (\ref{eq:prel}) to give
\begin{align}
\mathcal{M}_{Ch} &= \frac{\mathcal{M}_{3} \sqrt{3\pi}}{2} \left(\frac{\hbar c}{G}\right)^{\frac{3}{2}} \left(\frac{1}{\mu_{e} m_{u}}\right)^{2}\\
       &= 5.7157 \mu_{e}^{-2}  \mathcal{M}_{\odot}
\end{align}
where $\mathcal{M}_{3} = 2.0182$ is a solution from the Lane-Emden equation. For helium, carbon, oxygen or neon WDs, $\mu_{e} \approx 2$. This gives the critical mass as $1.43\msolar$.

\subsubsection{Atmosphere and Spectral Classification}
\label{sec:atmosphere_and_spectral_type}
After the formation of a WD, there are two main evolution loci, depending on the atmosphere -- hydrogen rich (DA) or hydrogen poor but helium and/or carbon--nitrogen--oxygen rich\,(DO/DB) (see Table \ref{table:spectraltype} for the spectral classification). DAs always outnumber DO/DBs, this is more obvious for hot and warm WDs. However, below $30,000$\,K, the non-DA to DA ratio increases steadily from $\sim 0.1$ to $\sim 0.4$~\citep{2013ASPC..469...77K}. This is because the helium convection zone has grown big enough and the convective velocities are high enough for overshooting to mix the hydrogen atmosphere with helium, provided that the superficial hydrogen layer is sufficiently thin ($\mathcal{M}_{\mathrm{H}}/\mathcal{M}_{\mathrm{WD}} \sim 10^{-15}$, \citealp{1991ApJ...371..719M}). This accounts for the sudden increase in the non-DA to DA ratio~(\citealp{1984ApJ...282..612S}; \citealp{2008ApJ...672.1144T}). This mixing also gives rise to a small number of DBAs, where both Balmer lines and He-I lines can be seen. When the effective temperature falls below $\sim 12,000\,$K, there is another decrease in the ratio. The convection zone of the surface hydrogen layer grows, as a consequence of the decreasing temperature, deep enough to mix with the interior helium. The thicker the envelope, the lower the mixing temperature, but if the hydrogen layer is more massive than $ \mathcal{M}_{\mathrm{H}}/\mathcal{M}_{\mathrm{WD}} \sim 10^{-6} $ then mixing will never occur~(Fontaine 2001); see Figure 1 in \citet{2008ApJ...672.1144T}. This mixing continues until $T_{\mathrm{eff}} \sim 6,000\,$K where there is a ``non-DA'' gap, as defined by Bergeron et al.~(1997), that very few non-DA WDs are found. Attempts were made on the theoretical front to explain the gap (\citealp{1997ApJS..108..339B}; \citealp{1999ApJ...520..680H}; \citealp{2001ApJS..133..413B}; \citealp{2012ApJ...753L..16C}) but there is still no convincing explanation for such a phenomenon. 

\begin{table}[h]
\begin{center}
    \begin{tabular}{ | c | p{10cm} | }
    \hline
    Spectral type & Characteristics \\ \hline
    DA & Only Balmer lines; no He-I or metals present \\
    DB & He-I  lines; no H or metals present \\
    DC & Continuous spectrum, no lines deeper than 5\% in any part of the spectrum \\
    DO & He-II strong; He-I or H present \\
    DZ & Metal lines only; no H or He lines \\
    DQ & Carbon features, either atomic or molecular in any part of the spectrum\\
    \hline
    \multicolumn{2}{ | c | }{Extra designation}\\
    \hline
    P  & Magnetic WDs with detectable polarization\\
    H  & Magnetic WDs without detectable polarization\\
    X  & Peculiar or unclassifiable spectrum\\
    E  & Emission lines are present\\
    ?  & Uncertain assigned classification; a colon~(:) may also by used\\
    V  & Optional symbol to denote variability\\
    \hline
    \end{tabular}
\end{center}
\caption[Definition of Primary Spectral Symbols for WDs]{Definition of Primary Spectral Symbols for WDs~(\citealp{1999ApJS..121....1M} and references therein)}
\label{table:spectraltype}
\end{table}

\subsection{Thermal Evolution}
\label{sec:thermalevolution}
The Chandrasekhar model discussed above provides a first insight into the structure of WDs. However, real WDs are not zero temperature stars. In fact, they are observed to have surface temperatures from the ultracool at $T_{\mathrm{eff}}<4,000$\,K to the new borns up to $\mathcal{O}(10^{6})$\,K. This implies that there exists a temperature gradient in the interior, as the surface loses heat, the core remains hot. Thus, WDs are not static but instead they continue to evolve through cooling. An early attempt to study the evolution was done by Mestel in 1952. The model was first to determine the total energy content and then to determine the rate at which energy is radiated away. This kind of simple treatment is possible due to the fact that the mechanical and thermal properties of degenerate materials are decoupled from each other.

\subsubsection*{Mechanical Properties}
Both degeneracy and gas pressure can contribute to balancing the gravitational contraction, but in a degenerate configuration the degeneracy pressure dominates. This can be easily demonstrated by comparing the electron degeneracy pressure with the ideal gas pressure. Assuming a pure carbon WD with an electron density of $ n_{e} = 12 n_{i} $, the average ion number density of $ n_{i} = 10^{35}\,m^{-3} $, at an effective temperature of $20,000$\,K, the ratio between the two pressures is
\begin{equation}
\frac{P_{e}}{P_{i}} = \frac{\frac{1}{20} \left(\frac{3}{\pi}\right)^{\frac{2}{3}}\frac{h^{2}}{m_{e}}n_{e}^{\frac{5}{3}}}{n_{i}k_{B}T} \approx 5 \times 10^{-18} \, n_{i}^{\frac{2}{3}} = \mathcal{O}(10^{6})
\end{equation}

\subsubsection*{Thermal Properties}
The heat capacities at constant volume can be written as $ C_{v} = (\partial u/ \partial T)_{v} $. From equation (\ref{eq:pnonrel}), $ u_{e} \propto P_{e} $, the energy density of electron degeneracy does not have dependency on temperature. However, for an ideal gas, $ u_{i} = \frac{3}{2} n_{i} k_{B} T $, by solving the partial derivatives,
\begin{align}
C_{v,e} &= (\partial u_{e}/ \partial T)_{v} = 0 \text{ and}\\
C_{v,i} &= (\partial u_{i}/ \partial T)_{v} = \frac{3}{2} n_{i} k_{B} ,
\end{align}
it is obvious that the heat capacity of a WD comes solely from gas in this simple treatment. Degenerate electrons are very good conductors of heat, so it is common to assume an isothermal core in crude models and the energy transfer only happens at the non-degenerate atmosphere. By further assuming that energy is only transported by radiation, the energy transfer is equal to the radiative transfer as described by the photon diffusion equation
\begin{equation} \label{eq:radiative}
\frac{\upd   T}{\upd   r} = -\frac{3}{4ac} \frac{\kappa \rho}{T^{3}}\frac{L_{r}}{4 \pi r^{2}} \text{,}
\end{equation}
where $ L_{r} $ is the luminosity at radius r, and $ \kappa $ is the radiative opacity which can be approximated by the Kramer's Law for bound-free and free-free process, $ \kappa = \kappa_{0} \rho T^{-\frac{7}{2}} $. Combining with the hydrostatic equilibrium equation and by assuming a thin envelope such that $ m \approx \mathcal{M} $ and $ L_{r} \approx L $, one arrives at
\begin{align}
\frac{\upd   P}{\upd   T} &= \frac{16 \pi ac}{3 \kappa_{0}} \frac{G\mathcal{M}T^{\frac{13}{2}}}{\rho L} \\
&=  \frac{16 \pi ac}{3 \kappa_{0}} \frac{\mathcal{R} G \mathcal{M} }{\mu L} \frac{T^{\frac{15}{2}}}{P} \text{,}
\end{align}
which can be integrated from the surface of the atmosphere with $ P = 0 $ and $ T = 0 $, to the base where $ P = k_{B}\, \rho_{b}\, T_{c} / A\, m_{u}\, \mu $ and $ T = T_{c} $, where the subscript $b$ denotes the base of the atmosphere. The choice of $ T_{c} $ is based on the assumption of an isothermal core. Substituting in the ideal gas equation, it can be rearranged to give
\begin{align}
\rho_{b} &= \overbrace{ \left( \frac{4}{17}\frac{32 \pi ac}{3 \kappa_{0}} \frac{\mu G }{\mathcal{R}} \right)^{\frac{1}{2}} }^{\alpha} T_{c}^{\frac{13}{4}} \mathcal{M}^{\frac{1}{2}} L^{-\frac{1}{2}} \\
&= \alpha T_{c}^{\frac{13}{4}} \mathcal{M}^{\frac{1}{2}} L^{-\frac{1}{2}} \text{.}
\end{align}
A second density relation can be found by equating the ideal gas pressure to the degeneracy pressure described by Equation \ref{eq:degeneracy_pressure_nonrel} and substitute in Equation \ref{eq:numberdensity} at the base of the atmosphere to get
\begin{align}
\frac{k_{B} \rho_{b} T_{c}}{\mathcal{A} m_{\mu} \mu} &= \frac{1}{20} \frac{3}{\pi}^{\frac{2}{3}} \frac{h^{2}}{m_{e}} \left( \frac{\rho_{b}}{\mu_{e} m_{u}} \right)^{\frac{5}{3}} \\
\rho_{b} &= \overbrace{ \left( 20 \frac{k_{B}}{\mu \mathcal{A} m_{u}} \frac{m_{e}}{h^{2}} \right)^{\frac{3}{2}} \frac{3}{\pi} \left( \mu_{e} m_{u} \right)^{\frac{5}{2}} }^{\beta} T_{c}^{\frac{3}{2}} \\
&= \beta T_{c}^{\frac{3}{2}} .
\end{align}
Finally, by equating the two density relations, we arrive at a power law relating the surface luminosity to the central temperature of a WD,
\begin{align}
\alpha T_{c}^{\frac{13}{4}} L^{-\frac{1}{2}} \mathcal{M}^{\frac{1}{2}} &= \beta T_{c}^{\frac{3}{2}} \\
L &= \left( \frac{\alpha}{\beta} \right)^{2} T_{c}^{\frac{7}{2}} \mathcal{M} .\label{eq:lum1}
\end{align}
Luminosity is the rate of change of energy, it can also be written as 
\begin{align}
L &= -\frac{\upd   E}{\upd   t}\\
  &= -\frac{\partial E}{\partial T_{c}}\frac{\upd T_{c}}{\upd t}\\
  &= -C_{v} \mathcal{M} \frac{\upd T_{c}}{\upd t} . \label{eq:lum2}
\end{align}
By equating the two luminosity relations,
\begin{align}
\left( \frac{\alpha}{\beta} \right)^{2} T_{c}^{\frac{7}{2}} \mathcal{M} &= -C_{v} \mathcal{M} \frac{\upd T_{c}}{\upd t} \\
\int_{0}^{t_{\mathrm{cool}}} \upd t &= - C_{v} \left( \frac{\beta}{\alpha} \right)^{2} \int_{T_{c,0}}^{T_{c,\mathrm{cool}}} T_{c}^{-\frac{7}{2}} \upd T_{c} \\
t_{\mathrm{cool}} &= \frac{2}{5} C_{v} \left( \frac{\beta}{\alpha} \right)^{2} \left[ T_{c,\mathrm{cool}}^{-\frac{5}{2}} - T_{c,0}^{-\frac{5}{2}} \right] ;
\end{align}
with a long cooling time, $T_{c,\mathrm{cool}} << T_{c,0}$, hence the second term is negligible and the relation becomes
\begin{equation}
t_{\mathrm{cool}} = \frac{2}{5} C_{v} \left( \frac{\beta}{\alpha} \right)^{2} T_{c,\mathrm{cool}}^{-\frac{5}{2}} .
\end{equation}
Substitute Equation \ref{eq:lum1} to express the relation in mass and luminosity
\begin{align}
t_{\mathrm{cool}} &= \frac{2}{5} C_{v} \left( \frac{\beta}{\alpha} \right)^{2} \left[ \frac{L}{M} \left(\frac{\beta}{\alpha}\right)^{2} \right]^{\frac{2}{7} \times -\frac{5}{2}} \\
                            &= \frac{2}{5} C_{v} \left( \frac{\beta}{\alpha} \right)^{\frac{4}{7}} \frac{\msolar}{L_{\odot}}^{\frac{5}{7}} \left( \frac{\mathcal{M}/\msolar}{L/L_{\odot}} \right)^{\frac{5}{7}}.
\end{align}
The constants used in the above derivation are:
\begin{itemize}
\item[] the radiation density constant $a = 7.5657 \times 10^{-16}$ J m$^{-3}$ K$^{-4}$
\item[] the speed of light $c = 2.9979 \times 10^{8}$ m s$^{-1}$
\item[] the opacity constant $\kappa_{0} = 10^{22}$ m$^{2}$ kg$^{-1}$
\item[] the gravitational constant $G = 6.6741 \times 10^{-11}$ m$^{3}$ kg$^{-1}$ s$^{-2}$
\item[] the ionisation fraction $\mu_{e} = 2$
\item[] the mean molecular mass $\mu = 1.35$~(using well-mixed 50/50 carbon and oxygen)
\item[] the Boltzmann constant $k_{B} = 1.3806 \times 10^{-23}$ J K$^{-1}$
\item[] the mass of electron $m_{e} = 9.1094 \times 10^{-31}$ kg
\item[] the atomic mass unit $m_{u} = 1.6605 \times 10^{-27}$ kg
\item[] the Planck's constant $h = 6.6261 \times 10^{-34}$ J s
\item[] the solar mass $\msolar = 1.9886 \times 10^{30}$ kg
\item[] the solar luminosity $L_{\odot} = 3.828 \times 10^{26}$ W
\end{itemize}
and the heat capacity can be expressed as  $ C_{v}= 3 k_{B} / 2 \mathcal{A} m_{u}$. By applying all the constants, the expression for the Mestel's cooling time becomes
\begin{align}
t_{\mathrm{cool}} &\approx 6.60 \times 10^{6} \left( \frac{\mathcal{M}/\msolar}{L/L_{\odot}}\right)^{\frac{5}{7}} \text{\, years} \\
                             &\approx 10^{7} \left( \frac{\mathcal{M}/\msolar}{L/L_{\odot}}\right)^{\frac{5}{7}} \text{\, years.} \label{eq:tcool}
\end{align}

This is the Mestel's model of WD evolution~\citep{1952MNRAS.112..583M}. It provides a simple power law relation between the cooling time, mass, luminosity and core chemical composition of WDs. This is the simplest picture of WD evolution, but it captures some of the most essential physics of the process. For a typical WD with $0.6\msolar$ and $0.1\%$ the solar luminosity, the cooling times is $\mathcal{O}(10^{9})$ years. The most important implications from this model include [i] the cooling time depends inversely on the core chemical composition, for example an oxygen WD is expected to cool faster than a carbon one; [ii] more massive WDs have larger thermal content but smaller radii so they are expected to cool slower; [iii] the cooling rate decreases with luminosity over time. Since \citet{1952MNRAS.112..583M}, the evolution model has been improved significantly. The major refinements include:

\begin{enumerate}
\item Neutrino cooling, which increases the cooling rate at high temperatures~($T>10^{9}$\,K).
\item Coulomb interactions, which reduces the cooling rate.
\item Crystallisation occurs as a consequence of coulomb interactions: it releases latent heat which acts as an extra source of heat and hence increases the cooling time.
\item Debye cooling which increases the cooling rate of very cool WDs which have very low surface luminosities.
\item At low luminosity, convection reaches the degenerate core, which lowers the central temperature and hence reduces the cooling rate.
\end{enumerate}

See Chapter \ref{sec:wd_heat_source} for more details. The luminosity function\,(LF) predicted by Mestel's Law agrees well with the observed LF, but it does not predict the sudden down turn at the faint end. This is now known to be a consequence of the finite age of the Galaxy where the oldest WDs have not had time to cool down to such temperatures.

\subsubsection*{Kelvin-Helmholtz Timescale}
The Kelvin-Helmholtz timescale is defined as
\begin{equation}
\tau_{\mathrm{KH}} = \frac{E_{g}}{L}.
\end{equation}
In the WD phase of the stellar evolution, there is little energy source, thus its cooling time can be estimated by the Kelvin-Helmholtz timescale that assumes all the energy remained in a WD comes from the gravitational contraction,
\begin{equation}
E_{g} \approx \frac{\mathrm{G}\mathcal{M}}{2R}.
\end{equation}
Hence, the timescale is
\begin{equation}
\tau_{\mathrm{KH}} \approx \frac{\mathrm{G}\mathcal{M}}{2RL}
\end{equation}
\citep{2012sse..book.....K}. For a $0.6\msolar$ young WD of the size of the Earth with solar luminosity, $\tau_{\mathrm{KH}} = \mathcal{O}(10^{9})$ years, comparable to the Mestel cooling timescale. However, if this is applied to the Sun, because of the much larger radius, its Kelvin-Helmholtz timescale is much short at $\tau_{\mathrm{KH}} = \mathcal{O}(10^{7})$ years. In addition, the Sun is fusing hydrogen in the core to replenish the energy radiated away, the the Kelvin-Helmholtz timescale is three orders of magnitude shorter than the MS timescale of a solar type star. Further discrepancy comes from the choice of static radii and luminosities used in the estimation, for example, the Sun will evolve to a red giant that has a radius hundreds times that of the Sun; or the luminosity of a WD reaches $10^{-4} L_{\odot}$ after $10\gyr$.

\subsection{Ultracool White Dwarfs}
``Ultracool'' is the temperature at which the absorption features from collisionally induced absorption of hydrogen~(H2CIA) become observable. This absorption is due to close-range H-H$_{2}$ collisions that strongly perturb the bound states of the atoms leading to the broadening of the Lyman-$\alpha$ line centred at $1,215.67\angstrom$~\citep{2006ApJ...651L.137K}. For example, at $T_\mathrm{eff} \sim 5,000\,\mathrm{K}$, the absorption broadens to $\sim4,500\angstrom$; and at $T_\mathrm{eff} \sim 3,000\,\mathrm{K}$, the absorption reaches $\sim5000\angstrom$~\citep{2010BAAA...53..111R}. Thus, the colour of WDs turns blue in the cooling sequence below such temperatures. This is a well-known phenomenon in the study of globular clusters where a ``blue hook'' appears on the WD sequence in the HR diagram. On the contrary, the helium CIA occurs during a ternary collision of helium and increases the opacity in the infrared. However, for an object at that age, it is unlikely that an object can retain a pure helium atmosphere because of the Bondi--Hoyle accretion of interstellar hydrogen~\citep{2001ApJS..133..413B}. Ultracool WDs~(UCWDs) are of great interest because they are the oldest members of the Galaxy. They have fossilised important information, for example the kinematics and past star formation history~(SFH) of the Galaxy from the oldest time. Culling a larger sample can help with the understanding of the early Galaxy as well as WD Physics.

\subsection{Helium White Dwarfs}
The core of stars less than $\sim0.5\msolar$ will never be hot and dense enough to fuse helium. Eventually, most of the hydrogen envelope will be consumed leaving a helium core WD with a hydrogen atmosphere. However, the timescale for such stellar evolution pathways is larger than the Hubble timescale implying the existing helium core WDs cannot be the products of singly evolved stars. The only possible way to explain the existence of helium WDs in the observed sample is with binary systems: when the more massive member has reached later stage of stellar evolution and inflated a few hundred times, the envelope overflows the gravitational equipotential of the system, the Roche Lobe, and leads to a mass transfer to the MS companion. When the envelope is exhausted, the evolved star becomes a helium WD and the main sequence star is slightly rejuvenated.

\subsection{Oxygen--Neon White Dwarfs}
The exact upper boundary of the initial mass of the progenitor for the formation of an Oxygen--Neon~(O/Ne) WD is still unclear. MS stars in the mass range between $7-10\msolar$ can end up as an O/Ne WD depending on the stellar model. Although, the WD mass range in which neon has to exist in the core is better known: WDs more massive than $1\msolar$ contain neon in the core. Using the Kroupa or Chabrier IMF with an exponent of $-2.3$, there are few massive stars compared to solar-type stars which means the contribution of O/Ne WD to the total WD population is low. Furthermore, even if there exists a bulk population of O/Ne WDs, their luminosities are much smaller as a consequence of the higher cooling rates and small radii, they would not be detectable with the current observational facilities~\citep{2007A&A...471..151C}. Through their Monte Carlo simulations, it was also found that they would contribute to only $\sim 2\%$ of the Massive Compact Halo Objects~(MACHOs). However, when the cooling time of a WD is concerned, the choice of a Carbon--Oxygen~(C/O) and an O/Ne core would lead to very different estimations. There are two major influences to the time: 1.) the higher atomic mass means that the size of an O/Ne WD is smaller than that of a C/O WD, so is has larger surface gravity. If a C/O composition is assumed, the derived mass would be higher than that assumed with an O/Ne composition. Therefore, the energy stored would be overestimated leading to a lower cooling rate; 2.) the crystallisation of neon occurs at higher temperature because of the higher atomic mass, it enters the Debye cooling phase quicker than a core with mostly oxygen. This further underestimates the cooling rate. The combined effect is that a WD would appear older than it is if an O/Ne WD is assumed to be a C/O WD. This can cause some complication in constraining the initial-final mass relation where the progenitor mass would be overestimated.

\subsection{Selection of White Dwarfs}
Hot WDs have UV excess compared to the MS stars. However, warm and cool WDs overlap with the MS stars in any colour combination so it is difficult to distinguish them in colour-colour space. In the ultracool regime, the H2CIA makes them blue and so they deviate from the MS colour; however, they are intrinsically too faint to be found in most surveys. To date, there are 19,712 WDs in the catalogue of spectroscopically confirmed isolated WD from Sloan Digital Sky Survey~(SDSS) DR7~\citep{2013ApJS..204....5K} and an addition of 9,088 from SDSS DR10~\citep{2015MNRAS.446.4078K}. Most of them are either false positives of follow up observations or from the BOSS ancillary science programs that has very strict colour selections~(see Appendix B2 of \citealp{2013AJ....145...10D}). Hence, the sample is biased towards hot and warm WDs~(typically $T_{\mathrm{eff}}>14,000$\,K for DAs, $T_{\mathrm{eff}}>8,000$\,K for DBs; and a minimum of $T_{\mathrm{eff}}=6,000$\,K). Thus, these catalogues are of little use when it comes to the faint end of the WDLF which reveals the star formation scenario of the Galaxy at early times. The use of reduced proper motion~(RPM) as a proxy-absolute magnitude can separate WDs from the MS stars in a RPM diagram, which resembles an HR diagram where the WDs are a few magnitudes fainter than the MS stars. It is only possible in this high speed digital imaging era to scan through the sky rapidly and to detect objects below the sky brightness, such that the survey volume is greatly increased for the search of these faint objects. This selection method has been proven to be efficient in identifying WD candidates~(e.g. \citealp{1992MNRAS.255..521E}; \citealp{1999MNRAS.306..736K}; H06 and RH11). Although this technique gives more leverage to separate WDs from MS stars, it is more difficult to combat between the completeness and contaminations because of the introduction of an extra parameter -- proper motion (see Chapter \ref{ch:ps_3pi} for more discussion).

\section{Science with White Dwarfs}
\subsection{White Dwarf Luminosity Functions}
The WDLF is an important tool for deriving important properties of the solar neighbourhood. In particular, the age of each component of the Galaxy, or with a sufficiently large sample, the SFH~(i.e.\ time resolved star formation rate). The WD number density and the age of the solar neighbourhood\footnote{The solar neighbourhood in this context is at a maximum of $\sim500\pc$.} are most studied; all published works point towards the solution of $8-10\gyr$ with $1-2\gyr$ of uncertainties~(see Table \ref{table:number_density}). While most studies focused on the Galactic discs~\citep{1992ApJ...386..539W,1995LNP...443...24O,1998ApJ...497..294L,1999MNRAS.306..736K,2012ApJS..199...29G}, some worked with open clusters~\citep{2000ApJ...529..318R}, globular clusters~\citep{2002ApJ...574L.155H,2009ApJ...705..408K,2010ApJ...708L..32B}, the stellar halo~(\citealp{1989LNP...328...15L}; \citealp{1999MNRAS.306..736K}; H06; RH11) and the Galactic bulge~\citep{2015ApJ...810....8C}.

Other works focus on cosmochronology -- the SFH of the Galaxy and the Physics of WD cooling regarding crystallisation~\citep{1968ApJ...151..227V,1975ApJ...200..306L}, C/O phase separation~\citep{1980JPhys..41C..61S,1983A&A...122..212M,1988A&A...193..141G,1993A&A...271L..13S,1994ApJ...434..652H}, chemical settling~\citep{1997ApJ...486..413S,2002ApJ...580.1077D,2008ApJ...677..473G}, atmospheric properties~(\citealp{1992ApJ...386..539W}, \citeyear{1995LNP...443...41W}; \citealp{1997ASSL..214..173F}; F01), initial-final mass relation~\citep{2008MNRAS.387.1693C,2009ApJ...692.1013S}, theoretical luminosity functions~\citep{1989ApJ...341..312I,1990ApJ...352..605N,1998ApJ...497..870W}, O/Ne WD~\citep{2007A&A...471..151C,2014A&A...569A..42V}, statistics with density estimator(s)~(\citealp{2006MNRAS.369.1654G}; \citealp{2007MNRAS.378.1461T}; \citealp{2015MNRAS.450.4098L}, hereafter LRH15) and helium WDs~\citep{2015ASPC..493..343K}.

\subsubsection{Different Density Estimators}
There are various density estimators that can be used to construct a WDLF but the maximum-volume density estimator~(\citealp{1968ApJ...151..393S}, 1/V$_{\mathrm{max}}$, see Chapter 4) is the only one that has been applied on observational data. To investigate the merits and drawbacks of this overwhelmingly popular estimator, \citet{2006MNRAS.369.1654G} and \citet{2007MNRAS.378.1461T} conducted in-depth statistical studies to compare the Cho{\l}oniewski~\citep{1986MNRAS.223....1C}, the parametric maximum likelihood method~(STY method, \citealp{1979ApJ...232..352S}) and the 1/V$_{\mathrm{max}}$ methods. It was found that for a small number of objects ($N<300$), Cho{\l}oniewski outperform 1/V$_{\mathrm{max}}$. Even with almost 10 times more data from Gaia, as simulation suggests~\citep{2014A&A...565A..11C}, each of the half-magnitude bins beyond $\mathrm{M}_{\mathrm{V}}=15.0$ will have less than 100 objects, that is exactly the regime where the 1/V$_{\mathrm{max}}$--type density estimator has the worst performance. Furthermore, \citet{2007MNRAS.378.1461T} reported that Cho{\l}oniewski density estimator was the most useful method in dealing with contaminated samples from other kinematic population because of its insensitivity to small amount of contamination~($<10\%$). However, this estimator has to be generalised over a proper motion limited sample before it can be used properly. To date, the 1/V$_{\mathrm{max}}$ is the only method that can be corrected for the incompleteness arising from the proper motion limits of the survey~(LRH15). The STY method, however, has the worst performance in every estimation among the three.

\subsubsection{Number Density}
\label{sec:past_number_density}
Integrating the LFs provide the measure of the number densities of WDs. The total WD number density is quite well measured; a unit weighted arithmetic mean of the post-1990s works is $3.71 \times 10^{-3}\pc^{-3}$ with a standard deviation of $0.74 \times 10^{-3}\pc^{-3}$. The individual thin disc and thick disc densities were only measured once by RH11. The stellar halo density was measured with SDSS~(H06) and SSS~(RH11) catalogued data in two occasions, and one of which with two different density estimators~(ie.\ the maximum volume and the effective volume methods), the solutions span two orders of magnitude from $\mathcal{O}(10^{-6})$ to $\mathcal{O}(10^{-4})$. See Table \ref{table:number_density} for the list of number densities found in published works in the last twenty years.

\begin{table}[h]
\begin{center}
    \begin{tabular}{  c  c  c  }
    Population & Number Density / pc$^{-3}$ & Reference\\
    \hline
    \hline
    \multirow{13}{*}{Total}
    & $ 3.2           \times 10^{-3}$ & \citealp{1988ApJ...332..891L}\\
    & $ >2.0          \times 10^{-3}$ & \citealp{1989LNP...328...15L}$^{\mathrm{a}}$\\
    & $ 5.0           \times 10^{-3}$ & \citealp{1991ASIC..336...67W}\\
    & $ 3.39          \times 10^{-3}$ & \citealp{1998ApJ...497..294L}\\
    & $ 4.16          \times 10^{-3}$ & \citealp{1999MNRAS.306..736K}\\    
    & $ 5.0  \pm 0.7  \times 10^{-3}$ & \citealp{2002ApJ...571..512H}\\    
    & $ 4.6           \times 10^{-3}$ & H06\\
    & $ 2.51          \times 10^{-3}$ & \citealp{2007AnA...466..627H}\\
    & $ 3.19 \pm 0.09 \times 10^{-3}$ & RH11\\
    & $ 4.39          \times 10^{-3}$ & \citealp{2012ApJS..199...29G}\\
    & $ 0.83          \times 10^{-3}$ & \citealp{2015MNRAS.450..743R}$^{\mathrm{b}}$\\
    \hline
    Thin Disc & $ 3.1 \times 10^{-3}$ & RH11$^{\mathrm{c}}$\\
    \hline
    Thick Disc & $ 6.4 \times 10^{-4}$ & RH11$^{\mathrm{c}}$\\
    \hline
    \multirow{4}{*}{Stellar Halo}
    & $ 2.6 \pm 1.2 \times 10^{-5}$ & \citealp{1989LNP...328...15L}\\
    & $ 4.0         \times 10^{-5}$ & H06\\
    & $ 4.4 \pm 1.3 \times 10^{-6}$ & RH11\\
    & $ 1.9         \times 10^{-4}$ & RH11$^{\mathrm{c}}$\\
    \hline
    \end{tabular}
\end{center}
\caption[Number density of WDs]{Number density of WDs measured in the last 20 years, uncertainties are shown if available}
\label{table:number_density}
\begin{enumerate}
\itemsep0em
\item[a] Without completeness correction, so the given value is only a lower limit
\item[b] DA-only
\item[c] Effective volume method
\end{enumerate}
\end{table}

\subsubsection{Galactic Star Formation History}
The algorithm for dating with the WDLF was rapidly evolved to allow the recovery of the SFH~(\citealp{1990ApJ...352..605N}; \citealp{2001ASPC..245..328I}; \citealp{2013MNRAS.434.1549R}, hereafter R13). For example, a short burst of increased star formation would appear as a bump in the WDLF. The use of WDLF inversion to derive the SFH is still in its infancy. R13 developed an inversion algorithm that requires input WDLF and WD atmosphere evolution models. It is similar to other inversion algorithms applied on colour-colour diagrams. However, there is some debate over the smoothing and possible amplification of noise during the application of Richardson-Lucy algorithm~\citep{1972JOSA...62...55R,1974AJ.....79..745L} and the determination of the point of convergence. \citet{2014ApJ...791...92T} used a set of spectroscopically confirmed WDs with well determined distance, temperature and surface gravity, hence the mass and radius, to derive the age of each individual WD. In their case, the derived SFH was mostly consistent with R13 but it lacks a peak in the SFH at recent times which they claim that as noise being amplified by the algorithm developed by R13. Overall, the results are broadly consistent with each other as well as those derived from the inversion of colour-colour diagrams with different algorithms~(\citealp{2002A&A...390..917V}; \citealp{2006A&A...459..783C}).

\subsection{White Dwarfs as Standard Candles}
The Chandrasekhar mass limit is the maximum mass in which a WD can exist. Beyond this, the degeneracy pressure in the centre of the WDs cannot support the gravity of the overlying layers. Upon the collapse of the core, the temperature at the core reaches the ignition temperature for carbon fusion as first proposed by \citet{1973ApJ...186.1007W}. In the degenerate state, there is no thermostatic control of the burning, this nuclear burning is a runaway reaction releasing $\mathcal{O}(10^{43})$\,J of energy in a Type Ia supernova. Because of the precise ignition scenario, they show similar luminosity profile, making them a good standard candle for distance measurement. Furthermore, because of their extreme brightness, they can be detected up to gigaparsec-scale. In 1998, they became the first observational evidence of an accelerating universe~\citep{1998AJ....116.1009R} opening a new chapter of modern cosmology. Soon after, a number of super-Chandrasekhar mass progenitors were identified, for example SN2003fg~\citep{2006Natur.443..308H}, SN2006gz~\citep{2007ApJ...669L..17H}, SN2007if~\citep{2007CBET.1059....2A} and SN2009dc~(\citealp{2010ApJ...714.1209T}; \citealp{2009ApJ...707L.118Y}). While some suggest a super-Chandrasekhar condition is due to optically thick winds, mass stripping from the binary companion star by the WD winds or differential rotation of the WD~\citep{2012ApJ...744...69H}, some suggest that a modified theory of gravity can provide a link between sub- and super-Chandrasekhar Type Ia supernova~\citep{2015JCAP...05..045D}. It is more commonly believed that they are the product of a binary WD merger~\citep{2007ApJ...669L..17H}.

\subsection{White Dwarf Planetary System}
WDs are small, with a size comparable to that of the Earth. Any companion passing through the line of sight between us and the WD can produce a deep or even total eclipse. However, there was little success in detecting such systems because WDs are scarce and their small projected areas give very low transit probabilities. These do not stop scientists because their detections can yield a huge amount of information that is usually not possible with stellar planetary systems. WDs are $\sim1,000$ times fainter than MS stars so the brightness contrast between a WD and a planet is much smaller than that between a MS star and a planet. For planets in the close proximity of a WD~($<2$\,a.u.), \citet{2001PASP..113.1227I} suggested that the infrared excess from the planet would be detectable for a WD hotter than $T_{\mathrm{eff}}=10,000$\,K. \citet{2001ApJ...546L..61C} suggested that a Jovian planet in the proximity of a UV-bright WD produced variable hydrogen recombination lines. For more widely separated systems the DODO survey~(\citealp{2008MNRAS.386L...5B}; \citealp{2009MNRAS.396.2074H}) reported that $<5\%$ of WDs have substellar companions with $T_{\mathrm{eff}}\geq500$\,K with projected separation of $60-200$\,a.u. based on non-detection from multi-epoch imaging of 23 WDs with Gemini North and the Spitzer Infrared Array Camera~(IRAC). Although planets have not yet been found around any WD, there have been some success in this area: a dusty disc was discovered from infrared excess around G19--38~\citep{1987Natur.330..138Z} and a metal rich gas disc was found rotating around TON\,345~\citep{2008MNRAS.391L.103G}. A number of young WDs show metal contamination in the photosphere suggesting they may possess terrestrial planets~(\citealp{2014A&A...566A..34K}; \citealp{2014MNRAS.440.1607B}). Most recently with the K2 mission~\citep{2014PASP..126..398H}, the first WD transit was reported on WD1145+047~\citep{2015Natur.526..546V}, which is believed to be due to the transit of a disintegrating minor planet based on the shape of the light curve. 

\subsection{White Dwarfs as Baryonic Dark Matter}
The 38 cool WDs with no spectral lines found from the SSS has sparked a controversy of whether WDs, with a typical mass of $\sim0.6\msolar$, have significant contribution towards the dark matter~\citep{2001Sci...292..698O}. At the time, the MACHO project estimated that those compact objects in the mass range $0.15-0.9\msolar$ contribute $20\%$ of the local dark matter at $95\%$ confidence~(Alcock et al. 2000). A similar project, Exp\'{e}rience pour la Recherche d'Objets Sombres~(EROS) places this limit at a maximum of $35\%$~\citep{2000A&A...355L..39L}. The EROS data towards the Small Magellanic Cloud has suggested a higher upper limit at $25\%$, and the objects have masses $2\times10^{-7}-1\msolar$~\citep{2003A&A...400..951A}. From the second phase, EROS2, based on non-detection of halo WDs, they found a $5\%$ upper limit of WD contribution to the halo mass at $95\%$ confidence. The absence of events with crossing time shorter than 10 days and the lack of low-mass MS stars in the Hubble Deep Field~(HDF) ruled out planet-sized objects~(\citealp{1996ApJ...466L..55F}; \citealp{2000A&A...355L..39L}). The microlensing events also ruled out more massive objects such as neutron stars and black holes as dark matter candidates. However, WDs were ruled out as a significant contributor to the dark matter halo after \citet{2001ApJ...559..942R} argued that the halo WD sample from \citet{2001Sci...292..698O} was misinterpreted and most of the WDs actually belong to the thick disc instead of the halo. This claim was further supported by \citet{2004ApJ...609..766K} who failed to find WDs in the 7-year baseline HDF proper motion studies, where there were 5 halo WD candidates with the formerly 2-year baseline HDF study~\citep{1999ApJ...524L..95I}.

\subsection{White Dwarfs as Sources of Gravitational Wave Radiation}
The first detection of gravitational waves was from a merger event of two stellar mass blackholes~(GW150914, \citealp{2016PhRvL.116f1102A}). However, a merger event with such massive objects is very rare. Binary stars in the Milky Way are, on the other hand, much more common. Strong sources of gravitational wave radiation include binary neutron stars, binary WDs, and binaries with neutron star and/or WD. Currently, the strongest known source is a pair of close binary WDs, WD~$1242-105$~\citep{2015AJ....149..176D}, with a strain amplitude of $h=1.66\times10^{-21}$, which is $\sim60\%$ larger than that of the GW150914 event. However, the orbital period for a pair of binary is much longer than the duration of a merger event~(0.1 second above the detection threshold for the GW150914 event) so the frequency of the ripples are in the mHz regime which is well outside the detection threshold of any existing gravitational wave detectors~(e.g. LIGO, Virgo). The Evolved Laser Interferometer Space Antenna~(eLISA, \citealp{2012CQGra..29l4016A}) is operating in the mHz frequency, but the sensitivity is expected to be at $h<10^{-20}$ so it will not be possible to confirm the gravitational waves from WD~$1242-105$. Many WDs which were thought to be in isolation are now found to be in binaries, so it is possible that there are even closer binary compact objects that are to be identified. Nevertheless, eLISA is scheduled to be operational in the 2030s, and it is still years before a direct detection of constant ripples in the fabric of space-time will be made even if there are stronger sources of this kind in the vicinity.

\section{Thesis Organisation}
In this thesis, I will focus on the generalisation of the $1/\mathrm{V}_{\mathrm{max}}$ estimator over a proper motion-limited sample and its application to produce the observed white dwarf luminosity functions and compare these against the theoretical ones. In Chapter 2, I will describe the properties of Pan--STARRS1, data reprocessing and data selection based on photometry and astrometry. The construction of WDLF and the relevant WD sciences are discussed in Chapter 3: comparison of WDLFs with four difference cooling models, three metallicities and six initial-final mass relations are performed towards to end of the chapter. Chapter 4 describes how the maximum-volume density estimator is generalised over a proper motion-limited sample, and WDLFs for different density profiles and kinematic properties are studied. The method is then applied to the Pan--STARRS1 data in Chapter 5 to study the Galactic WD populations. The final chapter is to conclude the thesis and discuss future extension work.

%% file: 2_ps_3pi.tex
\chapter{White Dwarfs from the Pan--STARRS1\\ 3$\pi$ Steradian Survey}
\label{ch:ps_3pi}
The Pan--STARRS 1~(PS1) system\footnote{\scriptsize{The PS1 Surveys have been made possible through contributions of the Institute for Astronomy, the University of Hawaii, the Pan--STARRS Project Office, the Max--Planck Society and its participating institutes, the Max Planck Institute for Astronomy, Heidelberg and the Max Planck Institute for Extraterrestrial Physics, Garching, The Johns Hopkins University, Durham University, the University of Edinburgh, Queen's University Belfast, the Harvard--Smithsonian Center for Astrophysics, the Las Cumbres Observatory Global Telescope Network Incorporated, the National Central University of Taiwan, the Space Telescope Science Institute, the National Aeronautics and Space Administration under Grant No.\ NNX08AR22G issued through the Planetary Science Division of the NASA Science Mission Directorate, the National Science Foundation under Grant No.\ AST-1238877, the University of Maryland, and Eotvos Lorand University~(ELTE) and the Los Alamos National Laboratory.}} is a wide-field optical imager devoted to survey operations~\citep{2010SPIE.7733E..0EK}. The telescope has a 1.8\,m diameter primary mirror, located on the peak of Haleakal$\bar{\mathrm{a}}$ on Maui~\citep{2004SPIE.5489..667H}. The site and optics deliver a point spread function~(PSF) with a full-width at half-maximum~(FWHM) of $\sim1"$ over a seven square degree field of view. The focal plane of the telescope is equipped with the Gigapixel Camera 1, an array of sixty $4800\times4800$ pixels orthogonal transfer array~(OTA) CCDs~\citep{2009amos.confE..40T,2008SPIE.7014E..0DO}. Each OTA CCD is further subdivided into an $8\times8$\,array of independently addressable detector regions, which are individually read out by the camera electronics through their own on-chip amplifier. Most of the PS1 observing time is dedicated to two surveys: the 3$\pi$ Sterdian Survey, that covers the entire sky north of declination $-30^{\circ}$, and the Medium-Deep Survey~(MDS), a deeper, multi-epoch survey of 10 fields, each of $\sim7$ square degrees in size~\citep{2012AAS...22010704C}. Each survey is conducted in five broadband filters, denoted g$_{\mathrm{P}1}$, r$_{\mathrm{P}1}$, i$_{\mathrm{P}1}$, z$_{\mathrm{P}1}$ and y$_{\mathrm{P}1}$, that span over the range of $400-1000$\,nm. These filters are similar to those used in the SDSS, except the g$_{\mathrm{P}1}$ filter extends $20$\,nm redward of g$_{\mathrm{SDSS}}$ while the z$_{\mathrm{P}1}$ filter is cut off at $920$\,nm. The y$_{\mathrm{P}1}$ filter covers the region from $920$ to $1030$\,nm where SDSS does not have an equivalent one. These filters and their absolute calibration in the context of PS1 are described in \citet{2012ApJ...750...99T}, \citet{2012ApJ...756..158S} and \citet{2013ApJS..205...20M}. The PS1 images are processed by the PS1 Image Processing Pipeline~(IPP; \citealp{2006amos.confE..50M}). This pipeline performs automatic bias subtraction, flat fielding, astrometry~\citep{2008IAUS..248..553M}, photometry~\citep{2007ASPC..364..153M}, and image stacking and differencing for every image taken by the system.

\section{3$\pi$ Steradian Survey}
The PS1 is comprised of six surveys where the $3\pi$ Steradian Survey~(3SS) is allocated with $56\%$ of all observing time. It is a survey of the entire sky north of declination $-30^\circ$ in the five broadband filters described above. The exposure times were changed since the beginning of the survey, but the majority of the programme used exposure times of $43, 40, 45, 30$ and $30$ seconds for the five filters respectively. Each observation of the 3SS visits a patch of sky two times with an interval of 15 minutes in between, which make a transit-time-interval\,(TTI) pair~\citep{2012AAS...22010704C}. These observations are used primarily to search for high proper-motion solar system objects~(asteroids and Near-Earth-Objects). As part of the nightly processing these TTI pairs are mutually subtracted and objects detected in the difference image are reported to the Moving Object Pipeline Software. Each of the TTI pairs are taken at exactly the same pointing and rotation angle so that the fill factor for searching for asteroids is not compromised. However, the other TTI pairs will be taken at a different rotation angle and centre offsets such that the stack should fill in the gaps and masked regions of the focal plane. The g$_{\mathrm{P}1}$, r$_{\mathrm{P}1}$ and i$_{\mathrm{P}1}$ bands are observed close to opposition to enable asteroid discovery while the z$_{\mathrm{P}1}$ and y$_{\mathrm{P}1}$ bands are scheduled as far from opposition as feasible in order to enhance the parallax factors of faint, low-mass objects in the solar neighbourhood. Each year, the field is then observed a second time in the same filter with an additional TTI pair of images, making four images of each part of the sky, in each of the five PS1 filters giving an average of $20$ images on $3\pi$ steradian of the sky per year. Over the 3-year period of the survey, this frequent imaging using the same response system allows the discovery of many proper motion objects that were not possible to be found in the past. However, the repeated imaging leaves many small gaps across the sky, which leads to wide distributions of detection limits in each filter. The median $10\sigma$ point source detection limits are at $21.55, 21.35, 20.95, 20.35$ and $19.35\magnitude$ in the five filters respectively~(Farrow et al. 2014).

\section{Data Access}
Data are stored in two formats, the Desktop Virtual Observatory~(DVO) database and the Published Science Products Subsystem~(PSPS). The DVO system consists of several programs which can insert, extract or manipulate data in the database. The PSPS hosts the databases where users can access through the web application Pan--STARRS Science Interface~(PSI). Proprietary Data access was granted to the PS1 contributing institutes. 

\subsection{DVO}
As of Processing Version 2.0~(PV2), the DVO databases are stored in form of Flexible Image Transport System~(FITS) binary files. One way to use the database is through DVO shell, which provides a command-line driven, programmable user interface to the astronomical objects and measurements in the database. It adds to this basic command set a collection of functions which provide direct access to the contents of the DVO database tables. Alternatively, common well-known packages, eg. \textsc{cfitsio} and \textsc{pyfits}, can be used to access the database directly. In this work, all data were accessed with \textsc{pyfits}. The DVO contains the following database tables:
\begin{table}[H]
\begin{center}
\begin{tabular}{ l p{8cm} }
SkyTable.fits    & Defines the organisation of the data, which is compartmentalised into $\sim10^{4}$ ``skycells''\\
Photcodes.dat    & Defines the photometric systems and transformations\\
Images.dat       & Describes all the images taken by the PS1\\
$[$skycell-ID$]$.cpt & Average properties of the objects, excluding average photometry ~(one row per object)\\
$[$skycell-ID$]$.cps & Average photometric properties for objects~(one row per filter per object)\\
$[$skycell-ID$]$.cpm & Photometry and relative astrometry of each detection\\
$[$skycell-ID$]$.cpn & Non-detections of known objects\\
$[$skycell-ID$]$.cpx & Lensing smearing and shearing of detections\\
$[$skycell-ID$]$.cpy & Average lensing smearing and shearing of objects\\
\end{tabular}
\end{center}
\end{table}

In the following work, data were compiled from SkyTable.fits, Photcodes.dat, .cpt, .cps and .cpm files with {\sc pyfits} package. The combination of catalogue ID~(CAT\_ID) and object ID~(OBJ\_ID) provides an unique ID in the database.

\subsection{PSI}
The PSI provides a login mechanism by which users become authenticated to the PSPS and thus able to access PS1 data products. Users can supply query scripts written in Structured Query Language~(SQL) directly within the interface, or with appropriate login details the query can be done via the Virtual Observatory~(VO). The outputs for the results can be delivered (1) directly to the web interface, (2) as a file with an option of several formats~(e.g. CSV, FITS), or (3) to a table in a personal database~(MyDB). Other functions include tracking the status of queries, graphing and plotting tools, managing the MyDB and obtaining Postage Stamps.

\section{Data Properties}
\subsection{Faint Magnitude Limits}
\label{sec:faint_magnitude_limits}
\begin{figure}
\includegraphics[width=0.9\textwidth]{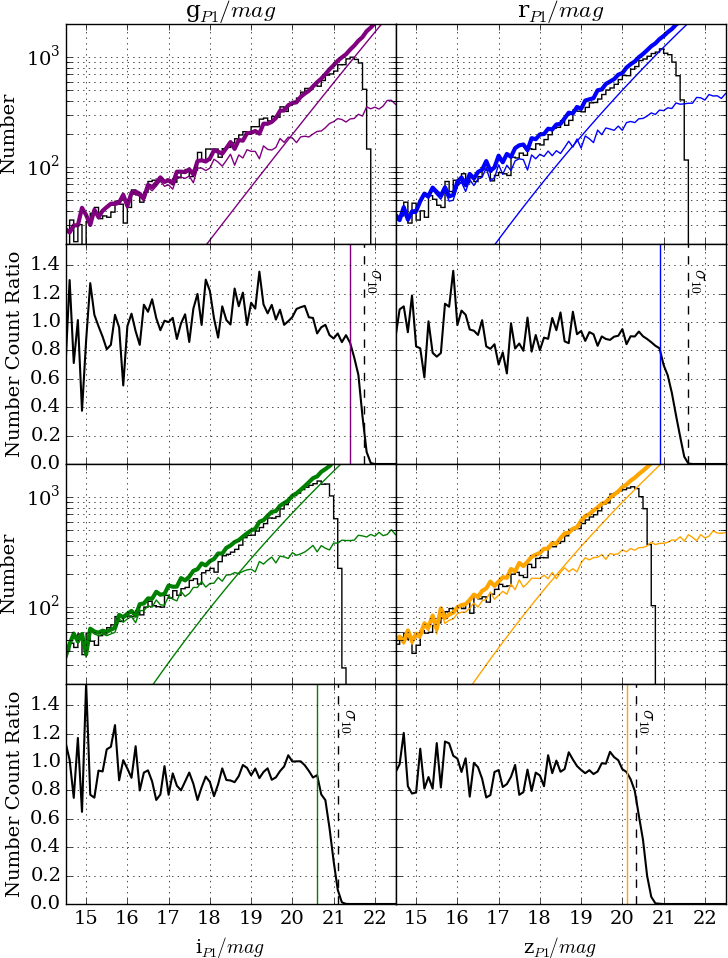}
\caption[Observed star and galaxy counts compared with model counts]{The star and galaxy counts in g$_{\mathrm{P}1}$~(purple), r$_{\mathrm{P}1}$~(blue), i$_{\mathrm{P}1}$~(green) and z$_{\mathrm{P}1}$~(yellow) filters. The odd rows show the star and galaxy counts with thin colour lines and the combined star and galaxy counts in thick colour line. The black lines are the observed number counts. The even rows show the ratios between the model and observation, the dashed lines show the 10$\sigma$ photometric limits and the colour lines mark the completeness limit~(see Chapter \ref{sec:completeness})}
\label{fig:star_counts}
\end{figure}

In order to find the faint magnitude limits at which data are complete, the object counts are compared against synthetic star and galaxy counts in g$_{\mathrm{P}1}$, r$_{\mathrm{P}1}$, i$_{\mathrm{P}1}$ and z$_{\mathrm{P}1}$ filters in 15 fields at high galactic latitudes to avoid interstellar dust hampering the completeness. We have chosen a field of view of $\sim3.4$ square degrees~(a HEALPix tile size of $N_{\mathrm{side}} = 32$, see the last section of this chapter), a size that is large enough for sufficient star counts and to smooth out inhomogeneity of galaxies, while at the same time small enough to limit variations in data quality across the field~(see Fig. \ref{fig:star_counts}).

\subsection*{Stars}
Differential star counts along the line of sight to each field are obtained using the Besan\c{c}on Galaxy model~(\citealp{2003A&A...409..523R}, \citeyear{2004A&A...416..157R}). This employs a population synthesis approach to produce a self-consistent model of the Galactic stellar populations, which can be ``observed'' to obtain the theoretical star counts. It is a useful tool to test various Galactic structure and formation scenarios although we have adopted all the default input physical parameters except the latest spectral type is DA9 instead of the default DA5. There are only two photometric systems available, the Johnson--Cousins and the CFHTLS-Megacam systems. Since there is only a small difference between the PS1 and Megacam filters, the g', r', i' and z' are used to approximate the g$_{\mathrm{P}1}$, r$_{\mathrm{P}1}$, i$_{\mathrm{P}1}$ and z$_{\mathrm{P}1}$ in this work. The faint magnitude limits of the model are set at $25\magnitude$ to guarantee that the model is always complete as compared to the data.

\subsection*{Galaxies}
Fainter than $\mathrm{M}_{\mathrm{bol}}\sim19\magnitude$, galaxies become unresolved~(i.e.\ point-like) and have photometric parameters that overlap with stars. Therefore, it is necessary to include galaxies in the synthetic number counts. Galaxy counts to faint magnitudes have been determined in many independent studies. The Durham Cosmology Group has combined their own results (\citealp{1991MNRAS.249..481J}; \citealp{1991MNRAS.249..498M}) with many other authors. These are available online along with transformations to different photometric bands\footnote{http://astro.dur.ac.uk/$\sim$nm/pubhtml/counts/counts.html}. They are provided in terms of log-number counts per square degree per half-magnitude unless specified otherwise. A cubic spline was fitted over all available observations to obtain the galaxy counts as functions of magnitude in each band.

\subsection{Bright Magnitude Limits}
The PS1 saturation limit is at 15\,mag, so this defines the bright limit in all filters.

\subsection{Proper Motion Recalculations}
Without parallaxes, WDs can only be found efficiently with proper motions. Therefore, on top of magnitude limits, a good knowledge of the proper motions and their associated uncertainties are needed to apply a completeness correction to a proper motion-limited sample. Beyond ~70\,pc, the parallax solution is only noise being amplified~\citep{2008IAUS..248..553M}. Yet, the major drawback is not the unreliable parallax, instead, when a parallax is found with its associated uncertainty, the uncertainty in the proper motion would be inaccurate since part of the uncertainty is propagated to the parallactic uncertainty. Thus, with the given proper motion solutions found with the IPP, where 5 parameters were solved simultaneously~(the pair of zero-point in the right ascension~($\alpha$) and declination~($\delta$) directions, the pair of proper motions and the parallax), the propagation of errors would lead to some unquantifiable over/underestimation in the proper motion uncertainties. Since the given set of astrometric solution is only good up to a few tens of parsecs, even for the study of nearby WDs, most of them lie outside the range where the parallactic solutions are meaningful. Therefore, I decided to compute my own set of 4-parameter solution where parallax is not solved for.

The direct recomputation can be expressed in the following matrix form which is equivalent to solving with the method of least squares, 
\begin{equation}
\left(
\begin{array}{cccc}
\frac{1}{w_{0}} & 0 & \frac{t_{0}}{w_{0}} & 0 \\
0 & \frac{1}{w_{0}} & 0 & \frac{t_{0}}{w_{0}} \\
\cdot & \cdot & \cdot & \cdot \\
\cdot & \cdot & \cdot & \cdot \\
\frac{1}{w_{n}} & 0 & \frac{t_{n}}{w_{n}} & 0 \\
0 & \frac{1}{w_{n}} & 0 & \frac{t_{n}}{w_{n}}
\end{array}
\right)
\left(
\begin{array}{c}
\alpha_{\mathrm{ZP}}\\
\delta_{\mathrm{ZP}}\\
\mu_{\alpha}\\
\mu_{\delta}
\end{array}
\right)
=
\left(
\begin{array}{c}
\frac{\Delta\alpha_{0}}{w_{0}}\\
\frac{\Delta\delta_{0}}{w_{0}}\\
\cdot\\
\cdot\\
\frac{\Delta\alpha_{n}}{w_{n}}\\
\frac{\Delta\delta_{n}}{w_{n}}
\end{array}
\right)
\end{equation}
with
\begin{equation}
w_{i} = \sqrt{ \sigma_{i}^{2} + 0.015^{2} }
\end{equation}
where $w_{i}$ is the weight~(the photometric uncertainty), $\sigma_{i}\magnitude$ is the random photometric uncertainty and the $0.015\magnitude$ comes from the systematic photometric uncertainty. The non-Gaussian PSF used in PS1 has the FWHM matches with the $\sigma_{\mathrm{PSF}}$ through $\mathrm{FWHM} = \sigma_{\mathrm{PSF}}*2*\sqrt{2}$, which is a factor of $0.8326$ of the standard Gaussian profile where $\mathrm{FWHM} = \sigma_{\mathrm{Gauss}}*2*\sqrt{2\ln{2}}$. By using an average FWHM of $1.2''$, the astrometric uncertainties, $\sigma_{x}$, can be written as a function of photometric uncertainties through \citep{1985MNRAS.214..575I} 
\begin{align}
\sigma_{x} &= \sigma_{\mathrm{Gauss}} \times w_{i} \\
           &= 0.8326 \times \mathrm{FWHM} \times w_{i} \\
           &\approx w_{i}.
\end{align}
$t_{i}$ is the epoch of the measurement, $\alpha_{\mathrm{ZP}}$ is the zero point of $\alpha_{i}$, $\delta_{\mathrm{ZP}}$ is the zero point of $\delta_{i}$, $\Delta\alpha_{i}$ is the $\alpha$-offset from the $\alpha_{\mathrm{ZP}}$, $\Delta\delta_{i}$ is that for $\delta$-offset, $\sigma_i$ is the photometric uncertainty and in the middle brackets, the $\alpha$, $\delta$, $\mu_{\alpha}$ and $\mu_{\delta}$ are the solution. The associated uncertainties is the dot product of the inverse of the first matrix, {\bf A}, with itself,
\begin{equation}
\left(
\begin{array}{c}
\sigma_{\alpha}^{2}\\
\sigma_{\delta}^{2}\\
\sigma_{\mu_{\alpha}}^{2}\\
\sigma_{\mu_{\delta}}^{2}
\end{array}
\right)
=
{\bf A}^{\mathrm{T}}{\bf A} \mathrm{.}
\end{equation}

For a uniformly sampled object, the proper motion uncertainty is inversely proportional to the maximum epoch difference and the square of the number of epochs. 3SS has a large number of observations and a short time baseline~(a maximum of 3.5 years) compared to other proper motion surveys that typically have either longer time baselines~(eg. USNO-B1.0, PPMXL and SSS) or many more epochs~(eg. Kepler, CoRoT, Catalina Surveys and PS1-MDS). However, the 3SS is the best survey for WDs because it has deeper and more accurate photometry than the old surveys, while it has a much larger footprint area than the surveys with hundreds or thousands of epochs. In order to further improve the quality of the astrometric solutions, four robust refitting algorithms were investigated to remove outliers.

\begin{enumerate}[label={\roman*}]

\item{\textbf{\textit{\textcolor{mediumelectricblue}{Fitting by minimising absolute deviation}}}}\\
Instead of maximising the likelihood estimator by minimising the mean squared deviation~(i.e. $\chi^{2}$), this method works by minimising the absolute deviation from the model. Consider a straight line $y(x;a,b) = a + bx$. To minimise the sum of the absolute deviation means that the function to be minimised is
\begin{equation}
\label{eq:mad}
\sum^{N}_{i=1} |y_{i}-a-bx_{i}|.
\end{equation}
This function may look slow to be solved, however, it can be simplified as the minimisation of this function is equivalent to minimising
\begin{equation}
\sum_{i} |c_{i}-c_{M}|
\end{equation}
where $c_{M}$ is the median of the set of $c_{i}$~(see Appendix A). By setting $a = \mathrm{median}\left[y_{i}-bx_{i}\right]$, Eq. \ref{eq:mad} would become an equation with a single variable which can be solved numerically. The use of a median in this method can reduce the effect of the outlying data. However, data points have to be unit weighted while the quality of the PS1 observations at different epoch is not similar throughout the 3.5 years period. Without appropriate weightings, the solutions are likely to be skewed by clustered data~(e.g. many observations over a short period of time, which is common at low latitude).\\

\item{\textbf{\textit{\textcolor{mediumelectricblue}{Jackknife resampling method}}}}\\
Jackknife is a resampling technique that takes the average value of the best fit solutions from all combinations of subsets of the data. A subset is formed by leaving out any number~(x>0) of data points. This method would be very effective against outliers, which is by definition, a small portion of data that does not follow the general behaviour. However, there is not enough computing power to go through this procedure. If 5(10) points out of the average of 60 epoch measurements were left out, the resampling of $10^{9}$ objects would require $\mathcal{O}(10^{15})(\mathcal{O}(10^{19}))$ calculations. Assume each operation takes 1 millisecond, it requires $10^{2}(10^{6})$ years on a single processor.\\

\item{\textbf{\textit{\textcolor{mediumelectricblue}{Iterative outlier rejection}}}}\\
This method identifies data points lying outside the 3$\sigma$ confidence limit from the best fit solution as outliers. A new weighted-least square solution will be solved. This process repeats until no more data points are rejected. This method assumes that the initial fit is not significantly affected by the outliers such that the best fit lines are not skewed towards the outliers instead of the true detections. Since the iteration lasts only a few times, it is orders of magnitudes faster than the jackknife method. However, it is more sensitive to data further away from the centroid of the solution. For a small change in the best-fit gradient, the absolute change in the ordinates far from the centroid would be much larger than those near the centroid.\\

\item{\textbf{\textit{\textcolor{mediumelectricblue}{Iterative outlier rejection with time dependent uncertainties}}}}\\
In order to solve the problem where data far from the centroid of the solution being more sensitive, an improved version has to propagate the uncertainties properly such that data points are further away from the centroid have higher tolerance levels. If one considers the best fit line described by the function
\begin{equation}
f = x + \mu \, t,
\end{equation}
the uncertainty given by the standard propagation of errors can be given by the covariance matrix $\left\langle \sigma_{i} \, \sigma_{j} \right\rangle$
\begin{align}
\begin{split}
\sigma^{2} &= \sum_{i} \sum_{j} \left\langle \sigma_{i} \, \sigma_{j} \right\rangle \frac{\partial f}{\partial \rho_{i}} \frac{\partial f}{\partial \rho_{j}} \\
&= \sigma^{2}_{x} \frac{\partial f}{\partial x}^{2} + 2 \sigma^{2}_{x \mu} \frac{\partial f}{\partial x} \frac{\partial f}{\partial \mu} + \sigma^{2}_{\mu} \frac{\partial f}{\partial \mu}^{2} \\
&= \sigma^{2}_{x} + \sigma^{2}_{\mu} t^{2}.
\end{split}
\end{align}
With the time-dependent term in the propagation of uncertainties from proper motions, data points that are further away from the centroid of the best-fit solution would have higher tolerance.
\end{enumerate}

The re-computation of proper motion is done to {\it all objects with detected proper motions from the IPP, regardless of the magnitude and the quality of the given proper motion}. This should, in theory, have removed all the quasars that are used to set the astrometric frame. The final catalogue has the proper motions fitted with the least squares method, where outliers are rejected with the improved outlier rejection method iteratively until no outlier is detected when solving for the solutions~\citep{2015ASPC..493..347L}. However, instead of simply reprocessing with all available epochs, data with poor observing conditions are rejected before solving for the initial astrometric solution.

\subsubsection*{Image quality in individual epoch}
\begin{figure}
\includegraphics[width=0.9\textwidth]{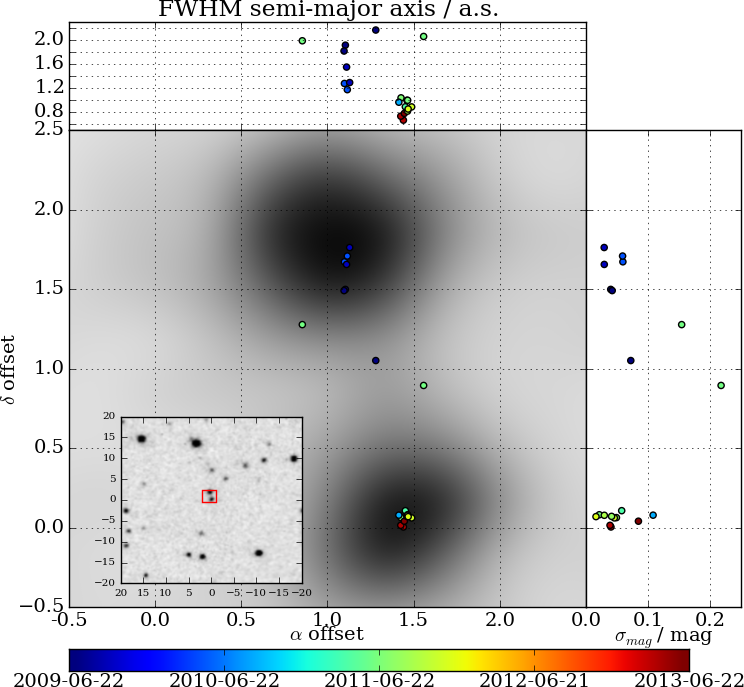}
\caption[Position of a high proper motion low temperature candidate at each epoch, supplied with FWHM and photometric uncertainties]{A high proper motion low temperature candidate at $\alpha = 19\mathrm{h} 19\mathrm{m} 34.67\mathrm{s}, \delta = 17^{\circ} 15' 9.92''$ ). When the positions, FWHM, photometric uncertainties and time are plotted against epoch, it appears that the early epochs are clustered about the object on the top, while the later epochs clustered about the counterpart. This is due to poor observing condition at the earliest epoch~(large FWHM, top panel) which was used to produce the source list where the two objects were blended as one. Subsequent pairings are then confused and sometimes they are measured as one object and sometimes as two. The photometric uncertainties which are used as weights in the astrometric solution, on the right panel, cannot reflect the astrometric quality of the given epochs}
\label{fig:multi-epoch}
\end{figure}

In order to confirm the quality of the proper motion recalculation, some supposedly high quality samples were drawn to check the validity. In one of the low temperature high proper motion candidates, $\alpha = 19\mathrm{h}\,19\mathrm{m}\,34.67\mathrm{s}, \delta = 17^{\circ}\,15'\,9.92''$, it appears that 2 objects that are separated by $\sim1.6''$ from centroid to centroid are confused in the early epoch with poor observing conditions. When the first epoch was used to produce the source list, the two objects were identified as one and the pair in the subsequent epochs were falsely matched. This produced large proper motions and with the iterative rejection, some good data points are rejected because there are a large number of outliers~(see Fig.~\ref{fig:multi-epoch}). As the ``outliers'' are rejected iteratively, the final solution is based on a cluster of poor data at early times and a cluster of good data at late times. Together they give a high proper motion signal with small uncertainties and small $\chi^{2}_{\nu}$ goodness-of-fit. In this case, the one at the bottom is classified as a galaxy based on the j-band morphology in VISTA Hemisphere Survey~(VHS). The counterpart above it appears to be a star in the VHS, from visual inspection, the total stack does not produce an object smeared in any direction so it cannot have carried a significant proper motion. Further inspection to digitised UKST~(Red) and POSS-I~(Red) photographic plate images, no moving object is identified in the $1'\times1'$ PS image upon blinking. When the declination offset and the size of the semi-major axis of the FWHM are plotted against epoch, a strong correlation between the offset in the declination position and the size of the FWHM is observed. From the plots, I deduced that a FWHM of 3 pixels~($\sim1.6''$) is the worst condition to be included in my recalculation.

\subsection{Lower Proper Motion Limits}
\label{sec:lower_pm_limit}
In order to select a clean sample of proper motion objects, we require our samples to have high signal to noise ratio in the proper motions. This excludes most of the non-moving objects from our catalogue and limits scatter in the RPM diagram\,(RPMD). The accurate description of total proper motion is very complex, especially with the multi-epoch nature of PS1. To simplify the problem, the first order approximation of the total proper motion uncertainty of an individual object can be found by the standard propagation of uncertainty, assuming the proper motion in the perpendicular directions are independent, which can be expressed as
\begin{equation}
\sigma_{a}^{2} = \sum_{i} \left[ \sigma_{i}^{2} \left( \frac{\partial a}{\partial x_{i}} \right) \right]
\end{equation}
where $a$ is the total proper motion, $\mu$, in this case, so
\begin{align}
\sigma_{\mu}^{2} &= \sigma_{\mu_{\alpha}\cos(\delta)}^{2} \left( \frac{\partial \mu}{\partial \mu_{\alpha}\cos(\delta)} \right)^{2} + \sigma_{\mu_{\delta}}^{2} \left( \frac{\partial \mu}{\partial \mu_{\delta}} \right)^{2} \\
\sigma_{\mu} &= \sqrt{\left(\frac{\mu_{\alpha}\cos(\delta)}{\mu}\right)^{2}\sigma_{\mu_{\alpha}\cos(\delta)}^{2}+\left(\frac{\mu_{\delta}}{\mu}\right)^{2}\sigma_{\mu_{\delta}}^{2}} .
\end{align}

\begin{figure}
\includegraphics[width=0.9\textwidth]{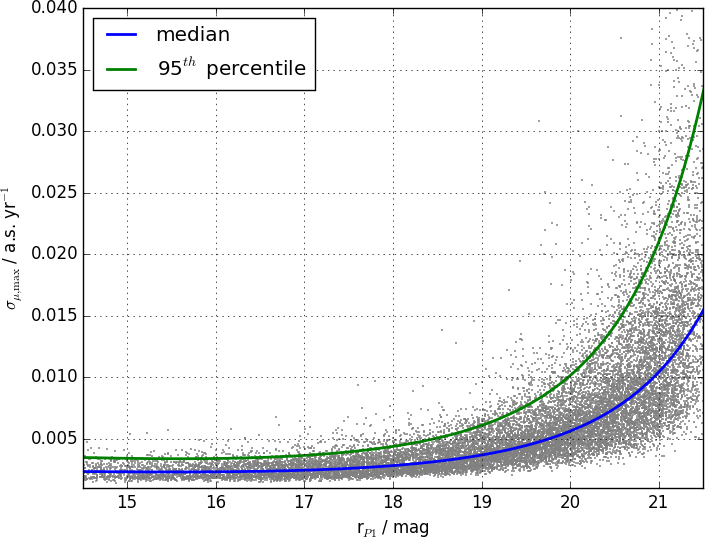}
\caption[Median and $95^{\mathrm{th}}$ percentile of and the median of $\sigma_{\mu}$ as a function of r$_{\mathrm{P}1}$]{The scatter plot with the median and the $95^{\mathrm{th}}$ percentile of $\sigma_{\mu}$ as a function of r$_{\mathrm{P}1}$, the curves are the cubic splines of the smoothed data}
\label{fig:low_pm_limit}
\end{figure}

However, when $\sigma_{\mu}$ is plotted against r$_{\mathrm{P}1}$, there is significant scatter at a given magnitude~(see Fig.~\ref{fig:low_pm_limit}). However, a well defined relation between the proper motion uncertainty and magnitude is needed for volume and completeness correction calculations~(see Chapter \ref{ch:vmax}). To overcome this problem I need to define the ``typical worst'' proper-motion uncertainty as a function of r$_{\mathrm{P}1}$ magnitude, and assign each object with the new semi-empirical proper motion uncertainty. This follows a similar algorithm that was designed by RH11. For the data with more than $1\sigma_{\mu}$ significance in proper motion, they were sorted by r$_{\mathrm{P}1}$. The median~($\sigma_{50}$) and the $95^{\mathrm{th}}$-percentile~($\sigma_{95}$) of the $\sigma_{\mu}$ within the moving bounding box with 500 objects are located, the 10 largest and smallest $\sigma_{\mu}$ in the bounding box were discarded to remove outliers. These smoothed data were then binned into $0.1\magnitude$. The median $\sigma_{50}$ and $\sigma_{95}$ of each bin were located and the corresponding magnitude was the arithmetic mean of r$_{\mathrm{P}1}$ in that bin. The 5 faintest bins were discarded because of the noisy $\sigma_{\mu}$--r$_{\mathrm{P}1}$ relation, 10 bins from 28 patches of the sky\footnote{HEALPix pixel ID: 392, 441, 1467, 1680, 2129, 2257, 2273, 2287, 2308, 2309, 2310, 2311, 2312, 2320, 2329, 2335, 2368, 2369, 2372, 2373, 2374, 2375, 2394, 2395, 2396, 2397, 2398 and 2399} were removed because of the significant noise at the faint end. Cubic splines were then fitted over the $\log(\sigma_{50})-\mathrm{r}_{\mathrm{P}1}$ and $\log(\sigma_{95})-\mathrm{r}_{\mathrm{P}1}$ relations. The fitting in logarithmic space was to serve as a simple positivity constrain at the bright end where the splines tend to turn negative. Fig.~\ref{fig:low_pm_limit} shows the splines overplotted on the data from one pixel. Despite a large number of points being used in each box, it was still too noisy to find a smooth function of the $99^{\mathrm{th}}$-percentile. Using the $\sigma_{50}$ and $\sigma_{95}$ I estimated the size of the distribution of $\sigma_{\mu}$ as a function of r$_{\mathrm{P}1}$ and reassigned each object with a new proper motion uncertainty, the $\sigma_{\mathrm{max}} = \sigma_{95}$, which followed a well defined function. The distribution of the $\sigma_{\mu}$ as a function of magnitude, $\sigma'(\mathrm{r}_{\mathrm{P}1})$, was estimated by $\sigma'=\frac{1}{2}(\sigma_{95}-\sigma_{50})$. Objects above the $\sigma_{50}+3\sigma'$ threshold were removed. This procedure was performed over a small area of sky to limit the effect of medium to large scale quality variations~(see the end of this chapter). The completeness of this outlier rejection based on proper motion uncertainty is $\sim99.87\%$.

\subsection{Upper Proper Motion Limit}
The upper proper motion limit is determined by both the matching radius, the matching algorithm and its efficiency.

\subsubsection*{Matching Radius}
The search radius for cross-matching between different epochs of PS1 data was $1''$. Although each part of the sky was imaged twelve times per year on average, some parts of the sky were limited by seasonal observability and weather. At low declinations, the sky could only be observed in a window of a few months every year so the maximum proper motion an object can carry is limited to roughly the size of the search radius per year, which is $1''\,yr^{-1}$.

\subsubsection*{Matching Algorithm}
For high proper motion objects, when they have moved for more than $1''$ throughout the survey period, they would be detected as 2 or more separate objects. The IPP solves for the 5-parameter astrometric solutions that include parallax. In order to break the degeneracy in the parallax and proper motion in the astrometric solution, a minimum epoch difference of 1.5 years is required. For objects that move faster than $0.66''\,yr^{-1}$, they would have moved outside the matching radius after 1.5 years. Otherwise, these objects would have either erroneously large proper motion with small parallax or vice versa. Although it is possible to ``stitch'' the multiple parts back together and recalculate the proper motions with the maximal use of data, this creates a completeness problem to the faint high proper motion objects. When objects close to the detection limits can only be observed under the best observing conditions, there are not enough epochs to solve for the astrometric solution when they are split into parts. For example, if an object has 10 evenly distributed measurements that are split into ``2 objects'' each with 5 measurements, the individual uncertainty would become $2\times\sqrt{2}\approx2.28$ times larger than that is solved as a single object where the $2$ comes from the ratio of the maximum epoch difference and $\sqrt{2}$ comes from the ratio of the number of epochs~\citep{2013ASPC..469..253H}.

\subsubsection*{Matching Efficiency}
\begin{figure}
\includegraphics[width=0.9\textwidth]{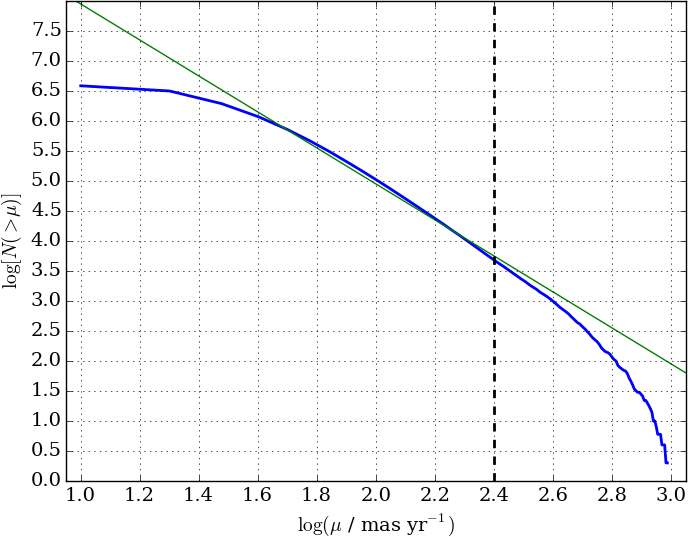}
\caption[Number of objects with proper motion larger than the given proper motion against proper motion]{The logarithm of the number of objects with proper motion larger than the given proper motion is plotted against proper motion~(blue). Due to the small distances at which the high proper motion objects are at, the relation has a gradient of $-3$ as shown by the green line. At $10^{2.4}=0.251''\,yr^{-1}$~(dashed line), the underestimation of the number of objects imply an incomplete matching of those high proper motion objects}
\label{fig:upper_pm_limit}
\end{figure}

The high proper motion population is in the immediate solar neighbourhood so the number density is uniform at this limit. Through using proper motion as a proxy-parallax~(like that in reduced proper motion), the number density follows
\begin{equation}
\log(N) \propto -3 \log(\mu)
\end{equation}
for a complete sample. In the 3SS, the gradient deviates from $-3$ at $0.251''\,yr^{-1}$, although the curve can be fitted with a steeper gradient, I believe this is an evidence of incompleteness of the apparently bright objects in the sample~(see Fig. \ref{fig:upper_pm_limit} and Chapter \ref{ch:wdlf}).

Combining the three cases, the matching efficiency gives the tightest limit among all, so the global upper proper motion limit is set at $0.251''\,yr^{-1}$.

\subsection{Object Morphology}
All objects were marked as good objects and were not flagged as extended, rock, ghost, trail, bleed, cosmic ray, asteroid or bad object.

\subsubsection*{Star-Galaxy Separation}
The 3SS catalogue has a star-galaxy separator~(SGS) entry for every object. The typical photometric limits are $\sim21\magnitude$ in the optical and a typical FWHM of $1.2''$ so at the faint end of the survey, the reliability of the SGS is limited to the observing conditions. Regardless of how sophisticated the IPP SGS algorithm is, when a galaxy is unresolved, there is little one can do about it. Therefore, I compared the SGS with the object classifier from the Canada--France--Hawaii' Telescope Lensing Survey~(CFHTLenS, \citealp{2012MNRAS.427..146H}; \citealp{2012AAS...21913009E} and \citealp{2012MNRAS.421.2355H}): ``CLASS\_STAR'', ``star\_flag'' and ``FITCLASS''. CFHTLenS is a 154 square degrees multi-colour optical survey with the Megacam u*, g', r', i' and z' filters incorporating all data collected over the five-year period on the CFHT Legacy Survey, which was optimised for weak lensing analysis. The deep photometry in i'-band was always taken at sub-arcsecond seeing condition. Both star\_flag and FITCLASS were optimised for galaxy selection, so the CLASS\_STAR provided by SExtractor was used for my analysis. Considering the superior quality in both photometry and observing conditions of CFHTLS, at the limit of i$_{\mathrm{P1}}\sim21\magnitude$, I assumed that CLASS\_STAR was completely reliable. The pairing criteria of the two catalogues were $2''$ matching radius and $5\sigma_{\mathrm{max}}$ proper motions.

\begin{figure}
\includegraphics[width=0.9\textwidth]{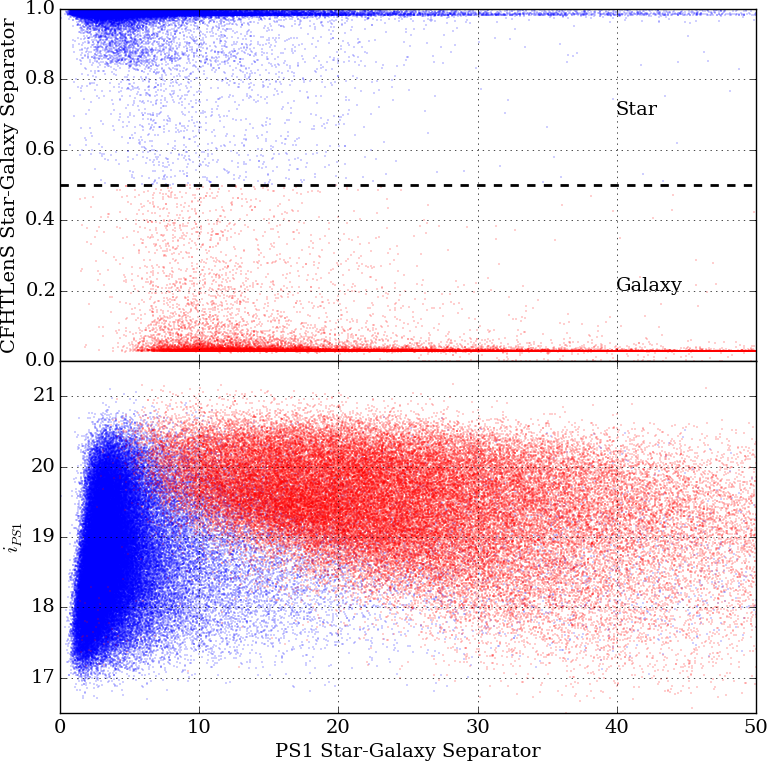}
\caption[CFHTLenS star galaxy separator, CLASS\_STAR, plotted against the PS1 star-galaxy separator]{Top: PS1 SGS is plotted against the CFHTLenS star galaxy separator, CLASS\_STAR. When the CLASS\_STAR is larger than 0.5, it is considered as a star; otherwise, a galaxy. Bottom: the i$_{\mathrm{P}1}$ is plotted against the PS1 SGS with the same colour code as the plot above}
\label{fig:sgs_compared}
\end{figure}

\begin{figure}
\includegraphics[width=0.9\textwidth]{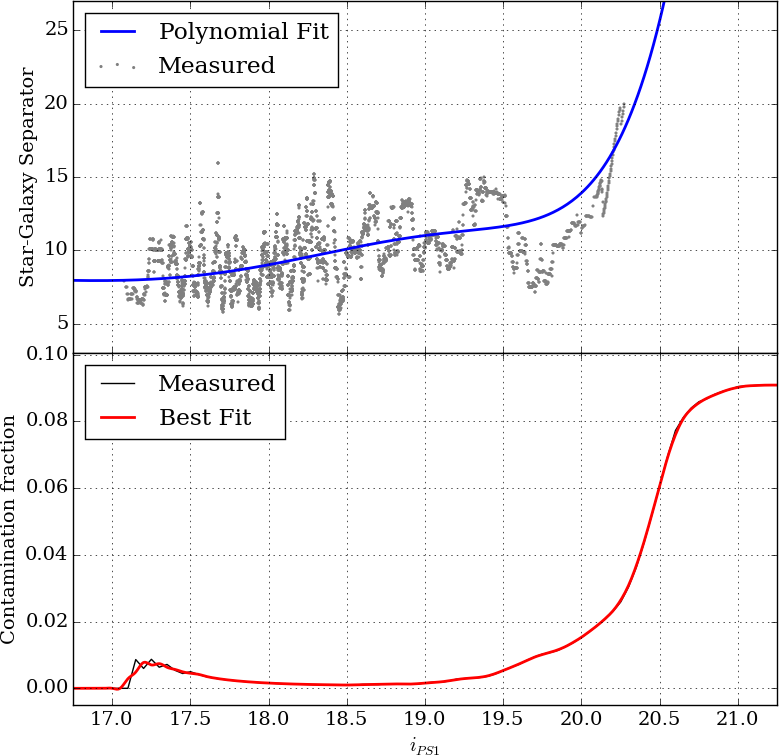}
\caption[Star-galaxy separator and contamination fraction as a function of i$_{\mathrm{P}1}$]{Top: the PS1 SGS as a function of i$_{\mathrm{P}1}$ at which the sample is 90\% complete. Bottom: contamination fraction as a function of i$_{\mathrm{P}1}$ at a constant completeness at 90\%}
\label{fig:sgs}
\end{figure}

In Fig.~\ref{fig:sgs_compared}, the PS1 SGS is plotted against CLASS\_STAR. I defined an object as a star when CLASS\_STAR $> 0.5$, as a contaminant otherwise. To find a relation of SGS as a function of magnitude at a constant completeness, the data was first sorted by i$_{\mathrm{P}1}$. With a moving box car of the size of $2\%$ of the data, the SGS at which the data was $90\%$ complete was found as a function of the median i$_{\mathrm{P}1}$ of the box. A $10^{\mathrm{th}}$ order polynomial was then fitted over the relation after being smoothed by a median filter~(upper panel of ~\ref{fig:sgs}). To account for the contamination, the sample of objects was first filtered by the SGS(i$_{\mathrm{P}1}$) function, then the fraction of the number of objects with CLASS\_STAR $<$ 0.5 was found as a function of i$_{\mathrm{P}1}$~(lower panel of Fig.~\ref{fig:sgs}).

\subsection{Reduced Proper Motion}
\begin{figure}
\includegraphics[width=\textwidth]{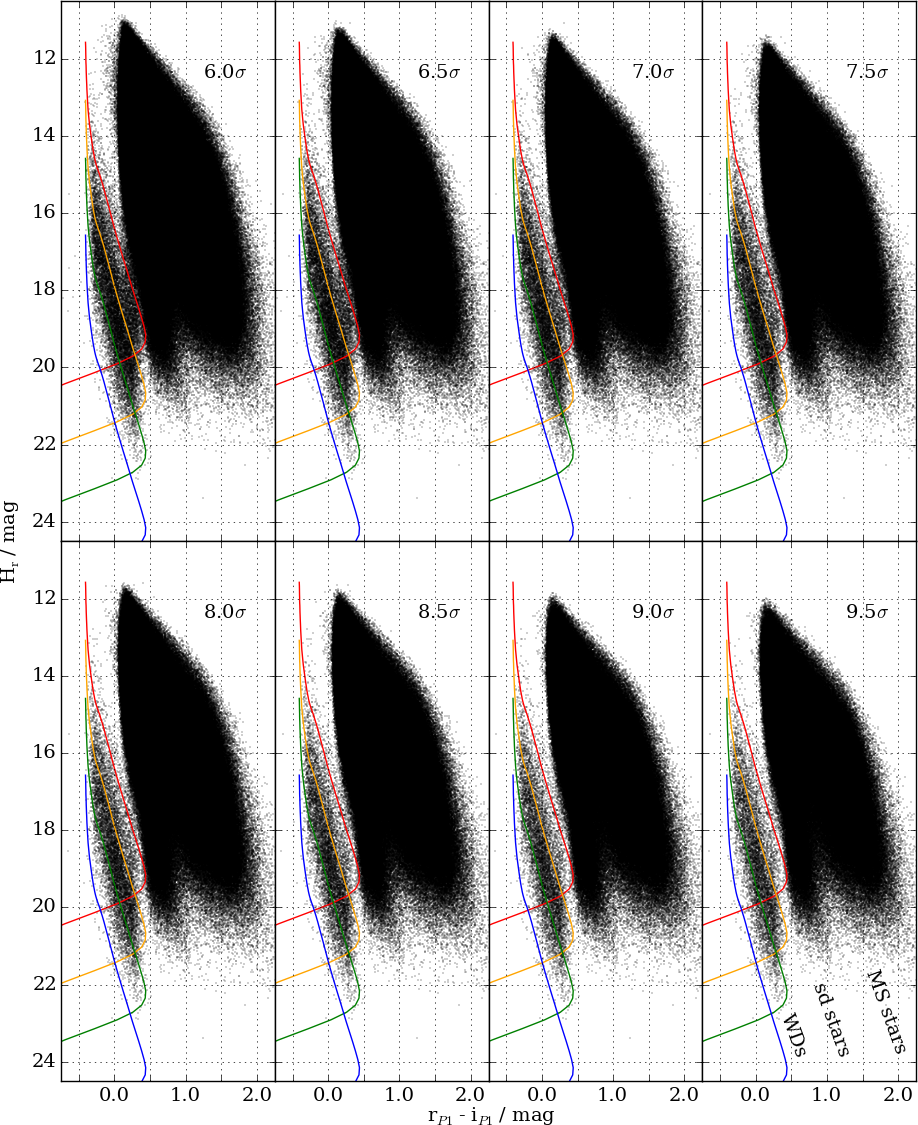}
\caption[RPM diagrams with $\mu > 5.0 - 9.5\sigma_{\mathrm{max}}$]{RPM diagrams with $\mu > 6.0 - 9.5\,\sigma_{\mathrm{max}}$ from top left to bottom right. The red, yellow, green and blue lines are the cooling sequence for DA WDs with tangential velocity at $20, 40, 80$ and $200\kmps$. Most contaminants appear as vertical scatter with neutral colour~($0.0<$r$_{\mathrm{P}1}-$i$_{\mathrm{P}1}<0.5$)}
\label{fig:rpmd_compared}
\end{figure}

There exists a correlation between proper motions and distance of nearby objects, since closer objects are more likely to show large proper motions. RPM, $H$, combines the proper motion with apparent magnitude, to provide a crude estimate of the absolute magnitude. Thus, the RPM equation has a close resemblance to the absolute-apparent magnitude relation,
\begin{align}
H_{m} &= m + 5\log(\mu) + 5 \\
      &= M + 5\log(v_{\mathrm{tan}}) - 3.38
\label{eq:rpm}
\end{align}
where $\mu$ is the proper motion in arcseconds per year, $m$ is the apparent magnitude, $M$ is the absolute magnitude and $v_{\mathrm{tan}}$ is the tangential velocity in kilometers per second. The RPM of WDs are a few magnitudes fainter than MS and subdwarfs stars of the same colour. Therefore, the WD locus is separated from other objects in the RPM diagram. This has been proved to be an efficient way to obtain a clean sample (H06 and RH11). In this work, we use the r$_{\mathrm{P}1}$ apparent magnitude for our RPM, denoted as $H_{r}$; see Fig.~\ref{fig:rpmd_compared} for RPM diagram at different confidence levels.

\subsection{Tangential Velocity Projection}
\label{sec:vtan_distribution}
Low-velocity WDs have similar RPMs to those of high velocity subdwarfs from the Galactic halo. Although RPMs and colours alone can clean up these contaminants, it is necessary to know how many WDs are removed so that a completeness correction can be applied. In order to obtain a clean sample of WDs, a tangential velocity limit is introduced to define a well known selection function in the RPM diagram. This can be seen from Equation \ref{eq:rpm} that there is a one to one relation between $H_{r}$ and $v_{\mathrm{tan}}$ at a given magnitude for a fixed model atmosphere. Through the application of a minimum tangential velocity, $v_{\mathrm{min}}$, a WD cooling curve is translated into the line of minimum RPM.

\begin{figure}
\includegraphics[width=0.9\textwidth]{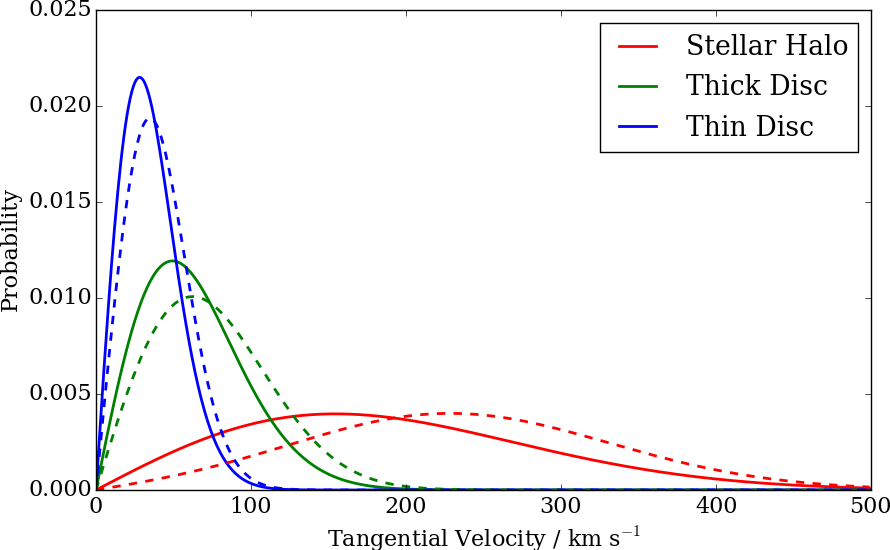}
\caption[The tangential velocity distribution in the direction of the North Galactic Pole and the Galactic Anti-Center]{The tangential velocity distribution of the thin disc, thick disc and stellar halo in the direction of the North Galactic Pole~(solid line) and the Galactic Anti-Center~(dashed line) based on the kinematic information from Table~\ref{table:galactic_kinematics}}
\label{fig:vtan_distribution}
\end{figure}

The problem of incompleteness was first identified by \citet{1986ApJ...308..347B}, and a Monte Carlo simulation was used to correct for such incompleteness by comparing with star count studies. This correction, known as the discovery fraction, $\chi$, was then applied by H06. Instead of using a simulation, \citet{2003MNRAS.344..583D} arrived at the discovery fractions by integrating over the Schwarzschild distribution functions to give the tangential velocity distribution, $\mathrm{P}(v_{\mathrm{tan}}, \alpha, \delta)$. This was done by projecting the velocity ellipsoid of the Galactic populations on to the tangent plane of observation, correcting for the mean motion relative to the Sun, and marginalising over the position angle to obtain the distribution in tangential velocity~(see Fig.~\ref{fig:vtan_distribution} and Appendix B for the mathematical construction). The values adopted for the mean reflex motions and velocity dispersion tensors are listed in Table~\ref{table:galactic_kinematics}. They were obtained from the \citet{2009AJ....137..266F} study of SDSS M dwarfs, with values taken from their $0-100\pc$ bin that is least affected by the problems associated with the deprojection of proper motions away from the plane~\citep{2009MNRAS.400L.103M}. RH11 further generalised the technique to cope with an all sky survey in a field by field basis. However, there were some discrepancies between the parameter space in which the volume and the discovery fractions were integrated over in all these cases. In order to generalise over a proper motion limited sample properly, the effects of the tangential velocity limits and the proper motion limits have to be considered simultaneously at each distance interval~(LRH15; see Chapter 4 for details).

\section{Derived Products}
The construction of a WDLF depends on the distance, luminosity and atmosphere type of the WDs, as well as the physical properties of the host population. Since most WDs are within a few hundred parsecs from us, the scalelength, in the order of kiloparsecs, of all Galactic components are not considered in this work. Interstellar reddening is corrected with the use of a three dimensional dust map when solving for the photometric parallax.

\subsection{WD Atmosphere Type}
On the theoretical front, WD atmospheres have been studied in detail. In recent years, with the abundant spectroscopic data available from SDSS, there were significant improvements in the understanding in the atmospheres. In addition to the conventional DA~($1,500$\,K $< T_{\mathrm{eff}}<120,000$\,K) models, synthetic photometry is available for 9 different hydrogen-helium mass ratios in the range $2,000$\,K $<T_{\mathrm{eff}}<12,000$\,K~(\citealp{2006AJ....132.1221H}; \citealp{2006ApJ...651L.137K}; \citealp{2011ApJ...730..128T}; and \citealp{2011ApJ...737...28B}\footnote{http://www.astro.umontreal.ca/$\sim$bergeron/CoolingModels}). The mass ratio of helium to hydrogen is denoted by $\mathcal{Y} = \log(\frac{\mathcal{M}_{\mathrm{He}}}{\mathcal{M}_{\mathrm{H}}})$. All models were provided in the PS1 filters by Dr. Bergeron upon request. The cooling tracks of different chemical compositions are very similar above $T_{\mathrm{eff}}\sim10,000$\,K~(i.e.\ $\mathrm{M}_{\mathrm{bol}} < 12.0$). See Appendix D for the bolometric correction.

\subsection{Interstellar Reddening}
A three dimensional map of interstellar dust reddening was produced using 800 million stars with PS1 photometry of which 200 million also have 2MASS photometry~\citep{2015ApJ...810...25G}. Although there is a health warning that the reddening is {\it best determined by using the representative samples, rather than the best-fit relation}, with $\sim$20,000 known WDs over the whole sky, the only way to deredden our samples is to use the given best-fit solution. In order to convert the reddening values of $E(B-V)$ to extinction in the PS1 photometric systems, the values on Table 6 of \citet{2011ApJ...737..103S} were used. I have adopted the values from the column $R_{v} = 3.1$ for this work~(see Table \ref{table:reddening}).

\begin{table}
\begin{tabular}{cccccc}
filter~(x) & g & r & i & z & y\\
\hline
\hline
$A_{x}/E(B-V)_{SFD}$       & 3.172 & 2.271 & 1.682 & 1.322 & 1.087\\
\hline
\end{tabular}
\caption[$A_{x}/E(B-V)_{SFD}$ in Pan--STARRS1 filters]{$A_{x}/E(B-V)_{\mathrm{SFD}}$ in different filters, x, evaluated according to a \citet{1999PASP..111...63F} reddening law with $R_{v} = 3.1$ using a $7,000$\,K source spectrum}
\label{table:reddening}
\end{table}
The reddening information along the line of sight was given between distance modulus 4.0 and 19.0 in 0.5 intervals. Each line of sight was interpolated with a cubic spline between the given points in order to compute the reddening at any given distance.

\subsection{Photometric Parallax}
The surface gravities of WDs are narrowly distributed at about
\begin{equation}
\label{eq:logg}
\left<\log\,g\right>=7.937\pm0.012
\end{equation}
with SDSS DR10~\citep{2015MNRAS.446.4078K}. Thus, by assuming a constant surface gravity at $\log(g)=8.0$, distance and temperature of an object can be determined simultaneously. However, in doing so, extra scatter is introduced to the solution statistics. The goodness-of-fit $\chi^{2}_{\nu}$ would not be at $\sim1$. Therefore, a simple Monte Carlo method was used to produce a catalogue of WDs following the distribution of equation \ref{eq:logg}. The standard deviations in magnitudes in each of the filters were found as a function of temperature for each of the DA and DB models; for the mixed hydrogen/helium atmosphere, they were found as functions of temperature and composition. The size of these standard deviations were comparable to the size of the uncertainties in photometry, so adding them to the photometric uncertainties in quadrature could reduce the $\chi^{2}_{\nu}$ by roughly $50\%$. With these relations, it was possible to propagate the uncertainties arisen from adopting constant surface gravity into my final parallactic solutions.

\begin{figure}
\includegraphics[width=0.9\textwidth]{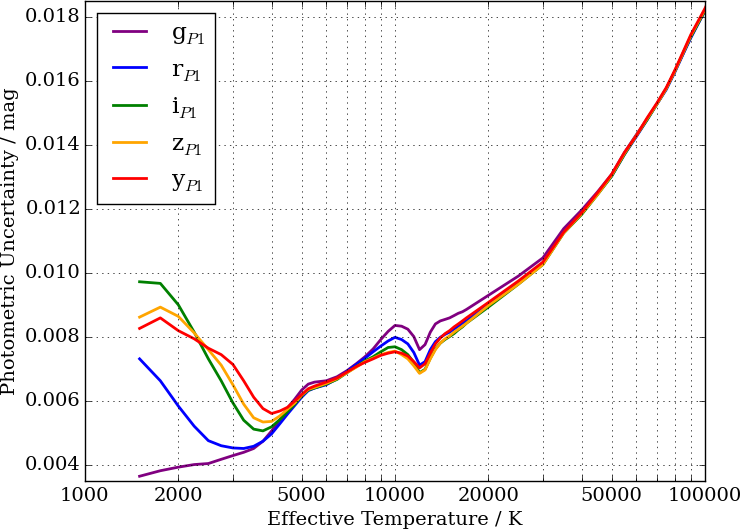}
\caption[The standard deviations in magnitude in each filter when a population is assumed to have fixed surface gravity of $\log(g)=8.0$ for DA WDs]{The standard deviations in magnitude in each filter when a population following the distribution described by equation \ref{eq:logg} is assumed to have fixed surface gravity of $\log(g)=8.0$ for DA WDs}
\label{fig:fixed_logg}
\end{figure}

\begin{figure}
\includegraphics[width=0.8\textwidth]{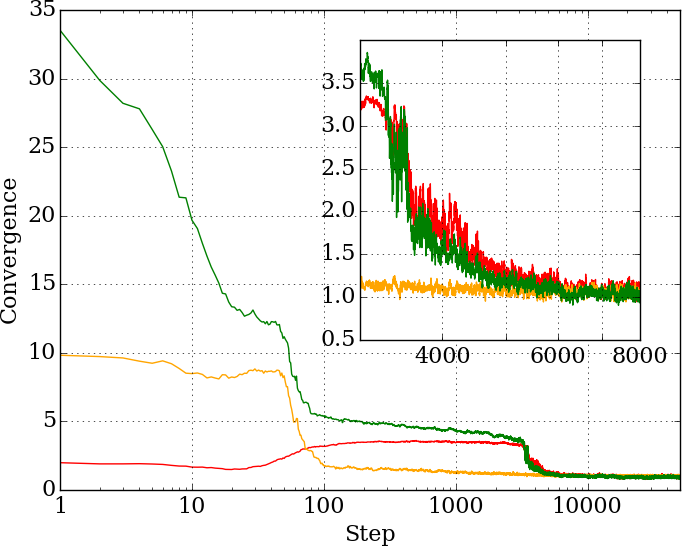}
\caption[The inter-quartile of the chains is plotted against the step number]{The interquartile ranges of the chains is plotted against the step number and rescaled by the overall interquartile range of the values from the last $20\%$ of the chains. The interquartile range ratios converge to better than $10\%$ after the first $6,000$ steps, so this defines the burn-in distance. The chains for temperature, distance and hydrogen/helium ratio are in red, yellow and green respectively}
\label{fig:convergence}
\end{figure}

\begin{figure}
\includegraphics[width=1.0\textwidth]{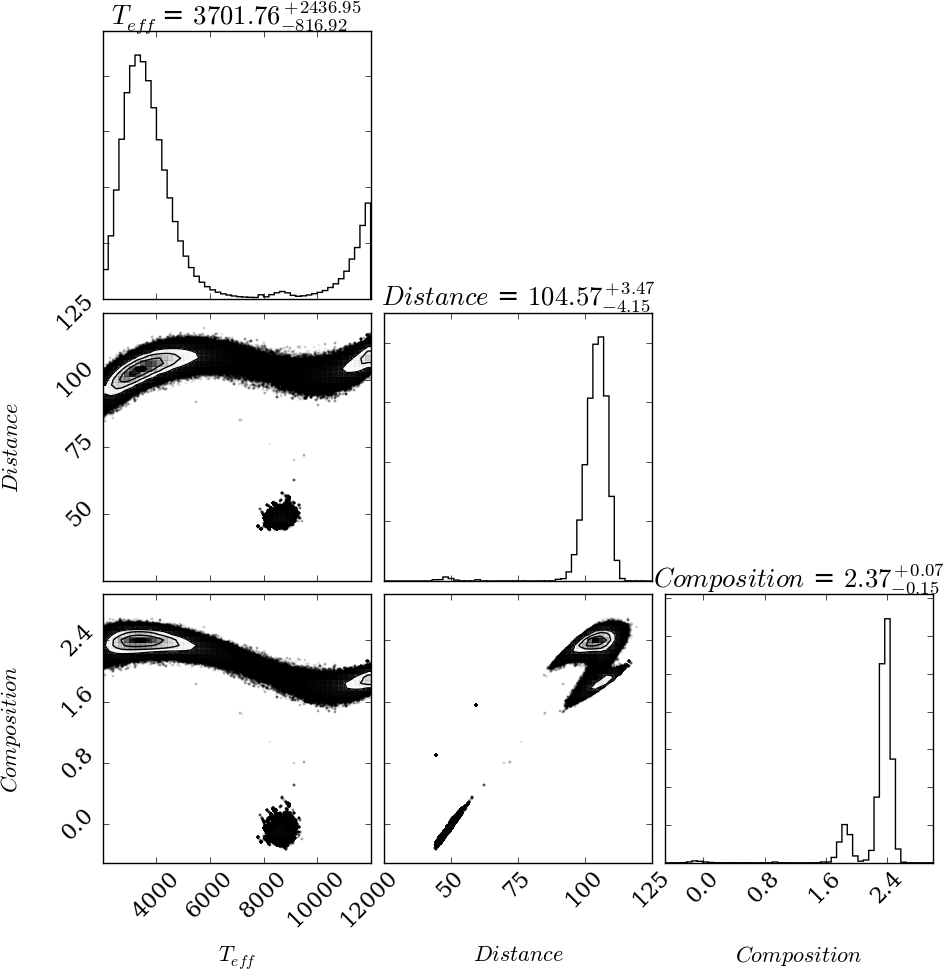}
\caption[Corner plot for the parallactic solution fitted with mixed atmosphere models with $1,000$ chains of size $50,000$]{The corner plot for the parallactic solution fitted with mixed atmosphere models with a burn-in distance of $6,000$ steps for $1,000$ chains of size $50,000$, solving for effective temperature, distance and hydrogen-helium mass ratio simultaneously}
\label{fig:parallax_long}
\end{figure}

\begin{figure}
\includegraphics[width=1.0\textwidth]{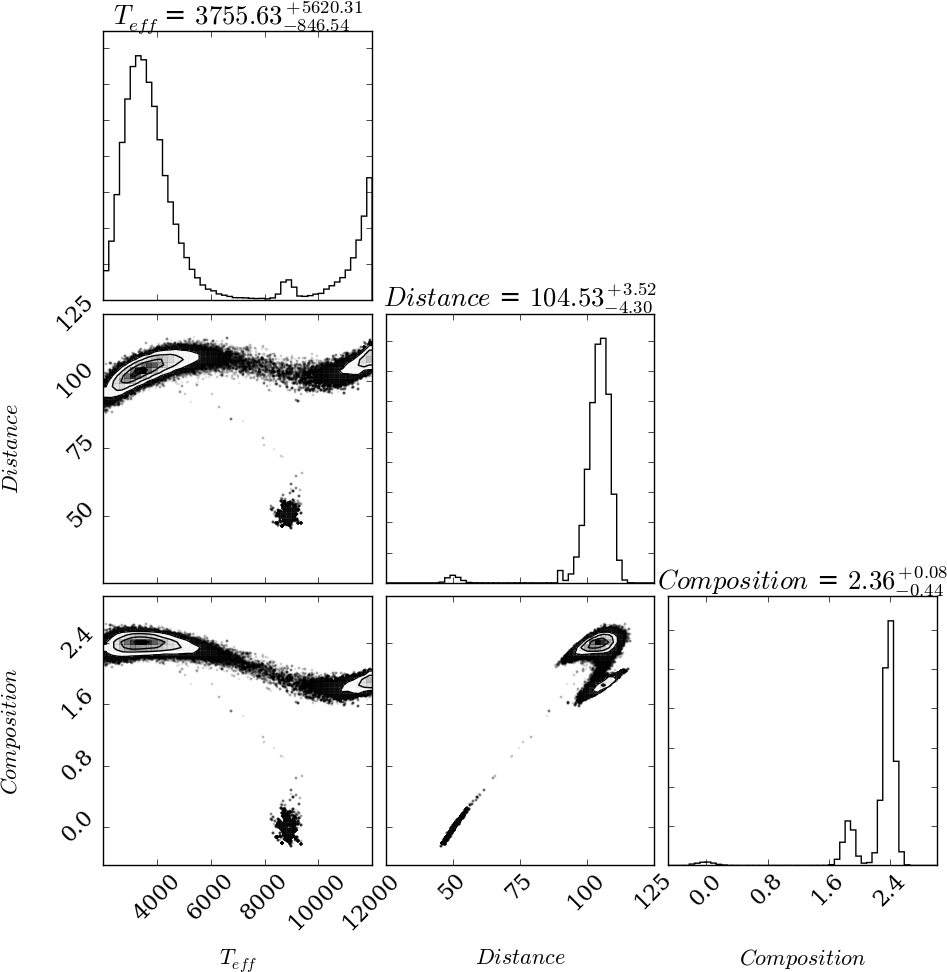}
\caption[Corner plot for the parallactic solution fitted with mixed atmosphere models with 100 chains of size 10,000]{The corner plot for the parallactic solution fitted with mixed atmosphere models with a burn-in distance of 6000 steps for 100 chains of size 10,000, solving for effective temperature, distance and hydrogen-helium mass ratio simultaneously}
\label{fig:parallax_short}
\end{figure}

The best-fit solutions in the 10 atmosphere types were found with the least-square method in the distance-temperature space sampled by a Markov chain Monte Carlo~(MCMC) algorithm \textsc{emcee}\footnote{http://dan.iel.fm/emcee}~\citep{2013PASP..125..306F}. These models are interpolated with a bicubic spline as functions of temperature and hydrogen-helium mass fraction~(distance alters mostly the overall normalisation, although the interstellar extinction changes the shape of the observed spectral energy distribution~(SED) slightly, the effects are typically much smaller than the photometric uncertainties). There are some degeneracies in the solution, and the most notable one is that between a cool WD at small distance and a hot WD at large distance because interstellar reddening alters the shape of the model SED. A simple minimisation technique sometimes cannot always arrive at the right solution. This was tested with a synthetic catalogue of WDs that a cool WD is more likely to be fitted as a hot WD with Nelder--Mead method\,(also known as downhill simplex method) when photometric noise is added. However, objects in the synthetic sample can be classified correctly by using \textsc{emcee}. A sufficient range of burn-in is extremely important because of multiple local minima that the final probability distributions of the solutions can span a huge range~(see Fig.~\ref{fig:parallax_long}). In order to find a good burn-in distance, $1,000$ excessively long chain of size $50,000$ are used to investigate the effect. The inter-quartile ranges of the chains is plotted against the step number and scaled by the overall inter-quartile range of the values from the last $20\%$ of the chains, this inter-quartile range ratio defines the convergence. Taking this pathological case where multiple minima exist, I deduce that the chains converge to better than $10\%$ after $6,000$ steps (see Figure~\ref{fig:convergence}). Because of the possible degenerate solutions, I adopt a conservative number and length of chains: 100 and 10,000 respectively~(which gives $100$ $\times$ 4,000 steps after burn-in, see Fig.~\ref{fig:parallax_short}). A second and third local minima are obvious in the temperature plots, the minimum at $T_{\textrm{eff}}\approx8,500$\,K has a very low signal to noise ratio, so it can be ruled out easily; the minimum at the boundary of the model grid ($T_{\textrm{eff}}\approx12,000$\,K) carries, however, a much large signal to noise ratio compared to the best solution. It is possible that there is a better fit at higher temperature, but this is limited to the available model. Instead of extrapolating the grid, in view of the similar colours in all chemical compositions above $8,000$\,K, objects are all fitted with pure hydrogen models~(available in the range of $1,500-120,000$\,K) and the reduced $\chi^{2}$ of the two fits are compared to identify the best fit solution. Without UV filter in the PS1, it is difficult to constrain the high temperature solution especially when a WD turns blue when it falls below $4,000$\,K. The strength of PS1 is the availability of y$_{\mathrm{P}1}$ filter which gives good constrain for the cool white dwarfs where the near infrared is suppressed due to collisionally induced absorption of hydrogen and helium molecules, which causes rapid change in the optical-infrared colour in this temperature range.

When interstellar reddening is included in the calculation, the likelihood function to be maximised is
\begin{equation}
\sum_{i} \left\{ \frac{ \left[ m_{i} - \mu_{D} - m_{\mathrm{model},i}(\mathrm{T}_{\mathrm{eff}}) - A_{i}(D) \right]^{2} }{ \sigma_{i}^{2} + \sigma_{log(g),i}^{2} } + \log\left[2\pi(\sigma_{i}^{2} + \sigma_{log(g),i}^{2})\right] \right\}
\end{equation}
where $m_{i}$ is the magnitude filter $i$, $\mu_{D} = 5 \log(D) - 5$ is the distance modulus with subscript $D$ to distinguish it from the symbol for proper motion, $m_{\mathrm{model},i}(T_{\mathrm{eff}})$ is the magnitude of a given model which depends only on the effective temperature and $A_{i}(D)$ is the total extinction at distance $D$. In the case of the mixed atmosphere models, the model magnitude becomes $m_{\mathrm{model},i}(T_{\mathrm{eff},\mathcal{Y}})$.

\subsection{Calibration of Photometric Parallax}
\begin{figure}
\includegraphics[width=0.9\textwidth]{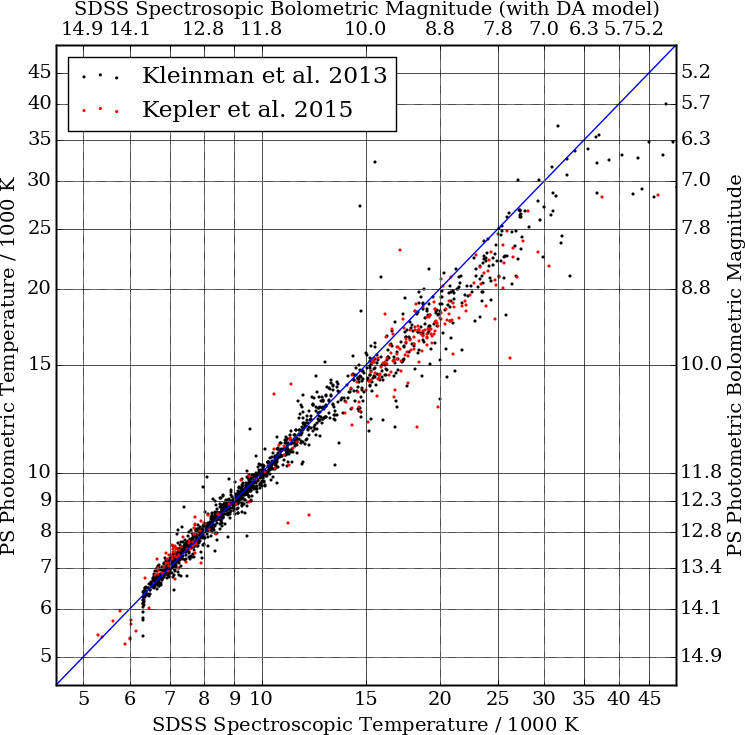}
\caption[WD candidates are crossmatched with confirmed DAs to compare the photometric temperatures against the spectroscopic temperatures]{WD candidates are crossmatched with confirmed DAs to compare the photometric temperatures against the spectroscopic temperatures. They agree well with each other}
\label{fig:spec_vs_phot}
\end{figure}

The lack of UV filters in PS1 means that when the SED of the hot WDs peak bluer than the coverage of the g$_{\mathrm{P}1}$, only the tail of the SED is fitted which may lead to a systematic bias. Therefore, the PS1 WD candidates were crossmatched with the confirmed DAs from \citet{2013ApJS..204....5K} and \citet{2015MNRAS.446.4078K} where spectroscopic temperatures were available. Above $T_{\mathrm{eff}}=15,000$\,K, the solutions were only slightly underestimated~(See Figure \ref{fig:spec_vs_phot}) so recalibration was not needed. At low temperatures, the solutions agreed very well.

\section{Survey Volume Maximisation with HEALPix}
\begin{figure}
\includegraphics[width=0.75\textwidth]{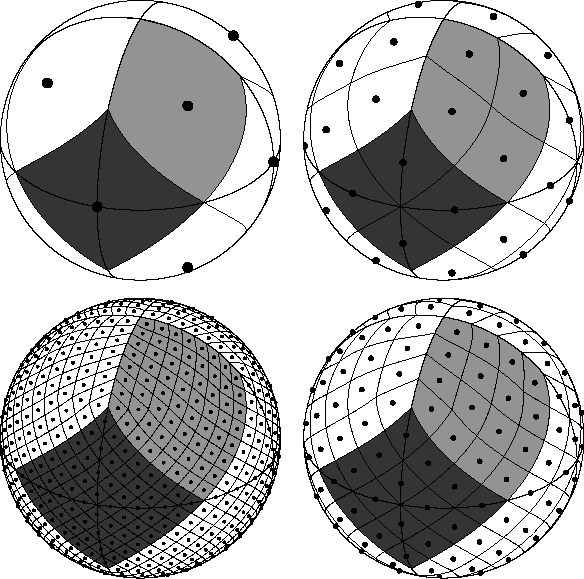}
\caption[The pixelisation of a sphere with HEALPix]{In the clockwise direction from the top left, the number of sides are $N_{\mathrm{side}} = 1, 2, 4, 8$ correspond to total number of pixels $N_{\mathrm{pix}} = 12 N_{\mathrm{side}}^{2} = 12, 48, 192, 768$. Figure is taken from \citet{2005ApJ...622..759G}}
\label{fig:healpix_sphere}
\end{figure}

HEALPix is an acronym for \underline{H}ierarchical \underline{E}qual \underline{A}rea iso\underline{L}atitude \underline{Pix}elization of a sphere~\citep{2005ApJ...622..759G}. This pixelisation routine produces a subdivision of a spherical surface in which all pixels cover the same surface area. All pixel centers are placed on rings of constant latitude, and are equidistant in azimuth~(on each ring). However, the pixels are not regular in shape. A HEALPix map has $N_{\mathrm{pix}} = 12 N_{\mathrm{side}}^{2}$ pixels each with the same area $\Omega = \frac{\pi}{3N_{\mathrm{side}}^{2}}$, where $N_{\mathrm{side}}$ is the square root of the number of division of the base pixel~(See Fig. \ref{fig:healpix_sphere}) and it can be any value with a base of 2~(ie. $2^{x}$ for any non-negative integer $x$) as long as the computer has sufficient memory to do so. This pixelisation routine is used in finding the faint magnitude limit, lower proper motion limit and completeness.

\subsection{Survey Depth}
\label{sec:completeness}
\begin{figure}
\includegraphics[width=0.9\textwidth]{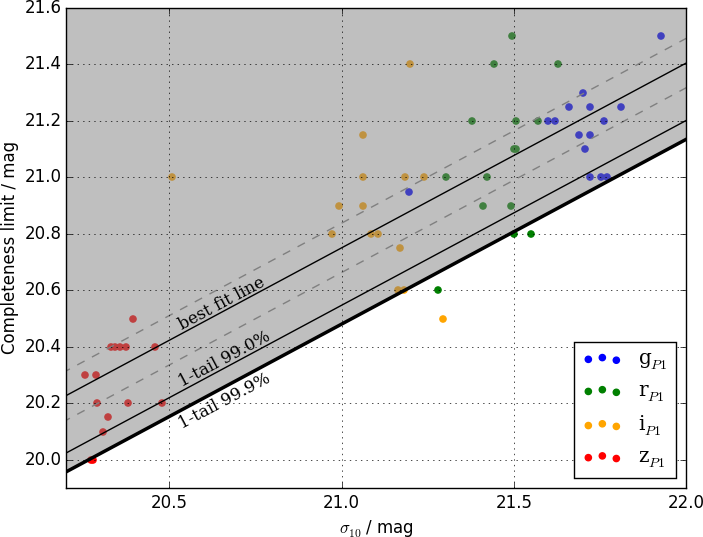}
\caption[The completeness limits against $10\sigma$ point source detection limits]{The completeness limits are plotted against the $10\sigma$ point source detection limits. The solid line is the best fit linear relation, dashed line is the vertical offset of the best fit solution that has covered $68.2\%$ of all data~(i.e. $1\sigma$ in each direction)}
\label{fig:completeness_relation}
\end{figure}

PS1 has a very complex variation in the data quality as a function of position. If the small/medium scale variations in the survey depth are not considered, the survey volume would be limited to the shallowest parts of the sky, which would be more than a magnitude brighter than the deepest parts. In order to take into account of these small scale effects, a linear relationship between the completeness magnitude~(Chapter \ref{sec:faint_magnitude_limits}) and the detection depth map~\citep{2014MNRAS.437..748F}, $D(\alpha,\delta)$, was found empirically; see Fig.~\ref{fig:completeness_relation}. Since the given depth maps are the $10\sigma$ detection limit in a fiducial $3''$ aperture, I had to convert it to FWHM magnitude by accounting for the flux included in the PSF, so the limiting magnitude was corrected with a linear transformation $D' = D - 2.5\times\log(2)$. The characteristics in the y$_{\mathrm{P}1}$ band were assumed to have similar behaviours as other filters. Since the magnitude at which the number counts departed from the model was judged by eye, and the $3.4$ square degrees field of view is not small compared to the size of the inhomogeneity of the survey quality, there is some scatter in the completeness-depth relation. In order to account for these variations, instead of choosing the best fit straight line, which has half the data points above the line and the other half below it, a straight line that would have covered 99.9\% of all data was used. The scatter of these points was measured from the median absolute deviation~(MAD) to minimize the effect from outliers, where $\sigma_{\mathrm{measured}} = 1.48\times \mathrm{MAD}$. Human reading error, $\sigma_{\mathrm{human}}$, was assumed to be $0.1\magnitude$, while the variation of the depth, $\sigma_{\mathrm{depth}}$, was accounted for by sampling the field at 400 equally spaced points in the given pixel to find the standard deviation in depth. Therefore, the corrected completeness limit is
\begin{equation}
C' = C - 3\sigma_{c},
\end{equation}
with
\begin{equation}
\sigma_{c} = \sqrt{ \sigma_{\mathrm{measured}}^{2} - \sigma_{\mathrm{human}}^{2} - \sigma_{\mathrm{depth}}^{2} }
\end{equation}

By applying this relation to the photometric depth maps, the completeness maps in the five PS1 filters were produced. However, the resolution in which the maps were provided was too high, so the completeness map used a degraded resolution of $N_{\mathrm{side}}=16$ to match the resolution of the tangential velocity completeness correction and to keep the computing time manageable.

\newpage
\begin{figure*}[ht!]
\includegraphics[width=0.85\textwidth]{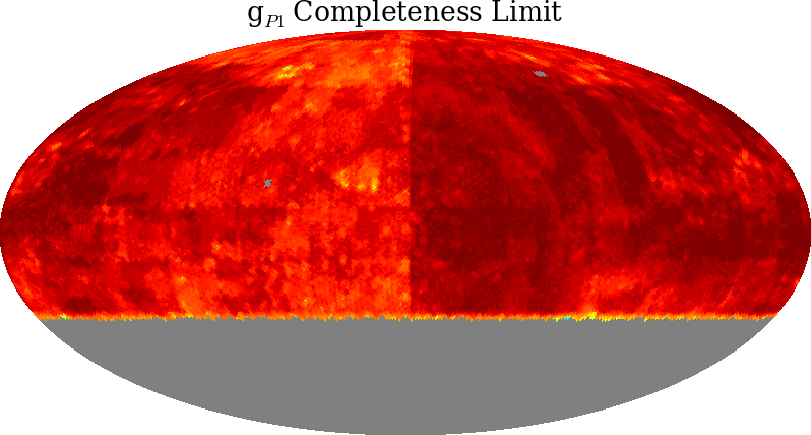}
\includegraphics[width=0.85\textwidth]{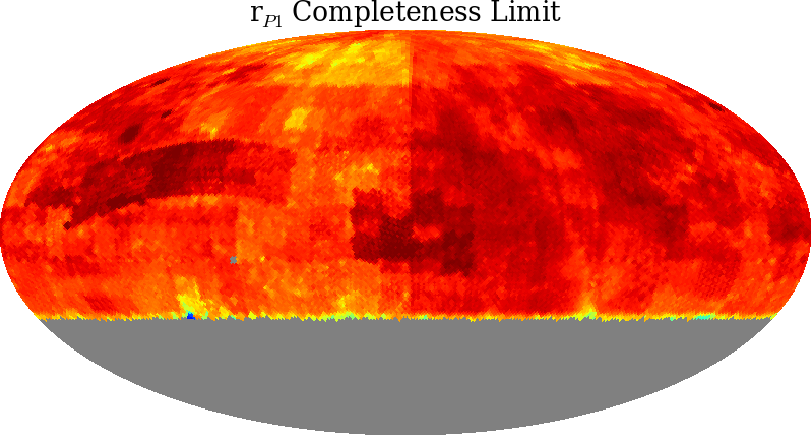}
\includegraphics[width=0.85\textwidth]{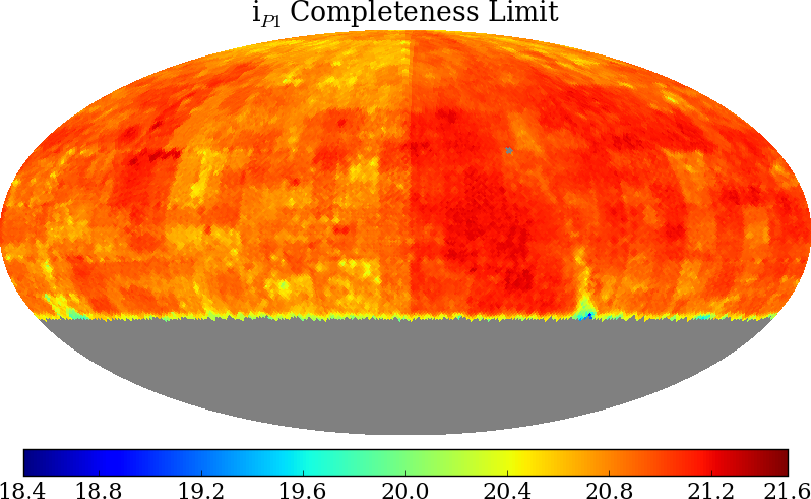}\\
\begin{flushright}
\text{(cont.)}
\end{flushright}
\end{figure*}

\begin{figure}[ht!]
\includegraphics[width=0.85\textwidth]{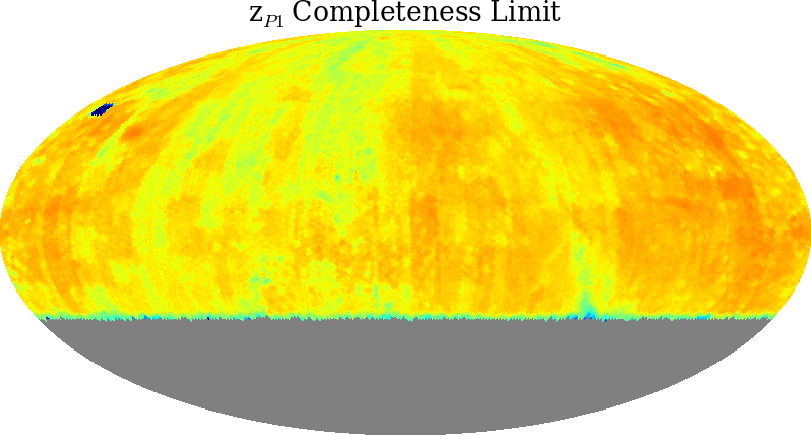}
\includegraphics[width=0.85\textwidth]{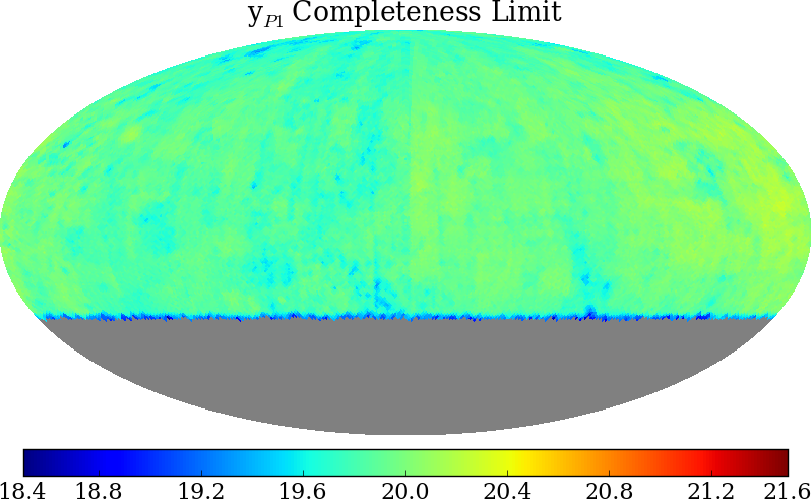}\\
\caption[Completeness map in g$_{\mathrm{P1}}$, r$_{\mathrm{P1}}$, i$_{\mathrm{P1}}$, z$_{\mathrm{P1}}$ and y$_{\mathrm{P1}}$ filters]{Completeness map in g$_{\mathrm{P1}}$, r$_{\mathrm{P1}}$, i$_{\mathrm{P1}}$, z$_{\mathrm{P1}}$ and y$_{\mathrm{P1}}$ filters respectively, all magnitudes are with the same colour scale}
\label{fig:depth_map}
\end{figure}

\subsection{Lower Proper Motion Limits}
The treatment described in Chapter \ref{sec:lower_pm_limit} was performed for each small area with $N_{\mathrm{side}}=16$ such that each pixel had an angular size of $\sim53.7$ square degrees. Each area was assigned with the default (non-nested) pixel ID number, and the interpolated $\sigma_{50}$ and $\sigma_{95}$ were saved for faster volume integration~(Eq.~\ref{eq:volume_new}). When an area had less than $5,000$ objects, all objects from the adjacent 8 pixels were included to derive the $\sigma_{50}$ and $\sigma_{95}$ for the central pixel. This only happened in a small number of high Galactic latitude pixels. If a resolution of $N_{\mathrm{side}}=32$ is used, this would result in a large number of pixels, hence a lower resolution was chosen.

\subsection{Tangential Velocity Completeness Correction}
In order to clean up the sample of proper motion objects, a lower tangential velocity was applied to remove spurious sources. For example, $20, 30$ or $40\kmps$ are typical choices to obtain clean sample of the thin disc population, the precise choice of the value depends on the data quality, see Chapter 5; and $160, 200$ or $240\kmps$ are used to obtain stellar halo objects~(H06, RH11). However, this process removes genuine objects from the sample. With some knowledge of the kinematics of the solar neighbourhood, it is possible to model the fractions of objects that are removed in any line of sight. A resolution of $N_{\mathrm{side}}=16$ is used to pixelise the sky to account for the variation in the projected kinematics across the sky.

%% file: 3_theoretical_wdlf.tex
\chapter{Theoretical White Dwarf Luminosity Functions}

The WDLF is a standard diagnostic tool for investigating the SFH of an aged stellar population. It is an evolving function of time, with a precise morphology determined by a number of parameters. \citet{1987ApJ...315L..77W} compared an observed WDLF derived from the Luyten Half-Second~(LHS) catalogue with a theoretical WDLF to obtain an estimate of the age of the Galaxy for the first time with this technique. However, a constant rate of WD formation and an average progenitor lifetime were used to give a first order estimate. The importance of careful treatment of the initial mass function~(IMF) and the star formation rate~(SFR) was addressed by \citet{1989ApJ...341..312I}, who developed a model to include the IMF, SFR and the progenitor--WD mass relation~(the initial-final mass relation, IFMR). \citet{1990ApJ...352..605N} examined WDLFs with various SFH scenarios. They showed that the WDLF is a sensitive probe of the history of the SFR as it can provide signatures of any irregularities in the SFH such as bursts and lulls; on the other hand, the WDLF is insensitive to the adopted form of the IMF, with a different choice of IMF affecting the amplitude but not the position or form. F01 discovered the importance of using dynamic envelopes to decouple crystallisation and convective coupling\footnote{The effect of convective coupling in slowing down the cooling decreases with the temperature gradient between the core and the envelope which decreases with time. However, once the excess of energy is liberated, convection speeds up the cooling process as it is more effective than a radiative envelope to transfer energy from the core to the surface.}~(adopted in the model developed by the Montr\'{e}al group). LPCODE, another major WD cooling model, also adopted a dynamic envelope. However, BaSTI models used a static atmosphere~(see more in Chapter~\ref{sec:wd_cooling_models}). Through combining the SFR~($\psi$), IMF~($\phi$), IFMR~($\zeta$), the stellar evolution time and WD cooling time, the WDLF~(i.e.\ number density as a function of magnitude) can be written in the form
\begin{equation}
\label{eq:WDLFnumberdensity}
n(\mathrm{M}_{\mathrm{bol}}) = \int^{\mathcal{M}_{u}}_{\mathcal{M}_{l}}
\tau(\mathrm{M}_{\mathrm{bol}},m)
\psi(\mathrm{T}_{0},\mathrm{M}_{\mathrm{bol}},\mathcal{M},m,Z)
\phi(\mathcal{M}) \mathrm{d}\mathcal{M}
\end{equation}
with
\begin{align}
\tau(\mathrm{M}_{\mathrm{bol}},m) &= \frac{\mathrm{d}t_{\mathrm{cool}}}{\mathrm{d}\mathrm{M}_{\mathrm{bol}}}(\mathrm{M}_{\mathrm{bol}},m)\\
\psi(\mathrm{T}_{0},\mathrm{M}_{\mathrm{bol}},\mathcal{M},m,Z) &= \psi[T_{0}-t_{\mathrm{cool}}(\mathrm{M}_{\mathrm{bol}},m)-t_{\mathrm{MS}}(\mathcal{M},Z)]\\
m &= \zeta(\mathcal{M})
\end{align}
where $\frac{\mathrm{d}t_{\mathrm{cool}}}{\mathrm{d}\mathrm{M}_{\mathrm{bol}}}$ is the inverse cooling rate in unit of magnitude, $\mathrm{M}_{\mathrm{bol}}$ is the absolute bolometric magnitude, $\mathrm{T}_{0}$ is the total look back time~(i.e.\ the time since the star was formed), $t_{\mathrm{cool}}$ is the time taken for a WD with mass $m$ to cool down to $\mathrm{M}_{\mathrm{bol}}$, and $t_{\mathrm{MS}}$ is the total MS time of the progenitor which is a function of both the zero-age MS (ZAMS) mass and metallicity. The integral is integrated over the progenitor mass from the lowest mass star that has had time to reach the WD stage at a given bolometric magnitude~(see Fig.~\ref{fig:ms_lifetime}), to the highest mass star that can produce a WD. The lower limit $\mathcal{M}_{l}$ can be found by solving
\begin{equation}
\mathrm{T}_{0}-t_{\mathrm{cool}}(\mathrm{M}_{\mathrm{bol}},m)-t_{\mathrm{MS}}(\mathcal{M}_{l},Z) = 0,
\end{equation}
while the upper mass limit is model dependent, typically at $7-8\msolar$. Because of the non-analytical nature of $\tau$, $t_{\mathrm{cool}}$ and $t_{\mathrm{MS}}$, they were interpolated from grids of precalculated tables.

\section{Main Sequence Stars}
\subsection{Lifetime and Mass}
The stellar evolution model PARSEC~(\citealp{2012MNRAS.427..127B}, \citeyear{2013EPJWC..4303001B}) was used to provide the total progenitor lifetimes of stars as a function of ZAMS mass at fixed solar metallicity. The mass range $0.75\msolar<\mathcal{M}<10.0\msolar$ was used for interpolation\footnote{The model covers $0.09-350\msolar$. Below $0.75\msolar$ the MS lifetime is longer than $30\gyr$ and the evolution calculations are stopped.}. The MS lifetime of stars with $\mathcal{M}<0.96\msolar$ is longer than the Hubble timescale, so they do not contribute to the WDLF; stars more massive than $\sim7.0\msolar$ may yield O/Ne WDs, which have a different cooling profile~\citep{2007A&A...465..249A}. The inclusion of the low and high mass stars are for stability of interpolation at the boundaries. The total progenitor lifetime is the time it takes to evolve from the Pre-MS~(PMS) to the first thermal pulse or the onset of carbon burning, depending on their ZAMS mass. The maximum mass of which a star becomes a WD at the end of the stellar evolution is not well understood, but is generally accepted to be in the range of $7-8\msolar$. In this work the upper limit is set by the WD model limit, $7.3\msolar$, where the IFMR would return a WD that is more massive than that available from the model grid of C/O WD. The inclusion of up to $8\msolar$ would increase the total stellar mass by $0.5\%$, which is assumed to have negligible effect on the WDLF. The lower mass limit is set at $0.516\msolar$ where the IFMR would return a WD with the same mass. 

\subsection{Metallicity}
\begin{figure}
\includegraphics[width=0.9\textwidth]{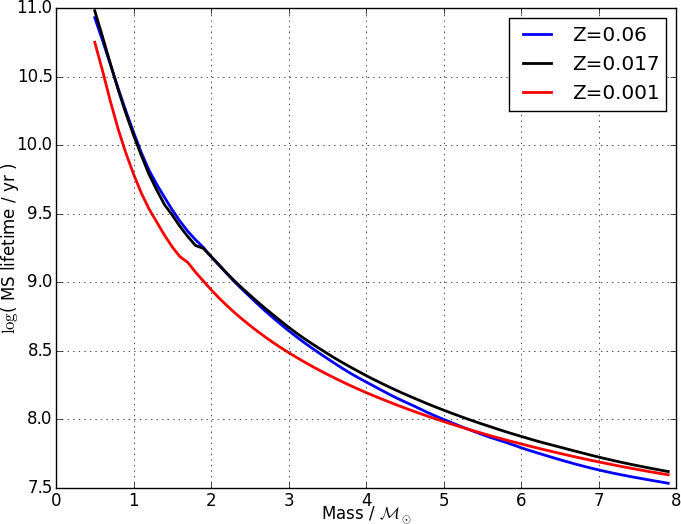}
\caption[The MS lifetimes from the PARSEC stellar evolution models]{The MS lifetimes of the PARSEC stellar evolution models with metallicities $Z=0.06$~(blue), $0.017$~(black) and $0.001$~(red). The most metal poor stars have the shortest MS lifetime below $5\msolar$, while extremely metal rich stars have similar MS lifetimes as solar type stars up to $3\msolar$. Metal rich stars burn faster than solar type ones above $3\msolar$ and metal poor ones above $5.5\msolar$}
\label{fig:ms_lifetime}
\end{figure}

The choice of metallicity is based on the available models. The disc metallicity is assumed to be solar, with the initial composition $Y_{\odot,\mathrm{init}}=0.279$ and $Z_{\odot,\mathrm{init}}=0.017$ which would evolve to the current day~($t=4.593\gyr$) values $Y_{\odot}=0.249$ and $Z_{\odot}=0.015$. These values differ from the measured values by $1.4\%$ and $9.8\%$ respectively~(\citealp{1997MNRAS.287..189B} and \citealp{2011SoPh..268..255C}). For the purpose of a statistical study, the differences have negligible effects. For the stellar halo, $Y=0.25$ and $Z=0.001$ are used. The main sequence lifetimes between the solar metallicity and metal poor stars differ by as much as $50\%$ at $1\msolar$, so the differences in the WDLFs are not negligible~(Fig. \ref{fig:ms_lifetime}). Old WDs are remnants of more massive progenitors and so the total age is dominated by the WD cooling time. Hence, the faint end of the WDLF is insensitive to metallicity. At the brighter parts, the MS lifetime has a non-negligible contribution to the total age so the difference in MS lifetime affects the shape of the WDLF. In the case of a population with a constant SFR, a shorter MS lifetime means that more stars would have become WDs at a given lookback time.

\subsection{Initial Mass Function $\phi$}
The IMF is well determined over the range of WD progenitor masses. The IMF without correcting for unresolved binaries of \citet{2001MNRAS.322..231K} is adopted, which is motivated by the fact that the stellar evolutionary model adopted used the same relations. It comprises a three-part power-law IMF of which the mass range of interest is
\begin{equation*}
\xi(m) \propto m^{-\alpha}\quad\text{where}\quad\alpha = 
\begin{cases}
 +2.3\pm0.3, \quad0.50\leq m/M_{\odot}\leq1.00 \\
 +2.3\pm0.7, \quad1.00\leq m/M_{\odot}.
\end{cases}
\end{equation*}
The only difference in the two relations is the size of the uncertainties.

\subsection{Star Formation Rate $\psi$}
The SFH is one of the most important quantities that can be extracted from a WDLF. However, in the forward modelling of a theoretical WDLF, a SFH must be supplied. The discs of the Milky Way are assumed to have a constant or exponentially decaying SFR, while the stellar halo has a single burst at constant amplitude for 1\,Gyr in the beginning.

\section{White Dwarfs}
As a general rule of thumb, more massive WDs have lower cooling rates because of their low luminosity, as a consequence of their small radii, compared to the less massive ones. However, the detailed treatments of the structural evolution of WDs have significant effects on the cooling rates as functions of composition, temperature and masses. Isolated stars predominantly end up as C/O WDs because stars less massive than $\sim 0.96 \msolar$ have lifetimes longer than the age of the Universe while stars only need to be more massive than $\sim 0.5\msolar$ to burn helium through either helium flash or stable helium burning. At the high mass end, the dividing properties for C/O and O/Ne WD production are still unclear. However, \citet{1997MNRAS.289..973G} found that for massive O/Ne WDs at $1.0$, $1.1$ and $1.2\msolar$, cooling rates are much larger than those of C/O cores due to the lack of latent heat of crystallisation. As neon has a higher nucleon to proton ratio, even at the late thermal pulsating stage~(i.e. the core reaches its highest temperature of its lifetime), the O/Ne mixture is still in solid state. In addition, the packing of the O/Ne crystals allows more efficient transfer of heat from the core to the surface.

Within a Hubble time, He WDs could only be formed in a binary system when the more massive member reached the late stages of stellar evolution and inflated a few hundred times and the envelope overflows the Roche Lobe leading to a mass transfer to the MS companion. When the envelope is exhausted, the system becomes a MS + He WD binary pair. To date, hydrodynamical simulations are the only way to study the extreme environment of the internal structure and chemistry of WDs; laboratories can only reproduce the atmospheric conditions of the first few meters under the photosphere. In order to construct a theoretical WDLF we require a good knowledge of the sources of energy and the cooling mechanisms of WDs.

\subsection{Heat Sources and Cooling of White Dwarfs}
\label{sec:wd_heat_source}
\begin{figure}
\includegraphics[width=0.9\textwidth]{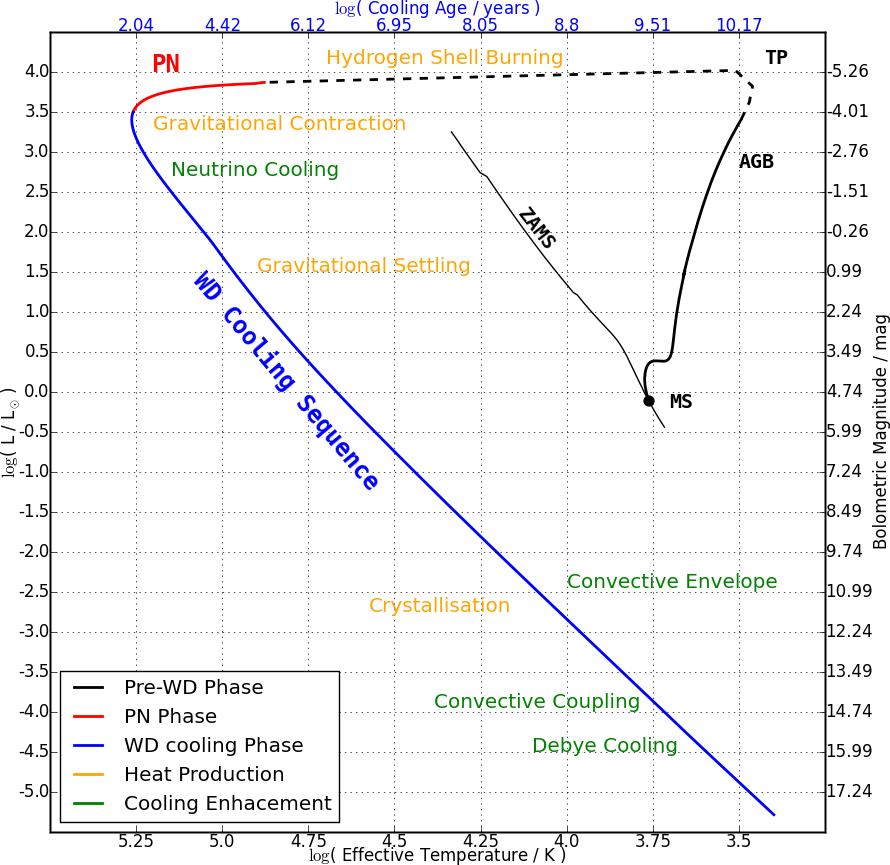}
\caption[An annotated stellar and WD evolution sequence]{An annotated stellar and WD evolution sequence. The progenitor evolution track is taken from PRASEC while the WD cooling track is of LPCODE. The evolution tracks of pre-WD, PN and WD phases are in black, red and blue respectively. The upper axis only applies to the WD cooling sequence, the other axes are universal. Heat production and cooling mechanisms are in yellow and green respectively, their positions correspond roughly to the onset of the mechanisms at those luminosities}
\label{fig:wd_evolution}
\end{figure}

Most of the internal energy of a WD is the remnant heat from the progenitor once it passed the planetary nebula~(PN) phase. However, there are various physical processes that can provide an appreciable amount of energy and slow down the cooling, while some do the opposite. The following outlines these processes of an isolated C/O WD in chronological order of their onsets~(see also Fig. \ref{fig:wd_evolution}), most of which is rearranged and simplified from~\citet{2010A&ARv..18..471A}.

\subsubsection*{Nuclear Burning}
Upon entering the PN phase, stable hydrogen shell burning via the CNO cycle and helium shell burning are the main source of luminosity. As a result of CNO burning, the hydrogen-rich envelope is consumed. The burning rate drops as the density and temperature at the base of the hydrogen-rich envelope become too low when its mass reaches $\mathcal{M}_{\mathrm{H}} \sim 10^{−4}\msolar$. They dominate the energy output in the early stage until $\sim10^{8}$ years when proton-proton chain hydrogen burning overtakes the luminosity. Recent calculation shows that for metal poor WD progenitors~($Z = 0.001, \mathcal{M}=0.593\msolar$), this hydrogen burning~(H-burning) can contribute $\sim30\%$ of the total luminosity at a cooling age of $\sim10^{8}-10^{9}$\,years~\citep{2010ApJ...717..183R}.

\subsubsection*{Gravitational energy}
The radius at the beginning of the WD phase can be up to twice the zero-temperature degenerate radius. Although being a slow process that lasts over $10^{10}$\,years, most of the contraction happens at early times. Thus, the contribution of compression to the energy output of the star is important in very hot WDs. In addition, changes in the internal density due to the increase in the core mass from H burning via the CNO cycle lead to an important release of gravitational potential energy from the core of very young WDs. Gravitational contraction would become important again in the very late stage~(beyond a Hubble time) when Debye cooling has already depleted the thermal energy content of the core. The contraction of the thin sub-atmospheric layers may provide up to $30\%$ of the stellar luminosity. However, gravitational energy barely contributes to the photo-luminosity for most of the WD evolution. This is because the energy released in the core is almost completely used by the degenerate electrons to increase their Fermi energy.

\subsubsection*{Neutrino losses}
Energy is not only radiated away from WDs in the form of photons, neutrino emission also plays an important part in the cooling at early stages. Neutrinos are created in the very deep interior of WDs and provide a main energy sink for hot WDs. Because of their extremely small interaction cross-section, even at a central density of $10^{6}$\,g\,cm$^{-3}$ typical of WDs, the mean free path of neutrinos is $\sim3,000\,\mathrm{R}_{\odot}$. Once created, neutrinos leave the WD without interacting with any overlying layers. In young WDs, neutrinos result from pure leptonic processes as a consequence of the electro-weak interaction. Under the conditions prevailing in hot WDs the plasma-neutrino process is usually dominant, but for massive WDs neutrino bremsstrahlung must also be taken into account~(\citealp{1994ApJ...425..222H}; \citealp{1996ApJS..102..411I}).

\subsubsection*{Gravitational settling}
The slow diffusion of $^{22}$Ne towards the centre of the liquid WD core releases enough gravitational potential energy to impact the cooling times of massive WDs~\citep{2002ApJ...580.1077D,2008ApJ...677..473G,2010ApJ...719..612A}. This is due to the fact that $^{22}$Ne, which is the product of helium burning~(He-burning) on $^{14}$N during prior evolution, has two extra neutrons as compared to nitrogen and oxygen. This results in an imbalance between the gravity and the electric fields compared to the matrix, leading to a slow gravitational settling of $^{22}$Ne towards the center. As predicted by \citet{2002ApJ...580.1077D}, the possible impact of $^{22}$Ne sedimentation on WD cooling could be better seen in old, metal-rich clusters, such as NGC~6791, where the $^{22}$Ne abundance expected in the cores of its WDs could be as high as $\sim$4\% by mass~\citep{2010Natur.465..194G}. They showed that the C/O phase separation upon crystallisation and $^{22}$Ne sedimentation in the core of cool WDs are associated with the slow down of the cooling rate that solved the long-standing age discrepancy in NGC~6791~\citep{2008ApJ...678.1279B}.

\subsubsection*{Convection}
\begin{figure}
\includegraphics[width=0.9\textwidth]{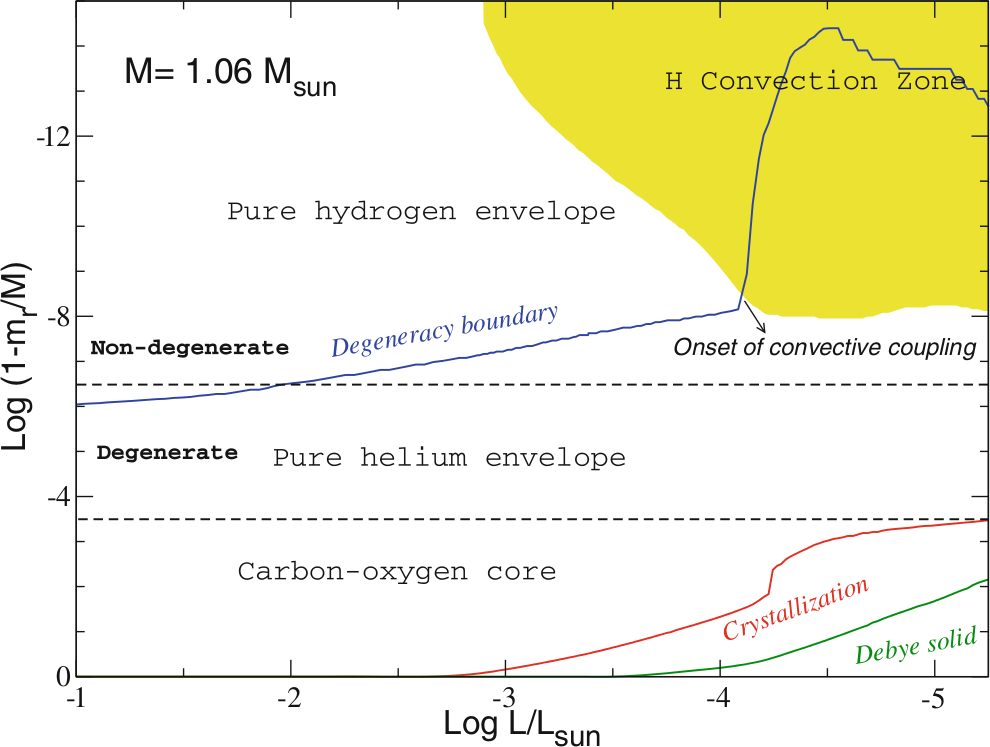}
\caption[The structural evolution of a WD]{The structural evolution of a WD as functions of mass fraction and luminosity. The blue, red and green lines are evolution of the boundary within which the material is degenerate, crystallised and behaving as Debye solid respectively. At the point where the degeneracy boundary touches the convection zone, it marks the onset of convective coupling and the degenerate materials are dredged up to almost the surface of the WD. Since heat can be transported much more efficiently with degenerate material, at the moment of coupling, the cooling rate of WD rapidly drops because of the increased rate of heat transfer from the interior. However, once the temperature gradient between the core and the surface has dropped sufficiently, the coupling enhances the cooling by bringing energy away from the core more effectively. The figure is taken from \citet{2010A&ARv..18..471A}}
\label{fig:wd_interior}
\end{figure}

In the envelope of cool WDs, energy transport is not only radiative but also through convection, which is a consequence of the increase in opacity due to the recombination of hydrogen and helium in the upper atmosphere. As temperature decreases, the convective zone grows deeper into the interior. At first, it reaches the partially ionised helium mantle, but the radiative temperature gradient remains, so the central temperature is independent of the outer boundary conditions. The heat flow depends on the opacity at the edge of the degenerate core. As the luminosity continues to decrease, the convective layer grows inward and eventually reaches the degenerate core and the radiative gradient is lost. The atmospheric conditions are therefore strongly correlated to those in the core, hence the thermal profile of an old WD can be well constrained. The degenerate core is roughly isothermal, while the convective zone is adiabatic which extends from the atmosphere to the outer edge of the core. This is known as the convective coupling, which modifies the relations between the WD luminosity and its core temperature~(Fig.~\ref{fig:wd_interior}), and hence the rate of cooling of cool WDs~(D'Antona \& Mazzitelli 1989; Fontaine et al. 2001). In particular, as the core temperature drops, convection becomes more and more efficient than radiation in energy transport. Although it causes a decrease in the cooling rate initially due to the steep temperature gradient, as the temperature gradient between the core and the envelope drops, the convective coupling facilitates cooling. Depending on the chemical stratification of the outer layer, convective coupling occurs at different times in the lives of WDs~\citep{1989ApJ...347..934D}. Convection may also play a key role in the interpretation of spectral evolution. This is particularly important if the WD is formed with a thin hydrogen envelope, and convection brings helium to the surface~(F01); it may explain the increase of non-DA to DA ratio from $100,000-30,000$\,K.

\subsubsection*{Crystallisation}
The degenerate electrons are already occupying their lowest energy states so they cannot be cooled. The non-degenerate ions, which are decoupled from the electrons, provide the thermal energy that slowly bring energy away from the core. As thermal energy is gradually lost from the star in the form of radiation, the kinetic motions of the ions lose amplitude and become correlated; the ionic state evolves from a gas to a fluid and eventually to a solid. The liquid-solid transition releases latent heat providing an energy input of the order of $1k_{B}T$ per ion, which is a significant amount of energy that slows down the cooling process. As a consequence of crystallisation, the core enters the Debye cooling regime where the specific heats deviate from the constant value given by the Dulong-Petit law. This accelerates the cooling and will be discussed further in the next section. The crystallisation couples with the release of gravitational energy associated with changes in the carbon-oxygen profile~\citep{1997ApJ...486..413S}. The phase diagram of a carbon-oxygen mixture is still poorly understood: the most recent calcuations take a spindle form~\citep{1993A&A...271L..13S} or a slightly positive azeptropic form with the azeotrope at $\mathrm{X}_{\mathrm{O}}=0.2$~\citep{2010PhRvL.104w1101H}. For a slowly cooling WD, both result in high oxygen content crystals forming in the matrix of C/O liquid mantle which slowly sinks to the core because of their larger density. The displacement of carbon due to the gravitational settling of oxygen-rich crystals further releases gravitational energy.

\subsubsection*{Coulomb Interaction and Debye Cooling}
In the dense WD interior, matter is completely pressure-ionised. As recognised by \citet{1961ApJ...134..669S}, Coulomb interactions modify the thermodynamical properties of the ion gas, in particular the specific heat. This, in turn, modifies the cooling times of WDs. The strength of Coulomb interactions, relative to the thermal kinetic energy, is determined by the Coulomb coupling parameter, $\Gamma$. For small $\Gamma$, Coulomb forces are weak relative to thermal motions and the ions behave like an ideal non-interacting gas. But, once $\Gamma$ approaches unity, ions begin to have short range interactions and eventually behave like a liquid. When $\Gamma$ approaches $180$, they form a lattice structure, experiencing a first order phase transition with the corresponding release of latent heat. This results in a new source of energy, which introduces an extra delay in the cooling of WDs. Its contribution to the total luminosity is small, $\sim$ 5\%, but not negligible~\citep{1976A&A....51..383S}. As the WD continues to cool, fewer modes of the lattice are excited, and the heat capacity drops according to the Debye law. This reduces the heat capacity once the temperature drops below the Debye temperature, $\theta_{D} = 4.0 \times 10^{3} \rho^{\frac{1}{2}}$, resulting in a fast cooling phase ($C_{v} \sim T^{3}$). As a result, the thermal content of a WD decreases with temperature. This fast cooling is expected to take place in very cool WDs, which take $10^{9}$ years to reach for a $1.0\msolar$ WD and longer than the Hubble time for one with $0.5\msolar$.

\subsection{WD Cooling models}
\label{sec:wd_cooling_models}
There are three major WD cooling models that are widely used: BaSTI~\citep{2010ApJ...716.1241S}, LPCODE~\citep{2010ApJ...717..183R} and the set provided by the Montr\'{e}al group~(\citealp{2006AJ....132.1221H}; \citealp{2006ApJ...651L.137K}; \citealp{2011ApJ...730..128T}; \citealp{2011ApJ...737...28B}). All of them agree with each other very well for the hot and warm WDs, but the predictions diverge beyond $\mathrm{M}_{\mathrm{bol}}=15\magnitude$ for DAs. The input Physics of the different models are detailed in this section; the missing Physics in the three models are summarised in Table \ref{table:cooling_physics}.

\subsubsection*{BaSTI}
\label{sec:basti}
BaSTI stands for a Bag of Stellar Tracks and Isochrones. It is a project aiming to deliver a homogeneous database of stellar evolution models, isochrones and integrated spectra for single-age, single-metallicity populations encompassing a large chemical composition range appropriate for stellar populations. In the WD cooling models, it has adopted most of the essential Physics and these are available at $\mwd=0.54, 0.55, 0.61, 0.68, 0.77, 0.87$ and $1.00 \msolar$. The treatment of neutrino production used that of \citet{1996ApJS..102..411I}. The initial chemical profile of the C/O core profile is taken from the pre-WD evolutionary computations in \citet{1997ApJ...486..413S} with $66-83\%$ oxygen depending on the progenitor mass. The equation of states~(EOS) for hydrogen and helium use \citep{1995ApJS...99..713S} and \citep{1993A&A...271L..13S} at the highest density, while that for the C/O mixture uses \citep{1988A&AS...76..157S} for gaseous phase and \citep{1994ApJ...434..641S} for liquid and solid phases; \citeauthor{1997ApJ...485..308I}~(\citeyear{1997ApJ...485..308I}; \citeyear{2000ApJ...528..397I}) was used to describe the C/O phase separation. Calculations include and exclude the release of gravitational energy associated with the phase separation and both are provided. The OPAL radiative opacities with zero metallicities are used for $T_{\mathrm{eff}} >10,000$\,K  in the hydrogen and helium envelopes~\citep{1993ApJ...412..752I}; for the temperatures and densities that are not covered by the OPAL tables, Rosseland mean opacities are taken from monochromatic opacities of \citet{1999ApJ...511L.107S}. Conductive opacities in liquid phase use \citet{1983ApJ...273..774I}, and \citet{1983ApJ...273..774I} for solid phase. Superadiabatic convection in the envelope was treated according to the ML2 parameterisation of the mixing-length theory~\citep{1992ApJ...387..288B}. The boundary conditions are taken from new sets of calculations of DA and DB atmospheres, the DA models have thick hydrogen layers, which means a mass fraction of $q_{\mathrm{H}} = 10^{-4} \mwd$ on top of an helium layer of mass $q_{\mathrm{H}} = 10^{-2} \mwd$. He-atmosphere models have an He-envelope with mass fraction $q_{\mathrm{He}} = 10^{-3.5} \mwd$. The surface boundary conditions required for integrating the stellar structure were obtained for $T_{\mathrm{eff}}<10,000$\,K from detailed non-gray model atmospheres. These include the latest physical improvements, CIA in particular, in the calculation of the chemistry, opacity, radiative transfer and EOS of dense hydrogen and helium in WD atmospheres~(\citealp{2004ApJ...607..970K}, \citeyear{2006ApJ...651L.137K}; \citealp{2006ApJ...641..488K}, \citeyear{2006ApJ...651.1120K}; \citealp{2007A&A...474..491K}).

{\bf Unaccounted Physics}: (1) the thickness of the envelope layers in the models is the same for the whole range of WD masses, and the possible dependence on the progenitor metallicity is disregarded. \citet{1986ApJ...301..164I} showed a dependence of $q_{\mathrm{H}}$ and $q_{\mathrm{He}}$ on progenitor mass and initial chemical composition; (2) for the massive WDs, H-burning at the bottom of the H envelope affects the cooling times by a few percent. In order to keep $q_{\mathrm{H}}$ at a strictly constant value for the whole model grid, nuclear burning is inhibited in all calculations; (3) the surface convective regions that develop during the cooling process are not able to cross the H-He interface in the DA or the He-C/O interface in the DB model~(i.e. convective coupling is not accounted for).

\subsubsection*{LPCODE}
\label{sec:lpcode}
LPCODE is a stellar evolutionary code developed at the University of La Plata, Argentina. It contains WD cooling sequences for progenitor stars with $Z=0.01$ and $Z=0.001$. The neutrino emission rates for pair, photo and bremsstrahlung processes were taken from \citet{1996ApJS..102..411I} while that from plasma processes are from \citet{1994ApJ...425..222H}. Since the LPCODE is a full stellar evolution simulation, the C/O core chemical profile is taken directly from the calculations of the progenitor evolution. Abundance changes resulting from residual nuclear burning and convective mixing were also taken into account; the release of energy by proton-burning was considered down to $\log(L/L_{\odot})=-4$. In the low-density regime, an updated version of the EOS of \citet{1979A&A....72..134M} was used; while in the high-density regime, the treatment of crystallisation follows the EOS of \citet{1994ApJ...434..641S} which accounts for both liquid and solid phases, with the C/O phase separation described by \citep{1997ApJ...485..308I} and \citep{2000ApJ...528..397I}. The change of chemical composition resulting from phase separation was computed using the spindle-form phase diagram of \citet{1993A&A...271L..13S}. The gravitational energy released from chemical settling and thermal and chemical diffusion of $^{1}$H, $^{3}$He, $^{4}$He, $^{12}$C, $^{14}$N and $^{16}$O follows that of \citet{2003A&A...404..593A}. Chemical rehomogenisation of the inner C/O profile induced by Rayleigh-Taylor instabilities follows \citet{1997ApJ...486..413S}. The radiative and conductive opacities are taken from the OPAL table by \citet{1993ApJ...412..752I} and \citet{2007ApJ...661.1094C} respectively. Below $10,000$\,K, the molecular opacities are found by assuming pure hydrogen composition with \citet{2009A&A...508.1539M}. Convection is treated with the ML2 parameterisation in \citet{1990ApJS...72..335T}; convective coupling is accounted for when the envelope becomes fully convective: the inner edge of the convective region reaches the boundary of the degenerate regions~(F01). Non-gray atmosphere models are used to derive the outer boundary conditions, which gives shallower outer convection zones as compared to the standard gray treatment of the atmosphere in \citet{1997ApJS..108..339B}, and this should be preferred as it provides a more accurate assessment of cooling times of cool white dwarfs~\citep{2007A&A...466.1043P}. The CIA treatments follow \citet{2002MNRAS.335..499R} and \citet{2006ApJ...651L.137K}; higher order corrections due to H-H and H-H2 are also considered~(\citealp{1965JChPh..43.2429K}; \citealp{1979JPhB...12L.501K}; \citealp{1986JChPh..84..822R} and \citealp{1988JChPh..89.6850P}).

{\bf Unaccounted Physics}: This model appears to have included all the well known WD Physics. However, this model is less used compared to the other two because of its young age.

\subsubsection*{Montr\'{e}al}
\label{sec:montreal}
The code handled the WD evolution in a robust way, which includes for the first time, at the time: (1) the crystallisation process (the moving boundary of the crystallisation front, the release of latent heat, and the redistribution of elements between the liquid/solid phases); (2) diffusion of the various atomic species; (3) convective mixing; and (4) residual thermonuclear burning and neutrino emission mechanisms. The key feature of the code is that the full structure of a model, from the central core region to the top of the atmosphere, is included in the evolutionary calculations. This is important for an accurate description of convective coupling and is essential for following the advancing crystallisation front in both the C/O core and the H/He envelope. Such an approach is also required for including residual gravitational contraction of the envelope, which is particularly important in the very low luminosity phases. The neutrino emissivities are taken from \citet{1996ApJS..102..411I} and the diffusion coefficients are of \citet{1986ApJS...61..197P}. Thermonuclear reaction rates for hydrogen residual burning at the base of the atmosphere are taken from \citet{1975ARA&A..13...69F}. The EOS of H/He, C and O are of \citet{1995ApJS...99..713S}, an unpublished version of \citet{1997ASSL..214..173F} and the solution of the Saha equations with a Coulomb term and a pressure ionisation term from \citet{1973A&A....23..325E} respectively. Upon crystallisation, the phase diagram of C/O is that of \citet{1993A&A...271L..13S}. The description of liquid and solid phases, including the calculation of the latent heat upon crystallisation uses the detailed thermodynamic data of \citet{1974PhDT........56L}. The OPAL radiative opacities are taken from \citet{1993ApJ...412..752I}; at low temperatures, where OPAL data are not available, the Rosseland opacities for pure H and pure He are computed by the authors, which takes into account the H quasi-molecular opacity of \citet{1994A&AS..108..417A} and CIA processes described in \citet{1995ApJ...443..764B}, \citet{1997ApJS..108..339B} and \citet{2001ApJS..133..413B}. Conductive opacities combine the low-density data of \citet{1969ApJS...18..297H} with the high-density results of \citet{1983ApJ...273..774I}, \citet{1984ApJ...285..758I}, \citet{1993ApJ...418..405I} and \citet{1993ApJ...404..268I} which cover both the liquid and solid phases. Convection is treated with the ML2 parameterisation in \citet{1990ApJS...72..335T}. The DA and DB models have H layers correspond to a mass fraction of $q_{\mathrm{H}} = 10^{-4}$ and $q_{\mathrm{H}} = 10^{-10} \mwd$ respectively on top of an He layer of mass $q_{\mathrm{H}} = 10^{-2} \mwd$~\citep{1995LNP...443...41W}.

{\bf Unaccounted Physics}: They provide three distinct families of evolutionary sequences but none of which is a realistic description of the core composition: (1) DA with pure C-core; (2) DA mixed C/O core composition (50/50 by mass fraction mixed uniformly); and (3) DB mixed C/O core composition. Note that the authors stress that the use of the pure C core would set a good upper limit of the derived age.

\begin{table}
\begin{tabular}{|l|c|c|c|}
\hline
Physics                          &  BaSTI & LPCODE & Montr\'{e}al\\
\hline
Neutrino                         & \cmark & \cmark &    \cmark    \\
Phase Separation                 & \cmark & \cmark &    \cmark    \\
Chemical Settling                & \cmark & \cmark &    \xmark    \\
EOS                              & \cmark & \cmark &    \cmark    \\
Radiative Opacities              & \cmark & \cmark &    \cmark    \\
Conductive Opacities             & \cmark & \cmark &    \cmark    \\
Molecular Opacities              & \xmark & \cmark &    \cmark    \\
Convective Coupling              & \xmark & \cmark &    \cmark    \\
Nuclear Burning                  & \xmark & \cmark &    \cmark    \\
Core and Atmosphere Growth       & \xmark & \cmark &    \cmark    \\
Corrections for Higher-order CIA & \cmark & \cmark &    \cmark    \\
\hline
\end{tabular}
\caption{WD Physics that are included in the cooling models}
\label{table:cooling_physics}
\end{table}

\section{Initial-Final Mass Relation $\zeta$}
Since it is impossible to compare the mass of the progenitor of a WD to that of itself directly given that the history of modern observational astronomy is no more than a few decades, the empirical form of this relation depends heavily on stellar models. There are a number of populations that can be used to constrain the IFMR, all of which depend heavily on stellar evolution models and WD cooling models; here is the \textit{complete} listing of IFMR studies:
\begin{enumerate}
\item[1.] globular clusters -- the age of a cluster is found by isochrone fitting of the main sequence turn off~(MSTO)~(1)  (\citealp{2004A&A...420..515M}, \citealp{2009ApJ...705..408K})
\item[2.] open clusters -- the age of a cluster is found by isochrone fitting of the MSTO~(\citealp{1993A&A...275..479K}, \citealp{1996A&A...313..810K}, \citealp{2001ApJ...563..987C}, \citealp{2004MNRAS.355L..39D}, \citealp{2004ApJ...615L..49W}, \citealp{2005ApJ...618L.123K}, \citealp{2006MNRAS.369..383D}, \citealp{2007AJ....133.1490W}, \citealp{2008ApJ...676..594K}, \citealp{2008AJ....135.2163R}, \citealp{2009MNRAS.395.2248D}, \citealp{2009ApJ...693..355W}, \citealp{2012MNRAS.423.2815D}, \citealp{2013MNRAS.428L..16D}, \citealp{2015ApJ...807...90C}, \citealp{2015MNRAS.451.4259C}, \citealp{2016ApJ...818...84C})
\item[3.] wide MS-WD binaries -- the age of the MS is found by isochrone fitting or gyrochronology~(\citealp{2008A&A...477..213C}, \citealp{2012ApJ...746..144Z})
\item[4.] wide WD-WD binaries -- the lifetime of the progenitors have to be found from stellar models~(\citealp{2015ASPC..493..325C}, \citealp{2015ApJ...815...63A}).
\end{enumerate}

Five empirical IFMRs from \citeauthor{2008ApJ...676..594K}~(\citeyear{2008ApJ...676..594K}, K08), \citeauthor{2008MNRAS.387.1693C}~(\citeyear{2008MNRAS.387.1693C}, C08), \citeauthor{2009ApJ...692.1013S}~(\citeyear{2009ApJ...692.1013S}, S09), \citeauthor{2009ApJ...693..355W}~(\citeyear{2009ApJ...693..355W}, W09) and \citeauthor{2009ApJ...705..408K}~(\citeyear{2009ApJ...705..408K}, K09) are demonstrated here. They have all chosen a slightly different approach in the analysis. In K08, each cluster was binned as a single data point before a single linear best fit solution of IFMR was found. In C08, wide MS-WD binaries were combined with open clusters to derive the IFMR, all data points were strictly WDs that have sufficient signal to noise ratio that data were never binned. A 2-part IFMR was also tested along with the conventional linear fit. Apart from adding new cluster data, S09 removed NGC~6791 from the analysis because of its high metallicity~([Fe/H]=+0.45). W09 had a complete list of open clusters. K09 also has a complete list of open clusters, but used various different models. A IFMR including a globular cluster was also available, but it was not advised to use because its low metallicity might have led to more mass loss and skewed the shape of the IFMR. The IFMR are plotted in Fig.~\ref{fig:ifmr_compared}.

\subsection[IFMR from \citet{2008ApJ...676..594K}]{IFMR from \citet{2008ApJ...676..594K} [reliable at $\mathcal{M}_{i}=1.1-6.5\msolar$]}
Three old open clusters, NGC~6791, NGC~6819 and NGC~7789, were used to calculate three mean IFMR data points at the low mass end down to a progenitor mass of $1.16\msolar$. By combining the mean IFMRs of WDs in each open cluster from previous works~(see Table~\ref{table:ifmr_objects}), they derived a linear IFMR of
\begin{equation}
m_{f} = (0.109 \pm 0.007)\mathcal{M}_{i} + 0.394 \pm 0.025;
\end{equation}
when two globular clusters NGC~6397 and NGC~6752~\citep{1996ApJ...465L..23R,2004A&A...420..515M} were included, a shallower IFMR was fitted as
\begin{equation}
m_{f} = (0.106 \pm 0.007)\mathcal{M}_{i} + 0.409 \pm 0.022;
\end{equation}
these two IFMRs were not included in the comparisons later because the authors have updated the relations in 2009. However, it can demonstrate that when metal poor globular clusters are included in the IFMR, they reduce the gradient of the relations.

\subsection[IFMR from \citet{2008MNRAS.387.1693C}]{IFMR from \citet{2008MNRAS.387.1693C} [reliable at $\mathcal{M}_{i}=1.5-6.4\msolar$]}
Instead of using averaged IFMR of each system, C08 recalculated the cooling age of all the known WDs in cluster and MS-WD binaries at the time with a single stellar evolution model and a single WD cooling model. It was the most homogeneous IFMR analysis at the time. The best fit IFMR was given by
\begin{equation}
m_{f} = (0.117 \pm 0.004)\mathcal{M}_{i} + 0.384 \pm 0.011;
\end{equation}
or when fitted with a 2-part IFMR with a break point at $2.7M_{\odot}$, it had a steeper relation at the high mass end that
\begin{equation}
m_{f} = 
\begin{cases}
(0.096 \pm 0.005)\mathcal{M}_{i} + 0.429 \pm 0.015, & \mathcal{M}_{i}<2.7M_{\odot} \\
(0.137 \pm 0.007)\mathcal{M}_{i} + 0.318 \pm 0.018, & \text{otherwise.}
\end{cases}
\end{equation}
The choice of having a break point was motivated by the possible lower efficiency of mass loss of low mass stars, but the choice of the value was not explained whether fitted as a free parameter or established from stellar evolution models.

\subsection[IFMR from \citet{2009ApJ...692.1013S}]{IFMR from \citet{2009ApJ...692.1013S} [reliable at $\mathcal{M}_{i}=1.7-8.5\msolar$]}
WDs from 10 open clusters were used to derive the IFMR for stars with progenitor mass above $1.7\msolar$. Due to the relative young age of open clusters, there was not enough time for lower mass stars to evolve into WDs hence it was unconstrained below $1.7\msolar$. A linear fit gives
\begin{equation}
m_{f} = 0.084\mathcal{M}_{i} + 0.466;
\end{equation}
or when it was fitted with two linear functions with a break point at $4\msolar$, as motivated by the kink in the stellar evolution calculations, it became
\begin{equation}
m_{f} = 
\begin{cases}
0.134\mathcal{M}_{i} + 0.331 , & 1.7M_{\odot} \leq \mathcal{M}_{i} \leq 4M_{\odot} \\
0.047\mathcal{M}_{i} + 0.679 , & 4M_{\odot} < \mathcal{M}_{i}.
\end{cases}
\end{equation}
This 2-part relation is significantly different from C08 at the high mass end where C08 finds an enhanced mass loss instead. The gradients of both IFMRs in this work are much shallower than all other works, and they did not estimate the uncertainties.

\subsection[IFMR from \citet{2009ApJ...693..355W}]{IFMR from \citet{2009ApJ...693..355W} [reliable at $\mathcal{M}_{i}=1.25-8.0\msolar$]}
Open cluster NGC 2287~(M41) was added to the collection of open clusters to derive the IFMR
\begin{equation}
m_{f} = (0.129 \pm 0.004)\mathcal{M}_{i} + 0.339 \pm 0.015;
\end{equation}
however, as noted by the authors, the \textit{``linear fit is formally inconsistent with the slopes of the fits given in \citet{2005MNRAS.361.1131F}, K08 and C08; the M35 WDs at the high-mass end of the relation prefer a steeper slope.''} Their IFMR has the largest gradient in all studies. The only exception is that the 2-part IFMR by C08 had a shallower gradient fitted over the low mass region and a much steeper relation was found at the high mass end, consistent with the argument of W09.

\subsection[IFMR from \citet{2009ApJ...705..408K}]{IFMR from \citet{2009ApJ...705..408K} [reliable at $\mathcal{M}_{i}=1.1-6.5\msolar$]}
They used all available open clusters to derive the IFMR for the lower mass end down to a progenitor mass of $0.8\msolar$ giving an IFMR of
\begin{equation}
m_{f} = (0.109 \pm 0.007)\mathcal{M}_{i} + 0.428 \pm 0.025;
\end{equation}
when the WDs from the globular cluster, M4, was included, a shallower IFMR was fitted as
\begin{equation}
m_{f} = (0.101 \pm 0.006)\mathcal{M}_{i} + 0.463 \pm 0.018;
\end{equation}
it was not explained why NGC~6397 and NGC~6752, which were used in K08, were excluded from this study for more leverage at the low mass end. Nevertheless, it was not advised to use the second IFMR because of an increasingly clear evidence of metallicity dependence on mass loss efficiency~(open clusters are more metal poor than open clusters). The first relation, is essentially an update of K08 with improved atmosphere models and included three more open clusters. It was merely an offset of $+0.034\msolar$ to the WD mass.

\begin{figure}
\includegraphics[width=0.9\textwidth]{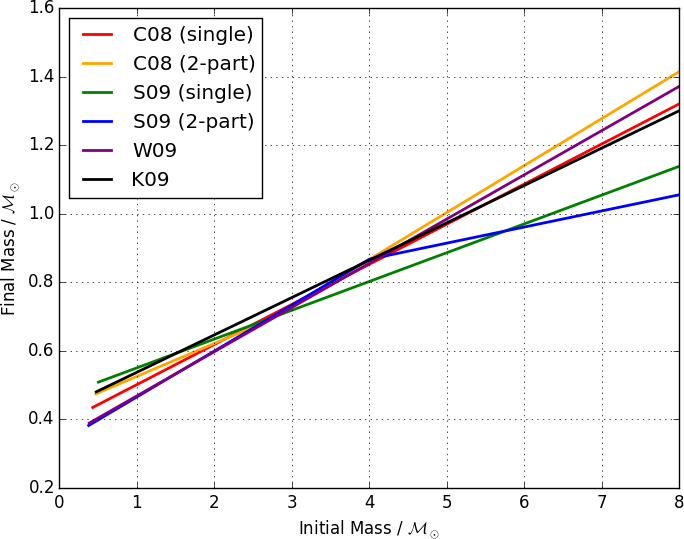}
\caption[The six initial-final mass relations]{The six IFMRs are plotted to illustrate the differences. The IFMRs from S09 are notably different. The differences of the rest are in the range of 10-20\% in final mass}
\label{fig:ifmr_compared}
\end{figure}

\begin{table}
\begin{tabular}{ | l | c | c | c | c | c | }
\hline
Object         & K08 & C08 & S09 & W09 & K09 \\
\hline
\multicolumn{6}{ | c | }{Stellar Systems} \\
\hline
Sirius~B       & \cmark & \cmark & \cmark & \cmark & \cmark \\
MS-WD binaries & \xmark & \cmark & \xmark & \xmark & \xmark \\
WD-WD binaries$^{(a)}$ & \xmark & \xmark & \xmark & \xmark & \xmark \\
\hline
\multicolumn{6}{ | c | }{Open Cluster} \\
\hline
Hyades (Caldwell 41)     & \cmark & \cmark & \cmark & \cmark & \cmark \\
Pleiades (M45)           & \cmark & \cmark & \cmark & \cmark & \cmark \\
Praesepe (NGC 2632, M44) & \cmark & \cmark & \cmark & \cmark & \cmark \\
NGC~1039 (M34)           & \xmark & \xmark & \cmark & \cmark & \cmark \\
NGC~2099 (M37)           & \cmark & \cmark & \cmark & \cmark & \cmark \\
NGC~2168 (M35)           & \cmark & \cmark & \cmark & \cmark & \cmark \\
NGC~2287 (M41)           & \xmark & \xmark & \xmark & \cmark & \cmark \\
NGC~2516                 & \cmark & \cmark & \cmark & \cmark & \cmark \\
NGC~3532                 & \xmark & \cmark & \cmark & \cmark & \cmark \\
NGC~6633                 & \cmark & \xmark$^{(b)}$ & \xmark$^{(b)}$ & \cmark & \cmark \\
NGC~6791                 & \cmark & \cmark & \xmark$^{(c)}$ & \cmark & \cmark \\
NGC~6819                 & \cmark & \cmark & \cmark & \cmark & \cmark \\
NGC~7789                 & \cmark & \cmark & \cmark & \cmark & \cmark \\
\hline
\multicolumn{6}{ | c | }{Globular Cluster$^{(d)}$} \\
\hline
NGC~6121 (M4)          & \xmark & \xmark & \xmark & \xmark & \cmark \\
NGC~6397 (Caldwell 86) & \cmark & \xmark & \xmark & \xmark & \xmark \\
NGC~6752               & \cmark & \xmark & \xmark & \xmark & \xmark \\
\hline
\end{tabular}
\begin{enumerate}
\itemsep0em
\item[a)] Not included in any homogeneous IFMR analysis yet.
\item[b)] Unclear why it is not included.
\item[c)] Ignored because of high metallicity.
\item[d)] K08 and K09 derived separate IFMRs with and without using globular clusters.
\end{enumerate}
\caption{Input data used for constructing IFMRs}
\label{table:ifmr_objects}
\end{table}

\section{White Dwarf Luminosity Functions}
\label{sec:theoretical_wdlf}
\begin{figure}
\includegraphics[width=0.9\textwidth]{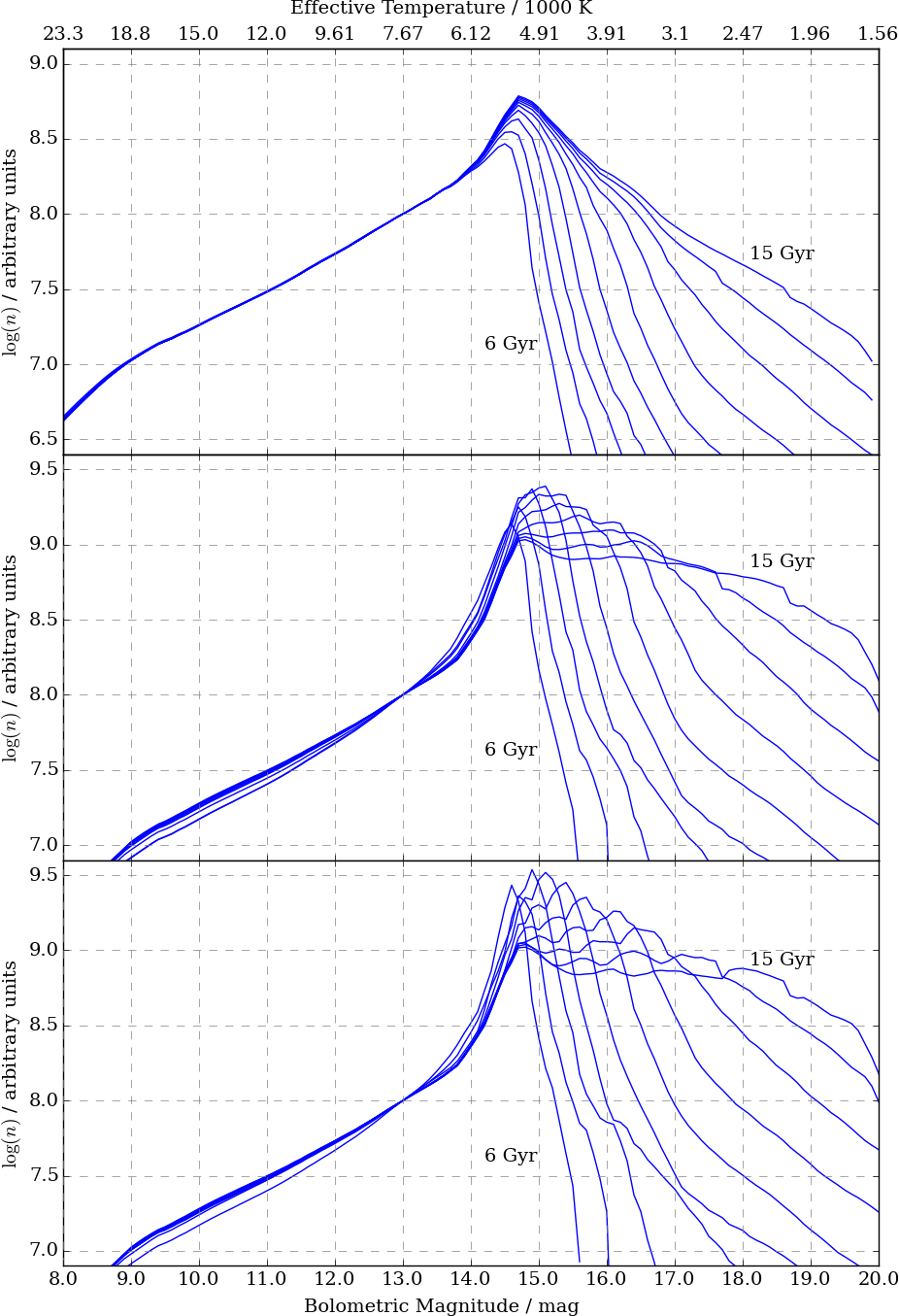}
\caption[Theoretical WDLFs with constant, exponential decay and burst star formation rate]{Top: theoretical WDLFs of $6-15\gyr$ old population in steps of $1\gyr$ with constant star formation rate. Middle: theoretical WDLFs of $6-15\gyr$ old population in steps of $1\gyr$ with a exponentially decaying star formation rate with a scale time of $1\gyr$. Bottom: theoretical WDLFs of $6-15\gyr$ old population in steps of $1\gyr$ with a constant star formation rate in the first $1\gyr$ and ceased afterwards}
\label{fig:wdlf_age_compared}
\end{figure}

When the models from the last three sections are in place, Eq.~\ref{eq:WDLFnumberdensity} can be integrated. I used the {\sc quad} function from {\sc SciPy} to perform the numerical integrations; the cooling models for different white dwarf masses are linearly interpolated and the choices of IMFR and SFR are described below. Five simple configurations are used to illustrate the effects of the most important parameter -- the SFR. In Fig.~\ref{fig:wdlf_age_compared}, theoretical WDLFs with constant SFR, exponentially decaying SFR with a decay constant of $1\gyr$, and a burst SFH with a constant SFR in the first $1\gyr$ and ceased afterwards are plotted for populations with age $6-15\gyr$ in step of $1\gyr$ in the top, middle and bottom panel respectively. In Fig.~\ref{fig:wdlf_burst_and_decay_compared}, exponentially decaying SFR with a decay constant of $1-10\gyr$ and burst SFH that last $1-10\gyr$ are plotted in steps of $1\gyr$ in both cases. The WDLFs are normalised at $\mathrm{M}_{\mathrm{bol}}=13.0\magnitude$. The most obvious feature from the five plots is the similarity in the shape of the WDLFs brighter than $\mathrm{M}_{\mathrm{bol}}\sim14.0\magnitude$. This means that WDLFs can only be used as cosmochronometer if the survey is deep enough to include the faintest objects, which was the case in \citet{1988ApJ...332..891L} where the downturn of the WDLF was first measured.

The following are used as our basic configuration: (1) solar metallicity, (2) Montr\'{e}al cooling model and (3) K09 IFMR; they will also be used for the rest of the thesis.

\subsection*{Constant SFR}
The bright edge of the WDLFs line up perfectly because WDs arrive at the corresponding bolometric magnitudes at a constant rate. The gradient is most sensitive to a time-dependent variable, for example, a change in SFR or metallicity with time. The sudden increase in the number density at $\mathrm{M}_{\mathrm{bol}}=14.0-15.0\magnitude$ is due to the slow down in cooling due to crystallisation which releases a significant amount of heat; convective coupling reduces the temperature gradient inside the WD, which slows down the cooling at early time.

\subsection*{Exponentially Decaying SFR}
The large contrast in the SFR between early and late times appears as a large step in the number density at $\mathrm{M}_{\mathrm{bol}}=14.0-15.0\magnitude$. This feature results from the fact that the rate at which WDs are formed is decreasing with time as a consequence of a lower SFR at late time. It has also allowed the faint objects to have major contributions to the WDLFs: a large number of faint WDs is sufficient to compensate the piling up of WDs at $\mathrm{M}_{\mathrm{bol}}\sim14.5\magnitude$ due to the decrease in cooling rate. The plateau at the faint end is because of a fairly constant cooling rate.

\subsection*{Burst SFR}
The WDLFs with burst SFR closely resemble those with exponential profile. They are both dominated by enhanced star formation at early times. The step at $\mathrm{M}_{\mathrm{bol}}=14.0-15.0\magnitude$ is even more prominent because of the abrupt change in the SFR. For the same reason, the plateau at the faint end is flatter than those produced from the SFH with exponential profiles.

\subsection*{Exponentially Decaying SFR with Different Decay Constant}
The top panel of Fig.~\ref{fig:wdlf_burst_and_decay_compared} shows the WDLFs with different SFR decay constants, from $1-10\gyr$ in step of $1\gyr$ for a $10\gyr$ old population. As explained earlier, the step at $\mathrm{M}_{\mathrm{bol}}=14.0-15.0\magnitude$ is a feature of the large change in SFR between early and late times. It is expected that when the decay is faster, the step is more prominent. The step for the $\tau_{D}=1\gyr$ is $\sim0.5$\,dex larger than that with $\tau_{D}=10\gyr$. As the decay constant increases, it approaches that with a constant SFR.

\subsection*{Burst SFR with Different Period}
\begin{figure}
\includegraphics[width=0.9\textwidth]{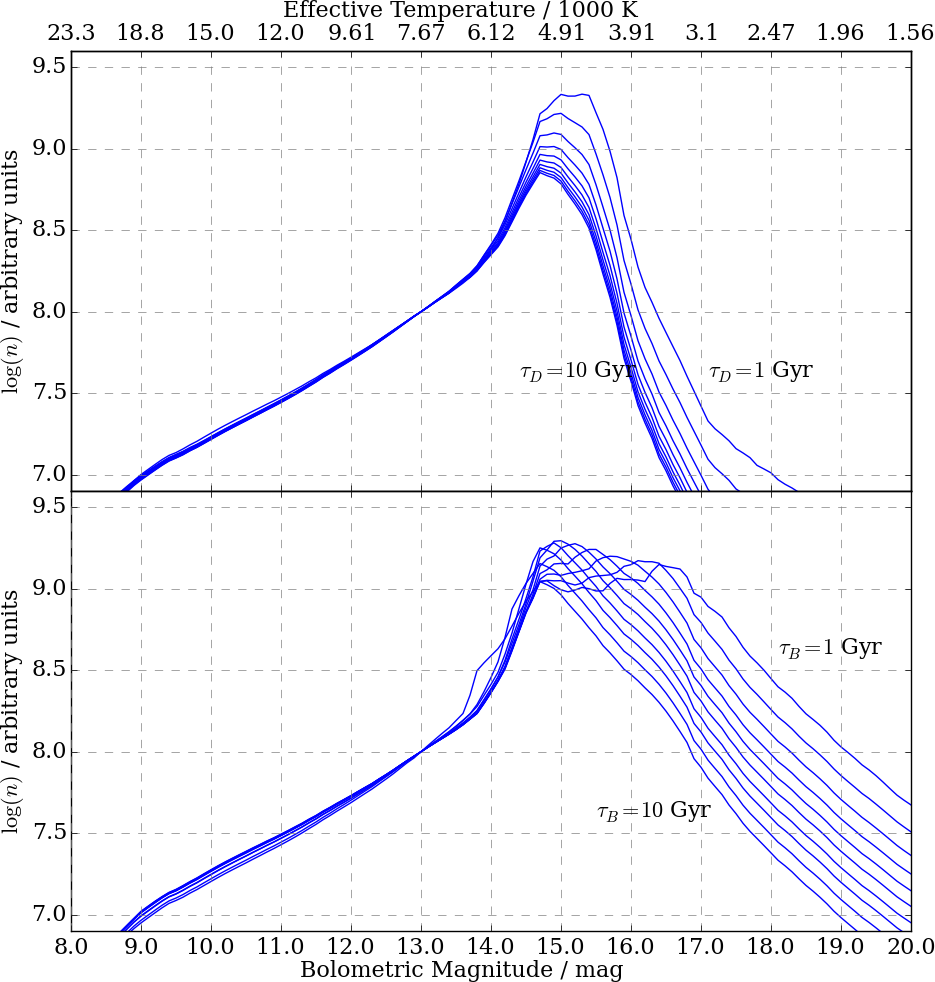}
\caption[Theoretical WDLFs with exponential decay and burst star formation rate with different decay constants and burst durations]{Top: theoretical WDLFs of $10\gyr$ old population with a exponentially decaying star formation rate with a scale time of $1-10\gyr$ in steps of $1\gyr$. Bottom: theoretical WDLFs of $13\gyr$ old population with a constant star formation rate in the first $1-10\gyr$  in steps of $1\gyr$ and ceased afterwards}
\label{fig:wdlf_burst_and_decay_compared}
\end{figure}

The bottom panel of Fig.~\ref{fig:wdlf_burst_and_decay_compared} shows the WDLFs with different period of star burst in the beginning, from $1-10\gyr$ in step of $1\gyr$ for a $13\gyr$ old population. The ``top hat''-like feature is the most prominent feature with the shortest burst. When the burst approaches a delta function, the WDLF is essentially the inverse-cooling rate as a function of magnitude. The plateau at the faint end of WDLFs with short burst, as well as the parallel edges at the faint end of all WDLFs are both the features of a roughly constant cooling rate at faint magnitudes. If the WDLFs are normalised at the parallel edges, for example $\mathrm{M}_{\mathrm{bol}}=18.0\magnitude$, it would mean that the WDLFs are normalised with the stars born in the first $1\gyr$. The vertical offsets in the WDLFs will then be transferred to the bright end, which is evidence of prolonged star formations. SFH with exponentially decaying profile can demonstrate the same effect but it is the most obvious with a bursty profile.

\section{Comparison of different models}
In order to compare different models, the metallicities, cooling models and IFMRs are varied one at a time in order to do valid comparisons.

\subsection*{Metallicity dependence of progenitor lifetimes}
\begin{figure}
\includegraphics[width=0.9\textwidth]{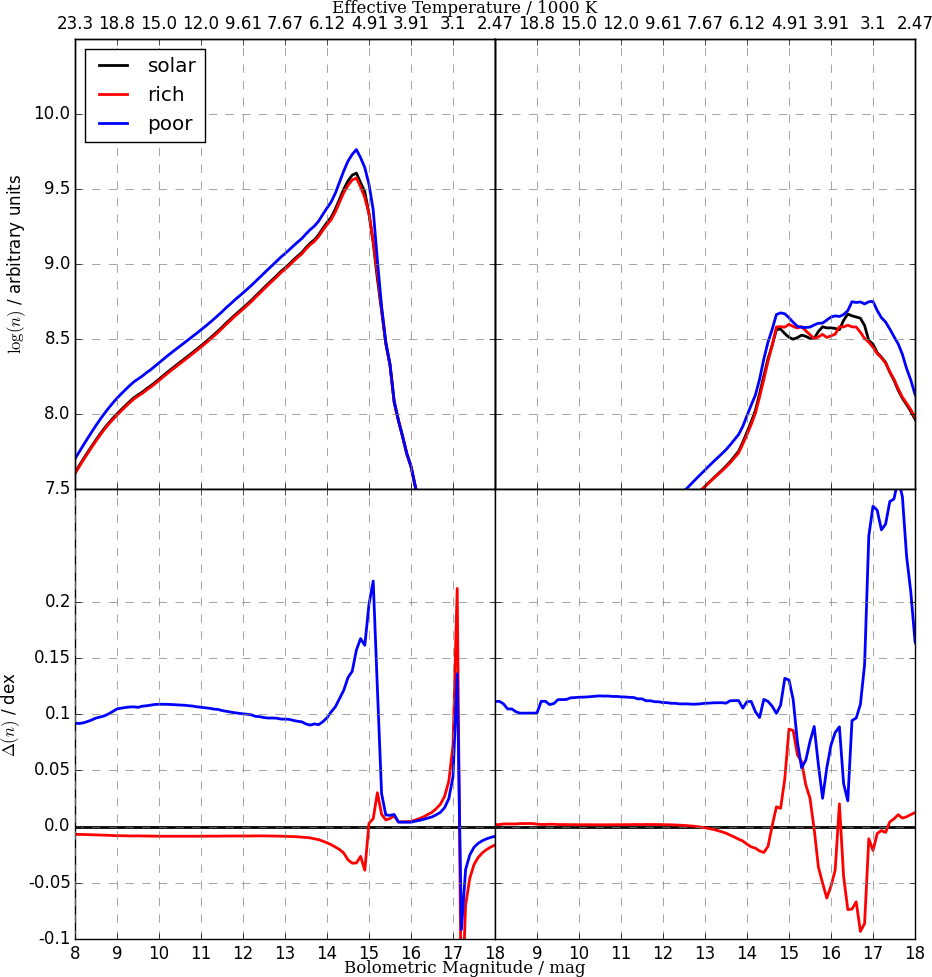}
\caption[Theoretical WDLFs with constant and burst star formation rate with different progenitor metallicities]{Top left: Theoretical WDLFs of a 8 billion years old population with constant star formation rate with $Z=0.06$~(blue), $0.017$~(black) and $0.001$~(red). Top right: Theoretical WDLFs for a 13 billion years old population with a constant star formation rate in the first billion years and ceased afterwards. Bottom: difference of WDLFs between different ZAM metallicity and the basic configuration}
\label{fig:wdlf_ms_metallicity_compared}
\end{figure}

Cool WDs have cooling time dominating their total ages; the choice of MS evolutionary models have small effects so different models are not compared, except for different metallicities. As WDs have little heat generation, they only evolve to lower bolometric magnitudes. The use of the most metal poor MS lifetime~(i.e. shortest MS lifetimes) gives a consistent enhanced density down to $\mathrm{M}_{\mathrm{bol}}\sim15.5\magnitude$ because stars are constantly formed at the same rate but they become WD quicker than the metal rich ones and rapidly replenish the WDs that have evolved to fainter magnitudes. For a population with constant SFR, this appears as WDs ``piling up'' on the WDLF as cooling rate decreases with temperature starting at $\mathrm{M}_{\mathrm{bol}}\sim14\magnitude$~(See Fig.~\ref{fig:wdlf_ms_metallicity_compared}). There is little effect on the shape of the WDLF of a burst population, the vertical offset continues down to $\mathrm{M}_{\mathrm{bol}}\sim15.5\magnitude$ which is the limit at which the WD cooling models are reliable. In R13, only the solar metallicity is used in the WDLF inversion algorithm so the SFH at early times is likely to be underestimated -- it would require a smaller normalisation~(i.e. number density) to compensate the vertical offset due to shorter MS lifetimes of metal poor stars formed in the early epoch. However, for the metal rich populations, there is negligible difference compared to a population with solar metallicity.

\subsection*{Cooling models}
\begin{figure}
\includegraphics[width=0.9\textwidth]{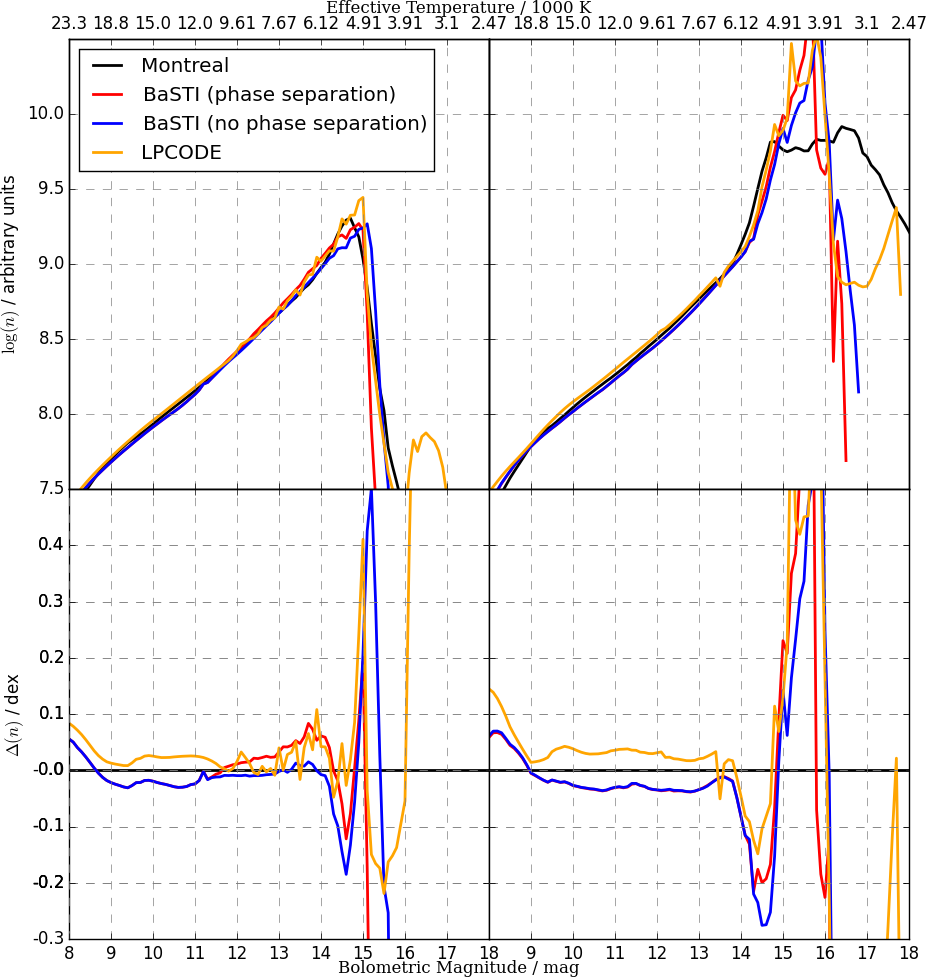}
\caption[Theoretical WDLFs with constant and burst star formation rate with different white dwarf cooling models]{Top left: Theoretical WDLFs of a 8 billion years old population with constant star formation rate with the Montr\'{e}al~(black), BaSTI with C/O phase separation~(red), BaSTI without C/O phase separation~(blue) and LPCODE~(orange) models. Top right: Theoretical WDLFs for a 13 billion years old population with a constant star formation rate in the first billion years and ceased afterwards for the four WD models. Models are extrapolated to cover the entire range of bolometric magnitudes when required to compute the density at the desired magnitude. The WDLFs are thus unreliable beyond $\mathrm{M}_{\mathrm{bol}}\sim15\magnitude$ and are completely dependent on extrapolated grid beyond $\mathrm{M}_{\mathrm{bol}}\sim16\magnitude$ for the BaSTIs and LPCODE. For the Montr\'{e}al model, the grid is available down to $\mathrm{M}_{\mathrm{bol}}\sim20\magnitude$. Bottom: difference of WDLFs between different WD models and the basic configuration}
\label{fig:wdlf_cooling_model_compared}
\end{figure}

The WDLFs constructed from the four sets of cooling models have small differences down to $\mathrm{M}_{\mathrm{bol}}\sim14\magnitude$. In the bottom left panel of Fig.~\ref{fig:wdlf_cooling_model_compared}, the BaSTI models with phase separation diverge at $\mathrm{M}_{\mathrm{bol}}\sim11.5\magnitude$, the temperature at which phase separation begins. The rapid divergence of all models at $\sim14\magnitude$ is due to the different treatment of the envelope/core boundary: in the BaSTI models, static envelopes are used so they lack convective coupling hence cooling is not slowed down due to envelop mixing; LPCODE uses a dynamic envelope like the Montr\'{e}al model. However, its inclusion of chemical settling means that it has a larger heat reservoir, hence it has a lower cooling rate. From the $13\gyr$ burst population on the right panel, it is obvious that the WD cooling rate beyond $\mathrm{M}_{\mathrm{bol}}\sim14.5\magnitude$ is much higher in the Montr\'{e}al model. This should be attributed to the adaptation of a well mixed 50/50 C/O core where energy released from chemical settling is not accounted for. Fainter than $\mathrm{M}_{\mathrm{bol}}\sim16\magnitude$, most calculations are based on extrapolated grids, with the exception of the Montr\'{e}al model, so the WDLFs are noisy and unreliable.

\subsection*{IFMRs}
\begin{figure}
\includegraphics[width=0.9\textwidth]{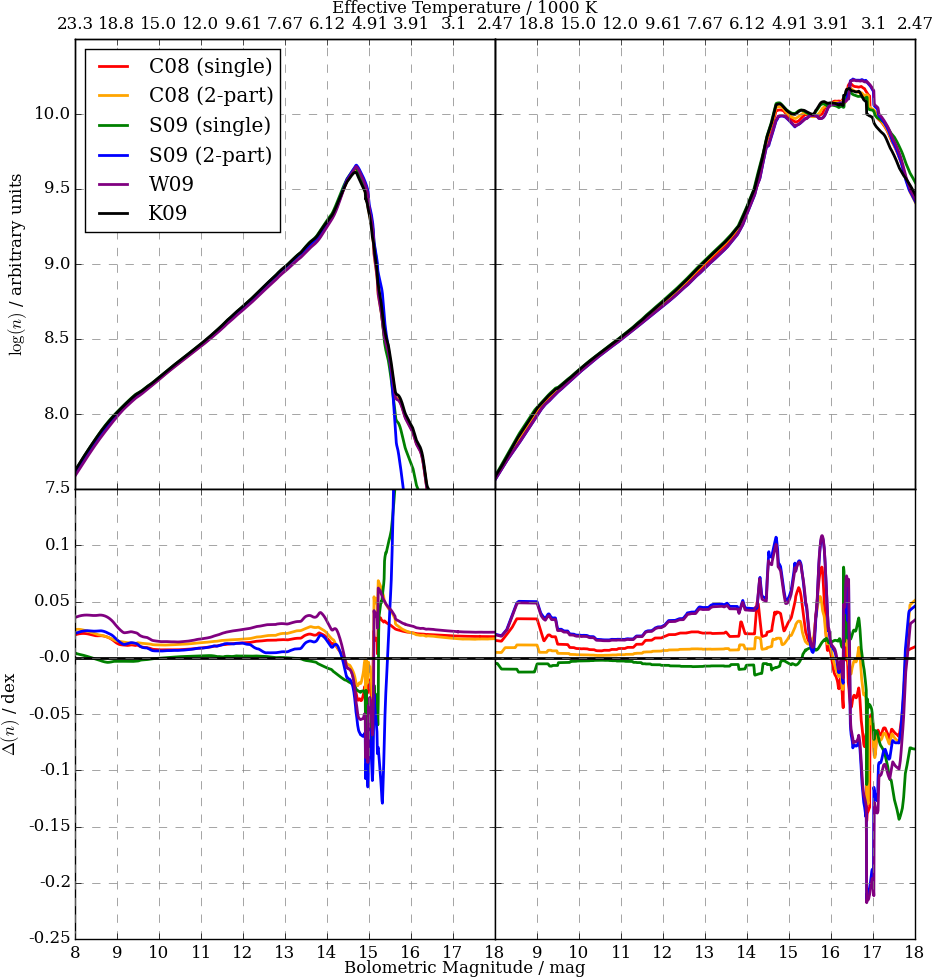}
\caption[Theoretical WDLFs with constant and burst star formation rate with different initial-final mass relations]{Top left: Theoretical WDLFs of a 8 billion years old population with constant star formation rate with the C08~(single), C08~(2-part), S09~(single), S09~(2-part), W09 and K09 IFMRs. Top right: Theoretical WDLFs for a 13 billion years old population with a constant star formation rate in the first billion years and ceased afterwards for the six IFMRs. Bottom: difference of WDLFs between different WD models and the basic configuration}
\label{fig:wdlf_ifmr_compared}
\end{figure}

Bridging between the MS and WD, the choice of IFMRs has negligible effect except at the faint end~($\mathrm{M}_{\mathrm{bol}}>15\magnitude$), where the uncertainties are dominated by WD atmosphere models and SFH. Nevertheless, their influence to the WDLF is illustrated in Fig.~\ref{fig:wdlf_ifmr_compared}. From the bottom right panel, the using of S09 single IFMR model leads to more WDs at the bright end as compared to other models, as evidenced from its position above all models below $\sim1.5\msolar$~(Fig.~\ref{fig:ifmr_compared}). The $\sim1\magnitude$ window at $\mathrm{M}_{\mathrm{bol}}\sim17\magnitude$ is the effect of the higher IFM ratio of the K09 of intermediate mass stars. At $4\msolar$, four IFMRs have higher IFM ratio than K09 which leads to the excess in the WDLF beyond $\mathrm{M}_{\mathrm{bol}}\sim17.5\magnitude$. The same effect can be seen from the left hand panel which begins at $\mathrm{M}_{\mathrm{bol}}\sim15\magnitude$. The significant differences in the WDLFs with S09 models are due to the high mass stars where S09 predicts a much lower IFM ratio than the other models.

\subsection{Summary}
This chapter has provided descriptions of the WD cooling models, choice of stellar evolution models and IFMRs and demonstrated how they affect the resulting WDLFs:
(1) The choice of WD cooling models have little effect on a young population~(e.g. thin disk) so there is little impact to the study of the solar neighbourhood. However, for an old population~(e.g. halo), the faint end of the WDLFs differ significantly between the Montr\'{e}al model and the rest. The major difference in the models are the treatment of the chemical composition where the Montr\'{e}al group uses a well-mixed 50/50 carbon-oxygen core, the others use more realistic description of the core chemistry. The exact reason of the diversion would require detailed code comparison. (2) The choice of IFMR has negligible effect on the WDLFs except at the faint end where the choice of S09 models diverges from the others, which is expected from the systematic inconsistency of the S09 models from other works. The difference between different WDLFs would otherwise be well within the uncertainties of each other. (3) The progenitor lifetime as a function of metallicity has little effect between solar metallicity and extremely metal rich cases. However, the metal poor population shows enhanced WDLFs because stars become WDs quicker. This means if one is to study a halo WDLF it is important to consider a different input metallicity such that the SFR at early time would not be overestimated. (4) The three forms of SFH show some broad differences: WDLFs with constant SFR are mostly featureless, except for the peak at $\textrm{M}_{\textrm{bol}} \sim 14.5 \magnitude$ which appears in any choice of models. The exponential decay and burst profiles give similar WDLFs when the decay timescale is comparable to the burst period. Otherwise, the burst profile is characterised by a broad plateau at faint magnitudes. (5) The bright ends~($8-13 \magnitude$) of the WDLFs reveal little information unless the population is younger than $\sim6\gyr$. Although at the even brighter magnitudes the gradient and shape can help to constrain exotic Astroparticle sciences. Since PS1 response system only reaches $400\,nm$ in the bluest filter, this cannot be inferred from my dataset.

%% file: 4_vmax_pm_limited.tex
\chapter{Maximum Volume Density Estimator}
\label{ch:vmax}
The use of a maximum volume as a density estimator began when \citet{1968ApJ...151..393S} introduced the V/V$_{\mathrm{max}}$ technique for analysing the luminosity function of quasars, $\phi_{s}$, where V and V$_{\mathrm{max}}$ are the volume enclosed by the object and the volume in which the object can be found at the given survey limits respectively and $\phi_{s}$ is the number density. The estimator can be used as a means of testing the completeness of a sample simultaneously. \citet{1976ApJ...207..700F} showed that V/V$_{\mathrm{max}}$ is unbiased and superior to the older ``classical'' estimator (N/V) when the magnitude or luminosity bins are not small. He further described the procedure of combining the density estimator from two non-overlapping areas.

\section{The Family of the V${\rm max}$ Density Estimator}
Information from a single survey is limited due to the relatively small number of objects. When several complete samples are combined, more fundamental parameters of the population of objects can be determined, and with smaller associated uncertainties. In the light of this problem, \citet{1980ApJ...235..694A} investigated different ways of combining catalogues, namely the incoherent region-independent method, the incoherent domain-independent method and the coherent method. All of these are superior over the original Schmidt method, with the coherent method being most accurate. When the effects of a space-density
gradient were corrected for \citep{1989MNRAS.238..709S, 1993ApJ...414..254T, 2015ASPC..493..347L} this method was extended to estimate stellar density where the density profile of the Galaxy varies significantly along different lines of sight. Because of the small distances probed, only the scale height effects are considered while the scale length is assumed to be constant.

In order to consider a sample of proper motion objects, \citet{1975ApJ...202...22S} extended his estimator to cope with both photometric and proper motion detection limits. The new estimator considers the tangential velocity as an intrinsic property of an object such that it can be kept as a constant. Then, the distance limits can be found easily by applying the upper and lower proper motion limits of the survey to a simple relation between tangential velocity, proper motion and distance: $v_{\mathrm{tan}} = 4.74\mu D\kmps$, where $\mu$ is the proper motion in arcseconds per year and $D$ is the distance to the object in parsecs.

Cool WDs and subdwarfs have similar optical colours to the MS stars while those of brown dwarfs (BDs) are similar to the giants. Therefore, it is difficult to distinguish them in colour-colour space. Due to their small radii, WDs, subdwarfs and BDs are located far from MS stars and giants in the HR diagram. However, when objects are only detected in a few broadband filters, it is impossible to classify them reliably which would lead to poor object selections and distance estimates. To overcome this problem, it is common to use RPM as a crude estimate of absolute magnitude to separate samples of subluminous objects from higher luminosity contaminants. In order to obtain a clean sample of WDs\,(for example an extreme subdwarf would easily be confused with a WDs with low tangential velocity) a lower tangential velocity has to be applied to remove the ambiguous objects. This procedure introduces an incompleteness which has to be corrected for. This problem was identified by \citet{1986ApJ...308..347B} and \citet{1992MNRAS.255..521E} separately. The former adopted a Monte Carlo\,(MC) simulation approach to correct for the incompleteness, while the latter was done analytically. However, the two methods evolved separately. In the simulation front, \citet{1999ASPC..169...51L} constructed some simulations based on different Galactic models to study the incompleteness due to proper motion selection after some strong arguments\,(\citealt{1995LNP...443...24O}, \citealt{1996Natur.382..692O}) pointing towards an incomplete LHS catalogue used in earlier studies\,(eg. \citealt{1987ApJ...315L..77W}, \citealt{1988ApJ...332..891L}). This correction, known as the discovery fraction, $\chi$, was then applied by H06. On the other hand, \citet{1999MNRAS.306..736K} used the analytical approach to correct for the incompleteness. Instead of calculating the discovery fractions from integrating over the density profile, \citet{2003MNRAS.344..583D} arrived at the discovery fractions by integrating over the Schwarzschild distribution functions. RH11 further generalised the technique to cope with an all sky survey as opposed to the individual fields of view employed in earlier works.

\subsection{Mathematical Constructions}
The classical estimator $\Phi = N/$V for a volume-limited sample is of little practical use for analysing small numbers of objects or strongly localised and kinematically biased groups of stars that are selected by proper motion and apparent magnitude. For these samples, the 1/V$_{\mathrm{max}}$ method is generally regarded as a superior estimator of the LF \citep{1976ApJ...207..700F}. The contribution of each object to the LF is weighted by the inverse of the maximum volume in which an object could be observed by the survey. For example, for a given bin of objects with index k, the space density is the sum of all 1/V$_{\mathrm{max}}$
\begin{equation}
\Phi_{k} = \displaystyle\sum\limits_{i=1}^{N_{k}} \frac{1}{\mathrm{V}_{\mathrm{max}, i}}
\end{equation}
for $N_{k}$ objects in the $k^{\mathrm{th}}$ bin. The uncertainty of each star's contribution is assumed to follow Poisson statistics. The sum of all errors in quadrature within a luminosity bin is therefore,
\begin{equation}
\sigma_{k} = \left[\displaystyle\sum\limits_{i=1}^{N_{k}} \left(\frac{1}{\mathrm{V}_{\mathrm{max}, i}}\right)^{2}\right]^{\frac{1}{2}}.
\end{equation}
The traditional 1/V$_{\mathrm{max}}$ technique assumes that objects are uniformly distributed in space. However, in reality, stars in the solar neighbourhood are concentrated in the plane of the disc. The effects of space-density gradient can be corrected by assuming a density law and defining a maximum generalised volume V$_{\mathrm{gen}}$ (\citealp{1989MNRAS.238..709S}; \citealp{1993ApJ...414..254T}) which is calculated by integrating the appropriate stellar density profile $\rho/\rho_{\odot}$ along the line of sight between the minimum distance, $d_{\mathrm{min}}$, and maximum distance, $d_{\mathrm{max}}$. This leads to the integral
\begin{equation}
\mathrm{V}_{\mathrm{gen,S}89} = \Omega \int^{d_{\mathrm{max}}}_{d_{\mathrm{min}}} \frac{\rho(r)}{\rho_{\odot}} \, r^{2}\,dr
\end{equation}
where $\Omega$ is the size of the solid angle of the survey. To minimise the contamination from extreme subdwarfs scattered into the WD regime in RPM-colour space, a lower tangential velocity limit, $v_{\mathrm{tan,lower}}$, is applied to remove most of the contaminants. Traditionally, the discovery fraction, $\chi_{v}$ which is the fraction of objects with tangential velocities larger than the lower tangential velocity limit, is only Galactic model and survey-footprint dependent (\citealp{1986ApJ...308..347B}, \citealp{1999ASPC..169...51L}, H06) such that the maximum volume density estimator can be written in the form
\begin{equation}
\mathrm{V}_{\mathrm{gen,H}06} = \chi_{v}(v_{\mathrm{tan,lower}}) \, \Omega \int^{d_{\mathrm{max}}}_{d_{\mathrm{min}}} \frac{\rho(r)}{\rho_{\odot}} \, r^{2} \, dr
\end{equation}
where the distance limits are derived from both photometric and proper motion limits of the survey by calculating
\begin{equation}
\label{eq:dmin}
d_{\mathrm{min}} = d \times \mathrm{max} \left[ 10^{\frac{(m_{\mathrm{min},i}-M_{i})}{5}}, \frac{\mu}{\mu_{\mathrm{max}}} \right]
\end{equation}
\begin{equation}
\label{eq:dmax}
d_{\mathrm{max}} = d \times \mathrm{min} \left[ 10^{\frac{(m_{\mathrm{max},i}-M_{i})}{5}}, \frac{\mu}{\mu_{\mathrm{min}}} \right]
\end{equation}
where $m_{\mathrm{min},i}$, $m_{\mathrm{max},i}$ and $M_{i}$ are the photometric limits and the absolute magnitudes of the object in filter $i$ respectively. The proper motion terms, $\mu/\mu_{\mathrm{max}}$ and $\mu/\mu_{\mathrm{min}}$, are rationalised by assuming an object would carry the same tangential velocity if it were placed closer to or farther from the observer (analogous to the absolute magnitude) and/or in an arbitrary line of sight.

\subsection{Attempt to Modify the Discovery Fraction}
RH11 extended the $\chi_{v}$ to include a directional dependence in order to account for the varying survey properties and stellar tangential velocity distribution across the sky. In RH11, the Schwarzschild distribution function is used instead to calculate the tangential velocity distribution, $P(v_{\mathrm{tan}})$, analytically. The discovery fraction can be found by projecting the velocity ellipsoid onto the tangent plane of observation \citep{1983veas.book.....M}. Thus it allows one to arrive at a precise $\chi(\alpha,\delta)$ without taking the average properties over a large area. Therefore, the volume integral can be modified to
\begin{equation}
\mathrm{V}_{\mathrm{gen,RH}11} = \sum_{i} \Omega_{i} \, \chi_{v}(i,v_{\mathrm{tan,lower}}) \,  \int^{d_{\mathrm{max}}}_{d_{\mathrm{min}}} \frac{\rho_{i}(r)}{\rho_{\odot}} \, r^{2} \, dr
\end{equation}
where $i$ denotes each sky cell covered by a Schmidt survey field employed in the production of the catalogue used by RH11 (hereafter the generalised method). This should, in theory, have taken into account all the small scale variations which a positional-independent $\chi_{v}$ would not be able to deal with.
\subsection{Problems in this framework}
We have identified a bias present in the {\it whole family} of maximum volume methods when calculating distance limits by holding the tangential velocity constant along a single line of sight \citep{1975ApJ...202...22S}; and the consequence to the discovery fraction applied in H06 and RH11 when the tangential velocity is further held constant across the sky. These can be described as follows:
\subsection*{Constant tangential velocity along line of sight}
The kinematics of an object is a property of the Galaxy. An object at a given magnitude at any given distance from the observer should not carry the same tangential velocity at a different line of sight distance when tested for observability. This assumption is only good over small field of views and small range of line of sight distances, while an acceptable size of the smallness is very difficult to assess if not unquantifiable.
\subsection*{Constant tangential velocity across the sky}
Consider a spatially uniform population like the stellar halo, and an all-sky survey where the proper motion limits are the same over the whole sky. The tangential velocity distribution varies along different lines of sight due to the solar motion. For stars at a given magnitude, a different fraction of the population will pass the proper motion limits along different lines of sight due to the differences in the tangential velocity distributions. In the most extreme cases, on average, a halo WD observed in the direction of the Anti-Galactic Center~(AGC) would appear to have a large velocity due to reflex motion imparted by the Sun in its orbit within the Galaxy. However, if one is observed in the direction to the solar apex instead, the motion would be much smaller on average. The consequence is that when proper motion limits and tangential velocity limits are applied, different numbers of stars would be detected in different regions of sky even for identical survey limits and spatial density. That is the motivation for RH11 to consider different tangential velocity distributions along different lines of sight. The method would be completely correct in the framework where {\it the tangential velocity is an intrinsic property of an object}. However, in the case where tangential velocity is not constant, there would be a mismatch between the parameter space which the discovery fraction and the maximum volume explore~(see Chapter~\ref{sec:new_approach} where we will demonstrate how the new approach can solve both problems).

\subsection{A Closer Look at the Discovery Fraction}
\label{sec:closer_look}
\begin{figure}
\begin{center}
\includegraphics[width=0.9\textwidth]{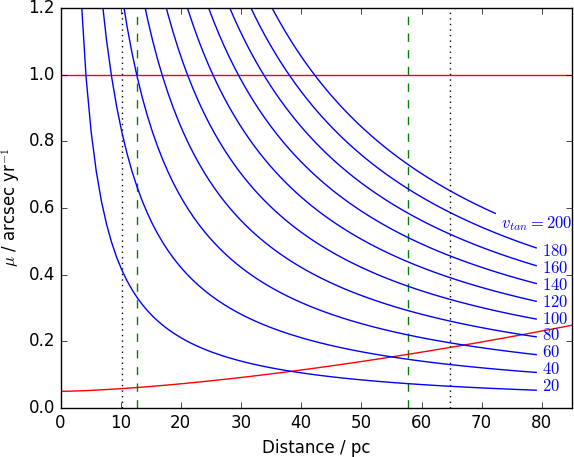}
\end{center}
\caption[The behaviours of tangential velocity and proper motion limits in the $\mu-D$ space]{This plot illustrates how tangential velocity and proper motion limits behave in the $\mu-D$ space. The solid lines are the contours of the tangential velocities from $20$ to $200\kmps$ in steps of $20\kmps$. The photometric distance limits mark the range of distances in which an object can be placed and stay within the detection limits. The dashed green lines are the distance limits calculated from Equations \ref{eq:dmin} and \ref{eq:dmax}, which is by fixing the tangential velocity of an object such that a proper motion limit correspond to a fixed distance limit. The dotted lines are the distance limits calculated from Equations \ref{eq:dmin_phot} and \ref{eq:dmax_phot}, where the distance limits are not functions of the kinematics. The red lines are the upper and lower proper motion limits; in the latter case, this limit rises as astrometric errors rise for increasingly faint flux levels as distance increases}
\label{fig:vtan_mu}
\end{figure}

\begin{figure}
\begin{center}
\includegraphics[width=0.9\textwidth]{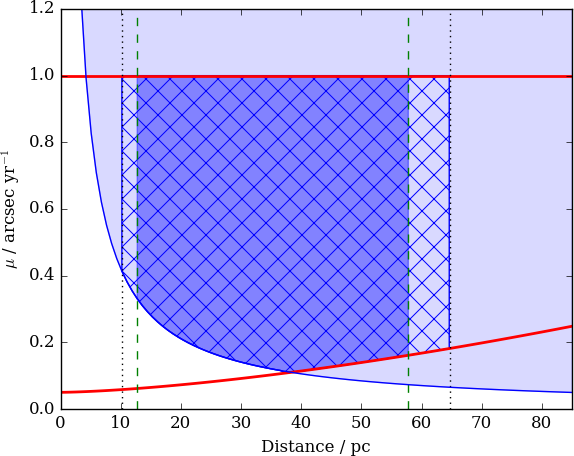}
\caption[The parameter space in which the discovery fraction, the generalised volume and the modified volume integrated over]{The area in light shade of blue~(which also includes the hatched and areas in dark shade of blue) is the parameter space where the discovery fraction is calculated when the proper motion limits and distance limits are not considered~(e.g. H06 and RH11). The area in dark shade of blue corresponds to the generalised volume where the distance limits~(Eq. \ref{eq:dmin} and \ref{eq:dmax}) are considered, it is clear that there are inconsistencies between the two parameter space. The hatched area is the parameter space where both the discovery fraction and the volume are integrated over in the modified method~(Please note that in this case the distance limits are found from Eq.~\ref{eq:dmin_phot} and \ref{eq:dmax_phot}). The weighted area gives the discovery fraction in the new method~(see Figure \ref{fig:discovery_fraction} for the weight maps). A treatment without considering the effects of the proper motion limits and distance limits on the parameter space in which the discovery fraction is integrated over would always overestimates the completeness, which translates to an underestimation in the number density}
\label{fig:different_discovery_fraction}
\end{center}
\end{figure}

\begin{figure}
\begin{center}
\includegraphics[width=0.75\textwidth]{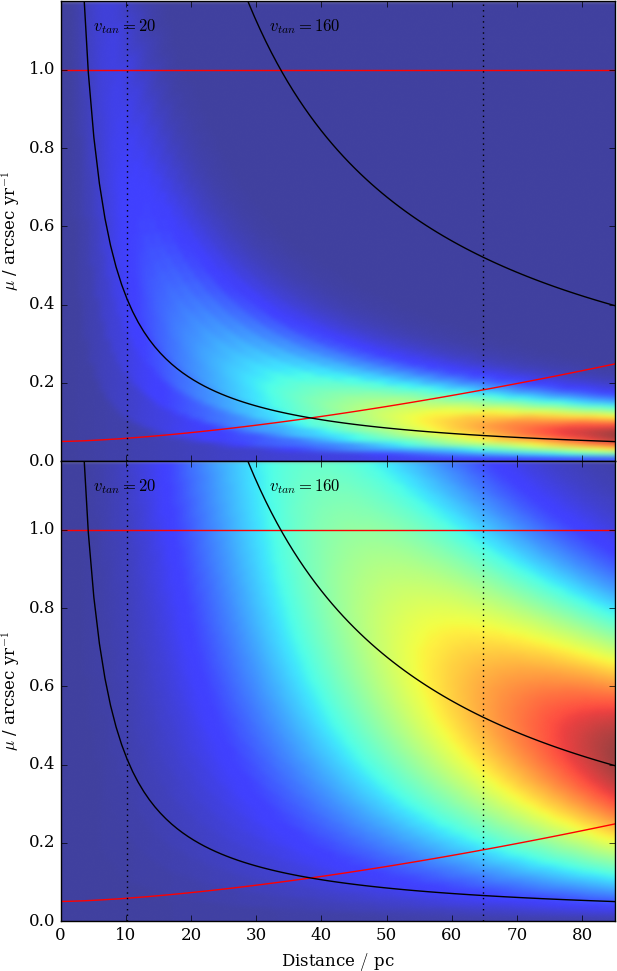}
\end{center}
\caption[The probability distribution as a function of proper motion and distance of a volume limited thin disc and halo population]{The top panel shows the probability distribution as a function of proper motion and distance of a volume limited thin disc population at a given luminosity in the direction of $(\alpha,\delta)=(0.0,0.0)$ over-plotted with the proper motions, tangential velocities and distance limits. The discovery fraction is the weighted area within the specified limits where the probability distribution function is used as the weights. The lower panel shows those of a halo population. The heaviest part of the thin disc weight function is excluded by the proper motion limit, but much more of the heavily weighted region of the stellar halo is within the allowed parameter space, so the tangential velocity limit at $\sim20\kmps$ has a much smaller effect in the thin disc than in the halo}
\label{fig:discovery_fraction}
\end{figure}

\begin{figure}
\begin{center}
\includegraphics[width=0.8\textwidth]{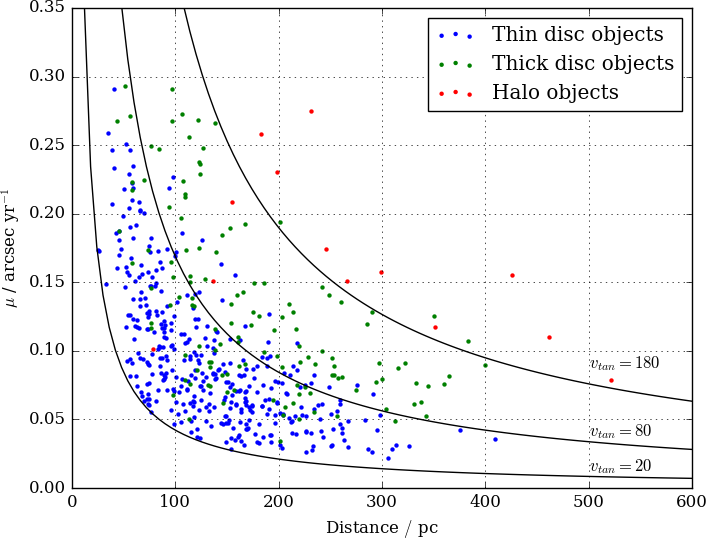}
\end{center}
\caption[Distribution of thin disc, thick disc and halo objects in proper motion and distance space]{Distribution of thin disc~(red), thick disc~(green) and halo~(blue) objects from Chapter \ref{sec:galaxy} in proper motion and distance space in the direction of $(\alpha,\delta)=(0.0,0.0)$ for an area of $30^{\circ} \times 30^{\circ}$. The $80$ and $180\kmps$ lines are the velocity limits adapted for this and next chapter in separating the disc and halo objects. At these velocity limits, the three components can be separated relatively cleanly, although putting a gap between $60$ and $100\kmps$ may suppress the confusion between thin and thick discs objects, likewise at $160-200\kmps$ for the thick disc and halo objects. It may be possible to classify objects probabilistically based on their tangential velocities}
\label{fig:model_dist_pm}
\end{figure}

Proper motions, tangential velocities and distances are related by a simple equation
\begin{equation}
\label{equation:vtan_mu_d}
v_{\mathrm{tan}} \approx 4.74\,\mu\,D.
\end{equation}
All the survey limits can be shown in the proper motion-distance space~(Fig.~\ref{fig:vtan_mu}). A valid approach should not impose any assumptions that restrict an object in this two dimensional space. The discovery fraction is equal to the weighted area restricted by the survey limits, where the weight map~(Fig.~\ref{fig:discovery_fraction}, and the true distribution of objects from the simulation in Fig.~\ref{fig:model_dist_pm} when the survey observability is taken into account) is generated from dividing the tangential velocity distribution by the distances,
\begin{equation}
\mathcal{W}(\mu,D)=\frac{P(v_{\mathrm{tan}})}{4.74 \times D}.
\end{equation}
In the generalised method, $v_{\mathrm{tan}}$ is fixed and is thus separable from $\mu$, which follow the lines of constant tangential velocity at $v_{\mathrm{tan}}=20-200\kmps$ in $20\kmps$ interval in Fig.\,\ref{fig:vtan_mu}. However, the two variables are related through distance as shown in Equation\,\ref{equation:vtan_mu_d}. This implies a selection criterion in one parameter would lead to a selection effect in the other. The generalised method only deals with the tangential velocity limits but ignores the consequential effect of the change in the proper motion limits bounding the discovery fractions. A similar effect also appears in the treatment of the distance limits. The volume integral is bounded by the maximum and minimum distances in which an object can be found but the discovery fraction in RH11 includes everything above the line $v_{\mathrm{tan}}=20\kmps$ as represented by the light blue area in Fig.\,\ref{fig:different_discovery_fraction}. The true discovery fraction should be bounded by the same limits as applied to the volume integral (hatched area). Thus, $\chi_{v}$ is always overestimated which translates to an underestimation in the luminosity function. This effect is stronger for:
\begin{itemize}
\item[i)] a survey with small upper proper motion limit because the lower this limit, the larger the overestimation of the discovery fraction. This can be seen from Fig. \ref{fig:different_discovery_fraction} where a smaller upper proper motion limit would lead to a smaller cross-hatched area, which means the discovery fraction would be overestimated.
\item[ii)] a population with large differences in the kinematics compared to the observer because objects tend to have large proper motions. This shifts the region with the highest probability in $\mathcal{W}(\mu,D)$ (Fig.~\ref{fig:discovery_fraction}) to larger proper motions. In the case of the Galaxy, the stellar halo is the most susceptible to this effect.
\item[iii)] intrinsically faint objects that carry small maximum observable distance. It is clear from Fig.\,\ref{fig:discovery_fraction} that the relative $\mathcal{W}(\mu,D)$ at low proper motion increases with distance when the $P(v_{\mathrm{tan}})/D$ peaks at smaller proper motion as distance increases. This affects the discovery fraction more severely when the lower proper motion limit is large since more heavy-weighted area would be included in, for example, both the H06 and RH11 methods.
\end{itemize}
\indent
As an example of a current survey which can be employed in WDLF studies \citep{2013ASPC..469..253H}, the Pan--STARRS upper proper motion limit is $\mathcal{O}(1)''$\,yr$^{-1}$ which means the first problem may affect the analysis of WDLF without proper treatment. Halo WDs have higher velocities and are older~(i.e.~fainter) than those of the discs and hence the effect on the halo population would be much larger than that in the discs. As it is of great interest to probe the halo WDLF at such low luminosity to explore the possible scenarios of the star formation history of the Galaxy, it is necessary to correct this bias.

\subsection{A New Approach}
\label{sec:new_approach}
In order to compute the discovery fraction properly, the parameter space in which an object could be observed by the survey has to be identical to that used in the discovery fraction integral. For each step of the numerical integration, it is necessary to calculate the instantaneous discovery fraction which is limited by the upper and lower proper motion limits, as well as the tangential velocity limits. It is worth mentioning that in the effective volume method in RH11 a similar approach was adopted that would have corrected for the bias noted in the last Section although the bias was not explicitly identified and discussed in that work. Instead of dealing in an object by object basis, their correction was applied statistically. The strength of that method is that the WDLFs of the three components could be untangled. However, binning objects by their bolometric magnitudes before membership association would lead to a loss of information. Furthermore, it loses the generality so that it cannot be applied to other luminosity estimators. In order to keep the analysis in an object by object basis, one should consider the {\it modified volume} integral. It is computed by integrating the physical survey volume along the line of sight, and considering at each distance step both the stellar density profile and the fraction of objects that pass the tangential velocity limits implied by the proper motion limits. The total generalised survey volume between $d_{\text{min}}$ and $d_{\text{max}}$ is therefore calculated by
\begin{equation}
\label{eq:volume_new}
V_{\mathrm{gen}}(r^\prime) = \Omega \int_{d_{\mathrm{min}}}^{d_{\mathrm{max}}} \frac{\rho(r)}{\rho_{\odot}} r^2 \left[ \int_{a(r)}^{b(r)} P(v_{\mathrm{tan}}) dv_{\mathrm{tan}} \right] dr,
\end{equation}
where the distance limits are solely determined by the photometric limits of the survey
\begin{equation}
\label{eq:dmin_phot}
d_{\mathrm{min}} = d \times \mathrm{max} \left[ 10^{\frac{(m_{\mathrm{min},i}-M_{i})}{5}} \right]
\end{equation}
\begin{equation}
\label{eq:dmax_phot}
d_{\mathrm{max}} = d \times \mathrm{min} \left[ 10^{\frac{(m_{\mathrm{max},i}-M_{i})}{5}} \right] ,
\end{equation}
$P(v_{\mathrm{tan}})$ is the tangential velocity distribution, $\rho(r)$ corrects for a non-uniform population density profile (i.e. disc populations), and $\Omega$ is the survey footprint area in steradians. Note that $P(v_{\mathrm{tan}})$ and $\rho(r)$ depend on the line of sight, so this model only holds for small fields. The limits on the tangential velocity integral are
\begin{align}
a(r)& = \mathrm{max}(v_{\mathrm{min}}, 4.74\mu_{\mathrm{min}}r)\\
b(r)& = \mathrm{min}(v_{\mathrm{max}}, 4.74\mu_{\mathrm{max}}r),
\end{align}
where $v_{\mathrm{min}}$ and $v_{\mathrm{max}}$ are any fixed tangential velocity limits applied to reduce contamination from other stellar populations, and $4.74\mu_{\mathrm{min}}r$ and $4.74\mu_{\mathrm{max}}r$ are the tangential velocity limits at distance $r$ arising from the proper motion limits. The appropriate limits on the integral are found by considering both of these effects.

The new volume integral has the distance limits decoupled from the kinematics, which are completely absorbed into the discovery fraction. The decoupling simultaneously means that regardless of how the kinematic behaviour changes with respect to the direction of observation, the entire $\mu-D-v_{\mathrm{tan}}$ parameter space in which an object can be found is explored. When the discovery fraction has a dependence on distance, it cannot be separated from the integrand. This means that the discovery fraction varies from object to object in any given direction, so it is sufficiently general to take a more realistic form of velocity distribution. This in turn allows for a distance dependent tangential velocity distribution that describes the Galaxy more realistically.

\subsection*{Survey volume generalised for kinematic selection}
It is instructive to examine how the survey volume as a function of distance is changed when the full effects of kinematic selection are taken into account, i.e. considering that proper motion limits result in implicit tangential velocity limits that vary as a function of distance. The generalised 1/V$_{\mathrm{max}}$ method does not consider this, and computes a constant discovery fraction only from any fixed, external tangential velocity limits that are applied e.g. to reduce contamination from certain types of object.

\begin{figure}
\begin{center}
\includegraphics[width=0.8\textwidth]{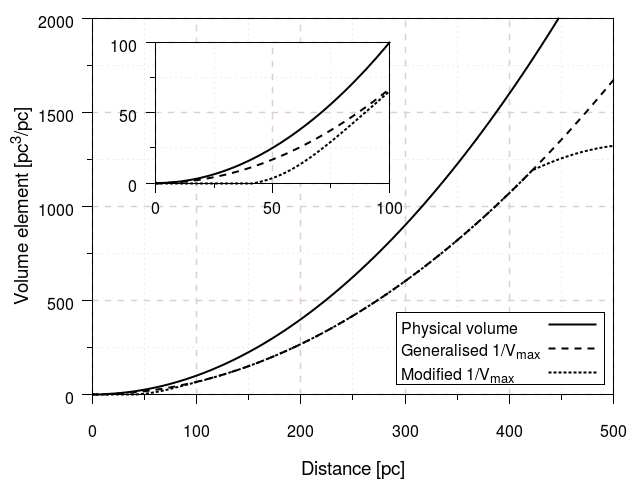}
\includegraphics[width=0.8\textwidth]{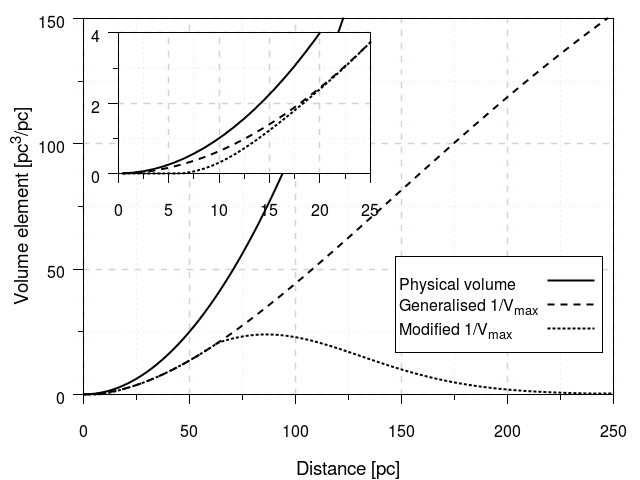}
\end{center}
\caption[The differential volume of the physical volume, generalised volume and modified volume as a function of line of sight distance for the thin disc and stellar halo]{Generalised survey volume as a function of line of sight distance, for the stellar halo~(top) and thin disc~(bottom). The lines represent the true physical volume~(solid), the generalised volume~(dashed), and the modified volume~(dotted). The functions diverge at large distances where $4.74\mu_{\mathrm{min}}$r $>$ $v_{\mathrm{min}}$~(see text) and the discovery fraction used by the generalised 1/V$_{\mathrm{max}}$ method breaks down. A similar effect operates at small distances~(inset), where the upper proper motion limit results in a small upper tangential velocity limit that excludes a large fraction of the population}
\label{fig:differential_volume}
\end{figure}

The differential survey volume as a function of distance, for the stellar halo and the thin disc, is presented in Fig.~\ref{fig:differential_volume}. These plots are computed for a line of sight towards the NGP, in order to exaggerate the effect of the non-uniform population density profile in the disc case, which falls off most rapidly in directions perpendicular to the plane. Halo and thin disc kinematics follow those in Table~\ref{table:galactic_kinematics}, with lower/upper proper motion limits of $0.1/1.0''$\,yr$^{-1}$ and a survey solid angle of 0.01\,sr.

In the case of the halo, a fixed lower tangential velocity limit of $v_{\mathrm{min}}=200\kmps$ has been applied, which is the usual way of reducing contamination from disc WDs. $v_{\mathrm{max}}$ has been left unconstrained; this corresponds to a constant discovery fraction of $\sim0.67$, which is the value used in the generalised 1/V$_{\mathrm{max}}$ method leading to a generalised volume that is a constant fraction of the physical volume.

In the case of the disc, a fixed lower limit of $v_{\mathrm{min}}=30\kmps$ has been applied, which is used to reduce contamination from high velocity sds. $v_{\mathrm{max}}$ is again unconstrained, leading to a constant discovery fraction of $\sim0.66$ used by the generalised 1/V$_{\mathrm{max}}$ method, although note that in the disc case the generalised volume diverges from the true physical volume due to the non-uniform population density profile.

In both cases, there is a range of intermediate distances over which the generalised and modified 1/V$_{\mathrm{max}}$ methods give the same result for the survey volume. This corresponds to the range over which the fixed external tangential velocity limits are active. Significant differences between the two methods arise at large distances, where the lower tangential velocity limit implied by the lower proper motion limit exceeds the fixed external limit. It is at these distances that the generalised 1/V$_{\mathrm{max}}$ method, which fails to consider this effect, overestimates the generalised survey volume and thus underestimates the number density. A similar effect arises at small distances~(inset), where the upper tangential velocity limit implied by the upper proper motion limit is greatly reduced, causing most of the population to be excluded. The generalised 1/V$_{\mathrm{max}}$ method, which considers only the fixed threshold $v_{\mathrm{max}}$, fails to account for this effect and again overestimates the survey volume.

\section{Monte Carlo Simulation}
\label{sec:mcsimulation}
\begin{figure}
\begin{center}
\includegraphics[width=0.85\textwidth]{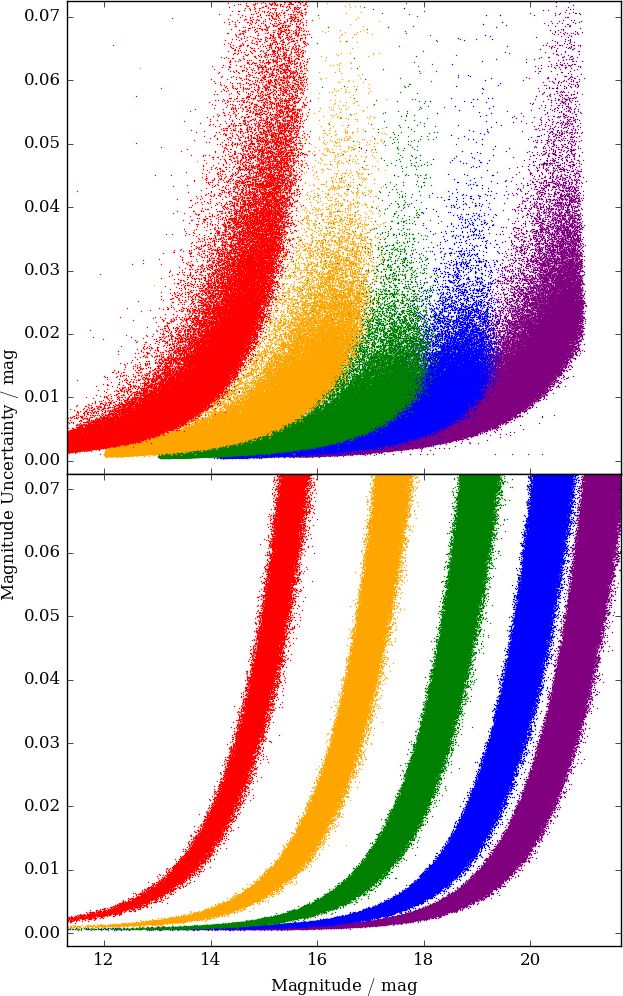}
\caption[The magnitudes and the associated uncertainties of point sources with $5\sigma$ proper motions in the PV1.1 of PS1 and in the simulations]{Top: the magnitudes and the associated uncertainties of point sources with $5\sigma$ proper motions in the PV1.1 of PS1, these measurements have not included the systematic uncertainties of $0.015$\,mag: from right to left, g$_{\mathrm{P}1}$~(purple), r$_{\mathrm{P}1}$~(blue), i$_{\mathrm{P}1}$~(green), z$_{\mathrm{P}1}$~(yellow) and y$_{\mathrm{P}1}$~(red) filters. Each successive filter is offset by one magnitude to the bright end. Bottom: the magnitudes and uncertainties distribution of WDs reproduced in the simulation}
\label{fig:simulated_noise}
\end{center}
\end{figure}

Monte Carlo simulations are used to produce snapshots of WD-only solar neighbourhoods which carry six dimensional phase space information. The volume probed in this work is assumed to be small such that the simulation is done in a Cartesian space, instead of a plane polar system centred at the Galactic Centre\,(GC). The Galaxy is further assumed to have three distinct kinematic components: a thin disc, a thick disc and a stellar halo, all of which have no density variations along the co-planar direction of the Galactic plane. All vertical structures follow exponential profiles, with scale height $H$. The velocity components, $\mathcal{U}$, $\mathcal{V}$ and $\mathcal{W}$, of each object are drawn from the Gaussian distributions constructed from the measured means and standard deviations of the three sets of kinematics that describes the three populations in the solar neighbourhood. The thin and thick disc populations are assigned with constant star formation rates since look back time, $\tau = 8\gyr$ and $\tau=10\gyr$ respectively while the halo has a star burst of duration $1\gyr$ at $\tau=12.5\gyr$. Theoretical WDLFs are constructed using the basic configuration as described in Chapter~\ref{sec:theoretical_wdlf} which can be used as the PDF in the Monte Carlo simulation. The normalisations of the PDFs are adopted from the WD densities found in RH11. The volume in which objects are distributed is limited to half a magnitude deeper than the maximum distance at which the survey can probe given its brightness. The half magnitude is to allow for random fluctuations near the detection limit after noise is added. The input parameters are assumed to be invariant with time and are summarised in Table~\ref{table:galactic_kinematics}.\\
\begin{table}
\centering
\caption{Physical properties of the Galaxy used in the Monte Carlo simulation}
\label{table:galactic_kinematics}
\begin{tabular}{ l *{3}{c} }
\hline
\hline
Parameter & Thin Disc & Thick Disc & Stellar Halo\\
\hline
$\langle\mathcal{U}\rangle$/km s$^{-1}$ & -8.62$^{a}$ & -11.0$^{d}$ & -26.0$^{d}$ \\
$\langle\mathcal{V}\rangle$/km s$^{-1}$ & -20.04$^{a}$ & -42.0$^{d}$ & -199.0$^{d}$ \\
$\langle\mathcal{W}\rangle$/km s$^{-1}$ & -7.10$^{a}$ & -12.0$^{d}$ & -12.0$^{d}$ \\
$\sigma_{\mathcal{U}}\kmps$ & 32.4$^{a}$ & 50.0$^{d}$ & 141.0$^{d}$ \\
$\sigma_{\mathcal{V}}\kmps$ & 23.0$^{a}$ & 56.0$^{d}$ & 106.0$^{d}$ \\
$\sigma_{\mathcal{W}}\kmps$ & 18.1$^{a}$ & 34.0$^{d}$ & 94.0$^{d}$ \\
H/pc & 250$^{b}$ & 780$^{e}$ & $\infty$ \\
n/pc$^{-3}$ & 0.00310$^{c}$& 0.00064$^{c}$& 0.00019$^{c}$ \\
\hline
\end{tabular}
\begin{enumerate}
\itemsep0em
\item[a)] \citealt{2009AJ....137..266F}
\item[b)] \citealt{1998A&A...333..106M}
\item[c)] RH11
\item[d)] \citealt{2000AJ....119.2843C}
\item[e)] \citealt{2006AJ....132.1768G}
\end{enumerate}
\end{table}
\indent
From the true distance and true bolometric magnitude drawn from the PDF, the true apparent magnitudes in the PS1 g$_{\mathrm{P}1}$, r$_{\mathrm{P}1}$, i$_{\mathrm{P}1}$, z$_{\mathrm{P}1}$ and y$_{\mathrm{P}1}$ filters are assigned \citep{2012ApJ...750...99T, 2012ApJ...756..158S, 2013ApJS..205...20M}. The uncertainties in those filters, $\sigma_{m_{i}}$, are assumed to scale exponentially with magnitude and  are described by
\begin{equation}
\label{eq:uncertainties}
\sigma_{m_{i}} = a_{i} \times \exp^{(m_{i}-15.0)} + b_{i}
\end{equation}
where $a_{i}$ and $b_{i}$ are constants measured from the PV 1.1 of PS1 and $m_{i}$ is the magnitude in filter $i$ (Fig. \ref{fig:simulated_noise}). Realistic dispersion is added to the uncertainties by resampling $\sigma_{m_{i}}$ with a Gaussian distribution with standard deviations of $0.1\times\sigma_{m_{i}}$ centred at the noiseless $\sigma_{m_{i}}$. The magnitudes in each filter are then drawn from a Gaussian distribution with a standard deviation $\sigma_{m_{i}}$. The proper motion uncertainty is based on the r$_{\rm P1}$ magnitude.
\begin{table}
\centering
\caption{Parameters for the photometric noise used in the MCMC simulation}
\begin{tabular}{ *{3}{c} }
\hline
\hline
Filter or Proper Motion & $a_{i}/mag$ & $b_{i}/mag$ \\
\hline
g         & 0.000125 & 0.00065 \\
r         & 0.000120 & 0.00075 \\
i         & 0.000175 & 0.00065 \\
z         & 0.000325 & 0.00089 \\
y         & 0.000750 & 0.00120 \\
$\mu$ & 0.000300 & 0.00050 \\
\hline
\end{tabular}
\end{table}

\section{Application to White Dwarf Luminosity Functions}
We have used the simulation outlined in Chapter \ref{sec:mcsimulation} to generate mock Pan--STARRS catalogues to compare the differences in applying the RH11 generalised volume and the new modified volume described in this work. The bright limits in all filters are set at $15.00\magnitude$, while the faint limits in g$_{\rm P1}$, r$_{\rm P1}$, i$_{\rm P1}$, z$_{\rm P1}$ and y$_{\rm P1}$ filters are at $21.50$, $21.00$, $20.50$, $20.00$ and $20.00\magnitude$ respectively. The lower proper motion limit is defined as five times the proper motion uncertainty at the given magnitude as defined in Equation \ref{eq:uncertainties}. The upper proper motion limit is set at $0.3''$\,yr$^{-1}$ unless specified otherwise. Photometric parallaxes are found by fitting the magnitudes to the WD synthetic atmosphere models at fixed surface gravity $\log{g}=8.0$. A lower tangential velocity limit is set at $20\kmps$.

\subsection{Stellar halo}
\begin{figure}
\begin{center}
\includegraphics[width=0.9\textwidth]{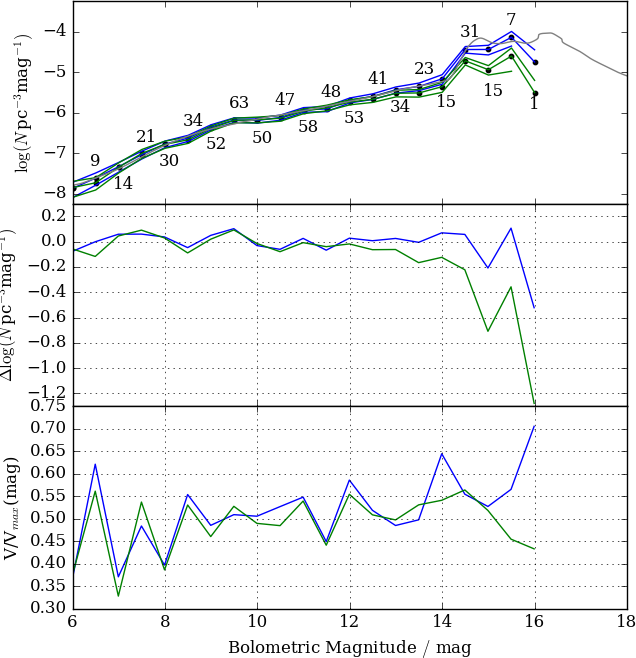}
\caption[WDLFs for a population with halo kinematics with generalised volume and modified volume methods]{Top: WDLFs for a population with halo kinematics with the measured halo WD density. The two WDLFs are derived from the new approach~(blue) and the RH11 method~(green) overplotted with the input true WDLF~(grey)}
\label{fig:wdlf_new_old}
\end{center}
\end{figure}

As discussed earlier, faint objects and objects from a population with large difference in the kinematics from the observers are most susceptible to underestimation of the number density when using the generalised volume technique. Thus, the largest systematic errors are expected to be found in the faint end of the stellar halo WDLF. In the top panel of Fig.\ref{fig:wdlf_new_old}, the differences in the WDLF constructed by the two methods are shown. The two LFs agree with each other up to M$_{\mathrm{bol}}\sim12.0\magnitude$. Beyond that, the generalised method consistently underestimates the number density and the deviation increases as the objects get fainter. The maximum difference is more than 1.0\,dex. The g$_{\mathrm{P}1}$ reaches $\sim400$\,nm at the blue edge, so it is expected that the photometric parallax solutions become unreliable at the bright end, while with small numbers of objects in the faint end the statistical noise is significant. Most of the $\langle$V/V$_{\mathrm{max}}\rangle$s in both cases are within $1\sigma$ from 0.5, which is a necessary condition for an unbiased sample.

\subsection{Stellar Halo with 100 times the Observed Density}
\label{sec:stellar_halo_100}
\begin{figure}
\begin{center}
\includegraphics[width=0.9\textwidth]{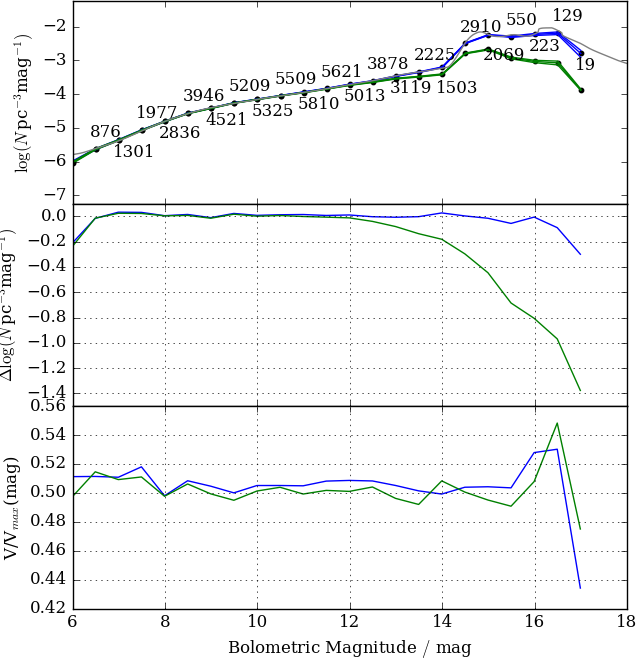}
\caption[WDLFs for a population with halo kinematics with generalised volume and modified volume methods for different upper tangential limits]{Same as Fig.\ref{fig:wdlf_new_old} except the WD density is artificially inflated by 100 times to reduce random noise. With the increase number density, the WDLF departs a magnitude fainter in both cases. The $\langle$V/V$_{\mathrm{max}} \rangle$ of the old method is still fluctuating about 0.5 even when the WDLF departs further and further away from the true LF. In the new method, the $\langle$V/V$_{\mathrm{max}} \rangle$ is consistent within statistical uncertainties}
\label{fig:wdlf_100_new_old}
\end{center}
\end{figure}

The stellar halo has a very low number density and it is well known that maximum volume estimators are prone to systematic bias with small number statistics. Thus, we have generated a stellar halo with 100 times the observed density to reduce such uncertainties. The WDLFs produced are more easily compared when the systematic uncertainties are much smaller and the corresponding plots are shown in Fig.~\ref{fig:wdlf_100_new_old}. The two LFs agree with each other up to M$_{\mathrm{bol}}\sim12.0\magnitude$ again. With 100 times the density, faint objects down to M$_{\mathrm{bol}}\sim17.0\magnitude$~($T_{\mathrm{eff}} \sim 3,000$\,K) can be generated easily. The LF produced with the new method agrees with the input LF to luminosities down to M$_{\mathrm{bol}}\sim16.0$\, while the RH11 method consistently underestimates the density, as predicted in Chapter \ref{sec:closer_look}. The $\langle$V/V$_{\mathrm{max}}\rangle$ shows an interesting behaviour - the distribution is fluctuating about 0.5 regardless of the bias in the WDLF. It implies that $\langle$V/V$_{\mathrm{max}}\rangle \approx 0.5$ is a necessary but not sufficient condition for both bias and completeness.

\subsection{Different upper proper motion limits}
\begin{figure}
\begin{center}
\includegraphics[width=0.8\textwidth]{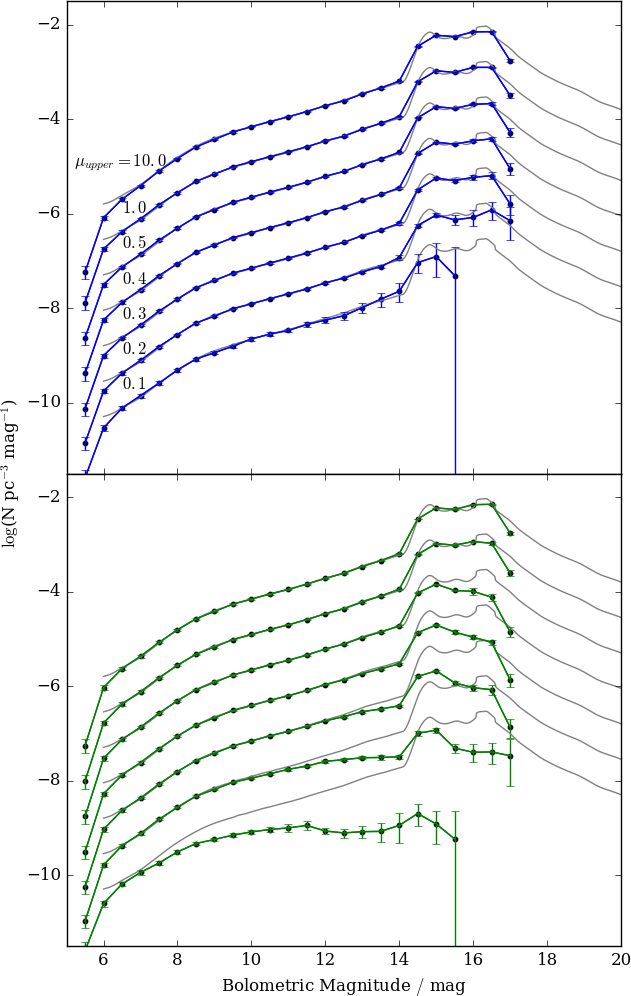}
\caption[WFLDs constructed with different upper proper motion limits with the modified volume method]{Top: WFLD constructed with different upper proper motion limits at 100 times the measured density with the new modified volume method. From top to bottom, the upper proper motion limit is set at $10.0, 1.0, 0.5, 0.4, 0.3, 0.2$ and $0.1''$\,yr$^{-1}$. Input WDLFs are shown in grey. Bottom: same procedure repeated with the generalised volume}
\label{fig:different_upper_limits}
\end{center}
\end{figure}

As mentioned in Chapter~\ref{sec:closer_look}, a small upper proper motion limit would lead to a strong bias. This is illustrated in Fig\,\ref{fig:different_upper_limits}: as the upper proper motion limit increases, the old and new approaches converge. However, when the limit decreases, the underestimation in the number density increases when adopting the old method. From the bottom panel, even at an upper limit of $0.5''$\,yr$^{-1}$, the generalised method fails to recover the WDLF beyond $\sim15.0\magnitude$. With the new method, the number density estimation is recovered under any restricted proper motion selection within the statistical noise. In the modern and future surveys with the rapid photometric systems, the upper proper motion limit would be in the order of arcseconds so this effect would be small. However, when the pairing of the bright objects includes older photographic plate data, the upper proper motion limit would be severely hampered due to a large maximum epoch difference (e.g.~\citealp{2003MNRAS.344..583D} has an upper limit of $0.18''$\,yr$^{-1}$ from pairing between POSS-I, POSS-II and SDSS). Another case is when samples at different velocity ranges are analysed separately (e.g.~the effective volume method in RH11 where the discovery fraction was treated correctly) the restricted velocity range would amplify the shortcoming of the old method.

\subsection{Thin Disc, Thick disc and Stellar Halo}
\label{sec:galaxy}
\begin{figure}
\begin{center}
\includegraphics[width=0.9\textwidth]{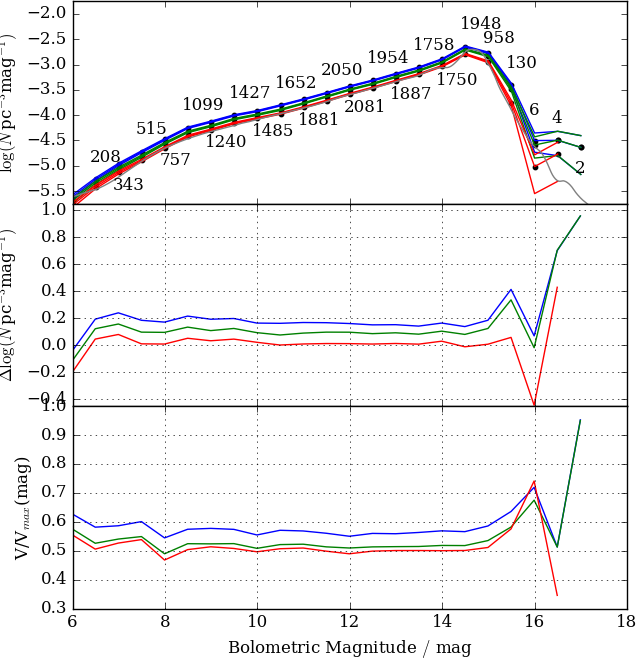}
\caption[WDLFs for a combined thin disc, thick disc and halo analysed with thin disc kinematics with generalised volume and modified volume methods]{Top: WDLFs of the Galaxy constructed by assuming negligible halo contamination~(blue), with an additional fixed upper tangential velocity limit at $80\kmps$~(green) and by picking up only objects tagged as ``thin disc object'' from the simulation~(red). Middle: the differences in the WDLF from the true LF as a function of magnitude. The overdensities are expected from the wrong association of kinematics model with the objects. Bottom: the $\langle$V/V$_{\mathrm{max}}\rangle$ distributions as a function of magnitude. $\langle$V/V$_{\mathrm{max}}\rangle$s are consistently larger than 0.5 implying objects tend to be found at larger distances than on average. This is expected from the objects with a uniform spatial density being associated with ones that have an exponential profile}
\label{fig:galaxy_upper_vtan}
\end{center}
\end{figure}

In the study of the Galactic WDLFs it is very difficult to assign population membership to individual objects. The consensus is that thin disc objects dominate the solar neighbourhood so that when studying the LF for the thin disc it is possible to assume a thin disc characteristic for all objects. This does not happen to be a good assumption as shown in the top panel of Fig. \ref{fig:galaxy_upper_vtan} where for the purposes of this exercise we apply only the new method. The solid line is the WDLF constructed from all stars regardless of the population. It is overestimated by $\sim0.2$\,dex at all magnitudes. An overestimation is expected but a consistent overdensity of $0.2$\,dex is not negligible. Due to the spatial density correction, the maximum volume of an object integrated over a disc profile has smaller volume than a halo profile. While objects are weighted by the inverse maximum volume, the apparently negligible contribution from the older populations would by amplified by density correction such that each contaminating object would have a volume underestimated by tens of percentage points. The discovery fraction for a disc population is always smaller than that of the halo, so this amplifies the 1/V$_{\mathrm{max}}$ by another few tens of percentage points. The V/V$_{\mathrm{max}}$ distributions are consistently at around $0.55$ as a consequence of the halo objects contribution. The ratio of the volume element between an exponential disc density profile and a uniform halo increases with distance. Therefore, for a group of uniformly distributed objects being assigned to follow an exponential profile in the integrals, the resulting $\langle$V/V$_{max}\rangle$ is expected to be larger than $0.5$. The dashed line is constructed with an addition of a fixed upper tangential velocity limit at $80\kmps$ which would eliminate most of the halo objects, but not the ones from the thick disc. This reduces the contamination down to roughly $0.1$\,dex at all magnitudes, with a smaller effect on the $\langle$V/V$_{\mathrm{max}}\rangle$ distribution. The dotted line is computed using only thin disc objects from that simulation for comparison. Everything agrees to within statistical uncertainties at all magnitudes in the thin disc-only analysis. Since the discovery fraction requires the kinematics of the populations, a good membership association is very important.

The aforementioned RH11 effective volume method in Chapter~\ref{sec:new_approach} was not dealt with in detail because of some fundamental differences in the method. That method can untangle the three components of the Galaxy. However, the binning of objects at an early stage means that it is not directly compatible with the framework in this work which, instead, bins objects in the final step. In dealing with observational data without the labels ``thin disc'', ``thick disc'' and ``halo'' tagged on the objects, the best one can do is to perform the membership assignment statistically. The effective volume method would provide a quick and easy way to do the job. However, if one would like to retain as much information as the maximum volume family would normally allow, the best way to do it is to find the probability of each object belonging to each of the components based on their observed and intrinsic properties, e.g. tangential velocities, metallicities, and/or distance from the Galactic Plane. Population membership assignment is beyond the scope of this work but if it were to be done in a maximum likelihood approach, readers are reminded that distributions are usually approximated as Gaussian functions. In the case of tangential velocity, the probability distribution function in fact follows a Rayleigh distribution in which case a Gaussian distribution cannot approximate the tail at high velocity well. It is important to assess whether this would lead to a bias in assigning objects to higher velocity populations. However, if one were to split the tangential velocity into ($l, b$) components, the distributions of the component velocities should be well approximated by Gaussian functions. As for the physical position, both the current distance from the plane and the maximum distance in which an object can reach based on the current kinematics, Z$_{\mathrm{max}}$, would be good parameters to test. Both suggested parameters have used the important piece of information that a halo object is much more likely to carry a larger velocity perpendicular to the plane than those of the discs.

\subsection{Different Galactic Models}
\begin{figure}
\begin{center}
\includegraphics[width=0.9\textwidth]{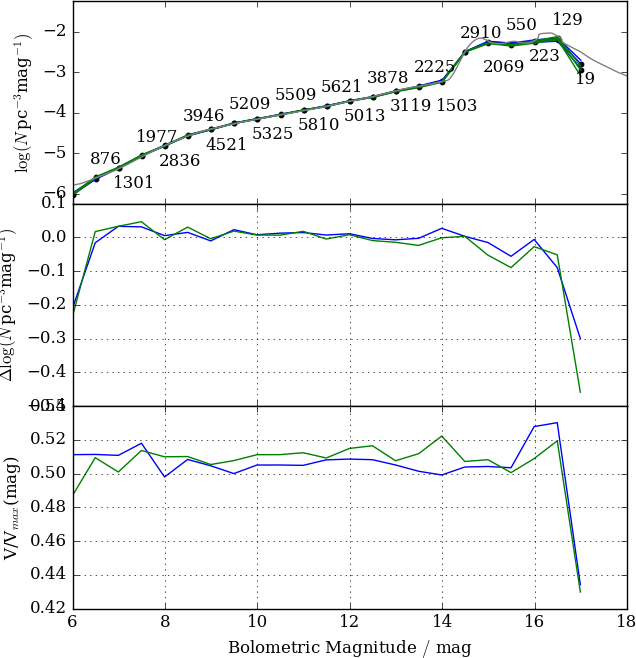}
\caption[WDLFs for the stellar halo with different input Galactic models]{Top: WDLFs for the stellar halo with different input Galactic models. The blue line is the original halo WDLF identical to the one in Fig\,\ref{fig:wdlf_100_new_old}, the green line is the one that follows a simplified Besan\c{c}on model but then analysed with the original set of kinematics. Middle and bottom panels are the same as before}
\label{fig:different_models}
\end{center}
\end{figure}

The accuracy of all LF estimators is sensitive to the assumed population kinematics and density profiles. Therefore it is important to investigate the effects of applying a different kinematic model in the analysis of a population simulated with another model. To demonstrate the effect, we have adopted a simplified version of the Besan\c{c}on galaxy model \citep{2003A&A...409..523R,2004A&A...416..157R}. A halo population at 100 times the measured density with $(\mathcal{U}, \mathcal{V}, \mathcal{W})=(0.0,-220.0,0.0)\kmps$ and velocity dispersion $(\sigma_{\mathcal{U}}, \sigma_{\mathcal{V}}, \sigma_{\mathcal{W}})=(131.0, 106.0, 85.0)\kmps$ was generated using the same recipe described in Chapter~\ref{sec:mcsimulation}. The halo population is then analysed using the original set of kinematic properties. The same WDLF from Chapter~\ref{sec:stellar_halo_100} is overplotted for comparison in Fig.~\ref{fig:different_models}. The two WDLFs agree at all magnitudes and even the trends in underdensities at faint magnitudes are very similar. The differences in the $\langle$V/V$_{\mathrm{max}}\rangle$ distribution at faint magnitudes can be explained by the same argument as in Chapter~\ref{sec:galaxy}. Due to the similar kinematic properties between the halo kinematic models applied in this work and the Besan\c{c}on model, the size of this effect is much smaller. From this simple experiment we conclude that small differences between the assumed and underlying kinematics results in little differences in both the derived WDLF and the $\langle$V/V$_{\mathrm{max}}\rangle$ distribution.

%% file: 5_wdlf_galaxy.tex
\chapter{White Dwarf Luminosity Functions of the Galaxy}
\label{ch:wdlf}

In this chapter, I will present the WDLFs from the 3SS showing the differences between the generalised and modified methods; and how interstellar reddening modifies the shape of the WDLFs. I refer the maximum volume method without taking into account of the proper motion and tangential velocity limits in the completeness correction as the \textit{generalised method}, V$_{\mathrm{gen}}$; the maximum volume with completeness corrected for the kinematic limits as the \textit{modified volume}, V$_{\mathrm{mod}}$; and the \textit{maximum volume}, V$_{\mathrm{max}}$, refers to the family of maximum volume density estimators~(i.e\ both methods).

\section{WDLFs with two Atmosphere Models}
\label{sec:two_atmosphere_models}
To limit contaminations, the maximum goodness-of-fit reduced chi-squared of the photometric parallax~($\chi^{2}_{\mu}$) is set at 10. The models are less reliable at low temperature, so for objects cooler than $5,000$\,K when colours start to diverge significantly, finder charts are inspected to remove spurious proper motion objects. Finder charts for objects with tangential velocity above $100\kmps$ are also inspected.

\begin{figure}
\includegraphics[width=0.9\textwidth]{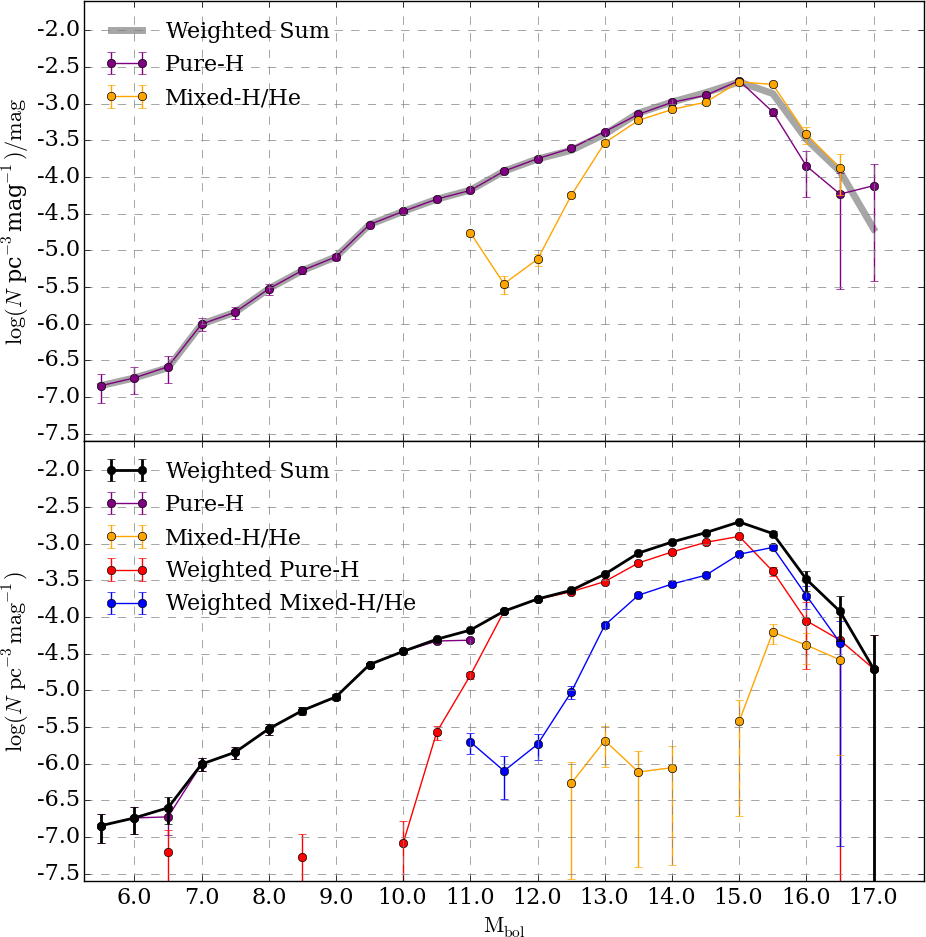}
\caption[Weighted WDLFs]{The top panel shows the WDLFs produced with the generalised volume method with $20\kmps<v_{\mathrm{tan}}<1,000\kmps$ when DA~(purple) and D-mixed~(yellow) are considered independently, the weight sum~(black line from the bottom panel) is plotted in grey for easier comparison; the bottom shows how the components contribute to the total WDLF}
\label{fig:da_dy_mixed}
\end{figure}

The mixed hydrogen-helium atmosphere model~(``D-mixed'') is only available below $12,000$\,K~($\sim11.5\magnitude$ depending on the atmospheric composition). Above this there is little difference in the models and only DA is considered in constructing the luminosity function. Because of the lack of available D-mixed models above $12,000$\,K, objects in the range of $11,500-12,000$\,K tend to have poor goodness of fit. To avoid systematic bias, the dividing temperature is set at $11,500$\,K. Objects are then divided into 3 groups and summed with appropriate weightings to give the total WDLFs :
\begin{itemize}
\item[i.] Best fit DA temperature above $11,500$\,K;
\item[ii.] Best fit D-mixed temperature below $11,500$\,K;
\item[iii.] Mutual to (i) and (ii), and everything else.
\end{itemize}
Objects in (i) and (ii) are unit-weighted but for those in (iii), they are weighted by the $\chi^{2}_{\mu}$ of the photometric parallax with the probability of the object having pure hydrogen atmosphere and mixed hydrogen-helium atmosphere being $P_{\mathcal{A}} \propto \exp(-0.5 \chi^{2}_{\mu,\mathcal{A}})$ and $P_{\mathcal{Y}} \propto \exp(-0.5 \chi^{2}_{\mu,\mathcal{Y}})$. The final weights are the ratio of the probability between the two. The total luminosity function is the weighted sum of the four component luminosity functions. In Fig.~\ref{fig:da_dy_mixed}, the top panel shows the WDLFs when DA~(purple) and D-mixed~(yellow) are considered independently, the weight sum~(black line from the bottom panel) is plotted in grey for easier comparison; the bottom shows how the components contribute to the total WDLF. From the top panel, it is obvious that the D-mixed is not reliable until $\sim13.0\magnitude$ and by comparing the DA with the weighted sum, it shows that D-mixed only dominates at  $\mathrm{M}_{\mathrm{bol}}>15.0\magnitude~(\sim 5,000\,\mathrm{K})$ when colours diverge significantly for different atmospheric composition~(See Appendix C). This method of weighting is used for both the generalised volume and modified volume method although it is only illustrated with the generalised method.

\section{The Solar Neighbourhood}
\begin{figure}
\includegraphics[width=0.75\textwidth]{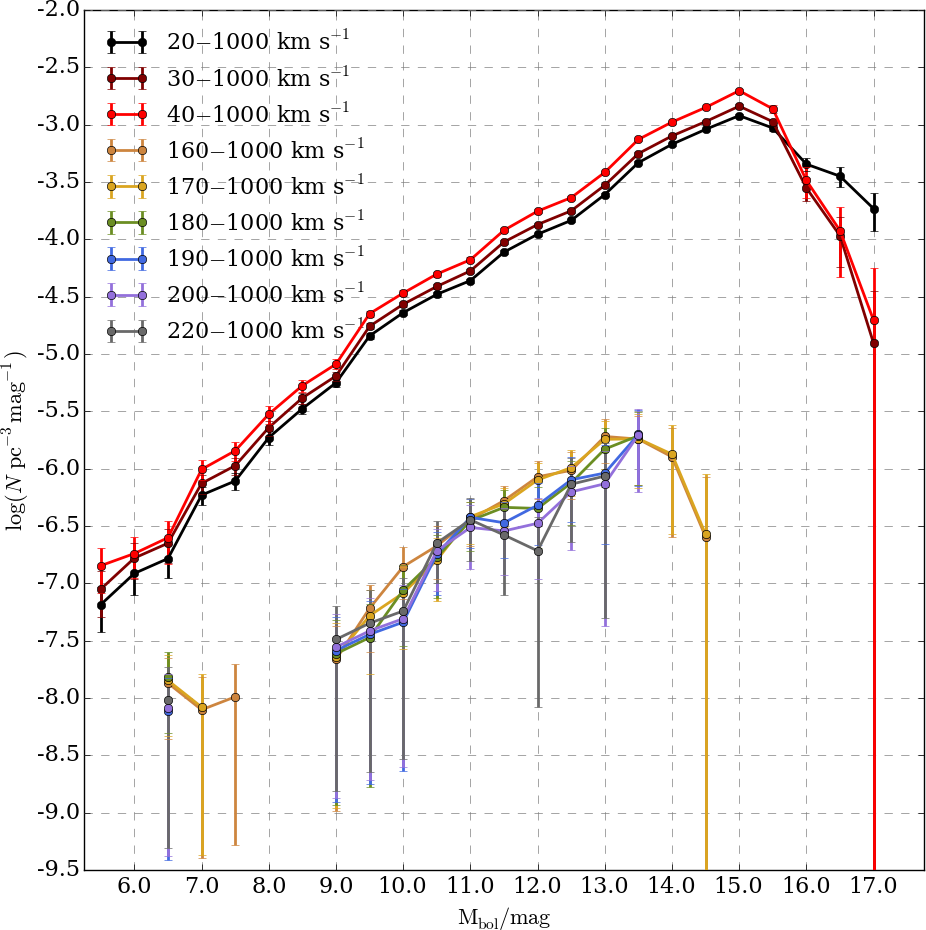}
\includegraphics[width=0.75\textwidth]{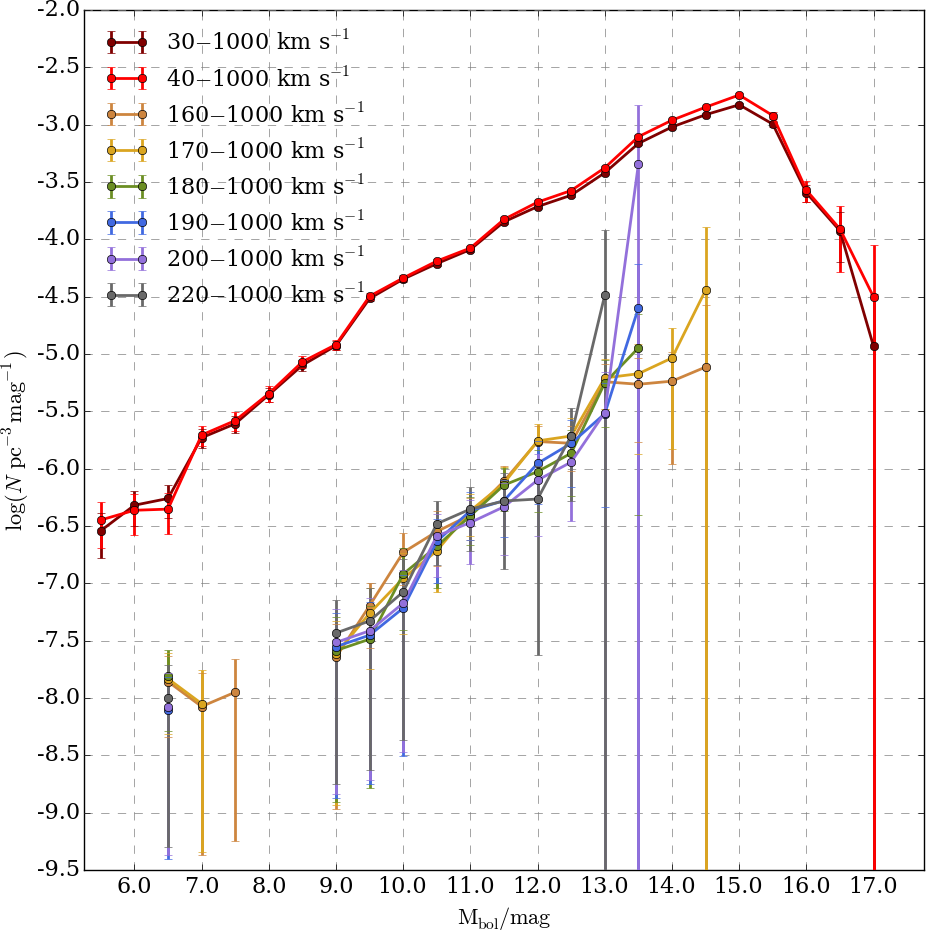}
\caption[WDLFs of the solar neighbourhood with the generalised and modified maximum volume density estimators]{Top: WDLFs of the solar neighbourhood with the generalised method. Bottom: WDLFs of the solar neighbourhood with the modified method}
\label{fig:wdlf_solar_neighbourhood}
\end{figure}

\begin{figure}
\includegraphics[width=0.8\textwidth]{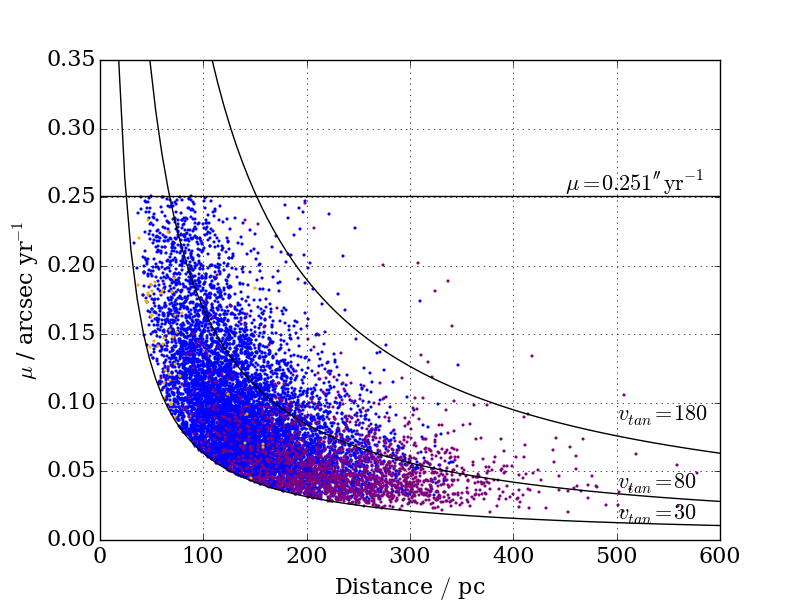}
\caption[Distribution of all objects in proper motion and distance space]{Distribution of all objects from the 3SS used in this chapter. Objects of type i, ii and iii (Chapter \ref{sec:two_atmosphere_models}) are in purple, yellow and blue respectively. The $80$ and $180\kmps$ lines are the velocity limits adapted the disc and halo objects. It is similar to the prediction from the Monte Carlo simulation~(Fig. \ref{fig:model_dist_pm})}
\label{fig:ps1_pm_dist}
\end{figure}

The total WD number density and LF of the solar neighbourhood have been traditionally studied by applying velocity limits at $20, 30 $ or $40\kmps$ depending on the quality of the astrometric solutions of the data. From the top panel of Fig.~\ref{fig:wdlf_solar_neighbourhood}, it shows that with a lower tangential velocity limit at $20\kmps$, there is significant contamination at the faintest magnitudes. Further analysis will exclude objects with tangential velocities below $30\kmps$. Brighter than $15.0\magnitude$, there is a roughly constant offset between the 3 WDLFs. The first instinct is to interpret it as contamination from the thick disc and halo: when more stringent cuts are applied to a sample, the proportional contributions from the older populations increases, as well as the completeness correction. Since the thin disc scale-height is only a fraction of those of the thick disc and halo, the maximum volume of an object corrected for the thin disc density profile is smaller than a volume that is corrected with larger scale-height. Thus, the inverse volume~(i.e.\ density) would be much larger. The combined effect is that the LF would be increasingly overestimated as the lower velocity limit increases. When the velocity limit is raised so much that the sample is no longer representative of the thin disc, the WDLF becomes completely meaningless, which would be the case if a $200\kmps$ limit is used in conjunction with a thin disc scale-height. However, this effect is in fact much smaller than expected as realised in the bottom panel of Fig.~\ref{fig:wdlf_solar_neighbourhood}. At the bright end, when the completeness correction is treated properly, the choice of the three velocities limits have small effect on the relative normalisations between the three. The absolute normalisation is, however, underestimated when the $\mathrm{V}_{\mathrm{gen}}$ is used instead of $\mathrm{V}_{\mathrm{mod}}$. This has the largest effect (1) at the bright end when objects can be seen from large distance where the proper motion uncertainties are not considered in the generalised method and (2) at the faint end when the approximation of constant tangential velocity is poor~(See Fig.~\ref{fig:different_discovery_fraction}). At $15.0\magnitude<\mathrm{M}_{\mathrm{bol}}<17.0\magnitude$, it is possible that the WDLF is dominated by thick disc WDs~(See Chapter~\ref{sec:normalisation_correction}). The total integrated number densities for the generalised volume method at the three velocity limits are $2.94\times10^{-3}, 3.18\times10^{-3}$ and $4.06\times10^{-3}$\,pc$^{-3}$; and the densities with $v_{\mathrm{tan}}>30$ and $40\kmps$ with the modified method are $3.48\times10^{-3}$ and $4.02\times10^{-3}$\,pc$^{-3}$ respectively. The $40\kmps$ samples agree with each other; with a $30\kmps$ selection, the number densities are similar between the two methods. The distribution of these objects in distance-proper motion space is shown in Fig.~\ref{fig:ps1_pm_dist}, which is similar to the prediction from the Monte Carlo simulation~(Fig. \ref{fig:model_dist_pm}).

The halo contribution to the solar neighbourhood is studied by applying high tangential velocity limits. The bright end of the LFs with both methods agree up to $13.5\magnitude$~(excluding the faintest bins with $200$ and $220\kmps$ selection, see below). At the faint end, the two methods estimate densities that are significantly different from each other: $\sim0.5$\,dex at M$_{\mathrm{bol}}=13.0\magnitude$ and $\sim2.0$\,dex at M$_{\mathrm{bol}}=15.0\magnitude$. The WDLFs with $v_{\mathrm{tan}}>200$ and $220\kmps$ have extremely large densities, but the small number of objects~(1 object at the respective faintest bin) gives rise to the huge uncertainties. At those bins, the densities are simply inflated by the input kinematics model. Similar effects happened at M$_{\mathrm{bol}}=14.0-14.5\magnitude$ for $v_{\mathrm{tan}}>160$ and $170\kmps$. Furthermore, a velocity cut at this level is likely to include the highest velocity WDs from the thick disc. Unfortunately, with the small upper proper motion limit in this work, there are few high velocity objects. The total integrated number densities for the generalised volume methods at the six velocity limits at $v_{\mathrm{tan}}>160, 170, 180, 190, 200$ and $220\kmps$ are $4.20\times10^{-6}, 4.13\times10^{-6}, 2.88\times10^{-6}, 2.60\times10^{-6}, 2.29\times10^{-6}$ and $1.39\times10^{-6}$\,pc$^{-3}$; and the corresponding densities with the modified volume method are $1.49\times10^{-5}, 3.17\times10^{-5}, 1.03\times10^{-5}, 1.62\times10^{-5}, 2.29\times10^{-4}$ and $1.83\times10^{-5}$\,pc$^{-3}$. 

\section{Velocity Limit Selection for the Thin Disc}
\begin{figure}
\includegraphics[width=0.9\textwidth]{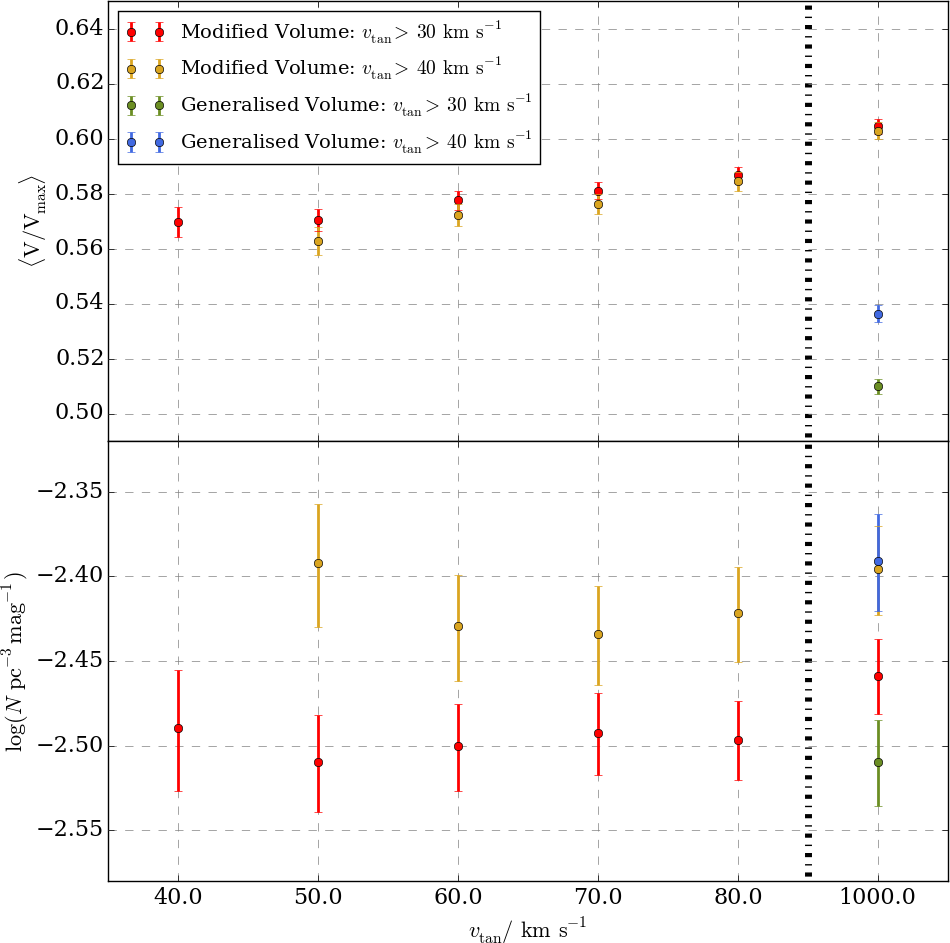}
\caption[The $\langle\frac{\mathrm{V}}{\mathrm{V}_{\mathrm{max}}}\rangle$s and number densities of the thin disc]{$\langle\frac{\mathrm{V}}{\mathrm{V}_{\mathrm{max}}}\rangle$ and the integrated number density as functions of upper tangential velocity limits with the thin disc scale-height correction}
\label{fig:vvmax_compared_thin_disc}
\end{figure}

In order to consider which velocity limits are the most appropriate for the respective samples, the $\langle\frac{\mathrm{V}}{\mathrm{V}_{\mathrm{max}}}\rangle$s and the number densities are plotted as a function of the upper tangential velocity limits. For comparison, the respective values with the generalised volume method with the lower limits at $30$ and $40\kmps$ are plotted on the same graph in Fig.~\ref{fig:vvmax_compared_thin_disc}. The number densities tend to be larger with more stringent velocity selections because the completeness correction amplified the contribution of contaminations from the older populations, even though they only contribute a minute fraction of the whole sample. These contaminants~(from populations with larger scale-heights) occupy space more uniformly compared to the thin disc, so when the upper velocity limit increases, more objects are likely to be found at large distance compared to a clean thin disc population. Hence, the $\langle\frac{\mathrm{V}}{\mathrm{V}_{\mathrm{max}}}\rangle$ increases with the upper limit.

From \ref{ch:vmax}, it is understood that with the inclusion of thick disc and halo objects, a velocity selection of $30\kmps<v_{\mathrm{tan}}<80\kmps$ would lead the $\langle\frac{\mathrm{V}}{\mathrm{V}_{\mathrm{max}}}\rangle$ to inflate to $\sim0.55$~(based on the RH11 number densities for the thin disc, thick disc and halo found with their effective volume method). This value depends on the input kinematics as well as the number density of the different populations. Only with more complex simulations or reliable population assignment~(which does not exist for white dwarfs to date) can this problem be tackled properly. The $\langle\frac{\mathrm{V}}{\mathrm{V}_{\mathrm{max}}}\rangle$s in this work are of similar size to the 0.55, and the number densities are slowly increasing as the upper tangential velocity limit increases from $60$ to $80$ to $1,000\kmps$, I deduce that a minimum of $50\kmps$ is needed for the $30\kmps$ sample, and a minimum of $70\kmps$ for the $40\kmps$ sample, where the number densities take the minimum values. When a $40\kmps$ limit is used, the number densities are larger than the $30\kmps$ counterparts at any choice of upper tangential velocity limits. This is due to the increased proportion of thick disc and halo objects as the most probable tangential velocity for a thin disc object is at $20-30\kmps$. For this reason, I believe $30\kmps$ is the more appropriate lower tangential velocity limit for the study of the thin disc.

\section{Velocity Limit Selection for the Halo}
\begin{figure}
\includegraphics[width=0.9\textwidth]{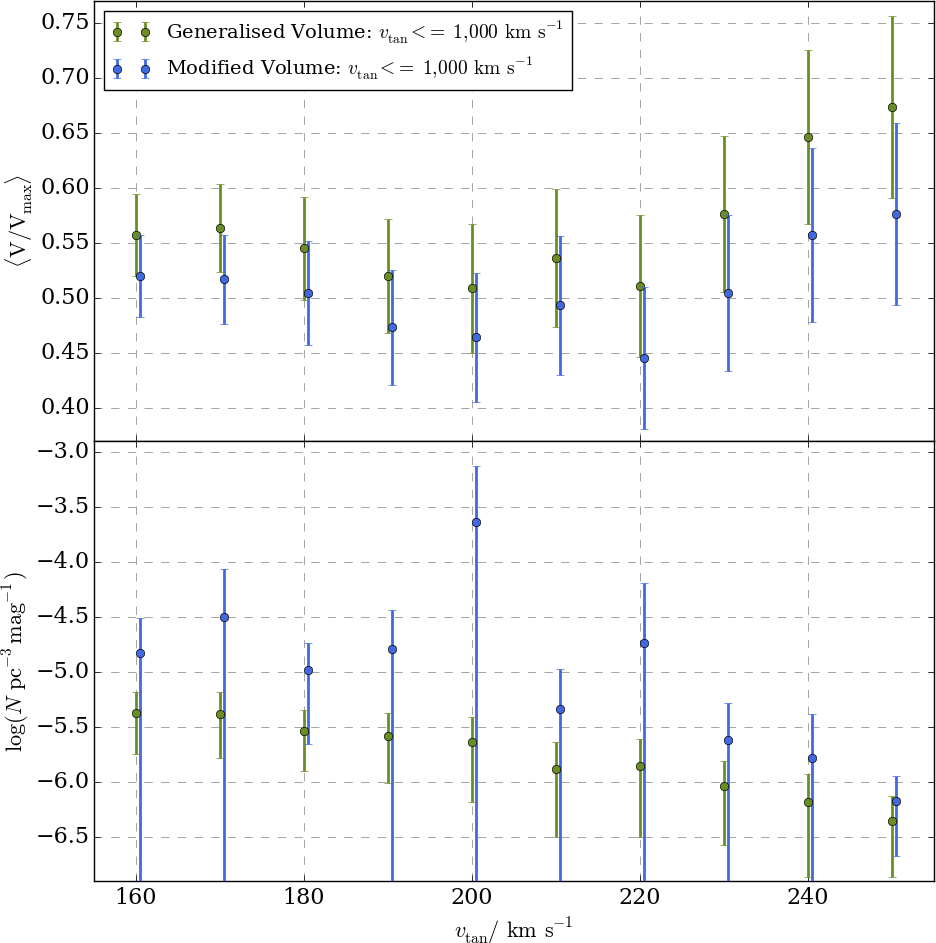}
\caption[The $\langle\frac{\mathrm{V}}{\mathrm{V}_{\mathrm{max}}}\rangle$s and number densities of the halo]{$\langle\frac{\mathrm{V}}{\mathrm{V}_{\mathrm{max}}}\rangle$ and the integrated number density as functions of lower tangential velocity limits with the halo scale-height correction}
\label{fig:vvmax_compared_halo}
\end{figure}

Since the halo sample is much smaller because of the large tangential velocity limits, it takes much less computation time so a wide range of values are investigated. However, having small numbers of objects mean there is large uncertainty. Despite the newly implemented completeness correction for the kinematic limits, the inherent limitation of the maximum volume density estimator is its poor estimation of the true density when the sample size is smaller than $\sim300$~\citep{2006MNRAS.369.1654G}. The $\langle\frac{\mathrm{V}}{\mathrm{V}_{\mathrm{max}}}\rangle$ fluctuate about $0.5$ with the $160$ to $240\kmps$ selection~(see Fig. \ref{fig:vvmax_compared_halo}). At $180\kmps$, $89\%$ of the sky have $<1\%$ of thick disc objects remaining according to the kinematic models I applied. At $160\kmps$, only $32\%$ of the sky has less than one percent of thick disc objects left. In comparison, at $190\kmps$, the maximum thick disc contribution is less than $0.8\%$. However, I believe at the faintest bins, the number densities with the $190$ and $200\kmps$ selections are inflated by the completeness corrections. If the velocity is set any higher, the number of objects remaining in the sample will become so low that most of the faint objects will no longer be present. Putting everything into consideration, the $180\kmps$ limit is the most representative for the halo population.

\section{WDLFs of the Thin Disc and the Halo}
The most representative velocity limits were identified in the last two sections. The thin disc is selected with tangential velocities in the range $v_{\mathrm{tan}}=30-80\kmps$ while the halo velocities are in the range $v_{\mathrm{tan}}=180-1,000\kmps$. The two WDLFs were integrated to give number densities of $3.188^{+0.171}_{-0.165}\times10^{-3}\pc^{-3}$ and $1.036^{+0.823}_{-0.813}\times10^{-5}\pc^{-3}$ respectively. The thin disc value is in good agreement with past works~(see Table~\ref{table:number_density}). The halo density is, however, difficult to compare with past works as values reported span three orders of magnitudes: it is within $1\sigma$ uncertainty of \citet{1989LNP...328...15L}, a quarter the size of that reported from H06, $10\%$ of what the effective volume method found from RH11; but larger than the RH11 halo density by a factor of $2.5$ when found with the generalised volume method.

\begin{figure}
\includegraphics[width=0.9\textwidth]{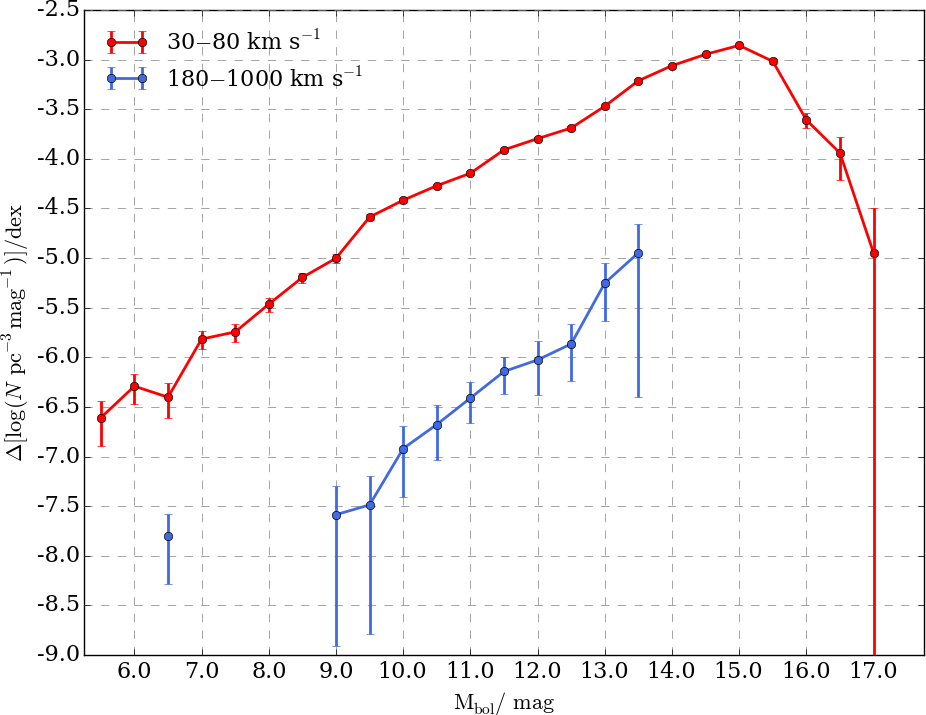}
\caption[WDLFs of the Thin Disc and the Halo from 3SS]{WDLFs of the thin disc~(red) and the halo~(blue) from the PS1 3SS, The two WDLFs were integrated to give number densities of $3.188^{+0.171}_{-0.165}\times10^{-3}\pc^{-3}$ and $1.036^{+0.823}_{-0.813}\times10^{-5}\pc^{-3}$ respectively}
\label{fig:wdlf_generalised_thin_disc_halo}
\end{figure}

\section{Interstellar reddening}
\label{sec:interstellar_reddening}
Hot WDs are blue objects, as they cool the colours turn red and become mostly neutral between $\mathrm{M}_{\mathrm{bol}}\sim12-16\magnitude$. When they become ultracool with $T_{\mathrm{eff}}<4,000$\,K, the H2CIA in the optical turns them blue again. Hence, the interstellar reddening has strongest effect towards hot WDs and the ultracool ones. For this reason, the total number density is a weak function of interstellar reddening as the number density is dominated by objects with neutral colours. The shape of the WDLF is, however, more strongly affected by the reddening at the bright end. Not only are these objects most affected by reddening because of their blue colours, they can be seen at much larger distances than the cooler counterparts so the light we see has gone through much more dust. On the other hand, the coolest objects can only be seen from a small distance, so in this magnitude-kinematics-limited sample, reddening has small effects on the WDLFs. This does not hold if one were to study a volume-limited sample. All in all, reddening has little effect on the study of the oldest objects of the Galaxy because of the observabilities, but at the bright end, which is used to constrain neutrino cooling and the possible WD cooling through axion emission,  the exclusion of extinction correction will lead to bias.

\begin{figure}
\includegraphics[width=0.9\textwidth]{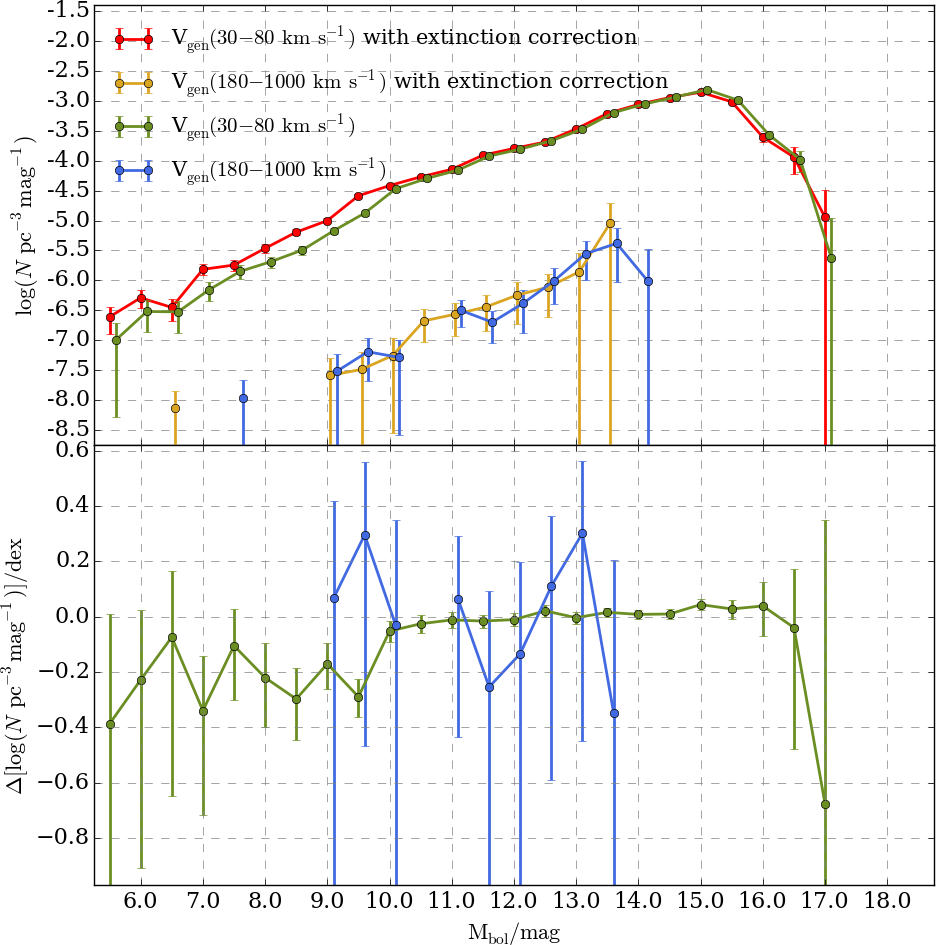}
\caption[Effects of interstellar reddening on WDLFs]{The effects of interstellar reddening on WDLFs are illustrated for the thin disc~(red and green) and the halo~(yellow and blue) populations. Without the proper treatment of extinction correction, the number densities are underestimated at the bright end and the faint end}
\label{fig:wdlf_reddening_compared}
\end{figure}

\section{Normalisation Correction and Star Formation Scenario}
\label{sec:normalisation_correction}
\begin{figure}
\includegraphics[width=0.9\textwidth]{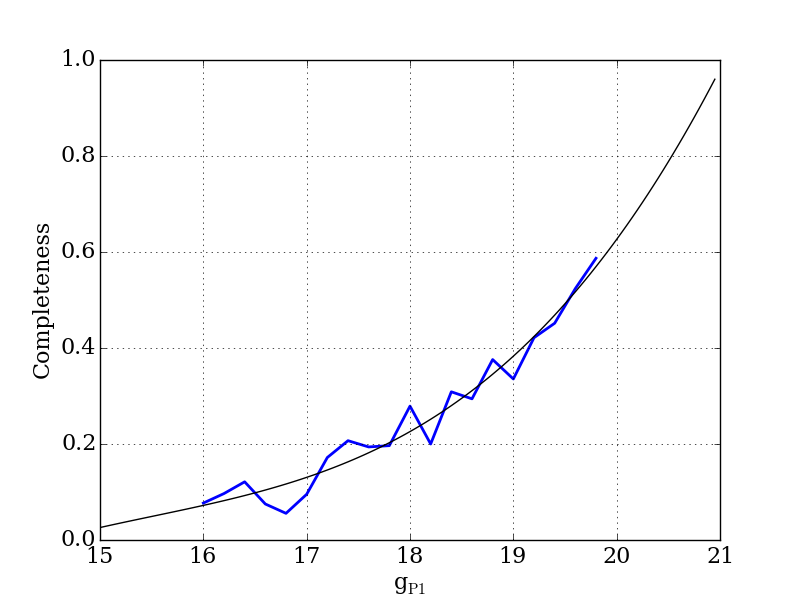}
\caption[Completeness correction as a function of g$_{P1}$]{The completeness correction for the WDLF as a function of g$_{P1}$. At the bright end, most objects are not recovered, while the completeness increases as the magnitude increases. The blue line is the measured completeness while the black line is the interpolated cubic-spline of the measurements}
\label{fig:wdlf_normalisation_correction}
\end{figure}

When the thin disc WDLF is overplotted with that from RH11, the WDLF I produced with the 3SS is underestimating the number density. Through pairing objects with the RH11 catalogue, it was found that apparently bright stars are less likely to be found from PS1. Most of the missing objects have a large number of epochs and good photometry, but manual astrometric fitting shows that they have poor astrometric solutions as indicated by the $\chi^{2}$. This is likely to be due to the effect of an underestimation of the photometric uncertainties. These uncertainties are used to weight the measurements for solving (1) astrometric solution and (2) photometric parallax. I require high signal to noise ratios in both to keep the sample clean. An underestimation in the photometric uncertainty will propagate through the data reduction pipeline and lead to large $\chi^{2}$ and hence the objects are removed. By correcting for the incompleteness with the g$_{P1}$ and B band from SSS because they have very similar response functions, the thin disc WDLF can be renormalised to match the RH11 WDLF. The high proper motion objects are used for matching two catalogues because the lower proper motion limits of the SSS are also very complex, so such selection can avoid complication with spurious matches and incompleteness in PS1 because of its higher proper motion uncertainties compared to SSS. The tangential velocities limits in both catalogues are set at $40$ and $1,000\kmps$, while the proper motions are in the range $100-250$ milliarcseconds per year. Objects within 15 degrees of the Galactic plane are disregarded. The completeness is then found as a function of g$_{P1}$~(Fig.~\ref{fig:wdlf_normalisation_correction}). The incompleteness is most severe at the bright end, while the completeness increases as the magnitude increases. The high correlation between apparent magnitude and completeness suggests that the effect is photometric rather than kinematic. One possible explanation of this incompleteness comes from the photometric reduction pipeline where the bright objects are known to carry unreliable photometry among the PS1 Science Consortium, thus the bright limits are usually set at $15\magnitude$ in all filters. It is possible that this effect is not completely accounted for above this magnitude limit and the effect gradually decrease as the magnitude increases. Another likely source of incompleteness is a bug in the proper motion recalculation.

The bright end is expected to be noisy as the completeness correction inflates the noise together with the signal; at the faint end, little correction is needed because most intrinsically faint objects are also apparently faint. The renormalised WDLF has a number density of $3.79\times10^{-3} \pc^{-3}$ which is very close to the weighted mean at $3.71\times10^{-3}\pc^{-3}$~(Chapter~\ref{sec:past_number_density}). It agrees with the RH11 WDLF up to M$_{\mathrm{bol}}=16.0\magnitude$; in the RH11, they found a plateau at fainter magnitudes. Such a feature is impossible for a population younger than $12\gyr$~(see Fig.~3.5) and only if it had much higher SFR at the earliest times. The most natural explanation is that those objects are either from the halo or they are not WDs. There is also a possibility of unknown WD physics and formation processes which could bring us a large number of ultracool WDs from the thin disc. The broad ``peak'' at M$_{\mathrm{bol}}=14.0-15.5\magnitude$ is not possible for any age of a population with constant, exponential or burst star formation history. For a WDLF to have the same number density at M$_{\mathrm{bol}}=14.0\magnitude$ and $15.5\magnitude$, the peak number density of the WDLF should be at least $0.5$~dex higher for a $10\gyr$ constant SFR, or $0.75$~dex higher for an exponential or burst SFH. In order to allow for such a broad feature in that magnitude range, the only possible explaination is a complex SFH. As a simple exercise, it can be easily demonstrated by superposing a $10\gyr$ exponential SFR population~(thick disc) over an $8\gyr$ constant SFR population~(thin disc), as motivated by earlier discussion in Chapter 1.1.1. In this simple case, the number density for the two populations are $2.54\times10^{-3}$ and $1.25\times10^{-3}\pc^{-3}$ respectively. This suggests a thin--thick disc number density ratio of $\sim1$ to $2.0$, which is much higher than what RH11 found from the effective volume method: $1$ to $4.8$. However, there are three factors that have to be taken into account: in RH11, their thick disc WDLF terminates at M$_{\mathrm{bol}}=15.25\magnitude$, but if their WDLF continues at the same density for another $0.5\magnitude$ bin, their ratio will go down to $1$ to $4.2$. The second point is the maximum volume density correction. The thick disc has a scale-height $\sim4$ times larger than that of the thin disc, so when a thick disc population is corrected for the density with thin disc density profile, the maximum volume would be smaller than its true value, hence the density is overestimated. Finally, the completeness correction based on thin disc kinematics would over-correct the number density. The tangential velocity distributions for the thick disc typically peak at $\sim60\kmps$ as compared to $\sim30\kmps$ for the thin disc. This means the correction factor with thin disc kinematics is smaller than the one that is appropriate for a thick disc object, and through the inverse volume weighting, the correction over-corrects the thick disc sample and provide a WDLF that is too large for a combined thin and thick disk sample.

\begin{figure}
\includegraphics[width=0.9\textwidth]{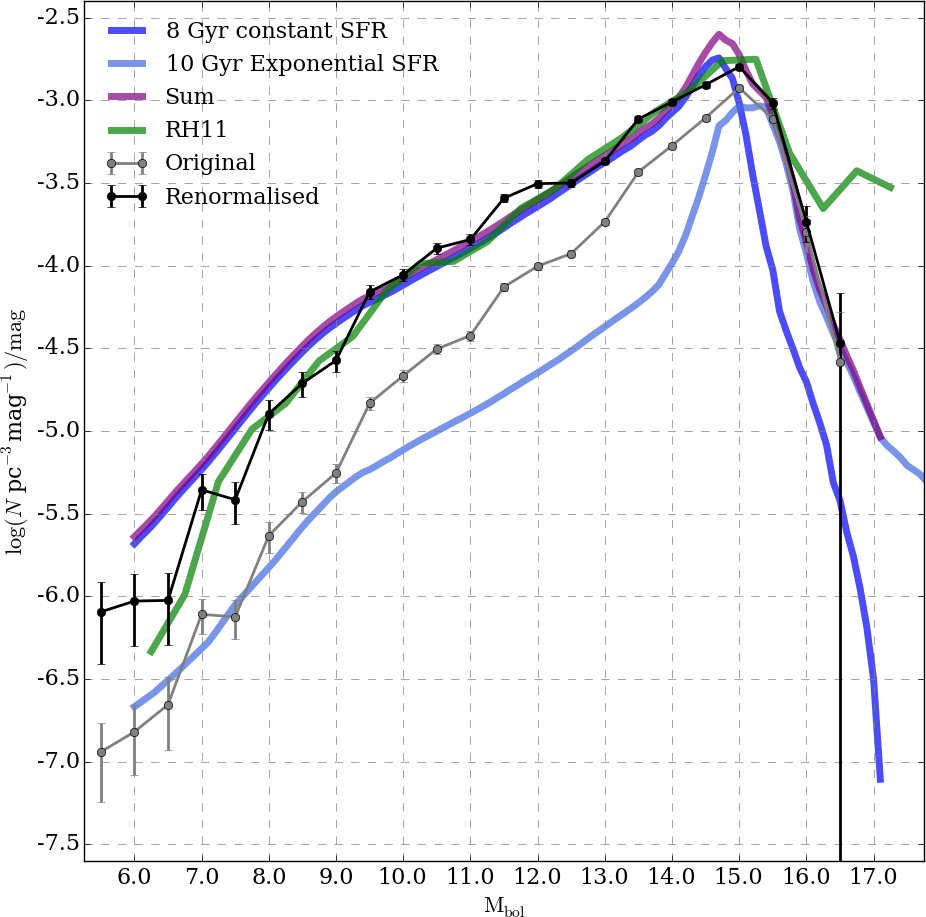}
\caption[WDLF normalisation correction]{The WDLF is renormalised as a function of apparent magnitude with the SSS data~(black). It is plotted against the WDLF from RH11 for comparison~(see text). The blue and purple lines are the theoretical WDLFs of an $8\gyr$ population with constant SFR, $10\gyr$ with exponential SFR and the two combined}
\label{fig:wdlf_normalisation}
\end{figure}

%% file: 6_conclusion.tex
\chapter{Conclusion}
In this thesis, I have (1)~compared the theoretical WDLFs with different input cooling models, IFMRs, SFHs and MS lifetimes for different initial metallicities; (2)~presented a new generalisation of the maximum volume density estimator over a tangential velocity and proper motion-limited sample; and (3)~applied the new density estimator, V$_{\mathrm{mod}}$, to the WD sample in the 3SS selected by colour and reduced proper motion and is compared this against the previous generation of estimator, V$_{\mathrm{gen}}$; the effect of interstellar reddening is also investigated.

\section{Theoretical White Dwarf Luminosity Functions}
I have compared the theoretical WDLFs for the first time by varying one parameter at a time. There is little variation in the shape of the WDLFs brighter than M$_{\mathrm{bol}}\sim14.0\magnitude$. The gradients from burst populations are larger than the ones with constant SFRs. However, most of the information is revealed in the faint end. The IFMRs and MS lifetimes have almost no effect until M$_{\mathrm{bol}}\sim15.0\magnitude$ at which the WDLF carries most information from the earliest epoch of star formation. They should be studied carefully for the WDLF inversion for the SFH with the future data~(e.g.\ PS1 PV3, Gaia and LSST) where objects fainter than M$_{\mathrm{bol}}\sim15.0\magnitude$ can be detected in bulk. Regarding different cooling models, they show large discrepancies in the faint end due to different treatments to the internal structures. At the end user level, there is little one can do apart from urging different groups to compare their models and work together~(e.g.\ \citealp{2013A&A...555A..96S}). The Montr\'{e}al model has the most updated atmosphere model but the least accurate description of the interior structure. The BaSTI and LPCODE models are the opposite; updates on the atmosphere models with different chemical compositions are needed at the faint end.

\section{Generalised Maximum Volume}
I have presented an improved treatment of the proper motion and tangential velocity limits to arrive at an unbiased luminosity function estimator for stellar populations selected on the basis of both magnitude and proper motion. My simulations have shown that the assumption of setting the tangential velocity of an object as its intrinsic property would be invalid when a sample is drawn from a survey that has restricted proper motion limits and for populations where the tangential velocity distribution function varies over the sky. In order to select high quality proper motion objects when analysing real survey data the lower proper motion limit and tangential velocity limit are usually set at such a level that the velocity space is very restricted. Therefore, it is necessary to correlate the tangential velocity with the proper motion depending on the properties of each individual object. Two biases were identified from dissecting the generalised V$_{\mathrm{max}}$ integral. The first one is an inconsistent density measurement of any non-uniform sample in which the density depends on the line of sight direction. The generalised method only considers the population spatial density and survey limits, so different densities would be measured at different line of sight directions as the distribution of proper motion and tangential velocity vary with line of sight. The other arises from the assumption that any given object would carry a constant tangential velocity. The choice of the upper proper motion limit has to be carefully determined from the properties of the survey. An arbitrarily large upper limit has negligible effects in the analysis as the survey volume is sensitive only to the lower proper motion limit. The discovery fraction is, however, sensitive to both upper and lower proper motion limits. From my simulations, it is found that surveys with upper proper motion limits smaller than $\sim0.5''$\,yr$^{-1}$ require more attention. This value seems small to the modern and future surveys but when a pairing is done with the early epoch Schmidt survey plate data for the bright objects for example the upper proper motion limits would shrink rapidly when the maximum epoch differences can be up to more than half a century (the pairing radius cannot be increased indefinitely when associating catalogues with very different depth and resolution without introducing many spurious matches).

Population membership association has to be done carefully because with a wrong set of kinematics the derived set of discovery fractions would become completely meaningless. In the case that a population completely dominates the density budget, for example, the thin disc objects in the solar neighbourhood, the wrong membership associations have small but non-negligible effects to the luminosity functions. Finally, I would like to stress the deductive meanings of $\langle$V/V$_{\mathrm{max}}\rangle$: (i)\,a flat V/V$_{\mathrm{max}}$ distribution with $\langle$V/V$_{\mathrm{max}}\rangle \sim 0.5$ only indicates that the sample is not biased towards distant or nearby objects, it gives no indication of the completeness of the sample. For a given flat distribution, one can conclude, at best, that the survey is equally sensitive to objects at any distance. However, (ii)\, given a complete unbiased sample, a flat distribution and $\langle$V/V$_{\mathrm{max}}\rangle \sim 0.5$ are expected.

\section{White Dwarf Luminosity Functions from the 3SS}
The most representative velocity limits were identified for the thin disc and the halo sample. The thin disc is selected with tangential velocities in the range $v_{\mathrm{tan}}=30-80\kmps$ while the halo velocities is in the range $v_{\mathrm{tan}}=180-1000\kmps$. The two WDLFs were integrated to give number densities of $3.188^{+0.171}_{-0.165}\times10^{-3}\pc^{-3}$ and $1.036^{+0.823}_{-0.813}\times10^{-5}\pc^{-3}$ respectively. The effect of interstellar reddening is investigated and demonstrated that it has small effect to the total number density of WDs, but it alters the shape of the WDLFs at the bright end and the extreme faint end. In the end, I illustrated that the overdensity of thin disc sample at the faint end can be explained by thick disc contamination. Combined theoretical WDLFs for a $10\gyr$ exponential SFR population~(thick disc) and an $8\gyr$ constant SFR population~(thin disc) can recover the shape of the WDLF. Although caution is needed for the respective number densities because an incorrect completeness correction can inflate the number density of the thick disc.

\section{Future Work}
\subsection*{Application to the PV3 of 3SS}
The final data release~(PV3) of PS1 will be available in late 2016. Once it is in place, the analysis can be applied directly onto the new data set. With the improved photometric and astrometric pipelines, major improvements are expected in both photometry and astrometry. With the improved photometric calibration, objects that were confused in the previous data release due to the buggy pipeline are likely to be recovered in PV3. This would allow the analysis to include lower Galactic latitude where the stellar density is much higher. High proper motion objects should be recovered more efficiently as well, which will lead to the discovery of more ultracool halo WDs.

\subsection*{Generalisation of other density estimators}
The treatment of detection limits due to proper motion selection have never been included in other density estimators. The modified Schmidt estimator is the only one that has been extended to incorporate such constraints. However, as detailed earlier, the treatment breaks down if we have restricted proper motion selection criteria. \citet{2006MNRAS.369.1654G} compared the properties of three different estimators: Generalised Schmidt, Stepwise Maximum Likelihood and Choloniewski method. It is clear that the treatment of the proper motion limits were only applied to the Schmidt estimator by considering the tangential velocity as an intrinsic property of an object, while the other two considered the case of a magnitude limited sample. Their analysis may be affected by the modifications to the density estimator made in this work because the sample adopted in their work has assumed a fixed tangential velocity along any line of sight. However, their work focused on the thin disc which is much less susceptible to the bias we identified, and by adopting a large upper proper motion limit in their data selection~($\mu \in \left[\,0.16,2.00\,\right]$), the effect should be tiny. Nonetheless I would like to remind readers to pay attention to the faint end of their analysis which is most affected if the bias is noticeable at all. It should be possible to generalise the discovery fraction (as a completeness correction arising from tangential velocity and/or proper motion limit) to other density estimators as the proper motion is now decoupled from the detection limit.

\subsection*{More detailed Galactic models for the generalisation of the density estimator}
The current Schwarzschild distribution assumes a Cartesian system centred at the Sun. This approximation is good for surveys that probe only small distances~(a few hundred parsecs). However, in the future surveys, most notably the Gaia and LSST, where the lower proper motion limits at the detection limits would be as low as $0.2$\,mas\,yr$^{-1}$~(G$_{\mathrm{Gaia}}\sim20\magnitude$), and $1$\,mas\,yr$^{-1}$~(r$_{\mathrm{LSST}}\sim24\magnitude$) respectively, an object with a tangential velocity of $20$\,km\,s$^{-1}$ could be detected with $5\sigma$ confidence at $\sim17$\,kpc and $\sim3.5$\,kpc respectively. A WD with $T_{\mathrm{eff}}\sim3000$\,K and M$_{\mathrm{bol}}\approx16\magnitude$, which translate to G$_{\text{Gaia}}\sim15.8\magnitude$ and g$_{\text{LSST}}\sim17.3\magnitude$ at $10$\,pc, would be detectable at $\sim1-2$\,kpc in both cases. The overdensity at the spiral arms is not accounted for in any work to date because the current surveys have not reached such distances where the nearest spiral arm lies. However, when the faintest objects can be detected at a distance of over a thousand parsecs, the brighter objects would lie easily inside the spiral arm region which is $\sim1$\,kpc from the Sun. In such a situation it is necessary to employ a more detailed density profile to account for the varying density in not just the vertical direction but also in the planar directions. With the decoupling of the proper motion limit from the photometric limits, the incorporation of a more sophisticated model would simply be a translation of the Galactic density and velocity profiles from the galactocentric frame to the geocentric frame. The key difference is that a Schwarzschild distribution function gives a constant tangential velocity distribution function as a function of the line of sight\,(i.e.\ $\chi = \chi(\alpha,\delta)$), but in the adoption of a complex model, the density and velocity distribution functions have to be found as functions of both distance and the direction of line of sight such that $\chi = \chi(\alpha,\delta,D)$.

%% file: appendix_A.tex
\chapter{Minimising Absolute Deviation}
Consider a function $\sum_{i} | c_{i} - c' |$ which has to be minimised by choosing the best $c'$ in the set of $c_{i}$. By setting the derivative of the function to zero,
\begin{equation}
\label{eq:mad_median}
\frac{\partial \sum_{i} | c_{i} - c' |}{\partial c'} = -\sum_{i} \frac{\partial | c_{i} - c' |}{\partial (c_{i}-c')}
= -\sum_{i} \frac{ c_{i} - c'}{ |c_{i}-c'|}
= 0
\end{equation}
as the derivative of the absolute value of a function can be given by
\begin{equation}
\frac{\partial |x|}{\partial x} = \frac{x}{|x|} 
  \begin{cases} 
   -1 & \text{if } x < 0 \\
   1   & \text{if } x > 0.
  \end{cases}
\end{equation}

In order for Eq. \ref{eq:mad_median} to be true, $c'$ has to be the median such that half the $c_{i}-c'$ will be equal to $-1$ and the other half will be equal to $1$. 

%% file: appendix_B.tex
\chapter{Tangential Velocity Distribution}

The following description follows the work of \citet{1987AJ.....93..864J,1983veas.book.....M}. For a population of stars, the peculiar velocities can be described by a single velocity ellipsoid. The peculiar velocity, $v$, of a given star can be described by a deviation from the mean, $\bar v$, of a residual velocity, $\eta$,:
\begin{equation}
\label{eq:peculiarvelocity}
v = \bar v + \eta
\end{equation}

It is assumed that the residual velocities follow the Schwarzschild distribution
\begin{equation}
S(\eta) = \left(\frac{1}{2\pi}\right)^{\frac{3}{2}} \left(\frac{1}{| \boldsymbol\Sigma |}\right)^{\frac{1}{2}} \exp(- \frac{1}{2}\, \underline{\eta}^{T}\,\boldsymbol\Sigma^{-1}\underline{\eta})
\end{equation}

where {$\boldsymbol\Sigma$} is the covariance matrix for the components of the residual velocity. Since the velocity ellipsoid of the Galactic components are always presented in the galactic frame, the most convenient set of triad of unit vectors points to the Galactic Centre, $\mathcal{U}$; in the direction perpendicular to $\mathcal{U}$ in the Galactic plane, $\mathcal{V}$; and in the direction of the North Galactic Pole~(NGP), $\mathcal{W}$. They form the Galactic triad \textbf{G}. Under the assumption of no vertex deviation, the covariance matrix can be written as
\begin{equation}
\boldsymbol\Sigma = \textbf{G}
\left(
\begin{matrix}
  \sigma_{\mathcal{U}}^{2} & 0 & 0 \\
  0 & \sigma_{\mathcal{V}}^{2} & 0 \\
  0 & 0 & \sigma_{\mathcal{W}}^{2}
\end{matrix}
\right)
\textbf{G}^{\mathrm{T}}.
\end{equation}

In the case of $\sigma_{U}^{2}=\sigma_{V}^{2}=\sigma_{W}^{2}$, the distribution is reduced to the Maxwell-Boltzmann distribution.

Since the distribution is a function of position, every object follows a different distribution. The natural solution is to express $\underline{\eta}$ and $\boldsymbol{\Sigma}$ in terms of the normal triad \textbf{R}, with $p$, $q$ and $r$ pointing in the direction of the right ascension, declination and line of sight respectively,
\begin{equation}
\underline{\eta} = \textbf{R}
\left(
\begin{matrix}
  \eta_{p}\\
  \eta_{q}\\
  \eta_{r}
\end{matrix}
\right) \mathrm{,}
\end{equation}
\begin{equation}
\boldsymbol\Sigma = \textbf{R}
\left(
\begin{matrix}
  \mathcal{C}_{pp} & \mathcal{C}_{pq} & \mathcal{C}_{pr} \\
  \mathcal{C}_{qp} & \mathcal{C}_{qq} & \mathcal{C}_{qr} \\
  \mathcal{C}_{rp} & \mathcal{C}_{rq} & \mathcal{C}_{rr}
\end{matrix}
\right)
\textbf{R}^{\mathrm{T}}
\end{equation}

and
\begin{equation}
\boldsymbol\Sigma^{-1} = \textbf{R}
\left(
\begin{matrix}
  c_{pp} & c_{pq} & c_{pr} \\
  c_{qp} & c_{qq} & c_{qr} \\
  c_{rp} & c_{rq} & c_{rr}
\end{matrix}
\right)
\textbf{R}^{\mathrm{T}}.
\end{equation}

Both $\boldsymbol\Sigma$ and $\boldsymbol\Sigma^{-1}$ are symmetric in the equatorial triad so the determinant $\mid \boldsymbol\Sigma \mid$ is invariant under any transformation,
$\boldsymbol\Sigma = \sigma^{2}_{\mathcal{U}}\sigma^{2}_{\mathcal{V}}\sigma^{2}_{\mathcal{W}}$.

The Schwarzschild distribution can then be written explicitly as
\begin{align*}
S(\eta) &=
\left(\frac{1}{2\pi}\right)^{\frac{3}{2}}
\left(\frac{1}{| \boldsymbol\Sigma |}\right)^{\frac{1}{2}}
\exp\left[- \frac{1}{2}
\left(
\begin{matrix}
  \eta_{p} & \eta_{q} & \eta_{r}
\end{matrix}
\right)
\left(
\begin{matrix}
  c_{pp} & c_{pq} & c_{pr} \\
  c_{qp} & c_{qq} & c_{qr} \\
  c_{rp} & c_{rq} & c_{rr}
\end{matrix}
\right)
\left(
\begin{matrix}
  \eta_{p}\\
  \eta_{q}\\
  \eta_{r}
\end{matrix}
\right)
\right]\\
&=  \left(\frac{1}{2\pi}\right)^{\frac{3}{2}}
 \left( \frac{1}{\mathcal{C}_{pp}\mathcal{C}_{qq}-\mathcal{C}_{pq}^{2}} \right)^{\frac{1}{2}} \times\\
&
 \exp\left[
 - \frac{\mathcal{C}_{qq}\eta_{p}^{2}+\mathcal{C}_{pp}\eta_{q}^{2}-2\mathcal{C}_{pq}\eta_{p}\eta_{q}}{2(\mathcal{C}^{2}_{pq}-\mathcal{C}_{pp}\times\mathcal{C}_{qq})} -\frac{c_{rr}}{2}\left( \eta_{r} + \frac{c_{qr}\eta_{q} + c_{pr}\eta_{p}}{c_{rr}} \right)^{2}
 \right]
\end{align*}

By integrating over the radial velocity $\eta_{r}$, and writing the determinant of the covariant matrix explicitly with the components, one can obtain the velocity ellipse in the tangent plane of observation
\begin{equation}
S_{pq} = \frac{1}{2\pi}
\left( \frac{1}{\mathcal{C}_{pp}\mathcal{C}_{qq}-\mathcal{C}_{pq}^{2}} \right)^{\frac{1}{2}}
\exp\left[ - \frac{\mathcal{C}_{qq}\eta_{p}^{2}+\mathcal{C}_{pp}\eta_{q}^{2}-2\mathcal{C}_{pq}\eta_{p}\eta_{q}}{2(\mathcal{C}^{2}_{pq}-\mathcal{C}_{pp}\times\mathcal{C}_{qq})} \right]
\end{equation}

Consider the solar neighbourhood, the observed velocity of a star can be expressed as $v = \eta - v_{\odot}$ using equation (\ref{eq:peculiarvelocity}), where $v_{\odot}$ is the solar motion. Expressing $v$ in terms of tangential velocity $v_{\mathrm{tan}}$ and position angle $\theta$, the velocity in the direction of right ascension and declination can be rewritten as
\begin{align*}
\eta_{p} &= v_{\mathrm{tan}}\sin\theta + v_{\odot, p}\\
\eta_{q} &= v_{\mathrm{tan}}\cos\theta + v_{\odot, q} .
\end{align*}

The distribution over $v_{\mathrm{tan}}$ can be obtained by marginalising over the position angle, $\theta$, giving the tangential velocity distribution
\begin{equation}
P(v_{\mathrm{tan}}) = \int^{2\pi}_{0} v_{\mathrm{tan}}S_{pq}(v_{\mathrm{tan}},\theta)\,d\theta.
\end{equation}

This integral is in general non-analytical and needs to be done numerically. To determine $S_{pq}$, the components $\mathcal{C}_{pp}$, $\mathcal{C}_{qq}$ and $\mathcal{C}_{pq}$ of $\boldsymbol\Sigma$ and the solar motion components $v_{\odot, p}$ and $v_{\odot, q}$ are required. The components $\mathcal{C}_{ij}$ of $\boldsymbol\Sigma$ are found from the following coordinate transformation,
\begin{equation}
\left(
\begin{matrix}
  \mathcal{C}_{pp} & \mathcal{C}_{pq} & \mathcal{C}_{pr} \\
  \mathcal{C}_{qp} & \mathcal{C}_{qq} & \mathcal{C}_{qr} \\
  \mathcal{C}_{rp} & \mathcal{C}_{rq} & \mathcal{C}_{rr}
\end{matrix}
\right)
= \mathrm{\textbf{R}}^{\mathrm{T}}\textbf{G}
\left(
\begin{matrix}
  \sigma_{\mathcal{U}}^{2} & 0 & 0 \\
  0 & \sigma_{\mathcal{V}}^{2} & 0 \\
  0 & 0 & \sigma_{\mathcal{W}}^{2}
\end{matrix}
\right)
\mathrm{\textbf{G}}^{\mathrm{T}}\textbf{R}
\end{equation}

where
\begin{align*}
\mathrm{\textbf{G}}^{\mathrm{T}} = \mathrm{\textbf{T}}_{1}\textbf{N}^{\mathrm{T}} \quad\quad\mathrm{and}\quad\quad \mathrm{\textbf{G}} = \textbf{N}\,\mathrm{\textbf{T}}_{1}^{\mathrm{T}}\\
\mathrm{\textbf{N}}^{\mathrm{T}} = \mathrm{\textbf{T}}_{2}\textbf{R}^{\mathrm{T}} \quad\quad\mathrm{and}\quad\quad \mathrm{\textbf{N}} = \textbf{R}\,\mathrm{\textbf{T}}_{2}^{\mathrm{T}}
\end{align*}

where \textbf{N} is the equatorial triad pointing towards the equator, the North Celestial Pole and the third one orthogonal to both; \textbf{T}$_{1}$ the transformation matrix from equatorial to galactic coordinates and \textbf{T}$_{2}$ from equatorial to orthonormal system in the direction of the line of sight.

The solar motion can be transformed from Galactic coordinates to the orthonormal system by
\begin{align*}
\langle v_{\odot}\rangle =
\mathrm{\textbf{R}}
\left(
\begin{matrix}
\langle v_{\odot}\rangle_{p}\\
\langle v_{\odot}\rangle_{q}\\
\langle v_{\odot}\rangle_{r}
\end{matrix}
\right)
= \mathrm{\textbf{G}}
\left(
\begin{matrix}
\langle v_{\odot}\rangle_{\mathcal{U}}\\
\langle v_{\odot}\rangle_{\mathcal{V}}\\
\langle v_{\odot}\rangle_{\mathcal{W}}
\end{matrix}
\right)
= \mathrm{\textbf{NT}}^{\mathrm{T}}_{1}
\left(
\begin{matrix}
\langle v_{\odot}\rangle_{\mathcal{U}}\\
\langle v_{\odot}\rangle_{\mathcal{V}}\\
\langle v_{\odot}\rangle_{\mathcal{W}}
\end{matrix}
\right)
 & = \mathrm{\textbf{RT}}_{2}\mathrm{\textbf{T}}^{\mathrm{T}}_{1}
\left(
\begin{matrix}
\langle v_{\odot}\rangle_{\mathcal{U}}\\
\langle v_{\odot}\rangle_{\mathcal{V}}\\
\langle v_{\odot}\rangle_{\mathcal{W}}
\end{matrix}
\right) \\
\left(
\begin{matrix}
\langle v_{\odot}\rangle_{p}\\
\langle v_{\odot}\rangle_{q}\\
\langle v_{\odot}\rangle_{r}
\end{matrix}
\right)
 & = \mathrm{\textbf{T}}_{2}\mathrm{\textbf{T}}^{\mathrm{T}}_{1}
\left(
\begin{matrix}
\langle v_{\odot}\rangle_{\mathcal{U}}\\
\langle v_{\odot}\rangle_{\mathcal{V}}\\
\langle v_{\odot}\rangle_{\mathcal{W}}
\end{matrix}
\right)
\end{align*}

The transformation matrices are given by
\begin{equation}
\mathrm{\textbf{T}}_{1} = 
\left(\,
\begin{matrix}
  \cos(\alpha_{c}) \cos(\delta_{c}) & \sin(\alpha_{c}) \cos(\delta_{c}) & \cos(\alpha_{p}) \cos(\delta_{p})\\
  \\
  \begin{matrix}
    \sin(\delta_{c}) \sin(\alpha_{p}) \cos(\delta_{p})\\
    -\sin(\delta_{p}) \sin(\alpha_{c}) \cos(\delta_{c})
  \end{matrix} & 
  \begin{matrix}
    \sin(\delta_{p}) \cos(\alpha_{c}) \cos(\delta_{c})\\
    -\cos(\alpha_{p}) \cos(\delta_{p}) \sin(\delta_{c})
  \end{matrix}
     & \sin(\alpha_{p}) \cos(\delta_{p})\\
     \\
  \cos(\alpha_{p}) \cos(\delta_{p}) & \cos(\delta_{c}) \cos(\delta_{p}) \sin(\alpha_{c}-\alpha_{p}) & \sin(\delta_{p})
\end{matrix}
\,\right)
\end{equation}

where $c$ and $p$ denotes the Galactic Centre and the NGP respectively; and

\begin{equation}
\mathrm{\textbf{T}}_{2} = 
\left(
\begin{matrix}
-\sin(\alpha) & -\sin(\delta)\cos(\alpha) & \cos(\delta)\cos(\alpha)\\
\cos(\alpha) & - \sin(\delta)\sin(\alpha) & \cos(\delta)\sin(\alpha)\\
0 & \cos(\delta) & \sin(\delta)
\end{matrix}
\right)
\end{equation}

with $\alpha$ and $\delta$ the position of the object.

%% file: appendix_C.tex
\chapter{Colour-Colour Diagrams for Different Hydrogen/Helium Mass Fraction}

\begin{figure}
\includegraphics[width=0.9\textwidth]{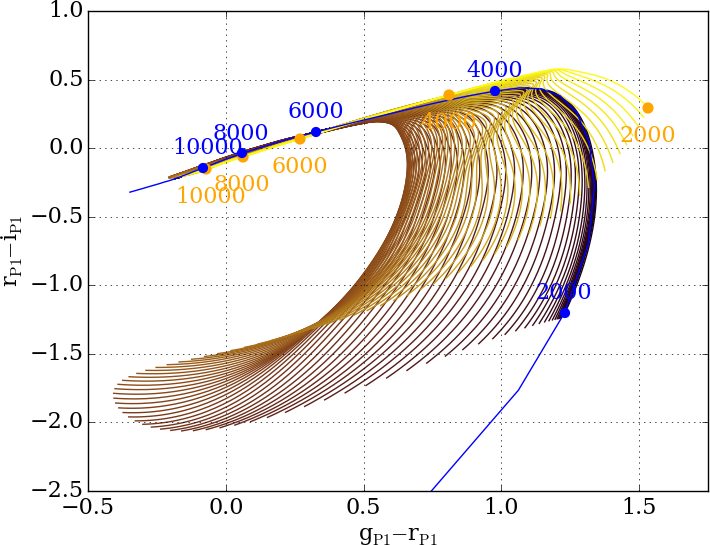}
\includegraphics[width=0.9\textwidth]{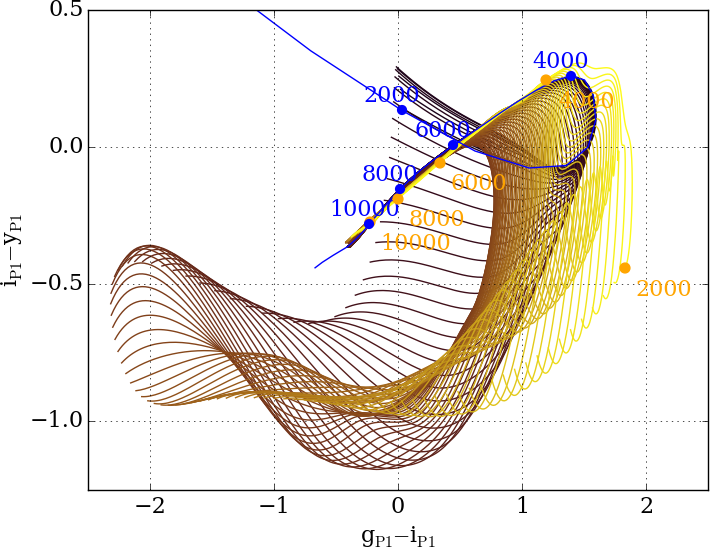}
\caption[Colour-colour diagrams for different hydrogen/helium mass fraction in the photosphere]{Colour-colour diagrams for different hydrogen/helium mass fraction in the photosphere~(Montr\`{e}al model). Blue line shows the cooling track of a pure hydrogen WD; from dark brown to yellow, they show the cooling tracks of WDs with gradual increase in helium mass fraction. Each successive line shows $+0.1$\,dex in the helium mass fraction, $\log[\frac{\mathcal{M}_{\mathrm{He}}}{\mathcal{M}_{\mathrm{H}}}]$, from $-2.0$ to $+8.0$}
\end{figure}

%% file: appendix_D.tex
\chapter{Table of Bolometric Correction for Mixed Hydrogen-Helium Atmosphere Model}

\begin{table}[h]
\begin{center}
\begin{tabular}{ccc ccc ccc ccc}
$T_{\mathrm{eff}}$ & $\mathcal{M}_{\mathrm{bol}}$ & $G_{\mathrm{P1}}$ & $R_{\mathrm{P1}}$ & $I_{\mathrm{P1}}$ & $Z_{\mathrm{P1}}$ & $Y_{\mathrm{P1}}$ \\\hline
  1500 & 20.108 & 19.728 & 18.917 & 21.266 & 20.828 & 20.631 \\
  1750 & 19.438 & 19.288 & 18.226 & 19.991 & 19.674 & 19.640 \\
  2000 & 18.858 & 18.925 & 17.695 & 18.894 & 18.683 & 18.760 \\
  2250 & 18.347 & 18.614 & 17.298 & 17.994 & 17.849 & 18.008 \\
  2500 & 17.889 & 18.347 & 17.009 & 17.290 & 17.170 & 17.366 \\
  2750 & 17.475 & 18.097 & 16.778 & 16.740 & 16.614 & 16.807 \\
  3000 & 17.097 & 17.855 & 16.581 & 16.323 & 16.172 & 16.323 \\
  3250 & 16.749 & 17.610 & 16.397 & 16.015 & 15.837 & 15.916 \\
  3500 & 16.427 & 17.358 & 16.216 & 15.784 & 15.589 & 15.594 \\
  3750 & 16.126 & 17.093 & 16.028 & 15.589 & 15.393 & 15.344 \\
  4000 & 15.845 & 16.795 & 15.819 & 15.403 & 15.216 & 15.144 \\
  4250 & 15.581 & 16.468 & 15.588 & 15.207 & 15.038 & 14.965 \\
  4500 & 15.332 & 16.124 & 15.340 & 15.000 & 14.854 & 14.790 \\
  4750 & 15.095 & 15.775 & 15.086 & 14.790 & 14.667 & 14.618 \\
  5000 & 14.870 & 15.439 & 14.837 & 14.584 & 14.485 & 14.450 \\
  5250 & 14.654 & 15.123 & 14.602 & 14.387 & 14.311 & 14.290 \\
  5500 & 14.447 & 14.835 & 14.386 & 14.206 & 14.151 & 14.143 \\
  6000 & 14.063 & 14.348 & 14.020 & 13.901 & 13.881 & 13.894 \\
  6500 & 13.712 & 13.953 & 13.718 & 13.648 & 13.657 & 13.688 \\
  7000 & 13.387 & 13.617 & 13.454 & 13.426 & 13.460 & 13.506 \\
  7500 & 13.084 & 13.321 & 13.215 & 13.223 & 13.279 & 13.340 \\
  8000 & 12.801 & 13.053 & 12.995 & 13.034 & 13.112 & 13.186 \\
  8500 & 12.535 & 12.807 & 12.790 & 12.858 & 12.956 & 13.044 \\
  9000 & 12.285 & 12.579 & 12.598 & 12.694 & 12.812 & 12.913 \\
  9500 & 12.047 & 12.366 & 12.419 & 12.541 & 12.678 & 12.791 \\
 10000 & 11.822 & 12.170 & 12.254 & 12.400 & 12.557 & 12.681 \\
 10500 & 11.607 & 11.993 & 12.107 & 12.276 & 12.450 & 12.585 \\
 11000 & 11.403 & 11.844 & 11.984 & 12.174 & 12.362 & 12.505 \\
 11500 & 11.208 & 11.719 & 11.883 & 12.090 & 12.290 & 12.439 \\
 12000 & 11.021 & 11.616 & 11.800 & 12.023 & 12.232 & 12.386 \\
 12500 & 10.841 & 11.539 & 11.740 & 11.975 & 12.189 & 12.346 \\
 13000 & 10.669 & 11.470 & 11.688 & 11.932 & 12.151 & 12.310 \\
 13500 & 10.503 & 11.398 & 11.631 & 11.885 & 12.109 & 12.270 \\
 14000 & 10.343 & 11.324 & 11.572 & 11.835 & 12.063 & 12.226
\end{tabular}
\end{center}
\caption{Spectral Energy Distribution of WDs with surface gravity of $\log(g)=8.0$ and a pure hydrogen atmosphere}
\end{table}

\begin{table}[h]
\begin{center}
\begin{tabular}{ccc ccc ccc ccc}
$T_{\mathrm{eff}}$ & $\mathcal{M}_{\mathrm{bol}}$ & $G_{\mathrm{P1}}$ & $R_{\mathrm{P1}}$ & $I_{\mathrm{P1}}$ & $Z_{\mathrm{P1}}$ & $Y_{\mathrm{P1}}$ \\\hline
 14500 & 10.188 & 11.252 & 11.513 & 11.784 & 12.016 & 12.181 \\
 15000 & 10.039 & 11.184 & 11.456 & 11.734 & 11.969 & 12.136 \\
 15500 &  9.895 & 11.119 & 11.402 & 11.685 & 11.923 & 12.092 \\
 16000 &  9.755 & 11.057 & 11.349 & 11.638 & 11.878 & 12.049 \\
 16500 &  9.619 & 10.997 & 11.298 & 11.592 & 11.834 & 12.007 \\
 17000 &  9.488 & 10.939 & 11.248 & 11.547 & 11.791 & 11.966 \\
 20000 &  8.770 & 10.626 & 10.974 & 11.295 & 11.551 & 11.735 \\
 25000 &  7.782 & 10.181 & 10.577 & 10.927 & 11.198 & 11.395 \\
 30000 &  6.972 &  9.765 & 10.198 & 10.568 & 10.851 & 11.058 \\
 35000 &  6.277 &  9.428 &  9.883 & 10.263 & 10.550 & 10.760 \\
 40000 &  5.679 &  9.216 &  9.682 & 10.068 & 10.357 & 10.569 \\
 45000 &  5.151 &  9.060 &  9.534 &  9.923 & 10.213 & 10.426 \\
 50000 &  4.676 &  8.931 &  9.410 &  9.802 & 10.094 & 10.307 \\
 55000 &  4.247 &  8.824 &  9.307 &  9.701 &  9.993 & 10.207 \\
 60000 &  3.854 &  8.726 &  9.213 &  9.609 &  9.901 & 10.115 \\
 65000 &  3.492 &  8.638 &  9.128 &  9.525 &  9.818 & 10.033 \\
 70000 &  3.156 &  8.557 &  9.050 &  9.448 &  9.741 &  9.956 \\
 75000 &  2.843 &  8.482 &  8.976 &  9.376 &  9.670 &  9.885 \\
 80000 &  2.550 &  8.411 &  8.907 &  9.308 &  9.602 &  9.817 \\
 85000 &  2.274 &  8.343 &  8.842 &  9.243 &  9.537 &  9.753 \\
 90000 &  2.013 &  8.279 &  8.779 &  9.181 &  9.476 &  9.692 \\
100000 &  1.530 &  8.158 &  8.661 &  9.064 &  9.360 &  9.576 \\
110000 &  1.091 &  8.045 &  8.551 &  8.955 &  9.251 &  9.468 \\
120000 &  0.686 &  7.937 &  8.445 &  8.851 &  9.147 &  9.364
\end{tabular}
\end{center}
\end{table}

\begin{table}[h]
\begin{center}
\begin{tabular}{ccc ccc ccc ccc}
$T_{\mathrm{eff}}$ & $\mathcal{M}_{\mathrm{bol}}$ & $G_{\mathrm{P1}}$ & $R_{\mathrm{P1}}$ & $I_{\mathrm{P1}}$ & $Z_{\mathrm{P1}}$ & $Y_{\mathrm{P1}}$ \\\hline
   2000 & 18.858 & 18.969 & 17.736 & 18.929 & 18.716 & 18.790 \\
   2250 & 18.347 & 18.667 & 17.350 & 18.076 & 17.921 & 18.068 \\
   2500 & 17.889 & 18.396 & 17.051 & 17.381 & 17.261 & 17.445 \\
   2750 & 17.475 & 18.144 & 16.815 & 16.823 & 16.705 & 16.894 \\
   3000 & 17.097 & 17.900 & 16.613 & 16.387 & 16.246 & 16.403 \\
   3250 & 16.749 & 17.649 & 16.424 & 16.056 & 15.884 & 15.975 \\
   3500 & 16.427 & 17.384 & 16.233 & 15.804 & 15.613 & 15.625 \\
   3750 & 16.126 & 17.103 & 16.034 & 15.597 & 15.402 & 15.357 \\
   4000 & 15.845 & 16.799 & 15.821 & 15.405 & 15.219 & 15.148 \\
   4250 & 15.581 & 16.471 & 15.589 & 15.208 & 15.039 & 14.966 \\
   4500 & 15.332 & 16.125 & 15.340 & 15.001 & 14.855 & 14.791 \\
   4750 & 15.095 & 15.778 & 15.088 & 14.792 & 14.669 & 14.620 \\
   5000 & 14.870 & 15.442 & 14.839 & 14.585 & 14.487 & 14.452 \\
   5250 & 14.654 & 15.126 & 14.603 & 14.389 & 14.313 & 14.292 \\
   5500 & 14.447 & 14.837 & 14.387 & 14.208 & 14.152 & 14.144 \\
   6000 & 14.063 & 14.349 & 14.020 & 13.900 & 13.879 & 13.892 \\
   6500 & 13.712 & 13.953 & 13.717 & 13.647 & 13.656 & 13.686 \\
   7000 & 13.387 & 13.617 & 13.454 & 13.425 & 13.459 & 13.505 \\
   7500 & 13.084 & 13.321 & 13.214 & 13.222 & 13.278 & 13.339 \\
   8000 & 12.801 & 13.053 & 12.994 & 13.033 & 13.110 & 13.185 \\
   8500 & 12.535 & 12.807 & 12.789 & 12.857 & 12.955 & 13.042 \\
   9000 & 12.285 & 12.578 & 12.597 & 12.692 & 12.810 & 12.910 \\
   9500 & 12.047 & 12.366 & 12.418 & 12.539 & 12.676 & 12.789 \\
 10000 & 11.822 & 12.169 & 12.252 & 12.398 & 12.554 & 12.679 \\
 10500 & 11.607 & 11.993 & 12.106 & 12.275 & 12.448 & 12.582 \\
 11000 & 11.403 & 11.843 & 11.982 & 12.172 & 12.360 & 12.502 \\
 11500 & 11.208 & 11.719 & 11.881 & 12.088 & 12.288 & 12.436 \\
 12000 & 11.021 & 11.615 & 11.799 & 12.021 & 12.230 & 12.383
\end{tabular}
\end{center}
\caption[Spectral Energy Distribution of WDs with surface gravity of $\log(g)=8.0$ and a mixed hydrogen-helium atmosphere with mass ratio of $\log(\frac{\mathcal{M}_{\mathrm{He}}}{\mathcal{M}_{\mathrm{H}}}) = -2.0$]{Spectral Energy Distribution of WDs with surface gravity of $\log(g)=8.0$ and a mixed hydrogen-helium atmosphere with mass ratio of $\log(\frac{\mathcal{M}_{\mathrm{He}}}{\mathcal{M}_{\mathrm{H}}}) = -2.0$ which is essentially a pure hydrogen atmosphere as helium has weak atmospheric features}
\end{table}

\begin{table}[h]
\begin{center}
\begin{tabular}{ccc ccc ccc ccc}
$T_{\mathrm{eff}}$ & $\mathcal{M}_{\mathrm{bol}}$ & $G_{\mathrm{P1}}$ & $R_{\mathrm{P1}}$ & $I_{\mathrm{P1}}$ & $Z_{\mathrm{P1}}$ & $Y_{\mathrm{P1}}$ \\\hline
  2000 & 18.858 & 18.889 & 17.694 & 18.942 & 18.836 & 18.906 \\
  2250 & 18.347 & 18.587 & 17.299 & 18.065 & 18.010 & 18.151 \\
  2500 & 17.889 & 18.318 & 16.996 & 17.359 & 17.327 & 17.506 \\
  2750 & 17.475 & 18.069 & 16.759 & 16.795 & 16.751 & 16.937 \\
  3000 & 17.097 & 17.827 & 16.557 & 16.356 & 16.271 & 16.431 \\
  3250 & 16.749 & 17.581 & 16.371 & 16.023 & 15.890 & 15.988 \\
  3500 & 16.427 & 17.324 & 16.185 & 15.772 & 15.603 & 15.626 \\
  3750 & 16.126 & 17.050 & 15.991 & 15.566 & 15.385 & 15.349 \\
  4000 & 15.845 & 16.757 & 15.786 & 15.378 & 15.200 & 15.136 \\
  4250 & 15.581 & 16.443 & 15.564 & 15.189 & 15.024 & 14.955 \\
  4500 & 15.332 & 16.109 & 15.325 & 14.989 & 14.845 & 14.784 \\
  4750 & 15.095 & 15.770 & 15.079 & 14.785 & 14.664 & 14.615 \\
  5000 & 14.870 & 15.439 & 14.835 & 14.582 & 14.484 & 14.450 \\
  5250 & 14.654 & 15.125 & 14.601 & 14.387 & 14.312 & 14.291 \\
  5500 & 14.447 & 14.836 & 14.385 & 14.206 & 14.151 & 14.143 \\
  6000 & 14.063 & 14.345 & 14.016 & 13.897 & 13.877 & 13.890 \\
  6500 & 13.712 & 13.950 & 13.714 & 13.644 & 13.654 & 13.684 \\
  7000 & 13.387 & 13.615 & 13.452 & 13.423 & 13.457 & 13.504 \\
  7500 & 13.084 & 13.320 & 13.214 & 13.221 & 13.278 & 13.338 \\
  8000 & 12.801 & 13.053 & 12.995 & 13.033 & 13.111 & 13.185 \\
  8500 & 12.535 & 12.809 & 12.791 & 12.859 & 12.956 & 13.043 \\
  9000 & 12.285 & 12.582 & 12.601 & 12.695 & 12.812 & 12.912 \\
  9500 & 12.047 & 12.372 & 12.423 & 12.543 & 12.679 & 12.792 \\
 10000 & 11.822 & 12.175 & 12.258 & 12.402 & 12.557 & 12.681 \\
 10500 & 11.607 & 11.998 & 12.109 & 12.276 & 12.449 & 12.583 \\
 11000 & 11.403 & 11.848 & 11.985 & 12.172 & 12.359 & 12.502 \\
 11500 & 11.208 & 11.722 & 11.882 & 12.087 & 12.286 & 12.436 \\
 12000 & 11.021 & 11.617 & 11.798 & 12.018 & 12.227 & 12.381
 \end{tabular}
\end{center}
\caption{Spectral Energy Distribution of WDs with surface gravity of $\log(g)=8.0$ and a mixed hydrogen-helium atmosphere with mass ratio of $\log(\frac{\mathcal{M}_{\mathrm{He}}}{\mathcal{M}_{\mathrm{H}}}) = -1.0$}
\end{table}

\begin{table}[h]
\begin{center}
\begin{tabular}{ccc ccc ccc ccc}
$T_{\mathrm{eff}}$ & $\mathcal{M}_{\mathrm{bol}}$ & $G_{\mathrm{P1}}$ & $R_{\mathrm{P1}}$ & $I_{\mathrm{P1}}$ & $Z_{\mathrm{P1}}$ & $Y_{\mathrm{P1}}$ \\\hline
  2000 & 18.871 & 18.510 & 17.592 & 19.018 & 19.691 & 19.636 \\
  2250 & 18.359 & 18.216 & 17.130 & 18.107 & 18.724 & 18.752 \\
  2500 & 17.901 & 17.954 & 16.788 & 17.326 & 17.857 & 18.009 \\
  2750 & 17.487 & 17.751 & 16.558 & 16.748 & 17.088 & 17.245 \\
  3000 & 17.109 & 17.517 & 16.343 & 16.281 & 16.497 & 16.656 \\
  3250 & 16.762 & 17.286 & 16.157 & 15.927 & 16.010 & 16.142 \\
  3500 & 16.440 & 17.052 & 15.983 & 15.659 & 15.631 & 15.709 \\
  3750 & 16.140 & 16.809 & 15.810 & 15.449 & 15.351 & 15.368 \\
  4000 & 15.859 & 16.554 & 15.631 & 15.270 & 15.141 & 15.116 \\
  4250 & 15.596 & 16.285 & 15.442 & 15.100 & 14.965 & 14.922 \\
  4500 & 15.347 & 16.005 & 15.242 & 14.929 & 14.803 & 14.756 \\
  4750 & 15.112 & 15.713 & 15.032 & 14.752 & 14.642 & 14.602 \\
  5000 & 14.888 & 15.414 & 14.814 & 14.571 & 14.479 & 14.450 \\
  5250 & 14.675 & 15.119 & 14.597 & 14.391 & 14.320 & 14.302 \\
  5500 & 14.472 & 14.839 & 14.391 & 14.218 & 14.167 & 14.161 \\
  6000 & 14.090 & 14.354 & 14.029 & 13.913 & 13.897 & 13.912 \\
  6500 & 13.740 & 13.963 & 13.732 & 13.664 & 13.675 & 13.708 \\
  7000 & 13.416 & 13.635 & 13.475 & 13.447 & 13.483 & 13.531 \\
  7500 & 13.114 & 13.345 & 13.243 & 13.249 & 13.307 & 13.369 \\
  8000 & 12.832 & 13.085 & 13.029 & 13.066 & 13.145 & 13.220 \\
  8500 & 12.567 & 12.847 & 12.831 & 12.896 & 12.994 & 13.082 \\
  9000 & 12.317 & 12.627 & 12.646 & 12.737 & 12.854 & 12.954 \\
  9500 & 12.080 & 12.422 & 12.472 & 12.588 & 12.723 & 12.835 \\
 10000 & 11.855 & 12.231 & 12.310 & 12.449 & 12.602 & 12.726 \\
 10500 & 11.641 & 12.056 & 12.161 & 12.323 & 12.492 & 12.626 \\
 11000 & 11.437 & 11.901 & 12.031 & 12.213 & 12.398 & 12.541 \\
 11500 & 11.242 & 11.772 & 11.924 & 12.124 & 12.321 & 12.472 \\
 12000 & 11.056 & 11.665 & 11.836 & 12.051 & 12.258 & 12.414 \\
 \end{tabular}
\end{center}
\caption{Spectral Energy Distribution of WDs with surface gravity of $\log(g)=8.0$ and a mixed hydrogen-helium atmosphere with mass ratio of $\log(\frac{\mathcal{M}_{\mathrm{He}}}{\mathcal{M}_{\mathrm{H}}}) = 0.0$}
\end{table}

\begin{table}[h]
\begin{center}
\begin{tabular}{ccc ccc ccc ccc}
$T_{\mathrm{eff}}$ & $\mathcal{M}_{\mathrm{bol}}$ & $G_{\mathrm{P1}}$ & $R_{\mathrm{P1}}$ & $I_{\mathrm{P1}}$ & $Z_{\mathrm{P1}}$ & $Y_{\mathrm{P1}}$ \\\hline
  2000 & 18.871 & 18.060 & 17.843 & 19.769 & 20.220 & 20.239 \\
  2250 & 18.359 & 17.760 & 17.215 & 18.701 & 19.509 & 19.408 \\
  2500 & 17.901 & 17.491 & 16.741 & 17.734 & 18.858 & 18.806 \\
  2750 & 17.487 & 17.236 & 16.382 & 16.923 & 17.818 & 18.037 \\
  3000 & 17.109 & 17.014 & 16.124 & 16.367 & 17.037 & 17.256 \\
  3250 & 16.762 & 16.809 & 15.921 & 15.952 & 16.402 & 16.598 \\
  3500 & 16.440 & 16.605 & 15.739 & 15.626 & 15.884 & 16.048 \\
  3750 & 16.140 & 16.417 & 15.580 & 15.374 & 15.482 & 15.594 \\
  4000 & 15.859 & 16.194 & 15.408 & 15.160 & 15.169 & 15.235 \\
  4250 & 15.596 & 15.964 & 15.236 & 14.979 & 14.933 & 14.959 \\
  4500 & 15.347 & 15.724 & 15.058 & 14.811 & 14.746 & 14.750 \\
  4750 & 15.112 & 15.479 & 14.876 & 14.650 & 14.583 & 14.578 \\
  5000 & 14.888 & 15.233 & 14.693 & 14.491 & 14.432 & 14.427 \\
  5250 & 14.675 & 14.987 & 14.511 & 14.335 & 14.288 & 14.289 \\
  5500 & 14.472 & 14.745 & 14.331 & 14.183 & 14.149 & 14.158 \\
  6000 & 14.090 & 14.322 & 14.021 & 13.919 & 13.913 & 13.938 \\
  6500 & 13.740 & 13.947 & 13.736 & 13.674 & 13.695 & 13.736 \\
  7000 & 13.416 & 13.627 & 13.483 & 13.459 & 13.505 & 13.561 \\
  7500 & 13.114 & 13.346 & 13.257 & 13.266 & 13.334 & 13.405 \\
  8000 & 12.832 & 13.093 & 13.049 & 13.090 & 13.177 & 13.261 \\
  8500 & 12.567 & 12.863 & 12.859 & 12.927 & 13.033 & 13.129 \\
  9000 & 12.317 & 12.651 & 12.682 & 12.776 & 12.899 & 13.006 \\
  9500 & 12.080 & 12.454 & 12.515 & 12.633 & 12.772 & 12.890 \\
 10000 & 11.855 & 12.269 & 12.357 & 12.497 & 12.652 & 12.779 \\
 10500 & 11.641 & 12.098 & 12.209 & 12.371 & 12.540 & 12.677 \\
 11000 & 11.437 & 11.943 & 12.077 & 12.257 & 12.440 & 12.585 \\
 11500 & 11.242 & 11.810 & 11.963 & 12.161 & 12.355 & 12.506 \\
 12000 & 11.056 & 11.699 & 11.868 & 12.081 & 12.285 & 12.441
 \end{tabular}
\end{center}
\caption{Spectral Energy Distribution of WDs with surface gravity of $\log(g)=8.0$ and a mixed hydrogen-helium atmosphere with mass ratio of $\log(\frac{\mathcal{M}_{\mathrm{He}}}{\mathcal{M}_{\mathrm{H}}}) = 1.0$}
\end{table}

\begin{table}[h]
\begin{center}
\begin{tabular}{ccc ccc ccc ccc}
$T_{\mathrm{eff}}$ & $\mathcal{M}_{\mathrm{bol}}$ & $G_{\mathrm{P1}}$ & $R_{\mathrm{P1}}$ & $I_{\mathrm{P1}}$ & $Z_{\mathrm{P1}}$ & $Y_{\mathrm{P1}}$ \\\hline
  2000 & 18.871 & 17.887 & 18.157 & 20.191 & 20.614 & 20.690 \\
  2250 & 18.359 & 17.561 & 17.430 & 19.158 & 19.795 & 19.736 \\
  2500 & 17.901 & 17.291 & 16.894 & 18.123 & 19.267 & 19.180 \\
  2750 & 17.487 & 17.059 & 16.494 & 17.361 & 18.363 & 18.357 \\
  3000 & 17.109 & 16.838 & 16.188 & 16.695 & 17.560 & 17.665 \\
  3250 & 16.762 & 16.667 & 15.968 & 16.167 & 16.721 & 16.859 \\
  3500 & 16.440 & 16.444 & 15.754 & 15.766 & 16.149 & 16.301 \\
  3750 & 16.140 & 16.225 & 15.563 & 15.457 & 15.682 & 15.823 \\
  4000 & 15.859 & 16.005 & 15.381 & 15.212 & 15.316 & 15.424 \\
  4250 & 15.596 & 15.783 & 15.203 & 15.011 & 15.036 & 15.105 \\
  4500 & 15.347 & 15.557 & 15.025 & 14.833 & 14.820 & 14.860 \\
  4750 & 15.112 & 15.331 & 14.846 & 14.668 & 14.641 & 14.666 \\
  5000 & 14.888 & 15.108 & 14.670 & 14.511 & 14.483 & 14.504 \\
  5250 & 14.675 & 14.890 & 14.498 & 14.361 & 14.340 & 14.362 \\
  5500 & 14.472 & 14.681 & 14.332 & 14.218 & 14.208 & 14.234 \\
  6000 & 14.090 & 14.400 & 14.114 & 14.028 & 14.034 & 14.070 \\
  6500 & 13.740 & 14.024 & 13.824 & 13.780 & 13.813 & 13.864 \\
  7000 & 13.416 & 13.682 & 13.555 & 13.547 & 13.605 & 13.673 \\
  7500 & 13.114 & 13.389 & 13.320 & 13.344 & 13.424 & 13.507 \\
  8000 & 12.832 & 13.137 & 13.115 & 13.167 & 13.267 & 13.361 \\
  8500 & 12.567 & 12.913 & 12.930 & 13.008 & 13.125 & 13.230 \\
  9000 & 12.317 & 12.709 & 12.760 & 12.863 & 12.994 & 13.109 \\
  9500 & 12.080 & 12.521 & 12.603 & 12.728 & 12.873 & 12.997 \\
 10000 & 11.855 & 12.348 & 12.457 & 12.603 & 12.762 & 12.893 \\
 10500 & 11.641 & 12.190 & 12.323 & 12.488 & 12.658 & 12.797 \\
 11000 & 11.437 & 12.048 & 12.202 & 12.384 & 12.563 & 12.708 \\
 11500 & 11.242 & 11.919 & 12.090 & 12.287 & 12.476 & 12.625 \\
 12000 & 11.056 & 11.802 & 11.989 & 12.199 & 12.395 & 12.549
 \end{tabular}
\end{center}
\caption{Spectral Energy Distribution of WDs with surface gravity of $\log(g)=8.0$ and a mixed hydrogen-helium atmosphere with mass ratio of $\log(\frac{\mathcal{M}_{\mathrm{He}}}{\mathcal{M}_{\mathrm{H}}}) = 2.0$}
\end{table}

\begin{table}[h]
\begin{center}
\begin{tabular}{ccc ccc ccc ccc}
$T_{\mathrm{eff}}$ & $\mathcal{M}_{\mathrm{bol}}$ & $G_{\mathrm{P1}}$ & $R_{\mathrm{P1}}$ & $I_{\mathrm{P1}}$ & $Z_{\mathrm{P1}}$ & $Y_{\mathrm{P1}}$ \\\hline
  2000 & 18.871 & 17.961 & 18.352 & 20.074 & 20.950 & 20.962 \\
  2250 & 18.359 & 17.620 & 17.585 & 19.205 & 19.920 & 19.997 \\
  2500 & 17.901 & 17.339 & 17.002 & 18.280 & 19.043 & 19.057 \\
  2750 & 17.487 & 17.105 & 16.584 & 17.446 & 18.306 & 18.417 \\
  3000 & 17.109 & 16.880 & 16.255 & 16.751 & 17.514 & 17.642 \\
  3250 & 16.762 & 16.668 & 16.006 & 16.207 & 16.806 & 16.965 \\
  3500 & 16.440 & 16.468 & 15.807 & 15.802 & 16.187 & 16.351 \\
  3750 & 16.140 & 16.267 & 15.628 & 15.496 & 15.696 & 15.834 \\
  4000 & 15.859 & 16.076 & 15.461 & 15.258 & 15.321 & 15.409 \\
  4250 & 15.596 & 15.860 & 15.289 & 15.068 & 15.057 & 15.102 \\
  4500 & 15.347 & 15.635 & 15.114 & 14.901 & 14.860 & 14.879 \\
  4750 & 15.112 & 15.408 & 14.934 & 14.742 & 14.696 & 14.706 \\
  5000 & 14.888 & 15.184 & 14.756 & 14.588 & 14.549 & 14.559 \\
  5250 & 14.675 & 14.965 & 14.582 & 14.439 & 14.412 & 14.427 \\
  5500 & 14.472 & 14.752 & 14.412 & 14.296 & 14.282 & 14.305 \\
  6000 & 14.090 & 14.350 & 14.092 & 14.024 & 14.037 & 14.077 \\
  6500 & 13.740 & 13.988 & 13.803 & 13.779 & 13.819 & 13.876 \\
  7000 & 13.416 & 13.670 & 13.549 & 13.561 & 13.626 & 13.697 \\
  7500 & 13.114 & 13.392 & 13.326 & 13.369 & 13.453 & 13.538 \\
  8000  & 12.832 & 13.145 & 13.128 & 13.197 & 13.299 & 13.395 \\
  8500  & 12.567 & 12.926 & 12.951 & 13.043 & 13.161 & 13.267 \\
  9000  & 12.317 & 12.731 & 12.793 & 12.906 & 13.039 & 13.154 \\
  9500  & 12.080 & 12.557 & 12.650 & 12.782 & 12.927 & 13.051 \\
 10000  & 11.855 & 12.398 & 12.517 & 12.668 & 12.824 & 12.955 \\
 10500  & 11.641 & 12.250 & 12.393 & 12.560 & 12.727 & 12.864 \\
 11000  & 11.437 & 12.111 & 12.276 & 12.459 & 12.635 & 12.778 \\
 11500  & 11.242 & 11.979 & 12.165 & 12.362 & 12.547 & 12.696 \\
 12000  & 11.056 & 11.854 & 12.058 & 12.269 & 12.463 & 12.616
 \end{tabular}
\end{center}
\caption{Spectral Energy Distribution of WDs with surface gravity of $\log(g)=8.0$ and a mixed hydrogen-helium atmosphere with mass ratio of $\log(\frac{\mathcal{M}_{\mathrm{He}}}{\mathcal{M}_{\mathrm{H}}}) = 3.0$}
\end{table}

\begin{table}[h]
\begin{center}
\begin{tabular}{ccc ccc ccc ccc}
$T_{\mathrm{eff}}$ & $\mathcal{M}_{\mathrm{bol}}$ & $G_{\mathrm{P1}}$ & $R_{\mathrm{P1}}$ & $I_{\mathrm{P1}}$ & $Z_{\mathrm{P1}}$ & $Y_{\mathrm{P1}}$ \\\hline
  2000 & 18.871 & 18.122 & 18.151 & 19.649 & 20.572 & 20.570 \\
  2250 & 18.359 & 17.794 & 17.482 & 18.776 & 19.697 & 19.675 \\
  2500 & 17.901 & 17.512 & 16.966 & 17.924 & 18.774 & 18.831 \\
  2750 & 17.487 & 17.267 & 16.590 & 17.177 & 17.955 & 18.014 \\
  3000 & 17.109 & 17.042 & 16.311 & 16.565 & 17.208 & 17.320 \\
  3250 & 16.762 & 16.857 & 16.099 & 16.089 & 16.492 & 16.620 \\
  3500 & 16.440 & 16.646 & 15.910 & 15.748 & 15.956 & 16.078 \\
  3750 & 16.140 & 16.433 & 15.735 & 15.493 & 15.548 & 15.628 \\
  4000 & 15.859 & 16.214 & 15.563 & 15.294 & 15.258 & 15.289 \\
  4250 & 15.596 & 15.982 & 15.386 & 15.124 & 15.051 & 15.049 \\
  4500 & 15.347 & 15.739 & 15.197 & 14.959 & 14.883 & 14.872 \\
  4750 & 15.112 & 15.497 & 15.006 & 14.796 & 14.730 & 14.722 \\
  5000 & 14.888 & 15.257 & 14.817 & 14.638 & 14.586 & 14.586 \\
  5250 & 14.675 & 15.020 & 14.629 & 14.480 & 14.446 & 14.455 \\
  5500 & 14.472 & 14.789 & 14.446 & 14.326 & 14.309 & 14.329 \\
  6000 & 14.090 & 14.365 & 14.107 & 14.040 & 14.053 & 14.093 \\
  6500 & 13.740 & 13.996 & 13.810 & 13.789 & 13.830 & 13.887 \\
  7000 & 13.416 & 13.677 & 13.554 & 13.571 & 13.636 & 13.708 \\
  7500 & 13.114 & 13.399 & 13.330 & 13.380 & 13.464 & 13.549 \\
  8000 & 12.832 & 13.154 & 13.132 & 13.209 & 13.311 & 13.407 \\
  8500 & 12.567 & 12.934 & 12.953 & 13.054 & 13.171 & 13.277 \\
  9000 & 12.317 & 12.733 & 12.789 & 12.911 & 13.043 & 13.157 \\
  9500 & 12.080 & 12.548 & 12.637 & 12.778 & 12.922 & 13.045 \\
 10000 & 11.855 & 12.375 & 12.494 & 12.653 & 12.809 & 12.939 \\
 10500 & 11.641 & 12.214 & 12.358 & 12.533 & 12.700 & 12.837 \\
 11000 & 11.437 & 12.063 & 12.230 & 12.420 & 12.596 & 12.740 \\
 11500 & 11.242 & 11.921 & 12.108 & 12.311 & 12.497 & 12.646 \\
 12000 & 11.056 & 11.789 & 11.992 & 12.208 & 12.401 & 12.555
\end{tabular}
\end{center}
\caption{Spectral Energy Distribution of WDs with surface gravity of $\log(g)=8.0$ and a mixed hydrogen-helium atmosphere with mass ratio of $\log(\frac{\mathcal{M}_{\mathrm{He}}}{\mathcal{M}_{\mathrm{H}}}) = 4.0$}
\end{table}

\begin{table}[h]
\begin{center}
\begin{tabular}{ccc ccc ccc ccc}
$T_{\mathrm{eff}}$ & $\mathcal{M}_{\mathrm{bol}}$ & $G_{\mathrm{P1}}$ & $R_{\mathrm{P1}}$ & $I_{\mathrm{P1}}$ & $Z_{\mathrm{P1}}$ & $Y_{\mathrm{P1}}$ \\\hline
  2000 & 18.871 & 18.363 & 17.899 & 19.118 & 20.027 & 20.040 \\
  2250 & 18.359 & 18.030 & 17.356 & 18.263 & 19.135 & 19.169 \\
  2500 & 17.901 & 17.744 & 16.951 & 17.506 & 18.292 & 18.342 \\
  2750 & 17.487 & 17.488 & 16.646 & 16.877 & 17.523 & 17.616 \\
  3000 & 17.109 & 17.281 & 16.408 & 16.368 & 16.783 & 16.898 \\
  3250 & 16.762 & 17.046 & 16.209 & 16.006 & 16.211 & 16.323 \\
  3500 & 16.440 & 16.817 & 16.025 & 15.735 & 15.773 & 15.842 \\
  3750 & 16.140 & 16.583 & 15.843 & 15.524 & 15.463 & 15.479 \\
  4000 & 15.859 & 16.340 & 15.656 & 15.343 & 15.241 & 15.223 \\
  4250 & 15.596 & 16.085 & 15.460 & 15.171 & 15.064 & 15.034 \\
  4500 & 15.347 & 15.825 & 15.259 & 15.003 & 14.907 & 14.879 \\
  4750 & 15.112 & 15.564 & 15.057 & 14.835 & 14.757 & 14.739 \\
  5000 & 14.888 & 15.302 & 14.851 & 14.664 & 14.606 & 14.600 \\
  5250 & 14.675 & 15.043 & 14.647 & 14.494 & 14.456 & 14.463 \\
  5500 & 14.472 & 14.799 & 14.453 & 14.332 & 14.313 & 14.331 \\
  6000 & 14.090 & 14.364 & 14.106 & 14.040 & 14.054 & 14.094 \\
  6500 & 13.740 & 13.995 & 13.810 & 13.789 & 13.831 & 13.888 \\
  7000 & 13.416 & 13.676 & 13.553 & 13.571 & 13.636 & 13.708 \\
  7500 & 13.114 & 13.395 & 13.324 & 13.376 & 13.461 & 13.546 \\
  8000 & 12.832 & 13.140 & 13.116 & 13.196 & 13.299 & 13.396 \\
  8500 & 12.567 & 12.905 & 12.921 & 13.027 & 13.146 & 13.253 \\
  9000 & 12.317 & 12.689 & 12.738 & 12.866 & 13.000 & 13.116 \\
  9500 & 12.080 & 12.490 & 12.568 & 12.715 & 12.862 & 12.986 \\
 10000 & 11.855 & 12.310 & 12.412 & 12.576 & 12.732 & 12.864 \\
 10500 & 11.641 & 12.146 & 12.271 & 12.448 & 12.614 & 12.751 \\
 11000 & 11.437 & 11.998 & 12.144 & 12.333 & 12.506 & 12.648 \\
 11500 & 11.242 & 11.863 & 12.028 & 12.229 & 12.408 & 12.555 \\
 12000 & 11.056 & 11.739 & 11.921 & 12.133 & 12.319 & 12.470
\end{tabular}
\end{center}
\caption{Spectral Energy Distribution of WDs with surface gravity of $\log(g)=8.0$ and a mixed hydrogen-helium atmosphere with mass ratio of $\log(\frac{\mathcal{M}_{\mathrm{He}}}{\mathcal{M}_{\mathrm{H}}}) = 5.0$}
\end{table}

\begin{table}[h]
\begin{center}
\begin{tabular}{ccc ccc ccc ccc}
$T_{\mathrm{eff}}$ & $\mathcal{M}_{\mathrm{bol}}$ & $G_{\mathrm{P1}}$ & $R_{\mathrm{P1}}$ & $I_{\mathrm{P1}}$ & $Z_{\mathrm{P1}}$ & $Y_{\mathrm{P1}}$ \\\hline
  2000 & 18.871 & 19.785 & 18.238 & 17.871 & 18.207 & 18.256 \\
  2250 & 18.359 & 19.325 & 17.861 & 17.376 & 17.521 & 17.577 \\
  2500 & 17.901 & 18.907 & 17.533 & 16.980 & 16.954 & 16.982 \\
  2750 & 17.487 & 18.513 & 17.233 & 16.658 & 16.514 & 16.492 \\
  3000 & 17.109 & 18.159 & 16.969 & 16.395 & 16.181 & 16.108 \\
  3250 & 16.762 & 17.785 & 16.696 & 16.160 & 15.931 & 15.836 \\
  3500 & 16.440 & 17.426 & 16.425 & 15.932 & 15.712 & 15.613 \\
  3750 & 16.140 & 17.062 & 16.158 & 15.718 & 15.520 & 15.431 \\
  4000 & 15.859 & 16.725 & 15.909 & 15.519 & 15.347 & 15.272 \\
  4250 & 15.596 & 16.362 & 15.640 & 15.302 & 15.160 & 15.102 \\
  4500 & 15.347 & 15.984 & 15.354 & 15.071 & 14.959 & 14.920 \\
  4750 & 15.112 & 15.632 & 15.086 & 14.851 & 14.768 & 14.746 \\
  5000 & 14.888 & 15.318 & 14.844 & 14.652 & 14.593 & 14.587 \\
  5250 & 14.675 & 15.036 & 14.624 & 14.470 & 14.433 & 14.440 \\
  5500 & 14.472 & 14.782 & 14.423 & 14.301 & 14.284 & 14.304 \\
  6000 & 14.090 & 14.338 & 14.065 & 13.996 & 14.011 & 14.052 \\
  6500 & 13.740 & 13.959 & 13.750 & 13.723 & 13.764 & 13.821 \\
  7000 & 13.416 & 13.629 & 13.473 & 13.479 & 13.540 & 13.611 \\
  7500 & 13.114 & 13.336 & 13.227 & 13.262 & 13.340 & 13.422 \\
  8000 & 12.832 & 13.072 & 13.006 & 13.067 & 13.162 & 13.253 \\
  8500 & 12.567 & 12.836 & 12.808 & 12.895 & 13.004 & 13.105 \\
  9000 & 12.317 & 12.625 & 12.634 & 12.744 & 12.866 & 12.976 \\
  9500 & 12.080 & 12.438 & 12.481 & 12.612 & 12.748 & 12.865 \\
 10000 & 11.855 & 12.270 & 12.345 & 12.495 & 12.642 & 12.767 \\
 10500 & 11.641 & 12.118 & 12.222 & 12.389 & 12.546 & 12.677 \\
 11000 & 11.437 & 11.980 & 12.109 & 12.292 & 12.458 & 12.595 \\
 11500 & 11.242 & 11.852 & 12.005 & 12.202 & 12.377 & 12.519 \\
 12000 & 11.056 & 11.734 & 11.908 & 12.118 & 12.300 & 12.448
\end{tabular}
\end{center}
\caption[Spectral Energy Distribution of WDs with surface gravity of $\log(g)=8.0$ and a mixed hydrogen-helium atmosphere with mass ratio of $\log(\frac{\mathcal{M}_{\mathrm{He}}}{\mathcal{M}_{\mathrm{H}}}) = 8.0$]{Spectral Energy Distribution of WDs with surface gravity of $\log(g)=8.0$ and a mixed hydrogen-helium atmosphere with mass ratio of $\log(\frac{\mathcal{M}_{\mathrm{He}}}{\mathcal{M}_{\mathrm{H}}}) = 8.0$ which is essentially a pure helium atmosphere}
\end{table}

%% file: mlam_thesis.bbl
\begin{thebibliography}{268}
\expandafter\ifx\csname natexlab\endcsname\relax\def\natexlab#1{#1}\fi

\bibitem[{{Abbott} {et~al}\mbox{.}(2016){Abbott}, {Abbott}, {Abbott},
  {Abernathy}, {Acernese}, {Ackley}, {Adams}, {Adams}, {Addesso}, {Adhikari},
  \& et~al.}]{2016PhRvL.116f1102A}
{Abbott} B.~P. {et~al.}, 2016, Physical Review Letters, 116, 061102

\bibitem[{{Afonso} {et~al}\mbox{.}(2003){Afonso}, {Albert}, {Andersen},
  {Ansari}, {Aubourg}, {Bareyre}, {Beaulieu}, {Blanc}, {Charlot}, {Couchot},
  {Coutures}, {Ferlet}, {Fouqu{\'e}}, {Glicenstein}, {Goldman}, {Gould},
  {Graff}, {Gros}, {Haissinski}, {Hamadache}, {de Kat}, {Lasserre}, {Le
  Guillou}, {Lesquoy}, {Loup}, {Magneville}, {Marquette}, {Maurice}, {Maury},
  {Milsztajn}, {Moniez}, {Palanque-Delabrouille}, {Perdereau}, {Pr{\'e}vot},
  {Rahal}, {Rich}, {Spiro}, {Tisserand}, {Vidal-Madjar}, {Vigroux}, \&
  {Zylberajch}}]{2003A&A...400..951A}
{Afonso} C. {et~al.}, 2003, \aap, 400, 951

\bibitem[{{Akerlof} {et~al}\mbox{.}(2007){Akerlof}, {Miller}, {Peters},
  {Thorstensen}, {Baltay}, {Bauer}, {Rabinowitz}, {Scalzo}, {Rigaudier},
  {Pecontal}, {Buton}, {Copin}, {Gangler}, {Smadja}, {Tao}, {Antilogus},
  {Bailey}, {Pain}, {Pereira}, {Wu}, {Aldering}, {Aragon}, {Bongard},
  {Childress}, {Loken}, {Nugent}, {Perlmutter}, {Runge}, {Thomas}, {Weaver},
  {Birchall}, {Cough}, {Holtzman}, {Rau}, {Kasliwal}, {Gal-Yam}, {Yuan}, \&
  {Quimby}}]{2007CBET.1059....2A}
{Akerlof} C. {et~al.}, 2007, Central Bureau Electronic Telegrams, 1059

\bibitem[{{Allard} {et~al}\mbox{.}(1994){Allard}, {Koester}, {Feautrier}, \&
  {Spielfiedel}}]{1994A&AS..108..417A}
{Allard} N.~F., {Koester} D., {Feautrier} N., {Spielfiedel} A., 1994, \aaps,
  108

\bibitem[{{Althaus} {et~al}\mbox{.}(2010{\natexlab{a}}){Althaus},
  {C{\'o}rsico}, {Isern}, \& {Garc{\'{\i}}a-Berro}}]{2010A&ARv..18..471A}
{Althaus} L.~G., {C{\'o}rsico} A.~H., {Isern} J., {Garc{\'{\i}}a-Berro} E.,
  2010{\natexlab{a}}, \aapr, 18, 471

\bibitem[{{Althaus} {et~al}\mbox{.}(2007){Althaus}, {Garc{\'{\i}}a-Berro},
  {Isern}, {C{\'o}rsico}, \& {Rohrmann}}]{2007A&A...465..249A}
{Althaus} L.~G., {Garc{\'{\i}}a-Berro} E., {Isern} J., {C{\'o}rsico} A.~H.,
  {Rohrmann} R.~D., 2007, \aap, 465, 249

\bibitem[{{Althaus} {et~al}\mbox{.}(2010{\natexlab{b}}){Althaus},
  {Garc{\'{\i}}a-Berro}, {Renedo}, {Isern}, {C{\'o}rsico}, \&
  {Rohrmann}}]{2010ApJ...719..612A}
{Althaus} L.~G., {Garc{\'{\i}}a-Berro} E., {Renedo} I., {Isern} J.,
  {C{\'o}rsico} A.~H., {Rohrmann} R.~D., 2010{\natexlab{b}}, \apj, 719, 612

\bibitem[{{Althaus} {et~al}\mbox{.}(2003){Althaus}, {Serenelli}, {C{\'o}rsico},
  \& {Montgomery}}]{2003A&A...404..593A}
{Althaus} L.~G., {Serenelli} A.~M., {C{\'o}rsico} A.~H., {Montgomery} M.~H.,
  2003, \aap, 404, 593

\bibitem[{{Amaro-Seoane} {et~al}\mbox{.}(2012){Amaro-Seoane}, {Aoudia},
  {Babak}, {Bin{\'e}truy}, {Berti}, {Boh{\'e}}, {Caprini}, {Colpi}, {Cornish},
  {Danzmann}, {Dufaux}, {Gair}, {Jennrich}, {Jetzer}, {Klein}, {Lang}, {Lobo},
  {Littenberg}, {McWilliams}, {Nelemans}, {Petiteau}, {Porter}, {Schutz},
  {Sesana}, {Stebbins}, {Sumner}, {Vallisneri}, {Vitale}, {Volonteri}, \&
  {Ward}}]{2012CQGra..29l4016A}
{Amaro-Seoane} P. {et~al.}, 2012, Classical and Quantum Gravity, 29, 124016

\bibitem[{{Anderson}(1929)}]{1929ZPhy...56..851A}
{Anderson} W., 1929, Zeitschrift fur Physik, 56, 851

\bibitem[{{Andrews} {et~al}\mbox{.}(2015){Andrews}, {Ag{\"u}eros}, {Gianninas},
  {Kilic}, {Dhital}, \& {Anderson}}]{2015ApJ...815...63A}
{Andrews} J.~J., {Ag{\"u}eros} M.~A., {Gianninas} A., {Kilic} M., {Dhital} S.,
  {Anderson} S.~F., 2015, \apj, 815, 63

\bibitem[{{Avni} \& {Bahcall}(1980)}]{1980ApJ...235..694A}
{Avni} Y., {Bahcall} J.~N., 1980, \apj, 235, 694

\bibitem[{{Bahcall} \& {Casertano}(1986)}]{1986ApJ...308..347B}
{Bahcall} J.~N., {Casertano} S., 1986, \apj, 308, 347

\bibitem[{{Barstow} {et~al}\mbox{.}(2014){Barstow}, {Barstow}, {Casewell},
  {Holberg}, \& {Hubeny}}]{2014MNRAS.440.1607B}
{Barstow} M.~A., {Barstow} J.~K., {Casewell} S.~L., {Holberg} J.~B., {Hubeny}
  I., 2014, \mnras, 440, 1607

\bibitem[{{Basu} \& {Antia}(1997)}]{1997MNRAS.287..189B}
{Basu} S., {Antia} H.~M., 1997, \mnras, 287, 189

\bibitem[{{Bedin} {et~al}\mbox{.}(2008){Bedin}, {King}, {Anderson}, {Piotto},
  {Salaris}, {Cassisi}, \& {Serenelli}}]{2008ApJ...678.1279B}
{Bedin} L.~R., {King} I.~R., {Anderson} J., {Piotto} G., {Salaris} M.,
  {Cassisi} S., {Serenelli} A., 2008, \apj, 678, 1279

\bibitem[{{Bedin} {et~al}\mbox{.}(2010){Bedin}, {Salaris}, {King}, {Piotto},
  {Anderson}, \& {Cassisi}}]{2010ApJ...708L..32B}
{Bedin} L.~R., {Salaris} M., {King} I.~R., {Piotto} G., {Anderson} J.,
  {Cassisi} S., 2010, \apjl, 708, L32

\bibitem[{{Beers} \& {Sommer-Larsen}(1995)}]{1995ApJS...96..175B}
{Beers} T.~C., {Sommer-Larsen} J., 1995, \apjs, 96, 175

\bibitem[{{Bergeron}, {Leggett} \& {Ruiz}(2001){Bergeron}, {Leggett}, \&
  {Ruiz}}]{2001ApJS..133..413B}
{Bergeron} P., {Leggett} S.~K., {Ruiz} M.~T., 2001, \apjs, 133, 413

\bibitem[{{Bergeron}, {Ruiz} \& {Leggett}(1997){Bergeron}, {Ruiz}, \&
  {Leggett}}]{1997ApJS..108..339B}
{Bergeron} P., {Ruiz} M.~T., {Leggett} S.~K., 1997, \apjs, 108, 339

\bibitem[{{Bergeron}, {Saumon} \& {Wesemael}(1995){Bergeron}, {Saumon}, \&
  {Wesemael}}]{1995ApJ...443..764B}
{Bergeron} P., {Saumon} D., {Wesemael} F., 1995, \apj, 443, 764

\bibitem[{{Bergeron} {et~al}\mbox{.}(2011){Bergeron}, {Wesemael}, {Dufour},
  {Beauchamp}, {Hunter}, {Saffer}, {Gianninas}, {Ruiz}, {Limoges}, {Dufour},
  {Fontaine}, \& {Liebert}}]{2011ApJ...737...28B}
{Bergeron} P. {et~al.}, 2011, \apj, 737, 28

\bibitem[{{Bergeron}, {Wesemael} \& {Fontaine}(1992){Bergeron}, {Wesemael}, \&
  {Fontaine}}]{1992ApJ...387..288B}
{Bergeron} P., {Wesemael} F., {Fontaine} G., 1992, \apj, 387, 288

\bibitem[{{Boyle}(1989)}]{1989MNRAS.240..533B}
{Boyle} B.~J., 1989, \mnras, 240, 533

\bibitem[{{Bressan} {et~al}\mbox{.}(2013){Bressan}, {Marigo}, {Girardi},
  {Nanni}, \& {Rubele}}]{2013EPJWC..4303001B}
{Bressan} A., {Marigo} P., {Girardi} L., {Nanni} A., {Rubele} S., 2013, in
  European Physical Journal Web of Conferences, Vol.~43, European Physical
  Journal Web of Conferences, p. 03001

\bibitem[{{Bressan} {et~al}\mbox{.}(2012){Bressan}, {Marigo}, {Girardi},
  {Salasnich}, {Dal Cero}, {Rubele}, \& {Nanni}}]{2012MNRAS.427..127B}
{Bressan} A., {Marigo} P., {Girardi} L., {Salasnich} B., {Dal Cero} C.,
  {Rubele} S., {Nanni} A., 2012, \mnras, 427, 127

\bibitem[{{Burleigh} {et~al}\mbox{.}(2008){Burleigh}, {Clarke}, {Hogan},
  {Brinkworth}, {Bergeron}, {Dufour}, {Dobbie}, {Levan}, {Hodgkin}, {Hoard}, \&
  {Wachter}}]{2008MNRAS.386L...5B}
{Burleigh} M.~R. {et~al.}, 2008, \mnras, 386, L5

\bibitem[{{Caffau} {et~al}\mbox{.}(2011){Caffau}, {Ludwig}, {Steffen},
  {Freytag}, \& {Bonifacio}}]{2011SoPh..268..255C}
{Caffau} E., {Ludwig} H.-G., {Steffen} M., {Freytag} B., {Bonifacio} P., 2011,
  \solphys, 268, 255

\bibitem[{{Calamida} {et~al}\mbox{.}(2015){Calamida}, {Sahu}, {Casertano},
  {Anderson}, {Cassisi}, {Gennaro}, {Cignoni}, {Brown}, {Kains}, {Ferguson},
  {Livio}, {Bond}, {Buonanno}, {Clarkson}, {Ferraro}, {Pietrinferni},
  {Salaris}, \& {Valenti}}]{2015ApJ...810....8C}
{Calamida} A. {et~al.}, 2015, \apj, 810, 8

\bibitem[{{Camacho} {et~al}\mbox{.}(2007){Camacho}, {Torres}, {Isern},
  {Althaus}, \& {Garc{\'{\i}}a-Berro}}]{2007A&A...471..151C}
{Camacho} J., {Torres} S., {Isern} J., {Althaus} L.~G., {Garc{\'{\i}}a-Berro}
  E., 2007, \aap, 471, 151

\bibitem[{{Carney} {et~al}\mbox{.}(1996){Carney}, {Laird}, {Latham}, \&
  {Aguilar}}]{1996AJ....112..668C}
{Carney} B.~W., {Laird} J.~B., {Latham} D.~W., {Aguilar} L.~A., 1996, \aj, 112,
  668

\bibitem[{{Carrasco} {et~al}\mbox{.}(2014){Carrasco}, {Catal{\'a}n}, {Jordi},
  {Tremblay}, {Napiwotzki}, {Luri}, {Robin}, \&
  {Kowalski}}]{2014A&A...565A..11C}
{Carrasco} J.~M., {Catal{\'a}n} S., {Jordi} C., {Tremblay} P.-E., {Napiwotzki}
  R., {Luri} X., {Robin} A.~C., {Kowalski} P.~M., 2014, \aap, 565, A11

\bibitem[{{Casewell} {et~al}\mbox{.}(2015){Casewell}, {Dobbie}, {Geier},
  {Lodieu}, \& {Hambly}}]{2015MNRAS.451.4259C}
{Casewell} S.~L., {Dobbie} P.~D., {Geier} S., {Lodieu} N., {Hambly} N.~C.,
  2015, \mnras, 451, 4259

\bibitem[{{Cassisi} {et~al}\mbox{.}(2007){Cassisi}, {Potekhin}, {Pietrinferni},
  {Catelan}, \& {Salaris}}]{2007ApJ...661.1094C}
{Cassisi} S., {Potekhin} A.~Y., {Pietrinferni} A., {Catelan} M., {Salaris} M.,
  2007, \apj, 661, 1094

\bibitem[{{Catal{\'a}n}(2015)}]{2015ASPC..493..325C}
{Catal{\'a}n} S., 2015, in Astronomical Society of the Pacific Conference
  Series, Vol. 493, 19th European Workshop on White Dwarfs, {Dufour} P.,
  {Bergeron} P., {Fontaine} G., eds., p. 325

\bibitem[{{Catal{\'a}n} {et~al}\mbox{.}(2008{\natexlab{a}}){Catal{\'a}n},
  {Isern}, {Garc{\'{\i}}a-Berro}, \& {Ribas}}]{2008MNRAS.387.1693C}
{Catal{\'a}n} S., {Isern} J., {Garc{\'{\i}}a-Berro} E., {Ribas} I.,
  2008{\natexlab{a}}, \mnras, 387, 1693

\bibitem[{{Catal{\'a}n} {et~al}\mbox{.}(2008{\natexlab{b}}){Catal{\'a}n},
  {Isern}, {Garc{\'{\i}}a-Berro}, {Ribas}, {Allende Prieto}, \&
  {Bonanos}}]{2008A&A...477..213C}
{Catal{\'a}n} S., {Isern} J., {Garc{\'{\i}}a-Berro} E., {Ribas} I., {Allende
  Prieto} C., {Bonanos} A.~Z., 2008{\natexlab{b}}, \aap, 477, 213

\bibitem[{{Chambers}(2012)}]{2012AAS...22010704C}
{Chambers} K.~C., 2012, in American Astronomical Society Meeting Abstracts,
  Vol. 220, American Astronomical Society Meeting Abstracts \#220, p. 107.04

\bibitem[{{Chandrasekhar}(1939)}]{1939isss.book.....C}
{Chandrasekhar} S., 1939, {An introduction to the study of stellar structure}

\bibitem[{{Chen} \& {Hansen}(2012)}]{2012ApJ...753L..16C}
{Chen} E.~Y., {Hansen} B.~M.~S., 2012, \apjl, 753, L16

\bibitem[{{Chiba} \& {Beers}(2000)}]{2000AJ....119.2843C}
{Chiba} M., {Beers} T.~C., 2000, \aj, 119, 2843

\bibitem[{{Chiba} \& {Yoshii}(1998)}]{1998AJ....115..168C}
{Chiba} M., {Yoshii} Y., 1998, \aj, 115, 168

\bibitem[{{Choloniewski}(1986)}]{1986MNRAS.223....1C}
{Choloniewski} J., 1986, \mnras, 223, 1

\bibitem[{{Chu} {et~al}\mbox{.}(2001){Chu}, {Dunne}, {Gruendl}, \&
  {Brandner}}]{2001ApJ...546L..61C}
{Chu} Y.-H., {Dunne} B.~C., {Gruendl} R.~A., {Brandner} W., 2001, \apjl, 546,
  L61

\bibitem[{{Cignoni} {et~al}\mbox{.}(2006){Cignoni}, {Degl'Innocenti}, {Prada
  Moroni}, \& {Shore}}]{2006A&A...459..783C}
{Cignoni} M., {Degl'Innocenti} S., {Prada Moroni} P.~G., {Shore} S.~N., 2006,
  \aap, 459, 783

\bibitem[{{Claver} {et~al}\mbox{.}(2001){Claver}, {Liebert}, {Bergeron}, \&
  {Koester}}]{2001ApJ...563..987C}
{Claver} C.~F., {Liebert} J., {Bergeron} P., {Koester} D., 2001, \apj, 563, 987

\bibitem[{{Cummings} {et~al}\mbox{.}(2015){Cummings}, {Kalirai}, {Tremblay}, \&
  {Ramirez-Ruiz}}]{2015ApJ...807...90C}
{Cummings} J.~D., {Kalirai} J.~S., {Tremblay} P.-E., {Ramirez-Ruiz} E., 2015,
  \apj, 807, 90

\bibitem[{{Cummings} {et~al}\mbox{.}(2016){Cummings}, {Kalirai}, {Tremblay}, \&
  {Ramirez-Ruiz}}]{2016ApJ...818...84C}
{Cummings} J.~D., {Kalirai} J.~S., {Tremblay} P.-E., {Ramirez-Ruiz} E., 2016,
  \apj, 818, 84

\bibitem[{{D'Antona} \& {Mazzitelli}(1989)}]{1989ApJ...347..934D}
{D'Antona} F., {Mazzitelli} I., 1989, \apj, 347, 934

\bibitem[{{Das} \& {Mukhopadhyay}(2015)}]{2015JCAP...05..045D}
{Das} U., {Mukhopadhyay} B., 2015, \jcap, 5, 045

\bibitem[{{Dawson} {et~al}\mbox{.}(2013){Dawson}, {Schlegel}, {Ahn},
  {Anderson}, {Aubourg}, {Bailey}, {Barkhouser}, {Bautista}, {Beifiori},
  {Berlind}, {Bhardwaj}, {Bizyaev}, {Blake}, {Blanton}, {Blomqvist}, {Bolton},
  {Borde}, {Bovy}, {Brandt}, {Brewington}, {Brinkmann}, {Brown}, {Brownstein},
  {Bundy}, {Busca}, {Carithers}, {Carnero}, {Carr}, {Chen}, {Comparat},
  {Connolly}, {Cope}, {Croft}, {Cuesta}, {da Costa}, {Davenport}, {Delubac},
  {de Putter}, {Dhital}, {Ealet}, {Ebelke}, {Eisenstein}, {Escoffier}, {Fan},
  {Filiz Ak}, {Finley}, {Font-Ribera}, {G{\'e}nova-Santos}, {Gunn}, {Guo},
  {Haggard}, {Hall}, {Hamilton}, {Harris}, {Harris}, {Ho}, {Hogg}, {Holder},
  {Honscheid}, {Huehnerhoff}, {Jordan}, {Jordan}, {Kauffmann}, {Kazin},
  {Kirkby}, {Klaene}, {Kneib}, {Le Goff}, {Lee}, {Long}, {Loomis}, {Lundgren},
  {Lupton}, {Maia}, {Makler}, {Malanushenko}, {Malanushenko}, {Mandelbaum},
  {Manera}, {Maraston}, {Margala}, {Masters}, {McBride}, {McDonald}, {McGreer},
  {McMahon}, {Mena}, {Miralda-Escud{\'e}}, {Montero-Dorta}, {Montesano},
  {Muna}, {Myers}, {Naugle}, {Nichol}, {Noterdaeme}, {Nuza}, {Olmstead},
  {Oravetz}, {Oravetz}, {Owen}, {Padmanabhan}, {Palanque-Delabrouille}, {Pan},
  {Parejko}, {P{\^a}ris}, {Percival}, {P{\'e}rez-Fournon},
  {P{\'e}rez-R{\`a}fols}, {Petitjean}, {Pfaffenberger}, {Pforr}, {Pieri},
  {Prada}, {Price-Whelan}, {Raddick}, {Rebolo}, {Rich}, {Richards}, {Rockosi},
  {Roe}, {Ross}, {Ross}, {Rossi}, {Rubi{\~n}o-Martin}, {Samushia},
  {S{\'a}nchez}, {Sayres}, {Schmidt}, {Schneider}, {Sc{\'o}ccola}, {Seo},
  {Shelden}, {Sheldon}, {Shen}, {Shu}, {Slosar}, {Smee}, {Snedden}, {Stauffer},
  {Steele}, {Strauss}, {Streblyanska}, {Suzuki}, {Swanson}, {Tal}, {Tanaka},
  {Thomas}, {Tinker}, {Tojeiro}, {Tremonti}, {Vargas Maga{\~n}a}, {Verde},
  {Viel}, {Wake}, {Watson}, {Weaver}, {Weinberg}, {Weiner}, {West}, {White},
  {Wood-Vasey}, {Yeche}, {Zehavi}, {Zhao}, \& {Zheng}}]{2013AJ....145...10D}
{Dawson} K.~S. {et~al.}, 2013, \aj, 145, 10

\bibitem[{{Debes} {et~al}\mbox{.}(2015){Debes}, {Kilic}, {Tremblay},
  {L{\'o}pez-Morales}, {Anglada-Escude}, {Napiwotzki}, {Osip}, \&
  {Weinberger}}]{2015AJ....149..176D}
{Debes} J.~H., {Kilic} M., {Tremblay} P.-E., {L{\'o}pez-Morales} M.,
  {Anglada-Escude} G., {Napiwotzki} R., {Osip} D., {Weinberger} A., 2015, \aj,
  149, 176

\bibitem[{{Dehnen} \& {Binney}(1998)}]{1998MNRAS.298..387D}
{Dehnen} W., {Binney} J.~J., 1998, \mnras, 298, 387

\bibitem[{{Deloye} \& {Bildsten}(2002)}]{2002ApJ...580.1077D}
{Deloye} C.~J., {Bildsten} L., 2002, \apj, 580, 1077

\bibitem[{{Digby} {et~al}\mbox{.}(2003){Digby}, {Hambly}, {Cooke}, {Reid}, \&
  {Cannon}}]{2003MNRAS.344..583D}
{Digby} A.~P., {Hambly} N.~C., {Cooke} J.~A., {Reid} I.~N., {Cannon} R.~D.,
  2003, \mnras, 344, 583

\bibitem[{{Dobbie} {et~al}\mbox{.}(2012){Dobbie}, {Day-Jones}, {Williams},
  {Casewell}, {Burleigh}, {Lodieu}, {Parker}, \&
  {Baxter}}]{2012MNRAS.423.2815D}
{Dobbie} P.~D., {Day-Jones} A., {Williams} K.~A., {Casewell} S.~L., {Burleigh}
  M.~R., {Lodieu} N., {Parker} Q.~A., {Baxter} R., 2012, \mnras, 423, 2815

\bibitem[{{Dobbie} {et~al}\mbox{.}(2013){Dobbie}, {K{\"u}lebi}, {Casewell},
  {Burleigh}, {Parker}, {Baxter}, {Lawrie}, {Jordan}, \&
  {Koester}}]{2013MNRAS.428L..16D}
{Dobbie} P.~D. {et~al.}, 2013, \mnras, 428, L16

\bibitem[{{Dobbie} {et~al}\mbox{.}(2006){Dobbie}, {Napiwotzki}, {Burleigh},
  {Barstow}, {Boyce}, {Casewell}, {Jameson}, {Hubeny}, \&
  {Fontaine}}]{2006MNRAS.369..383D}
{Dobbie} P.~D. {et~al.}, 2006, \mnras, 369, 383

\bibitem[{{Dobbie} {et~al}\mbox{.}(2009){Dobbie}, {Napiwotzki}, {Burleigh},
  {Williams}, {Sharp}, {Barstow}, {Casewell}, \&
  {Hubeny}}]{2009MNRAS.395.2248D}
{Dobbie} P.~D., {Napiwotzki} R., {Burleigh} M.~R., {Williams} K.~A., {Sharp}
  R., {Barstow} M.~A., {Casewell} S.~L., {Hubeny} I., 2009, \mnras, 395, 2248

\bibitem[{{Dobbie} {et~al}\mbox{.}(2004){Dobbie}, {Pinfield}, {Napiwotzki},
  {Hambly}, {Burleigh}, {Barstow}, {Jameson}, \&
  {Hubeny}}]{2004MNRAS.355L..39D}
{Dobbie} P.~D., {Pinfield} D.~J., {Napiwotzki} R., {Hambly} N.~C., {Burleigh}
  M.~R., {Barstow} M.~A., {Jameson} R.~F., {Hubeny} I., 2004, \mnras, 355, L39

\bibitem[{{Downes}(1986)}]{1986ApJS...61..569D}
{Downes} R.~A., 1986, \apjs, 61, 569

\bibitem[{{Eggen}, {Lynden-Bell} \& {Sandage}(1962){Eggen}, {Lynden-Bell}, \&
  {Sandage}}]{1962ApJ...136..748E}
{Eggen} O.~J., {Lynden-Bell} D., {Sandage} A.~R., 1962, \apj, 136, 748

\bibitem[{{Eggleton}, {Faulkner} \& {Flannery}(1973){Eggleton}, {Faulkner}, \&
  {Flannery}}]{1973A&A....23..325E}
{Eggleton} P.~P., {Faulkner} J., {Flannery} B.~P., 1973, \aap, 23, 325

\bibitem[{{Erben} \& {CFHTLenS Collaboration}(2012)}]{2012AAS...21913009E}
{Erben} T., {CFHTLenS Collaboration}, 2012, in American Astronomical Society
  Meeting Abstracts, Vol. 219, American Astronomical Society Meeting Abstracts
  \#219, p. 130.09

\bibitem[{{ESA}(1997)}]{1997ESASP1200.....E}
{ESA}, ed., 1997, ESA Special Publication, Vol. 1200, {The HIPPARCOS and TYCHO
  catalogues. Astrometric and photometric star catalogues derived from the ESA
  HIPPARCOS Space Astrometry Mission}

\bibitem[{{Evans}(1992)}]{1992MNRAS.255..521E}
{Evans} D.~W., 1992, \mnras, 255, 521

\bibitem[{{Farrow} {et~al}\mbox{.}(2014){Farrow}, {Cole}, {Metcalfe}, {Draper},
  {Norberg}, {Foucaud}, {Burgett}, {Chambers}, {Kaiser}, {Kudritzki},
  {Magnier}, {Price}, {Tonry}, \& {Waters}}]{2014MNRAS.437..748F}
{Farrow} D.~J. {et~al.}, 2014, \mnras, 437, 748

\bibitem[{{Felten}(1976)}]{1976ApJ...207..700F}
{Felten} J.~E., 1976, \apj, 207, 700

\bibitem[{{Ferrario} {et~al}\mbox{.}(2005){Ferrario}, {Wickramasinghe},
  {Liebert}, \& {Williams}}]{2005MNRAS.361.1131F}
{Ferrario} L., {Wickramasinghe} D., {Liebert} J., {Williams} K.~A., 2005,
  \mnras, 361, 1131

\bibitem[{{Fitzpatrick}(1999)}]{1999PASP..111...63F}
{Fitzpatrick} E.~L., 1999, \pasp, 111, 63

\bibitem[{{Flynn}, {Gould} \& {Bahcall}(1996){Flynn}, {Gould}, \&
  {Bahcall}}]{1996ApJ...466L..55F}
{Flynn} C., {Gould} A., {Bahcall} J.~N., 1996, \apjl, 466, L55

\bibitem[{{Fontaine}, {Brassard} \& {Bergeron}(2001){Fontaine}, {Brassard}, \&
  {Bergeron}}]{2001PASP..113..409F}
{Fontaine} G., {Brassard} P., {Bergeron} P., 2001, \pasp, 113, 409

\bibitem[{{Fontaine} \& {Wesemael}(1997)}]{1997ASSL..214..173F}
{Fontaine} G., {Wesemael} F., 1997, in Astrophysics and Space Science Library,
  Vol. 214, White dwarfs, {Isern} J., {Hernanz} M., {Garcia-Berro} E., eds., p.
  173

\bibitem[{{Foreman-Mackey} {et~al}\mbox{.}(2013){Foreman-Mackey}, {Hogg},
  {Lang}, \& {Goodman}}]{2013PASP..125..306F}
{Foreman-Mackey} D., {Hogg} D.~W., {Lang} D., {Goodman} J., 2013, \pasp, 125,
  306

\bibitem[{{Fowler}, {Caughlan} \& {Zimmerman}(1975){Fowler}, {Caughlan}, \&
  {Zimmerman}}]{1975ARA&A..13...69F}
{Fowler} W.~A., {Caughlan} G.~R., {Zimmerman} B.~A., 1975, \araa, 13, 69

\bibitem[{{Fuchs}, {Jahrei{\ss}} \& {Flynn}(2009){Fuchs}, {Jahrei{\ss}}, \&
  {Flynn}}]{2009AJ....137..266F}
{Fuchs} B., {Jahrei{\ss}} H., {Flynn} C., 2009, \aj, 137, 266

\bibitem[{{G{\"a}nsicke} {et~al}\mbox{.}(2008){G{\"a}nsicke}, {Koester},
  {Marsh}, {Rebassa-Mansergas}, \& {Southworth}}]{2008MNRAS.391L.103G}
{G{\"a}nsicke} B.~T., {Koester} D., {Marsh} T.~R., {Rebassa-Mansergas} A.,
  {Southworth} J., 2008, \mnras, 391, L103

\bibitem[{{Garc{\'{\i}}a-Berro} {et~al}\mbox{.}(2008){Garc{\'{\i}}a-Berro},
  {Althaus}, {C{\'o}rsico}, \& {Isern}}]{2008ApJ...677..473G}
{Garc{\'{\i}}a-Berro} E., {Althaus} L.~G., {C{\'o}rsico} A.~H., {Isern} J.,
  2008, \apj, 677, 473

\bibitem[{{Garcia-Berro} {et~al}\mbox{.}(1988){Garcia-Berro}, {Hernanz},
  {Mochkovitch}, \& {Isern}}]{1988A&A...193..141G}
{Garcia-Berro} E., {Hernanz} M., {Mochkovitch} R., {Isern} J., 1988, \aap, 193,
  141

\bibitem[{{Garcia-Berro}, {Isern} \& {Hernanz}(1997){Garcia-Berro}, {Isern}, \&
  {Hernanz}}]{1997MNRAS.289..973G}
{Garcia-Berro} E., {Isern} J., {Hernanz} M., 1997, \mnras, 289, 973

\bibitem[{{Garc{\'{\i}}a-Berro} {et~al}\mbox{.}(2010){Garc{\'{\i}}a-Berro},
  {Torres}, {Althaus}, {Renedo}, {Lor{\'e}n-Aguilar}, {C{\'o}rsico},
  {Rohrmann}, {Salaris}, \& {Isern}}]{2010Natur.465..194G}
{Garc{\'{\i}}a-Berro} E. {et~al.}, 2010, \nat, 465, 194

\bibitem[{{Geijo} {et~al}\mbox{.}(2006){Geijo}, {Torres}, {Isern}, \&
  {Garc{\'{\i}}a-Berro}}]{2006MNRAS.369.1654G}
{Geijo} E.~M., {Torres} S., {Isern} J., {Garc{\'{\i}}a-Berro} E., 2006, \mnras,
  369, 1654

\bibitem[{{Geisler} {et~al}\mbox{.}(2007){Geisler}, {Wallerstein}, {Smith}, \&
  {Casetti-Dinescu}}]{2007PASP..119..939G}
{Geisler} D., {Wallerstein} G., {Smith} V.~V., {Casetti-Dinescu} D.~I., 2007,
  \pasp, 119, 939

\bibitem[{{Giammichele}, {Bergeron} \& {Dufour}(2012){Giammichele}, {Bergeron},
  \& {Dufour}}]{2012ApJS..199...29G}
{Giammichele} N., {Bergeron} P., {Dufour} P., 2012, \apjs, 199, 29

\bibitem[{{Girard} {et~al}\mbox{.}(2006){Girard}, {Korchagin},
  {Casetti-Dinescu}, {van Altena}, {L{\'o}pez}, \&
  {Monet}}]{2006AJ....132.1768G}
{Girard} T.~M., {Korchagin} V.~I., {Casetti-Dinescu} D.~I., {van Altena} W.~F.,
  {L{\'o}pez} C.~E., {Monet} D.~G., 2006, \aj, 132, 1768

\bibitem[{{G{\'o}rski} {et~al}\mbox{.}(2005){G{\'o}rski}, {Hivon}, {Banday},
  {Wandelt}, {Hansen}, {Reinecke}, \& {Bartelmann}}]{2005ApJ...622..759G}
{G{\'o}rski} K.~M., {Hivon} E., {Banday} A.~J., {Wandelt} B.~D., {Hansen}
  F.~K., {Reinecke} M., {Bartelmann} M., 2005, \apj, 622, 759

\bibitem[{{Green} {et~al}\mbox{.}(2015){Green}, {Schlafly}, {Finkbeiner},
  {Rix}, {Martin}, {Burgett}, {Draper}, {Flewelling}, {Hodapp}, {Kaiser},
  {Kudritzki}, {Magnier}, {Metcalfe}, {Price}, {Tonry}, \&
  {Wainscoat}}]{2015ApJ...810...25G}
{Green} G.~M. {et~al.}, 2015, \apj, 810, 25

\bibitem[{{Green}(1980)}]{1980ApJ...238..685G}
{Green} R.~F., 1980, \apj, 238, 685

\bibitem[{{Hachisu} {et~al}\mbox{.}(2012){Hachisu}, {Kato}, {Saio}, \&
  {Nomoto}}]{2012ApJ...744...69H}
{Hachisu} I., {Kato} M., {Saio} H., {Nomoto} K., 2012, \apj, 744, 69

\bibitem[{{Haft}, {Raffelt} \& {Weiss}(1994){Haft}, {Raffelt}, \&
  {Weiss}}]{1994ApJ...425..222H}
{Haft} M., {Raffelt} G., {Weiss} A., 1994, \apj, 425, 222

\bibitem[{{Hambly} {et~al}\mbox{.}(2013){Hambly}, {Rowell}, {Tonry}, {Magnier},
  \& {Stubbs}}]{2013ASPC..469..253H}
{Hambly} N., {Rowell} N., {Tonry} J., {Magnier} E., {Stubbs} C., 2013, in
  Astronomical Society of the Pacific Conference Series, Vol. 469, 18th
  European White Dwarf Workshop., p. 253

\bibitem[{{Hambly} {et~al}\mbox{.}(2001{\natexlab{a}}){Hambly}, {Davenhall},
  {Irwin}, \& {MacGillivray}}]{2001MNRAS.326.1315H}
{Hambly} N.~C., {Davenhall} A.~C., {Irwin} M.~J., {MacGillivray} H.~T.,
  2001{\natexlab{a}}, \mnras, 326, 1315

\bibitem[{{Hambly}, {Irwin} \& {MacGillivray}(2001){Hambly}, {Irwin}, \&
  {MacGillivray}}]{2001MNRAS.326.1295H}
{Hambly} N.~C., {Irwin} M.~J., {MacGillivray} H.~T., 2001, \mnras, 326, 1295

\bibitem[{{Hambly} {et~al}\mbox{.}(2001{\natexlab{b}}){Hambly}, {MacGillivray},
  {Read}, {Tritton}, {Thomson}, {Kelly}, {Morgan}, {Smith}, {Driver},
  {Williamson}, {Parker}, {Hawkins}, {Williams}, \&
  {Lawrence}}]{2001MNRAS.326.1279H}
{Hambly} N.~C. {et~al.}, 2001{\natexlab{b}}, \mnras, 326, 1279

\bibitem[{{Hambly}, {Smartt} \& {Hodgkin}(1997){Hambly}, {Smartt}, \&
  {Hodgkin}}]{1997ApJ...489L.157H}
{Hambly} N.~C., {Smartt} S.~J., {Hodgkin} S.~T., 1997, \apjl, 489, L157

\bibitem[{{Hansen}(1999)}]{1999ApJ...520..680H}
{Hansen} B.~M.~S., 1999, \apj, 520, 680

\bibitem[{{Hansen} {et~al}\mbox{.}(2002){Hansen}, {Brewer}, {Fahlman},
  {Gibson}, {Ibata}, {Limongi}, {Rich}, {Richer}, {Shara}, \&
  {Stetson}}]{2002ApJ...574L.155H}
{Hansen} B.~M.~S. {et~al.}, 2002, \apjl, 574, L155

\bibitem[{{Harris} {et~al}\mbox{.}(2006){Harris}, {Munn}, {Kilic}, {Liebert},
  {Williams}, {von Hippel}, {Levine}, {Monet}, {Eisenstein}, {Kleinman},
  {Metcalfe}, {Nitta}, {Winget}, {Brinkmann}, {Fukugita}, {Knapp}, {Lupton},
  {Smith}, \& {Schneider}}]{2006AJ....131..571H}
{Harris} H.~C. {et~al.}, 2006, \aj, 131, 571

\bibitem[{{Hernanz} {et~al}\mbox{.}(1994){Hernanz}, {Garcia-Berro}, {Isern},
  {Mochkovitch}, {Segretain}, \& {Chabrier}}]{1994ApJ...434..652H}
{Hernanz} M., {Garcia-Berro} E., {Isern} J., {Mochkovitch} R., {Segretain} L.,
  {Chabrier} G., 1994, \apj, 434, 652

\bibitem[{{Heymans} {et~al}\mbox{.}(2012){Heymans}, {Van Waerbeke}, {Miller},
  {Erben}, {Hildebrandt}, {Hoekstra}, {Kitching}, {Mellier}, {Simon},
  {Bonnett}, {Coupon}, {Fu}, {Harnois D{\'e}raps}, {Hudson}, {Kilbinger},
  {Kuijken}, {Rowe}, {Schrabback}, {Semboloni}, {van Uitert}, {Vafaei}, \&
  {Velander}}]{2012MNRAS.427..146H}
{Heymans} C. {et~al.}, 2012, \mnras, 427, 146

\bibitem[{{Hicken} {et~al}\mbox{.}(2007){Hicken}, {Garnavich}, {Prieto},
  {Blondin}, {DePoy}, {Kirshner}, \& {Parrent}}]{2007ApJ...669L..17H}
{Hicken} M., {Garnavich} P.~M., {Prieto} J.~L., {Blondin} S., {DePoy} D.~L.,
  {Kirshner} R.~P., {Parrent} J., 2007, \apjl, 669, L17

\bibitem[{{Hildebrandt} {et~al}\mbox{.}(2012){Hildebrandt}, {Erben}, {Kuijken},
  {van Waerbeke}, {Heymans}, {Coupon}, {Benjamin}, {Bonnett}, {Fu}, {Hoekstra},
  {Kitching}, {Mellier}, {Miller}, {Velander}, {Hudson}, {Rowe}, {Schrabback},
  {Semboloni}, \& {Ben{\'{\i}}tez}}]{2012MNRAS.421.2355H}
{Hildebrandt} H. {et~al.}, 2012, \mnras, 421, 2355

\bibitem[{{Hodapp} {et~al}\mbox{.}(2004){Hodapp}, {Siegmund}, {Kaiser},
  {Chambers}, {Laux}, {Morgan}, \& {Mannery}}]{2004SPIE.5489..667H}
{Hodapp} K.~W., {Siegmund} W.~A., {Kaiser} N., {Chambers} K.~C., {Laux} U.,
  {Morgan} J., {Mannery} E., 2004, in \procspie, Vol. 5489, Ground-based
  Telescopes, {Oschmann} Jr. J.~M., ed., pp. 667--678

\bibitem[{{Hogan}, {Burleigh} \& {Clarke}(2009){Hogan}, {Burleigh}, \&
  {Clarke}}]{2009MNRAS.396.2074H}
{Hogan} E., {Burleigh} M.~R., {Clarke} F.~J., 2009, \mnras, 396, 2074

\bibitem[{{Holberg} \& {Bergeron}(2006)}]{2006AJ....132.1221H}
{Holberg} J.~B., {Bergeron} P., 2006, \aj, 132, 1221

\bibitem[{{Holberg}, {Oswalt} \& {Sion}(2002){Holberg}, {Oswalt}, \&
  {Sion}}]{2002ApJ...571..512H}
{Holberg} J.~B., {Oswalt} T.~D., {Sion} E.~M., 2002, \apj, 571, 512

\bibitem[{{Horowitz}, {Schneider} \& {Berry}(2010){Horowitz}, {Schneider}, \&
  {Berry}}]{2010PhRvL.104w1101H}
{Horowitz} C.~J., {Schneider} A.~S., {Berry} D.~K., 2010, Physical Review
  Letters, 104, 231101

\bibitem[{{Howell} {et~al}\mbox{.}(2006){Howell}, {Sullivan}, {Nugent},
  {Ellis}, {Conley}, {Le Borgne}, {Carlberg}, {Guy}, {Balam}, {Basa},
  {Fouchez}, {Hook}, {Hsiao}, {Neill}, {Pain}, {Perrett}, \&
  {Pritchet}}]{2006Natur.443..308H}
{Howell} D.~A. {et~al.}, 2006, \nat, 443, 308

\bibitem[{{Howell} {et~al}\mbox{.}(2014){Howell}, {Sobeck}, {Haas}, {Still},
  {Barclay}, {Mullally}, {Troeltzsch}, {Aigrain}, {Bryson}, {Caldwell},
  {Chaplin}, {Cochran}, {Huber}, {Marcy}, {Miglio}, {Najita}, {Smith},
  {Twicken}, \& {Fortney}}]{2014PASP..126..398H}
{Howell} S.~B. {et~al.}, 2014, \pasp, 126, 398

\bibitem[{{Hu}, {Wu} \& {Wu}(2007){Hu}, {Wu}, \& {Wu}}]{2007AnA...466..627H}
{Hu} Q., {Wu} C., {Wu} X.-B., 2007, \aap, 466, 627

\bibitem[{{Hubbard} \& {Lampe}(1969)}]{1969ApJS...18..297H}
{Hubbard} W.~B., {Lampe} M., 1969, \apjs, 18, 297

\bibitem[{{Ibata} {et~al}\mbox{.}(1999){Ibata}, {Richer}, {Gilliland}, \&
  {Scott}}]{1999ApJ...524L..95I}
{Ibata} R.~A., {Richer} H.~B., {Gilliland} R.~L., {Scott} D., 1999, \apjl, 524,
  L95

\bibitem[{{Iben} \& {Laughlin}(1989)}]{1989ApJ...341..312I}
{Iben}, Jr. I., {Laughlin} G., 1989, \apj, 341, 312

\bibitem[{{Iben} \& {MacDonald}(1986)}]{1986ApJ...301..164I}
{Iben}, Jr. I., {MacDonald} J., 1986, \apj, 301, 164

\bibitem[{{Iben} \& {Tutukov}(1984)}]{1984ApJ...282..615I}
{Iben}, Jr. I., {Tutukov} A.~V., 1984, \apj, 282, 615

\bibitem[{{Iglesias} \& {Rogers}(1993)}]{1993ApJ...412..752I}
{Iglesias} C.~A., {Rogers} F.~J., 1993, \apj, 412, 752

\bibitem[{{Ignace}(2001)}]{2001PASP..113.1227I}
{Ignace} R., 2001, \pasp, 113, 1227

\bibitem[{{Irwin}(1985)}]{1985MNRAS.214..575I}
{Irwin} M.~J., 1985, \mnras, 214, 575

\bibitem[{{Isern} {et~al}\mbox{.}(2000){Isern}, {Garc{\'{\i}}a-Berro},
  {Hernanz}, \& {Chabrier}}]{2000ApJ...528..397I}
{Isern} J., {Garc{\'{\i}}a-Berro} E., {Hernanz} M., {Chabrier} G., 2000, \apj,
  528, 397

\bibitem[{{Isern}, {Garc{\'{\i}}a-Berro} \& {Salaris}(2001){Isern},
  {Garc{\'{\i}}a-Berro}, \& {Salaris}}]{2001ASPC..245..328I}
{Isern} J., {Garc{\'{\i}}a-Berro} E., {Salaris} M., 2001, in Astronomical
  Society of the Pacific Conference Series, Vol. 245, Astrophysical Ages and
  Times Scales, {von Hippel} T., {Simpson} C., {Manset} N., eds., p. 328

\bibitem[{{Isern} {et~al}\mbox{.}(1997){Isern}, {Mochkovitch},
  {Garc{\'{\i}}a-Berro}, \& {Hernanz}}]{1997ApJ...485..308I}
{Isern} J., {Mochkovitch} R., {Garc{\'{\i}}a-Berro} E., {Hernanz} M., 1997,
  \apj, 485, 308

\bibitem[{{Ishida} {et~al}\mbox{.}(1982){Ishida}, {Mikami}, {Noguchi}, \&
  {Maehara}}]{1982PASJ...34..381I}
{Ishida} K., {Mikami} T., {Noguchi} T., {Maehara} H., 1982, \pasj, 34, 381

\bibitem[{{Isobe}(1974)}]{1974A&A....36..333I}
{Isobe} S., 1974, \aap, 36, 333

\bibitem[{{Itoh}, {Hayashi} \& {Kohyama}(1993){Itoh}, {Hayashi}, \&
  {Kohyama}}]{1993ApJ...418..405I}
{Itoh} N., {Hayashi} H., {Kohyama} Y., 1993, \apj, 418, 405

\bibitem[{{Itoh} {et~al}\mbox{.}(1996){Itoh}, {Hayashi}, {Nishikawa}, \&
  {Kohyama}}]{1996ApJS..102..411I}
{Itoh} N., {Hayashi} H., {Nishikawa} A., {Kohyama} Y., 1996, \apjs, 102, 411

\bibitem[{{Itoh} \& {Kohyama}(1993)}]{1993ApJ...404..268I}
{Itoh} N., {Kohyama} Y., 1993, \apj, 404, 268

\bibitem[{{Itoh} {et~al}\mbox{.}(1984){Itoh}, {Kohyama}, {Matsumoto}, \&
  {Seki}}]{1984ApJ...285..758I}
{Itoh} N., {Kohyama} Y., {Matsumoto} N., {Seki} M., 1984, \apj, 285, 758

\bibitem[{{Itoh} {et~al}\mbox{.}(1983){Itoh}, {Mitake}, {Iyetomi}, \&
  {Ichimaru}}]{1983ApJ...273..774I}
{Itoh} N., {Mitake} S., {Iyetomi} H., {Ichimaru} S., 1983, \apj, 273, 774

\bibitem[{{Jahreiss} \& {Gliese}(1993)}]{1993IAUS..156..107J}
{Jahreiss} H., {Gliese} W., 1993, in IAU Symposium, Vol. 156, Developments in
  Astrometry and their Impact on Astrophysics and Geodynamics, {Mueller} I.~I.,
  {Kolaczek} B., eds., pp. 107--112

\bibitem[{{Johnson} \& {Soderblom}(1987)}]{1987AJ.....93..864J}
{Johnson} D.~R.~H., {Soderblom} D.~R., 1987, \aj, 93, 864

\bibitem[{{Jones} {et~al}\mbox{.}(1991){Jones}, {Fong}, {Shanks}, {Ellis}, \&
  {Peterson}}]{1991MNRAS.249..481J}
{Jones} L.~R., {Fong} R., {Shanks} T., {Ellis} R.~S., {Peterson} B.~A., 1991,
  \mnras, 249, 481

\bibitem[{{Kaiser} {et~al}\mbox{.}(2010){Kaiser}, {Burgett}, {Chambers},
  {Denneau}, {Heasley}, {Jedicke}, {Magnier}, {Morgan}, {Onaka}, \&
  {Tonry}}]{2010SPIE.7733E..0EK}
{Kaiser} N. {et~al.}, 2010, in \procspie, Vol. 7733, Ground-based and Airborne
  Telescopes III, p. 77330E

\bibitem[{{Kalirai} {et~al}\mbox{.}(2008){Kalirai}, {Hansen}, {Kelson},
  {Reitzel}, {Rich}, \& {Richer}}]{2008ApJ...676..594K}
{Kalirai} J.~S., {Hansen} B.~M.~S., {Kelson} D.~D., {Reitzel} D.~B., {Rich}
  R.~M., {Richer} H.~B., 2008, \apj, 676, 594

\bibitem[{{Kalirai} {et~al}\mbox{.}(2005){Kalirai}, {Richer}, {Reitzel},
  {Hansen}, {Rich}, {Fahlman}, {Gibson}, \& {von Hippel}}]{2005ApJ...618L.123K}
{Kalirai} J.~S., {Richer} H.~B., {Reitzel} D., {Hansen} B.~M.~S., {Rich} R.~M.,
  {Fahlman} G.~G., {Gibson} B.~K., {von Hippel} T., 2005, \apjl, 618, L123

\bibitem[{{Kalirai} {et~al}\mbox{.}(2009){Kalirai}, {Saul Davis}, {Richer},
  {Bergeron}, {Catelan}, {Hansen}, \& {Rich}}]{2009ApJ...705..408K}
{Kalirai} J.~S., {Saul Davis} D., {Richer} H.~B., {Bergeron} P., {Catelan} M.,
  {Hansen} B.~M.~S., {Rich} R.~M., 2009, \apj, 705, 408

\bibitem[{{Kepler} {et~al}\mbox{.}(2015){Kepler}, {Pelisoli}, {Koester},
  {Ourique}, {Kleinman}, {Romero}, {Nitta}, {Eisenstein}, {Costa},
  {K{\"u}lebi}, {Jordan}, {Dufour}, {Giommi}, \&
  {Rebassa-Mansergas}}]{2015MNRAS.446.4078K}
{Kepler} S.~O. {et~al.}, 2015, \mnras, 446, 4078

\bibitem[{{Kilic} {et~al}\mbox{.}(2004){Kilic}, {von Hippel}, {Mendez}, \&
  {Winget}}]{2004ApJ...609..766K}
{Kilic} M., {von Hippel} T., {Mendez} R.~A., {Winget} D.~E., 2004, \apj, 609,
  766

\bibitem[{{Kippenhahn}, {Weigert} \& {Weiss}(2012){Kippenhahn}, {Weigert}, \&
  {Weiss}}]{2012sse..book.....K}
{Kippenhahn} R., {Weigert} A., {Weiss} A., 2012, {Stellar Structure and
  Evolution}

\bibitem[{{Kleinman} {et~al}\mbox{.}(2013){Kleinman}, {Kepler}, {Koester},
  {Pelisoli}, {Pe{\c c}anha}, {Nitta}, {Costa}, {Krzesinski}, {Dufour},
  {Lachapelle}, {Bergeron}, {Yip}, {Harris}, {Eisenstein}, {Althaus}, \&
  {C{\'o}rsico}}]{2013ApJS..204....5K}
{Kleinman} S.~J. {et~al.}, 2013, \apjs, 204, 5

\bibitem[{{Knox}, {Hawkins} \& {Hambly}(1999){Knox}, {Hawkins}, \&
  {Hambly}}]{1999MNRAS.306..736K}
{Knox} R.~A., {Hawkins} M.~R.~S., {Hambly} N.~C., 1999, \mnras, 306, 736

\bibitem[{{Koester}(2009)}]{2009A&A...498..517K}
{Koester} D., 2009, \aap, 498, 517

\bibitem[{{Koester} \& {Chanmugam}(1990)}]{1990RPPh...53..837K}
{Koester} D., {Chanmugam} G., 1990, Reports on Progress in Physics, 53, 837

\bibitem[{{Koester}, {G{\"a}nsicke} \& {Farihi}(2014){Koester}, {G{\"a}nsicke},
  \& {Farihi}}]{2014A&A...566A..34K}
{Koester} D., {G{\"a}nsicke} B.~T., {Farihi} J., 2014, \aap, 566, A34

\bibitem[{{Koester} \& {Reimers}(1993)}]{1993A&A...275..479K}
{Koester} D., {Reimers} D., 1993, \aap, 275, 479

\bibitem[{{Koester} \& {Reimers}(1996)}]{1996A&A...313..810K}
{Koester} D., {Reimers} D., 1996, \aap, 313, 810

\bibitem[{{Ko{\l}os} \& {Wolniewicz}(1965)}]{1965JChPh..43.2429K}
{Ko{\l}os} W., {Wolniewicz} L., 1965, \jcp, 43, 2429

\bibitem[{{Kowalski}(2006{\natexlab{a}})}]{2006ApJ...641..488K}
{Kowalski} P.~M., 2006{\natexlab{a}}, \apj, 641, 488

\bibitem[{{Kowalski}(2006{\natexlab{b}})}]{2006ApJ...651.1120K}
{Kowalski} P.~M., 2006{\natexlab{b}}, \apj, 651, 1120

\bibitem[{{Kowalski}(2007)}]{2007A&A...474..491K}
{Kowalski} P.~M., 2007, \aap, 474, 491

\bibitem[{{Kowalski} \& {Saumon}(2004)}]{2004ApJ...607..970K}
{Kowalski} P.~M., {Saumon} D., 2004, \apj, 607, 970

\bibitem[{{Kowalski} \& {Saumon}(2006)}]{2006ApJ...651L.137K}
{Kowalski} P.~M., {Saumon} D., 2006, \apjl, 651, L137

\bibitem[{{Kroupa}(2001)}]{2001MNRAS.322..231K}
{Kroupa} P., 2001, \mnras, 322, 231

\bibitem[{{Krzesi{\'n}ski}(2013)}]{2013ASPC..469...77K}
{Krzesi{\'n}ski} J., 2013, in Astronomical Society of the Pacific Conference
  Series, Vol. 469, 18th European White Dwarf Workshop., p.~77

\bibitem[{{Krzesinski}, {Torres} \& {Garc{\'{\i}}a-Berro}(2015){Krzesinski},
  {Torres}, \& {Garc{\'{\i}}a-Berro}}]{2015ASPC..493..343K}
{Krzesinski} J., {Torres} S., {Garc{\'{\i}}a-Berro} E., 2015, in Astronomical
  Society of the Pacific Conference Series, Vol. 493, 19th European Workshop on
  White Dwarfs, {Dufour} P., {Bergeron} P., {Fontaine} G., eds., p. 343

\bibitem[{{Kulander} \& {Guest}(1979)}]{1979JPhB...12L.501K}
{Kulander} K.~C., {Guest} M.~F., 1979, Journal of Physics B Atomic Molecular
  Physics, 12, L501

\bibitem[{{Lam} \& {Hambly}(2015)}]{2015ASPC..493..347L}
{Lam} M.~C., {Hambly} N.~C., 2015, in Astronomical Society of the Pacific
  Conference Series, Vol. 493, 19th European Workshop on White Dwarfs, {Dufour}
  P., {Bergeron} P., {Fontaine} G., eds., p. 347

\bibitem[{{Lam}, {Rowell} \& {Hambly}(2015){Lam}, {Rowell}, \&
  {Hambly}}]{2015MNRAS.450.4098L}
{Lam} M.~C., {Rowell} N., {Hambly} N.~C., 2015, \mnras, 450, 4098

\bibitem[{{Lamb} \& {van Horn}(1975)}]{1975ApJ...200..306L}
{Lamb} D.~Q., {van Horn} H.~M., 1975, \apj, 200, 306

\bibitem[{{Lamb}(1974)}]{1974PhDT........56L}
{Lamb}, Jr. D.~Q., 1974, PhD thesis, THE UNIVERSITY OF ROCHESTER.

\bibitem[{{Larson}(1976)}]{1976MNRAS.176...31L}
{Larson} R.~B., 1976, \mnras, 176, 31

\bibitem[{{Lasserre} {et~al}\mbox{.}(2000){Lasserre}, {Afonso}, {Albert},
  {Andersen}, {Ansari}, {Aubourg}, {Bareyre}, {Bauer}, {Beaulieu}, {Blanc},
  {Bouquet}, {Char}, {Charlot}, {Couchot}, {Coutures}, {Derue}, {Ferlet},
  {Glicenstein}, {Goldman}, {Gould}, {Graff}, {Gros}, {Haissinski}, {Hamilton},
  {Hardin}, {de Kat}, {Kim}, {Lesquoy}, {Loup}, {Magneville}, {Mansoux},
  {Marquette}, {Maurice}, {Milsztajn}, {Moniez}, {Palanque-Delabrouille},
  {Perdereau}, {Pr{\'e}vot}, {Regnault}, {Rich}, {Spiro}, {Vidal-Madjar},
  {Vigroux}, {Zylberajch}, \& {EROS Collaboration}}]{2000A&A...355L..39L}
{Lasserre} T. {et~al.}, 2000, \aap, 355, L39

\bibitem[{{Lee} \& {Carney}(1999)}]{1999AJ....118.1373L}
{Lee} J.-W., {Carney} B.~W., 1999, \aj, 118, 1373

\bibitem[{{Leggett}, {Ruiz} \& {Bergeron}(1998){Leggett}, {Ruiz}, \&
  {Bergeron}}]{1998ApJ...497..294L}
{Leggett} S.~K., {Ruiz} M.~T., {Bergeron} P., 1998, \apj, 497, 294

\bibitem[{{Liebert} {et~al}\mbox{.}(1979){Liebert}, {Dahn}, {Gresham}, \&
  {Strittmatter}}]{1979ApJ...233..226L}
{Liebert} J., {Dahn} C.~C., {Gresham} M., {Strittmatter} P.~A., 1979, \apj,
  233, 226

\bibitem[{{Liebert} {et~al}\mbox{.}(1999){Liebert}, {Dahn}, {Harris}, \&
  {Legget}}]{1999ASPC..169...51L}
{Liebert} J., {Dahn} C.~C., {Harris} H.~C., {Legget} S.~K., 1999, in
  Astronomical Society of the Pacific Conference Series, Vol. 169, 11th
  European Workshop on White Dwarfs, {Solheim} S.-E., {Meistas} E.~G., eds.,
  p.~51

\bibitem[{{Liebert}, {Dahn} \& {Monet}(1988){Liebert}, {Dahn}, \&
  {Monet}}]{1988ApJ...332..891L}
{Liebert} J., {Dahn} C.~C., {Monet} D.~G., 1988, \apj, 332, 891

\bibitem[{{Liebert}, {Dahn} \& {Monet}(1989){Liebert}, {Dahn}, \&
  {Monet}}]{1989LNP...328...15L}
{Liebert} J., {Dahn} C.~C., {Monet} D.~G., 1989, in Lecture Notes in Physics,
  Berlin Springer Verlag, Vol. 328, IAU Colloq. 114: White Dwarfs, {Wegner} G.,
  ed., pp. 15--23

\bibitem[{{Lucy}(1974)}]{1974AJ.....79..745L}
{Lucy} L.~B., 1974, \aj, 79, 745

\bibitem[{{MacDonald} \& {Vennes}(1991)}]{1991ApJ...371..719M}
{MacDonald} J., {Vennes} S., 1991, \apj, 371, 719

\bibitem[{{Magni} \& {Mazzitelli}(1979)}]{1979A&A....72..134M}
{Magni} G., {Mazzitelli} I., 1979, \aap, 72, 134

\bibitem[{{Magnier}(2006)}]{2006amos.confE..50M}
{Magnier} E., 2006, in The Advanced Maui Optical and Space Surveillance
  Technologies Conference, p. E50

\bibitem[{{Magnier}(2007)}]{2007ASPC..364..153M}
{Magnier} E., 2007, in Astronomical Society of the Pacific Conference Series,
  Vol. 364, The Future of Photometric, Spectrophotometric and Polarimetric
  Standardization, {Sterken} C., ed., p. 153

\bibitem[{{Magnier} {et~al}\mbox{.}(2008){Magnier}, {Liu}, {Monet}, \&
  {Chambers}}]{2008IAUS..248..553M}
{Magnier} E.~A., {Liu} M., {Monet} D.~G., {Chambers} K.~C., 2008, in IAU
  Symposium, Vol. 248, A Giant Step: from Milli- to Micro-arcsecond Astrometry,
  {Jin} W.~J., {Platais} I., {Perryman} M.~A.~C., eds., pp. 553--559

\bibitem[{{Magnier} {et~al}\mbox{.}(2013){Magnier}, {Schlafly}, {Finkbeiner},
  {Juric}, {Tonry}, {Burgett}, {Chambers}, {Flewelling}, {Kaiser}, {Kudritzki},
  {Morgan}, {Price}, {Sweeney}, \& {Stubbs}}]{2013ApJS..205...20M}
{Magnier} E.~A. {et~al.}, 2013, \apjs, 205, 20

\bibitem[{{Majewski}(1992)}]{1992ApJS...78...87M}
{Majewski} S.~R., 1992, \apjs, 78, 87

\bibitem[{{Majewski}(1993)}]{1993ARA&A..31..575M}
{Majewski} S.~R., 1993, \araa, 31, 575

\bibitem[{{Marigo} \& {Aringer}(2009)}]{2009A&A...508.1539M}
{Marigo} P., {Aringer} B., 2009, \aap, 508, 1539

\bibitem[{{McCook} \& {Sion}(1999)}]{1999ApJS..121....1M}
{McCook} G.~P., {Sion} E.~M., 1999, \apjs, 121, 1

\bibitem[{{McMillan} \& {Binney}(2009)}]{2009MNRAS.400L.103M}
{McMillan} P.~J., {Binney} J.~J., 2009, \mnras, 400, L103

\bibitem[{{Mendez} \& {Guzman}(1998)}]{1998A&A...333..106M}
{Mendez} R.~A., {Guzman} R., 1998, \aap, 333, 106

\bibitem[{{Mestel}(1952)}]{1952MNRAS.112..583M}
{Mestel} L., 1952, \mnras, 112, 583

\bibitem[{{Metcalfe} {et~al}\mbox{.}(1991){Metcalfe}, {Shanks}, {Fong}, \&
  {Jones}}]{1991MNRAS.249..498M}
{Metcalfe} N., {Shanks} T., {Fong} R., {Jones} L.~R., 1991, \mnras, 249, 498

\bibitem[{{Mochkovitch}(1983)}]{1983A&A...122..212M}
{Mochkovitch} R., 1983, \aap, 122, 212

\bibitem[{{Moehler} {et~al}\mbox{.}(2004){Moehler}, {Koester}, {Zoccali},
  {Ferraro}, {Heber}, {Napiwotzki}, \& {Renzini}}]{2004A&A...420..515M}
{Moehler} S., {Koester} D., {Zoccali} M., {Ferraro} F.~R., {Heber} U.,
  {Napiwotzki} R., {Renzini} A., 2004, \aap, 420, 515

\bibitem[{{Murray}(1983)}]{1983veas.book.....M}
{Murray} C.~A., 1983, {Vectorial astrometry}

\bibitem[{{Newberg} {et~al}\mbox{.}(2002){Newberg}, {Yanny}, {Rockosi},
  {Grebel}, {Rix}, {Brinkmann}, {Csabai}, {Hennessy}, {Hindsley}, {Ibata},
  {Ivezi{\'c}}, {Lamb}, {Nash}, {Odenkirchen}, {Rave}, {Schneider}, {Smith},
  {Stolte}, \& {York}}]{2002ApJ...569..245N}
{Newberg} H.~J. {et~al.}, 2002, \apj, 569, 245

\bibitem[{{Noh} \& {Scalo}(1990)}]{1990ApJ...352..605N}
{Noh} H.-R., {Scalo} J., 1990, \apj, 352, 605

\bibitem[{{Nordstr{\"o}m} {et~al}\mbox{.}(2004){Nordstr{\"o}m}, {Mayor},
  {Andersen}, {Holmberg}, {Pont}, {J{\o}rgensen}, {Olsen}, {Udry}, \&
  {Mowlavi}}]{2004A&A...418..989N}
{Nordstr{\"o}m} B. {et~al.}, 2004, \aap, 418, 989

\bibitem[{{Norris} \& {Ryan}(1991)}]{1991ApJ...380..403N}
{Norris} J.~E., {Ryan} S.~G., 1991, \apj, 380, 403

\bibitem[{{Onaka} {et~al}\mbox{.}(2008){Onaka}, {Tonry}, {Isani}, {Lee},
  {Uyeshiro}, {Rae}, {Robertson}, \& {Ching}}]{2008SPIE.7014E..0DO}
{Onaka} P., {Tonry} J.~L., {Isani} S., {Lee} A., {Uyeshiro} R., {Rae} C.,
  {Robertson} L., {Ching} G., 2008, in \procspie, Vol. 7014, Ground-based and
  Airborne Instrumentation for Astronomy II, p. 70140D

\bibitem[{{Oppenheimer} {et~al}\mbox{.}(2001){Oppenheimer}, {Hambly}, {Digby},
  {Hodgkin}, \& {Saumon}}]{2001Sci...292..698O}
{Oppenheimer} B.~R., {Hambly} N.~C., {Digby} A.~P., {Hodgkin} S.~T., {Saumon}
  D., 2001, Science, 292, 698

\bibitem[{{Oswalt} \& {Smith}(1995)}]{1995LNP...443...24O}
{Oswalt} T.~D., {Smith} J.~A., 1995, in Lecture Notes in Physics, Berlin
  Springer Verlag, Vol. 443, White Dwarfs, {Koester} D., {Werner} K., eds.,
  p.~24

\bibitem[{{Oswalt} {et~al}\mbox{.}(1996){Oswalt}, {Smith}, {Wood}, \&
  {Hintzen}}]{1996Natur.382..692O}
{Oswalt} T.~D., {Smith} J.~A., {Wood} M.~A., {Hintzen} P., 1996, \nat, 382, 692

\bibitem[{{Paquette} {et~al}\mbox{.}(1986){Paquette}, {Pelletier}, {Fontaine},
  \& {Michaud}}]{1986ApJS...61..197P}
{Paquette} C., {Pelletier} C., {Fontaine} G., {Michaud} G., 1986, \apjs, 61,
  197

\bibitem[{{Peebles} \& {Dicke}(1968)}]{1968ApJ...154..891P}
{Peebles} P.~J.~E., {Dicke} R.~H., 1968, \apj, 154, 891

\bibitem[{{Petsalakis}, {Theodorakopoulos} \& {Wright}(1988){Petsalakis},
  {Theodorakopoulos}, \& {Wright}}]{1988JChPh..89.6850P}
{Petsalakis} I.~D., {Theodorakopoulos} G., {Wright} J.~S., 1988, \jcp, 89, 6850

\bibitem[{{Prada Moroni} \& {Straniero}(2007)}]{2007A&A...466.1043P}
{Prada Moroni} P.~G., {Straniero} O., 2007, \aap, 466, 1043

\bibitem[{{Preston}, {Shectman} \& {Beers}(1991){Preston}, {Shectman}, \&
  {Beers}}]{1991ApJ...375..121P}
{Preston} G.~W., {Shectman} S.~A., {Beers} T.~C., 1991, \apj, 375, 121

\bibitem[{{Quinn} \& {Goodman}(1986)}]{1986ApJ...309..472Q}
{Quinn} P.~J., {Goodman} J., 1986, \apj, 309, 472

\bibitem[{{Rebassa-Mansergas} {et~al}\mbox{.}(2015){Rebassa-Mansergas}, {Liu},
  {Cojocaru}, {Yuan}, {Torres}, {Garc{\'{\i}}a-Berro}, {Xiang}, {Huang},
  {Koester}, {Hou}, {Li}, \& {Zhang}}]{2015MNRAS.450..743R}
{Rebassa-Mansergas} A. {et~al.}, 2015, \mnras, 450, 743

\bibitem[{{Reid}(2005)}]{2005ARA&A..43..247R}
{Reid} I.~N., 2005, \araa, 43, 247

\bibitem[{{Reid}, {Sahu} \& {Hawley}(2001){Reid}, {Sahu}, \&
  {Hawley}}]{2001ApJ...559..942R}
{Reid} I.~N., {Sahu} K.~C., {Hawley} S.~L., 2001, \apj, 559, 942

\bibitem[{{Renedo} {et~al}\mbox{.}(2010){Renedo}, {Althaus}, {Miller
  Bertolami}, {Romero}, {C{\'o}rsico}, {Rohrmann}, \&
  {Garc{\'{\i}}a-Berro}}]{2010ApJ...717..183R}
{Renedo} I., {Althaus} L.~G., {Miller Bertolami} M.~M., {Romero} A.~D.,
  {C{\'o}rsico} A.~H., {Rohrmann} R.~D., {Garc{\'{\i}}a-Berro} E., 2010, \apj,
  717, 183

\bibitem[{{Renzini} {et~al}\mbox{.}(1996){Renzini}, {Bragaglia}, {Ferraro},
  {Gilmozzi}, {Ortolani}, {Holberg}, {Liebert}, {Wesemael}, \&
  {Bohlin}}]{1996ApJ...465L..23R}
{Renzini} A. {et~al.}, 1996, \apjl, 465, L23

\bibitem[{{Richardson}(1972)}]{1972JOSA...62...55R}
{Richardson} W.~H., 1972, Journal of the Optical Society of America
  (1917-1983), 62, 55

\bibitem[{{Richer} {et~al}\mbox{.}(2000){Richer}, {Hansen}, {Limongi},
  {Chieffi}, {Straniero}, \& {Fahlman}}]{2000ApJ...529..318R}
{Richer} H.~B., {Hansen} B., {Limongi} M., {Chieffi} A., {Straniero} O.,
  {Fahlman} G.~G., 2000, \apj, 529, 318

\bibitem[{{Richer} {et~al}\mbox{.}(1996){Richer}, {Harris}, {Fahlman}, {Bell},
  {Bond}, {Hesser}, {Holland}, {Pryor}, {Stetson}, {Vandenberg}, \& {van den
  Bergh}}]{1996ApJ...463..602R}
{Richer} H.~B. {et~al.}, 1996, \apj, 463, 602

\bibitem[{{Riess} {et~al}\mbox{.}(1998){Riess}, {Filippenko}, {Challis},
  {Clocchiatti}, {Diercks}, {Garnavich}, {Gilliland}, {Hogan}, {Jha},
  {Kirshner}, {Leibundgut}, {Phillips}, {Reiss}, {Schmidt}, {Schommer},
  {Smith}, {Spyromilio}, {Stubbs}, {Suntzeff}, \&
  {Tonry}}]{1998AJ....116.1009R}
{Riess} A.~G. {et~al.}, 1998, \aj, 116, 1009

\bibitem[{{Roach} \& {Kuntz}(1986)}]{1986JChPh..84..822R}
{Roach} A.~C., {Kuntz} P.~J., 1986, \jcp, 84, 822

\bibitem[{{Robin} {et~al}\mbox{.}(2003){Robin}, {Reyl{\'e}}, {Derri{\`e}re}, \&
  {Picaud}}]{2003A&A...409..523R}
{Robin} A.~C., {Reyl{\'e}} C., {Derri{\`e}re} S., {Picaud} S., 2003, \aap, 409,
  523

\bibitem[{{Robin} {et~al}\mbox{.}(2004){Robin}, {Reyl{\'e}}, {Derri{\`e}re}, \&
  {Picaud}}]{2004A&A...416..157R}
{Robin} A.~C., {Reyl{\'e}} C., {Derri{\`e}re} S., {Picaud} S., 2004, \aap, 416,
  157

\bibitem[{{Rohrmann}, {Althaus} \& {Kepler}(2010){Rohrmann}, {Althaus}, \&
  {Kepler}}]{2010BAAA...53..111R}
{Rohrmann} R.~D., {Althaus} L.~G., {Kepler} S.~O., 2010, Boletin de la
  Asociacion Argentina de Astronomia La Plata Argentina, 53, 111

\bibitem[{{Rohrmann} {et~al}\mbox{.}(2002){Rohrmann}, {Serenelli}, {Althaus},
  \& {Benvenuto}}]{2002MNRAS.335..499R}
{Rohrmann} R.~D., {Serenelli} A.~M., {Althaus} L.~G., {Benvenuto} O.~G., 2002,
  \mnras, 335, 499

\bibitem[{{Rosenberg} {et~al}\mbox{.}(1999){Rosenberg}, {Saviane}, {Piotto}, \&
  {Aparicio}}]{1999AJ....118.2306R}
{Rosenberg} A., {Saviane} I., {Piotto} G., {Aparicio} A., 1999, \aj, 118, 2306

\bibitem[{{Rowell}(2013)}]{2013MNRAS.434.1549R}
{Rowell} N., 2013, \mnras, 434, 1549

\bibitem[{{Rowell} \& {Hambly}(2011)}]{2011MNRAS.417...93R}
{Rowell} N., {Hambly} N.~C., 2011, \mnras, 417, 93

\bibitem[{{Rubin} {et~al}\mbox{.}(2008){Rubin}, {Williams}, {Bolte}, \&
  {Koester}}]{2008AJ....135.2163R}
{Rubin} K.~H.~R., {Williams} K.~A., {Bolte} M., {Koester} D., 2008, \aj, 135,
  2163

\bibitem[{{Saio} \& {Yoshii}(1979)}]{1979PASP...91..553S}
{Saio} H., {Yoshii} Y., 1979, \pasp, 91, 553

\bibitem[{{Salaris}, {Althaus} \& {Garc{\'{\i}}a-Berro}(2013){Salaris},
  {Althaus}, \& {Garc{\'{\i}}a-Berro}}]{2013A&A...555A..96S}
{Salaris} M., {Althaus} L.~G., {Garc{\'{\i}}a-Berro} E., 2013, \aap, 555, A96

\bibitem[{{Salaris} {et~al}\mbox{.}(2010){Salaris}, {Cassisi}, {Pietrinferni},
  {Kowalski}, \& {Isern}}]{2010ApJ...716.1241S}
{Salaris} M., {Cassisi} S., {Pietrinferni} A., {Kowalski} P.~M., {Isern} J.,
  2010, \apj, 716, 1241

\bibitem[{{Salaris} {et~al}\mbox{.}(1997){Salaris}, {Dom{\'{\i}}nguez},
  {Garc{\'{\i}}a-Berro}, {Hernanz}, {Isern}, \&
  {Mochkovitch}}]{1997ApJ...486..413S}
{Salaris} M., {Dom{\'{\i}}nguez} I., {Garc{\'{\i}}a-Berro} E., {Hernanz} M.,
  {Isern} J., {Mochkovitch} R., 1997, \apj, 486, 413

\bibitem[{{Salaris} {et~al}\mbox{.}(2009){Salaris}, {Serenelli}, {Weiss}, \&
  {Miller Bertolami}}]{2009ApJ...692.1013S}
{Salaris} M., {Serenelli} A., {Weiss} A., {Miller Bertolami} M., 2009, \apj,
  692, 1013

\bibitem[{{Salpeter}(1961)}]{1961ApJ...134..669S}
{Salpeter} E.~E., 1961, \apj, 134, 669

\bibitem[{{Sandage}, {Tammann} \& {Yahil}(1979){Sandage}, {Tammann}, \&
  {Yahil}}]{1979ApJ...232..352S}
{Sandage} A., {Tammann} G.~A., {Yahil} A., 1979, \apj, 232, 352

\bibitem[{{Saumon}, {Chabrier} \& {van Horn}(1995){Saumon}, {Chabrier}, \& {van
  Horn}}]{1995ApJS...99..713S}
{Saumon} D., {Chabrier} G., {van Horn} H.~M., 1995, \apjs, 99, 713

\bibitem[{{Saumon} \& {Jacobson}(1999)}]{1999ApJ...511L.107S}
{Saumon} D., {Jacobson} S.~B., 1999, \apjl, 511, L107

\bibitem[{{Schlafly} \& {Finkbeiner}(2011)}]{2011ApJ...737..103S}
{Schlafly} E.~F., {Finkbeiner} D.~P., 2011, \apj, 737, 103

\bibitem[{{Schlafly} {et~al}\mbox{.}(2012){Schlafly}, {Finkbeiner},
  {Juri{\'c}}, {Magnier}, {Burgett}, {Chambers}, {Grav}, {Hodapp}, {Kaiser},
  {Kudritzki}, {Martin}, {Morgan}, {Price}, {Rix}, {Stubbs}, {Tonry}, \&
  {Wainscoat}}]{2012ApJ...756..158S}
{Schlafly} E.~F. {et~al.}, 2012, \apj, 756, 158

\bibitem[{{Schmidt}(1959)}]{1959ApJ...129..243S}
{Schmidt} M., 1959, \apj, 129, 243

\bibitem[{{Schmidt}(1968)}]{1968ApJ...151..393S}
{Schmidt} M., 1968, \apj, 151, 393

\bibitem[{{Schmidt}(1975)}]{1975ApJ...202...22S}
{Schmidt} M., 1975, \apj, 202, 22

\bibitem[{{Searle}(1977)}]{1977egsp.conf..219S}
{Searle} L., 1977, in Evolution of Galaxies and Stellar Populations, {Tinsley}
  B.~M., {Larson} D.~Campbell R.~B.~G., eds., p. 219

\bibitem[{{Searle} \& {Zinn}(1978)}]{1978ApJ...225..357S}
{Searle} L., {Zinn} R., 1978, \apj, 225, 357

\bibitem[{{Segretain} \& {Chabrier}(1993)}]{1993A&A...271L..13S}
{Segretain} L., {Chabrier} G., 1993, \aap, 271, L13

\bibitem[{{Segretain} {et~al}\mbox{.}(1994){Segretain}, {Chabrier}, {Hernanz},
  {Garcia-Berro}, {Isern}, \& {Mochkovitch}}]{1994ApJ...434..641S}
{Segretain} L., {Chabrier} G., {Hernanz} M., {Garcia-Berro} E., {Isern} J.,
  {Mochkovitch} R., 1994, \apj, 434, 641

\bibitem[{{Shaviv} \& {Kovetz}(1976)}]{1976A&A....51..383S}
{Shaviv} G., {Kovetz} A., 1976, \aap, 51, 383

\bibitem[{{Sion}(1984)}]{1984ApJ...282..612S}
{Sion} E.~M., 1984, \apj, 282, 612

\bibitem[{{Stevenson}(1980)}]{1980JPhys..41C..61S}
{Stevenson} D.~J., 1980, Journal de Physique, 41, C2

\bibitem[{{Stobie}, {Ishida} \& {Peacock}(1989){Stobie}, {Ishida}, \&
  {Peacock}}]{1989MNRAS.238..709S}
{Stobie} R.~S., {Ishida} K., {Peacock} J.~A., 1989, \mnras, 238, 709

\bibitem[{{Stoner}(1930)}]{1930LEDPM...9..944S}
{Stoner} E., 1930, The London, Edinburgh, and Dublin Philosophical Magazine and
  Journal of Science: Series 7, Volume 9, Issue 60, p.~944-963, 9, 944

\bibitem[{{Straniero}(1988)}]{1988A&AS...76..157S}
{Straniero} O., 1988, \aaps, 76, 157

\bibitem[{{Tanaka} {et~al}\mbox{.}(2010){Tanaka}, {Kawabata}, {Yamanaka},
  {Maeda}, {Hattori}, {Aoki}, {Nomoto}, {Iye}, {Sasaki}, {Mazzali}, \&
  {Pian}}]{2010ApJ...714.1209T}
{Tanaka} M. {et~al.}, 2010, \apj, 714, 1209

\bibitem[{{Tassoul}, {Fontaine} \& {Winget}(1990){Tassoul}, {Fontaine}, \&
  {Winget}}]{1990ApJS...72..335T}
{Tassoul} M., {Fontaine} G., {Winget} D.~E., 1990, \apjs, 72, 335

\bibitem[{{Tinney}, {Reid} \& {Mould}(1993){Tinney}, {Reid}, \&
  {Mould}}]{1993ApJ...414..254T}
{Tinney} C.~G., {Reid} I.~N., {Mould} J.~R., 1993, \apj, 414, 254

\bibitem[{{Tonry} \& {Onaka}(2009)}]{2009amos.confE..40T}
{Tonry} J., {Onaka} P., 2009, in Advanced Maui Optical and Space Surveillance
  Technologies Conference, p. E40

\bibitem[{{Tonry} {et~al}\mbox{.}(2012){Tonry}, {Stubbs}, {Lykke}, {Doherty},
  {Shivvers}, {Burgett}, {Chambers}, {Hodapp}, {Kaiser}, {Kudritzki},
  {Magnier}, {Morgan}, {Price}, \& {Wainscoat}}]{2012ApJ...750...99T}
{Tonry} J.~L. {et~al.}, 2012, \apj, 750, 99

\bibitem[{{Torres}, {Garc{\'{\i}}a-Berro} \& {Isern}(2007){Torres},
  {Garc{\'{\i}}a-Berro}, \& {Isern}}]{2007MNRAS.378.1461T}
{Torres} S., {Garc{\'{\i}}a-Berro} E., {Isern} J., 2007, \mnras, 378, 1461

\bibitem[{{Tremblay} \& {Bergeron}(2008)}]{2008ApJ...672.1144T}
{Tremblay} P.-E., {Bergeron} P., 2008, \apj, 672, 1144

\bibitem[{{Tremblay}, {Bergeron} \& {Gianninas}(2011){Tremblay}, {Bergeron}, \&
  {Gianninas}}]{2011ApJ...730..128T}
{Tremblay} P.-E., {Bergeron} P., {Gianninas} A., 2011, \apj, 730, 128

\bibitem[{{Tremblay} {et~al}\mbox{.}(2014){Tremblay}, {Kalirai}, {Soderblom},
  {Cignoni}, \& {Cummings}}]{2014ApJ...791...92T}
{Tremblay} P.-E., {Kalirai} J.~S., {Soderblom} D.~R., {Cignoni} M., {Cummings}
  J., 2014, \apj, 791, 92

\bibitem[{{van Horn}(1968)}]{1968ApJ...151..227V}
{van Horn} H.~M., 1968, \apj, 151, 227

\bibitem[{{van Oirschot} {et~al}\mbox{.}(2014){van Oirschot}, {Nelemans},
  {Toonen}, {Pols}, {Brown}, {Helmi}, \& {Portegies
  Zwart}}]{2014A&A...569A..42V}
{van Oirschot} P., {Nelemans} G., {Toonen} S., {Pols} O., {Brown} A.~G.~A.,
  {Helmi} A., {Portegies Zwart} S., 2014, \aap, 569, A42

\bibitem[{{Vanderburg} {et~al}\mbox{.}(2015){Vanderburg}, {Johnson},
  {Rappaport}, {Bieryla}, {Irwin}, {Lewis}, {Kipping}, {Brown}, {Dufour},
  {Ciardi}, {Angus}, {Schaefer}, {Latham}, {Charbonneau}, {Beichman},
  {Eastman}, {McCrady}, {Wittenmyer}, \& {Wright}}]{2015Natur.526..546V}
{Vanderburg} A. {et~al.}, 2015, \nat, 526, 546

\bibitem[{{Vennes} {et~al}\mbox{.}(2002){Vennes}, {Smith}, {Boyle}, {Croom},
  {Kawka}, {Shanks}, {Miller}, \& {Loaring}}]{2002MNRAS.335..673V}
{Vennes} S., {Smith} R.~J., {Boyle} B.~J., {Croom} S.~M., {Kawka} A., {Shanks}
  T., {Miller} L., {Loaring} N., 2002, \mnras, 335, 673

\bibitem[{{Vergely} {et~al}\mbox{.}(2002){Vergely}, {K{\"o}ppen}, {Egret}, \&
  {Bienaym{\'e}}}]{2002A&A...390..917V}
{Vergely} J.-L., {K{\"o}ppen} J., {Egret} D., {Bienaym{\'e}} O., 2002, \aap,
  390, 917

\bibitem[{{Weidemann}(1991)}]{1991ASIC..336...67W}
{Weidemann} V., 1991, in NATO Advanced Science Institutes (ASI) Series C, Vol.
  336, NATO Advanced Science Institutes (ASI) Series C, {Vauclair} G., {Sion}
  E., eds., p.~67

\bibitem[{{Whelan} \& {Iben}(1973)}]{1973ApJ...186.1007W}
{Whelan} J., {Iben}, Jr. I., 1973, \apj, 186, 1007

\bibitem[{{White} \& {Rees}(1978)}]{1978MNRAS.183..341W}
{White} S.~D.~M., {Rees} M.~J., 1978, \mnras, 183, 341

\bibitem[{{Williams} \& {Bolte}(2007)}]{2007AJ....133.1490W}
{Williams} K.~A., {Bolte} M., 2007, \aj, 133, 1490

\bibitem[{{Williams}, {Bolte} \& {Koester}(2004){Williams}, {Bolte}, \&
  {Koester}}]{2004ApJ...615L..49W}
{Williams} K.~A., {Bolte} M., {Koester} D., 2004, \apjl, 615, L49

\bibitem[{{Williams}, {Bolte} \& {Koester}(2009){Williams}, {Bolte}, \&
  {Koester}}]{2009ApJ...693..355W}
{Williams} K.~A., {Bolte} M., {Koester} D., 2009, \apj, 693, 355

\bibitem[{{Winget} {et~al}\mbox{.}(1987){Winget}, {Hansen}, {Liebert}, {van
  Horn}, {Fontaine}, {Nather}, {Kepler}, \& {Lamb}}]{1987ApJ...315L..77W}
{Winget} D.~E., {Hansen} C.~J., {Liebert} J., {van Horn} H.~M., {Fontaine} G.,
  {Nather} R.~E., {Kepler} S.~O., {Lamb} D.~Q., 1987, \apjl, 315, L77

\bibitem[{{Wood}(1992)}]{1992ApJ...386..539W}
{Wood} M.~A., 1992, \apj, 386, 539

\bibitem[{{Wood}(1995)}]{1995LNP...443...41W}
{Wood} M.~A., 1995, in Lecture Notes in Physics, Berlin Springer Verlag, Vol.
  443, White Dwarfs, {Koester} D., {Werner} K., eds., p.~41

\bibitem[{{Wood} \& {Oswalt}(1998)}]{1998ApJ...497..870W}
{Wood} M.~A., {Oswalt} T.~D., 1998, \apj, 497, 870

\bibitem[{{Yamanaka} {et~al}\mbox{.}(2009){Yamanaka}, {Kawabata}, {Kinugasa},
  {Tanaka}, {Imada}, {Maeda}, {Nomoto}, {Arai}, {Chiyonobu}, {Fukazawa},
  {Hashimoto}, {Honda}, {Ikejiri}, {Itoh}, {Kamata}, {Kawai}, {Komatsu},
  {Konishi}, {Kuroda}, {Miyamoto}, {Miyazaki}, {Nagae}, {Nakaya}, {Ohsugi},
  {Omodaka}, {Sakai}, {Sasada}, {Suzuki}, {Taguchi}, {Takahashi}, {Tanaka},
  {Uemura}, {Yamashita}, {Yanagisawa}, \& {Yoshida}}]{2009ApJ...707L.118Y}
{Yamanaka} M. {et~al.}, 2009, \apjl, 707, L118

\bibitem[{{Zhao} {et~al}\mbox{.}(2012){Zhao}, {Oswalt}, {Willson}, {Wang}, \&
  {Zhao}}]{2012ApJ...746..144Z}
{Zhao} J.~K., {Oswalt} T.~D., {Willson} L.~A., {Wang} Q., {Zhao} G., 2012,
  \apj, 746, 144

\bibitem[{{Zuckerman} \& {Becklin}(1987)}]{1987Natur.330..138Z}
{Zuckerman} B., {Becklin} E.~E., 1987, \nat, 330, 138

\end{thebibliography}
